\documentclass[aps,pra,reprint,superscriptaddress,nofootinbib]{revtex4-2}
\makeatletter
\renewcommand{\frontmatter@abstractwidth}{410\p@}
\makeatother



\usepackage{comment}
\usepackage{cmap}  
\usepackage[utf8]{inputenc}
\usepackage[english]{babel}
\babelprovide[hyphenrules=english]{en}
\usepackage[T1]{fontenc}
\usepackage{appendix}
\usepackage{braket}
\usepackage{url, amsfonts, tikz, physics, listings, xcolor, graphicx, float}
\usepackage[shortlabels]{enumitem}
\usepackage{dsfont}
\usepackage{amsmath, amssymb, amstext, amscd, amsthm, makeidx, graphicx, url, mathrsfs, mathtools, longdivision, polynom, bbm, complexity}
\usepackage{nicefrac}
\usepackage[linesnumbered,ruled,vlined]{algorithm2e}
\usepackage{xcolor}
\usepackage[normalem]{ulem}
\usepackage{array}[=2016-10-06]
\usepackage{amsmath,tabularx,makecell, booktabs}
\usepackage{fancyhdr}

\fancypagestyle{aqsmain}{%
  \fancyhf{}%
  \fancyhead[R]{\thepage}%
}
\fancypagestyle{header}{%
  \fancyhf{}%
  \fancyhead[R]{\hyperref[apptoc]{\thepage}}%
}
\DeclareUnicodeCharacter{2009}{\,}
\DeclareUnicodeCharacter{202F}{\,}

\definecolor{blueviolet}{rgb}{0.2, 0.2, 0.6}
\definecolor{webgreen}{rgb}{0,.5,0}
\definecolor{webbrown}{rgb}{.6,0,0}
\usepackage[pdftex,
  bookmarks=false,
  colorlinks=true, 
  urlcolor=webbrown,
  linkcolor=blueviolet, 
  citecolor=webgreen,
  pdfstartpage=1,
  pdfstartview={FitH}, 
  bookmarksopen=false
  ]{hyperref}
\usepackage{cleveref}

\makeatletter
\newcommand\RedeclareMathOperator{%
  \@ifstar{\def\rmo@s{m}\rmo@redeclare}{\def\rmo@s{o}\rmo@redeclare}%
}

\newcommand\rmo@redeclare[2]{%
  \begingroup \escapechar\m@ne\xdef\@gtempa{{\string#1}}\endgroup
  \expandafter\@ifundefined\@gtempa
     {\@latex@error{\noexpand#1undefined}\@ehc}%
     \relax
  \expandafter\rmo@declmathop\rmo@s{#1}{#2}}

\newcommand\rmo@declmathop[3]{%
  \DeclareRobustCommand{#2}{\qopname\newmcodes@#1{#3}}%
}
\@onlypreamble\RedeclareMathOperator
\makeatother

\allowdisplaybreaks

\RedeclareMathOperator*{\E}{{\mathbb{E}}}
\DeclareMathOperator*{\Var}{\mathrm{Var}}

\newenvironment{appendixabstract}
  {\begin{center}\begin{minipage}{0.9\textwidth}\small}
  {\end{minipage}\end{center}}

\makeatletter
\DeclareFontFamily{OMX}{MnSymbolE}{}
\DeclareSymbolFont{MnLargeSymbols}{OMX}{MnSymbolE}{m}{n}
\SetSymbolFont{MnLargeSymbols}{bold}{OMX}{MnSymbolE}{b}{n}
\DeclareFontShape{OMX}{MnSymbolE}{m}{n}{
    <-6>  MnSymbolE5
   <6-7>  MnSymbolE6
   <7-8>  MnSymbolE7
   <8-9>  MnSymbolE8
   <9-10> MnSymbolE9
  <10-12> MnSymbolE10
  <12->   MnSymbolE12
}{}
\DeclareFontShape{OMX}{MnSymbolE}{b}{n}{
    <-6>  MnSymbolE-Bold5
   <6-7>  MnSymbolE-Bold6
   <7-8>  MnSymbolE-Bold7
   <8-9>  MnSymbolE-Bold8
   <9-10> MnSymbolE-Bold9
  <10-12> MnSymbolE-Bold10
  <12->   MnSymbolE-Bold12
}{}

\let\llangle\@undefined
\let\rrangle\@undefined
\DeclareMathDelimiter{\llangle}{\mathopen}%
                     {MnLargeSymbols}{'164}{MnLargeSymbols}{'164}
\DeclareMathDelimiter{\rrangle}{\mathclose}%
                     {MnLargeSymbols}{'171}{MnLargeSymbols}{'171}
\makeatother

\numberwithin{equation}{section}

\fancypagestyle{aqsequal}{%
  \fancyhf{}%
  \fancyhead[R]{\thepage}%
  \fancyfoot[L]{%
    \parbox[b]{\columnwidth}{%
      \raggedright
      \rule{0.4\linewidth}{0.4pt}\\[3pt]
      \footnotesize\textsuperscript{*}Equal contribution
    }%
  }%
}

\newtheorem{theorem}{Theorem}
\newtheorem{prop}{Proposition}
\newtheorem{observation}{Observation}
\newtheorem{lemma}{Lemma}
\newtheorem{corollary}{Corollary}
\newtheorem{definition}{Definition}

\newcommand{\indicator}{\mathds{1}}

\usepackage{needspace}
\usepackage{tcolorbox}
\tcbuselibrary{breakable}

\newtcolorbox{graydefinitionbox}{
    before={\Needspace{8\baselineskip}\par\medskip},
    after={\par\medskip},
    colback=gray!4,
    colframe=gray!80,
    boxrule=1.4pt,
    arc=6pt,
    left=8pt,
    right=8pt,
    top=8pt,
    bottom=8pt,
    breakable
}

\newcommand{\stkout}[1]{\ifmmode\text{\sout{\ensuremath{#1}}}\else\sout{#1}\fi}
\newif\ifverbose

\verbosefalse

\begin{document}
\title{Algorithmic Quantum Sensing}

\setlength{\skip\footins}{12pt}

\author{Ishaan Kannan}
\thanks{Equal contribution}
\affiliation{Department of Physics, Harvard University, Cambridge, MA 02138 USA}

\author{Aditi Venkatesh}
\thanks{Equal contribution}
\affiliation{Department of Physics, Harvard University, Cambridge, MA 02138 USA}

\author{Jordan Cotler}
\affiliation{Department of Physics, Harvard University, Cambridge, MA 02138 USA}

\begin{abstract}
We introduce a unified framework for sensing time-dependent fields through the lens of classical and quantum algorithms. First, we develop three pulse sequence protocols, namely randomized, matched, and pooled spectroscopy, which outperform established approaches such as dynamical decoupling across a wide range of tasks. These protocols attain the optimal sensitivity for linear and nonlinear AC sensing on conventional spin or qubit platforms, improving existing devices by changing only their pulse sequences. We then introduce quantum computational spectroscopy (QCS), which couples quantum sensors to a quantum processor. QCS combines two new primitives: a quantum computational frequency comb that coherently converts any classical signal into a quantum oracle, and a family of quantum algorithms that use the oracle to extract signal information in superposition. Variations can be implemented on current platforms with only one memory ancilla. We show that the ultimate limit imposed by quantum mechanics lies polynomially beyond the Heisenberg limit, and that QCS achieves it. Attaining this new limit requires quantum computation, establishing a fundamental separation between conventional and quantum computational sensing. Moreover, QCS retains beyond-Heisenberg scaling for broadband signals under constant-strength sensor depolarization, while we prove that the same noise limits any protocol lacking quantum computation to standard-quantum-limit scaling. Our result provides the first asymptotic, beyond-quadratic quantum advantage in AC metrology when noise is aligned with the signal, proving that Lindblad span conditions do not apply to broadband quantum sensing. Our work builds an algorithmic theory for designing sensing protocols and establishes quantum sensing as a new frontier for quantum algorithms, with quantum advantages that surpass the Heisenberg limit and that persist under noise.
\end{abstract}
\makeatletter
\renewcommand{\frontmatter@footnote@produce}{%
  \global\let\@FMN@list\@empty
}
\makeatother

\maketitle
\pagestyle{aqsmain}
\thispagestyle{aqsequal}

\makeatletter
\let\AQSsavedaddcontentsline\addcontentsline
\renewcommand{\addcontentsline}[3]{}%
\makeatother

\vspace{-2em}
{\renewcommand\addcontentsline[3]{} 
\section{Introduction}}
\vspace{-1.6mm}
\begingroup
\setlength{\parskip}{0pt}
\setlength{\baselineskip}{%
  \dimexpr\baselineskip\relax plus 0.2pt
}

Quantum sensors are among the most precise and diverse measurement devices used in modern science~\cite{giovannetti_quantum-enhanced_2004, degen_quantum_2017}. Today, platforms ranging from atomic clocks and spin defects to superconducting circuits and optical interferometers~\cite{ludlow_optical_2015, maze_nanoscale_2008, bylander_noise_2011, aasi_enhanced_2013} are used to probe phenomena spanning molecular composition~\cite{ernst_application_1966, glenn_high-resolution_2018}, protein structure~\cite{wuthrich_protein_1989, lovchinsky_nuclear_2016}, fundamental particles~\cite{budker_proposal_2014, backes_quantum_2021}, and gravity~\cite{kasevich_atomic_1991, abbott_observation_2016}. In many of these applications, the sensor detects a time-varying signal, and the information of interest lies in its frequency spectrum, phases, and correlations. Learning from time-dependent signals, broadly termed AC sensing, is a fundamental task across the sciences. Numerous application-specific techniques have been developed, often using pulse sequences derived from dynamical decoupling (DD) to selectively amplify signals at particular frequencies~\cite{carr_effects_1954, viola_dynamical_1998, de_lange_single-spin_2011, kotler_single-ion_2011}, while other approaches reconstruct entire waveforms with compressed sensing or optimize quantum control for specialized tasks~\cite{magesan_compressing_2013, magesan_reconstructing_2013, cooper_time-resolved_2014, poggiali_optimal_2018}. Nevertheless, a general understanding of how to design provably optimal strategies for AC sensing  remains incomplete.

\begin{figure*}
    \centering
      \includegraphics[width=\linewidth]{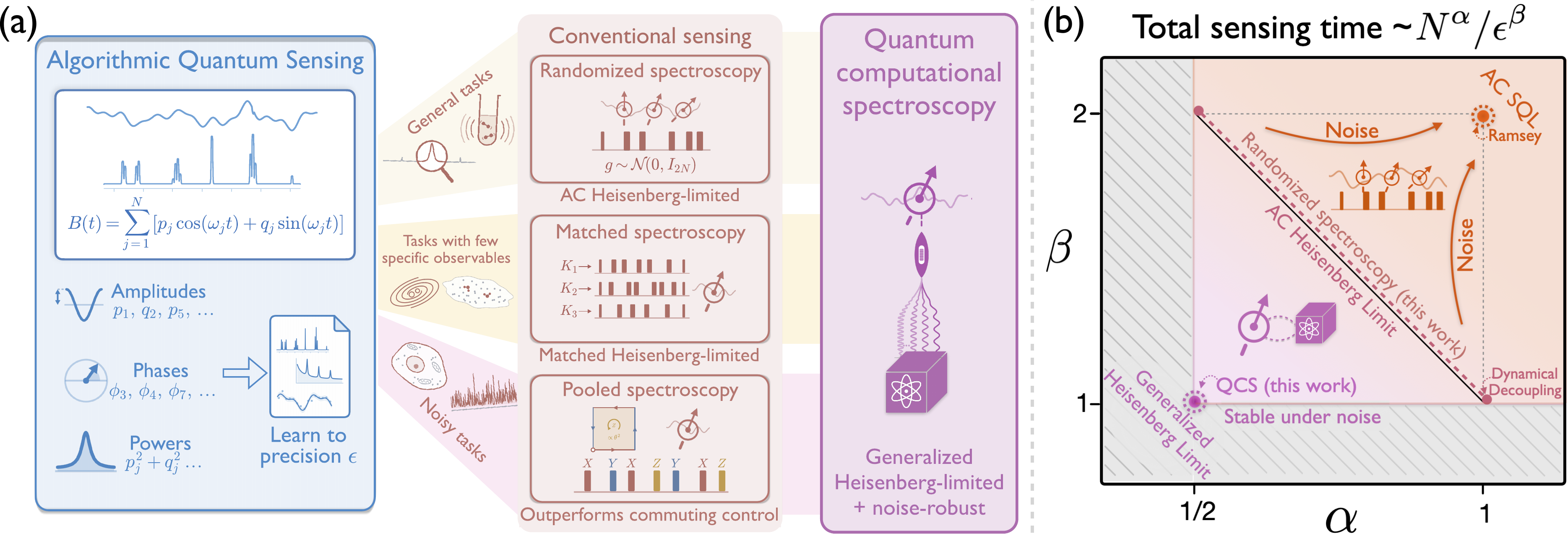}
    \caption{(a) \textit{Overview of protocols.} Algorithmic quantum sensing (AQS) frames AC sensing tasks as statistical learning problems, recasting sensor optimization as algorithm design and admitting lower bounds certifying the best achievable sensitivity.  With conventional sensor control, this formalism leads to three families of pulse sequence protocols: \textit{randomized} quantum spectroscopy (RQS) learns many observables simultaneously at the AC Heisenberg limit (ACHL, Theorem~\ref{thm:heisenberg_informal}); \textit{matched} spectroscopy achieves the \textit{matched} Heisenberg limit (Theorem~\ref{thm:matched_heisenberg_informal}), surpassing the ACHL when few observables are desired; pooled spectroscopy measures cumulative nonlinear observables asymptotically faster than commuting $\pi$-pulse sequences (Theorem~\ref{thm:pooled_adv}). With a quantum processor, quantum computational spectroscopy (QCS) solves these tasks with sensitivity beyond the Heisenberg limit. (b) \textit{AC sensing phase diagram}. $\epsilon$-accurate AC sensing of a signal with $N$ frequencies requires time $N^\alpha/\epsilon^\beta$, with $\alpha \text{ (x axis)}, \beta \text{ (y axis})$ determining a protocol's efficiency. Existing protocols lie at or above the ACHL, limited by $(\alpha=1, \beta=1)$ or $(\alpha=\tfrac{1}{2}, \beta=2)$ for specific sensing tasks, while RQS saturates both ends of this limit for any AQS task. Dephasing noise reverts \textit{conventional} protocols to the standard quantum limit $(\alpha=1,\beta=2)$, as in DC metrology~\cite{Tsang_2013, Demkowicz_Dobrza_ski_2017, zhou_achieving_2018, pradenne_restrictions_2026}. The ACHL can be surpassed if and only if ancillary quantum computation is available, and the resulting generalized Heisenberg limit (GHL) of $(\alpha=\tfrac{1}{2}, \beta = 1)$ is saturated by QCS. This performance is stable under noise, illustrating that the GHL, rather than the noise-sensitive Heisenberg limit, is the ultimate quantum-mechanical limit in a larger and previously unexplored phase diagram.}
    \vspace{-3.8mm}

    \label{fig:Fig_1}
\end{figure*}

We formulate AC sensing as a single statistical learning problem which we call \textit{Algorithmic Quantum Sensing} (AQS). Given a signal supported on at most $N$ frequencies, AQS asks for estimates of chosen functions of its spectrum to precision $\epsilon$. Through AQS, we find rigorous, saturable forms of the Heisenberg limit and Standard Quantum Limit (SQL) for AC sensing, establishing a unifying notion of optimality across tasks. We then introduce three families of AC sensing protocols, \textit{randomized}, \textit{matched}, and \textit{pooled} spectroscopy, which use only conventional pulse control of a single qubit or global control of a spin ensemble to achieve the optimal sensitivity (Fig.~\ref{fig:Fig_1}a).  Randomized quantum spectroscopy (RQS) uses random pulse sequences to collect a classical dataset from which many signal properties can be estimated downstream, and saturates the AC Heisenberg limit. Under noise, RQS continues to saturate the AC standard quantum limit.  Matched spectroscopy instead exploits advance knowledge of the observables of interest to tailor pulse sequences directly to the desired quantities, avoiding the cost of acquiring a general-purpose dataset. When only a few observables are needed, matched spectroscopy reduces interrogation time polynomially below that of any nonadaptive protocol and achieves the matched Heisenberg limit, which no conventional protocol can surpass. Under noise, both protocols continue to saturate their respective standard quantum limits, and together provide an optimal set of conventional protocols for AC sensing tasks requiring any number of observables and in any noise environment. Pooled spectroscopy, designed to estimate coarse-grained features when many noisy spectral lines are present, uses noncommuting pulse sequences to process signal information nonlinearly before readout, outperforming all commuting approaches for estimating quantities like the total power within a frequency band. All three protocols outperform established strategies such as DD and multiplexed Walsh-Hadamard sensing in tasks including broadband search, lock-in detection, and multidimensional spectroscopy through algorithmic design, without requiring additional hardware capabilities.

Recently, work on \textit{quantum computational sensing}~\cite{khan_quantum_2026} has considered coupling sensors to ancillary quantum memories that process the acquired information, often aiming to reach or surpass Heisenberg scaling~\cite{zhuang_physical-layer_2019, zheng_preparation_2022, allen_quantum_2025, cornell_all-mechanical_2025, rovny2025multiqubitnanoscalesensingentanglement, prabhu_quantum_2026, kannan_exponential_2026, sen_quantum_2026, sirotin2026quantumadvantagesensingimaging, liu2026optimalquantumsensingalgorithms}. Noise aligned with the signal sharply limits these gains: in single-parameter sensing, such noise provably returns the sensitivity to the standard quantum limit once the sensing time or number of sensors exceeds a scale set by the noise rate~\cite{Tsang_2013, Demkowicz_Dobrza_ski_2017, zhou_achieving_2018, pradenne_restrictions_2026}, and no quantum computational protocol has yet sustained Heisenberg scaling asymptotically under signal-aligned noise. 

We ask whether a quantum processor can provide an AC sensing advantage that conventional sensor control cannot replicate and that survives under realistic signal-aligned noise. We answer this question affirmatively by introducing \textit{quantum computational spectroscopy} (QCS). QCS solves AC sensing tasks polynomially faster than any protocol without ancillary quantum computation, surpassing the AC Heisenberg limit, and we prove that no protocol allowed by quantum mechanics can do better. Moreover, QCS can be realized on near-term platforms with as little as one ancilla qubit, while larger processors enable the algorithm to estimate more quantities simultaneously. This speedup arises from the ability of a quantum computer to process data in superposition, and is distinct from metrological enhancements arising from entangled sensors or coherent interrogation, which enables growing advantages without increasing sensor or ancilla count. Moreover, we introduce the notion of a \textit{quantum computational frequency comb}, which robustly converts ambient signals into quantum oracles and enables the broadband QCS speedup to persist under continuous, constant-strength depolarization of the sensor. To our knowledge, QCS is the first quantum computational protocol whose metrological advantage survives under constant-strength signal-aligned noise, and its performance shows that satisfying the HNLS condition of Ref.~\cite{zhou_achieving_2018} is not necessary to achieve asymptotic quantum advantage in broadband AC sensing.

While the AC Heisenberg limit on conventional protocols improves on the SQL quadratically in \textit{either} $N$ or $\epsilon$-dependence, the stronger limit saturated by QCS allows quadratic improvements simultaneously in \textit{both} parameters, a beyond-quadratic quantum speedup. This scaling is a special case of the \textit{generalized Heisenberg limit} (GHL) identified in Ref.~\cite{kannan_universal_2026}. Surprisingly, the GHL responds to noise very differently than the Heisenberg limit. Without quantum computation, we prove that any Heisenberg-limited sensing protocol degrades to the SQL under signal-aligned noise of any constant strength, establishing the multiparameter AC analog of no-go theorems for DC metrology~\cite{Demkowicz_Dobrza_ski_2017, zhou_achieving_2018, pradenne_restrictions_2026}. In contrast, the broadband GHL is unchanged under the same noise, and QCS continues to saturate this limit up to timescales far longer than the sensor's own coherence time. Physically, the observed linewidth is determined by the coherence time of a quantum memory, rather than that of the sensor itself. These observations reshape the canonical separation of AC quantum metrology, demonstrating that the frequency content of a signal is an essential axis of a larger metrological phase diagram in which beyond-quadratic speedups are both possible and practical. Within this diagram, the AC Heisenberg limit is an unstable optimum, while the separation between SQL and GHL is stable under noise and fundamentally enabled by quantum computation (Fig.~\ref{fig:Fig_1}b).
\par
\endgroup

\vspace{5mm}
{\renewcommand{\addcontentsline}[3]{}%
\section{Algorithmic formulation \\ of  AC sensing}}
\vspace{-2mm}

Consider two paradigmatic AC sensing tasks. In axionic dark matter detection, a very weak, narrowband signal is expected somewhere in a frequency band that may span hundreds of MHz, and the task is to locate this amplitude spike \cite{budker_proposal_2014, backes_quantum_2021}. In contrast, multidimensional NMR spectroscopy characterizes high-dimensional correlations between frequency tones emitted by a driven molecule \cite{bax_two-dimensional_1989, kay_three-dimensional_1990, kay_four-dimensional_1990}. Despite their different platforms and objectives, both tasks interrogate a time-varying field whose waveform decomposes into Fourier modes, $B(t) = \sum_{j=1}^N \left[p_j \cos(\omega_j t) + q_j \sin(\omega_j t)\right]$, and both estimate functions of the coefficients $p_j$ and $q_j$. Dark matter search looks for the index $j$ whose power $p_j^2 + q_j^2$ slightly exceeds that of the surrounding frequencies. Multidimensional NMR instead measures cross-peak correlations, which are linear combinations of products of coefficients from different frequencies, such as $p_i q_j$ or $p_i q_j p_k$.

We therefore formulate AC sensing as learning $M$ functions of a signal's vector, a task whose difficulty depends on the number of frequencies we wish to resolve, the target precision, and the observables of interest. We write a signal as $B(t) = A^T f(t)$, where $A=(p_1, ..., p_N, q_1,..., q_N)^T\in\mathbb{R}^{2N}$ is the unknown spectrum vector and $f(t)=\bigl( \cos(\omega_1t), ..., \allowbreak\cos(\omega_Nt),\allowbreak \sin(\omega_1t), \allowbreak...,\allowbreak \sin(\omega_N t)\bigr)^T$ arranges the known frequency modes. For a rank-$L$ tensor $K$, we use $K[A^{\otimes L}]$ to denote $K$ contracted against $L$ copies of $A$. The following definition extends the linear matched filtering formulation of Ref.~\cite{Cotler_2026}.
 
\vspace{1em}
\begin{definition}[Algorithmic quantum sensing problem]\label{def:aqs_problem}
    Fix a set of $N$ increasing frequencies $\{\omega_1,...,\omega_N\}$, a precision $\epsilon$, and a set of $M$ rank-$L$ tensors $\mathcal K = \{K_1,...,K_M\}$ with $\kappa = \max_j \|K_j\|_\infty$.  Given query access to a signal $B(t) = A^T f(t)$ with $\|A\|_2 \leq 1$, the algorithmic quantum sensing problem $\textnormal{AQS}[\{\omega_j\}, \epsilon, \mathcal 
    K]$ is to output estimates $\widehat o_1, ..., \widehat{o}_M$, such that
    \begin{equation}
        \left|\widehat o_j - K_j[A^{\otimes L}]\right| \leq \epsilon\,, \qquad \forall j \in \{1,..,M\} ,
    \end{equation}
    with high success probability.
\end{definition}

General rank-$L$ multilinear observables incorporate both linear and nonlinear tasks. Spectrum reconstruction involves $L=1$ with $K_j = e_j$ (the $j^{\rm th}$ standard basis vector), so that $K_j^T A = A_j$. Estimating signal power at every frequency requires $L=2$, with diagonal matrices $K_j$ that are zero except at the $j^{\rm th}$ and $(N+j)^{\rm th}$ diagonal entries, so that $A^T K_j A = p_j^2 + q_j^2$. In practice, the frequencies $\{\omega_j\}$ can be chosen based on priors or on the available $T_2$, which imposes a lower limit of $\sim 1/T_2$ on the minimum frequency spacing $\Delta_\omega$, and the choice $\|A\|_2 \leq 1$ simply normalizes the signal amplitude. The scale $\kappa$ is defined as the spectral norm of any $K$, and absorbs the normalization of the signal. Because $\kappa$ absorbs the signal's coupling strength and $\epsilon$ captures uncertainty in a physical amplitude, both have units of s$^{-1}$ when $\hbar = 1$, so $\epsilon/\kappa$ is unitless relative precision. For scale-invariance, our results are stated in terms of unitless time, and dividing by $\kappa$ recovers physical time. 

\begin{figure}
    \centering
    \includegraphics[width=\linewidth]{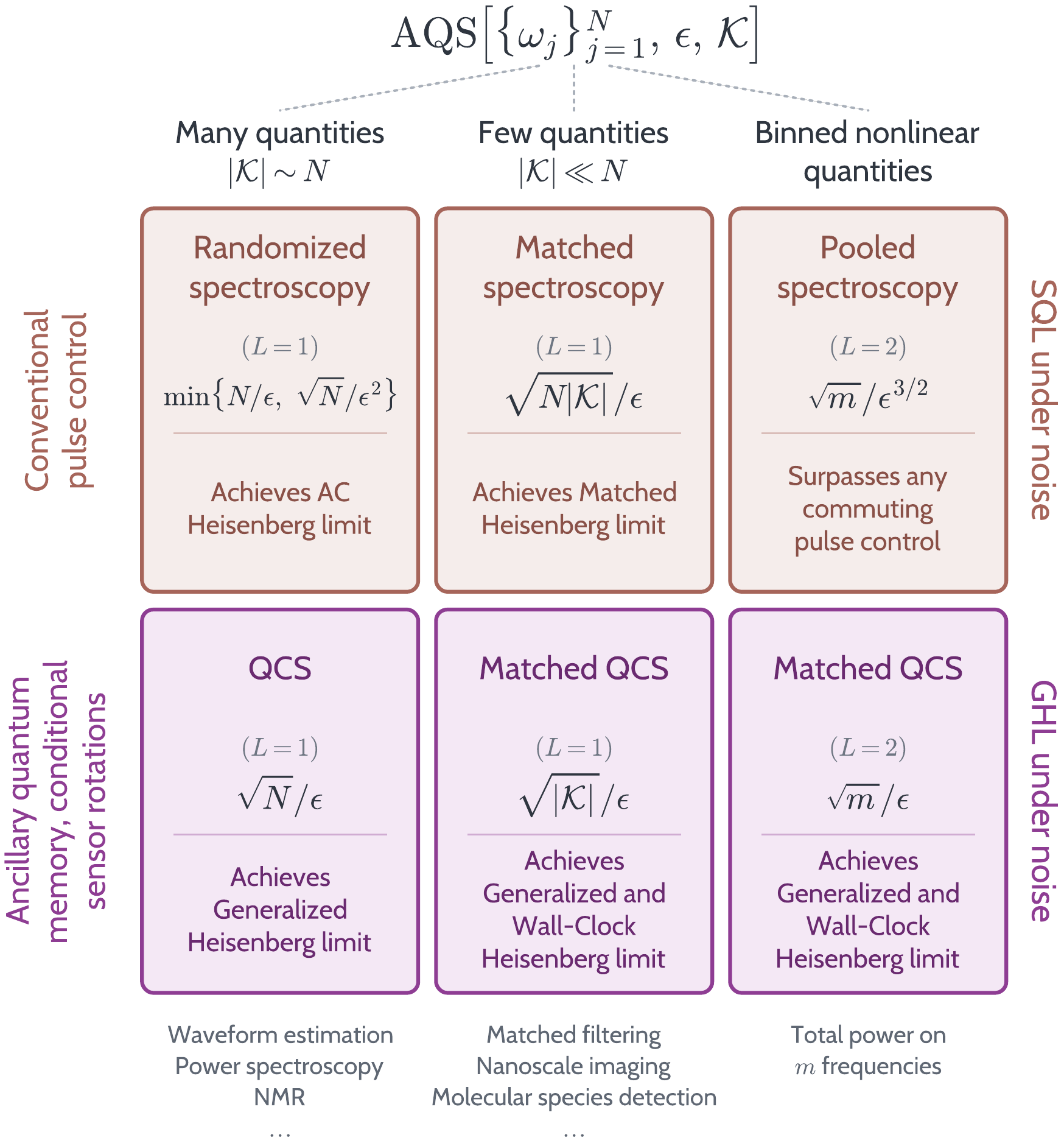}
    \par\vspace{-2mm}
    \caption{All protocols and their regimes of application are summarized. Simplified, noiseless scalings are depicted for linear (for pooled, quadratic) observables, alongside the relevant lower bounds. For greater detail see Table~\ref{tab:aqs_protocol_comparison}, and for general-$L$, noise-aware scalings see the Appendices.}
    \label{fig:protocol_grid}
    \vspace{-4mm}
\end{figure}

AQS is distinct from the AC detection task of Ref.~\cite{allen_quantum_2025}, which involves deciding whether a single tone of amplitude at least $\epsilon$ is present among otherwise empty frequency bins, and instead encompasses learning linear and nonlinear properties of general signals at a finite frequency resolution. The detection, search, and learning tasks of Ref.~\cite{liu2026optimalquantumsensingalgorithms} are each special cases of AQS with $L=1$. In the rest of this work, we present optimal conventional and quantum computational sensing protocols for every AQS instance (Fig.~\ref{fig:protocol_grid}).

\vspace{-1em}
{\renewcommand{\addcontentsline}[3]{}%
\section{Sensing with pulse sequences}}
\vspace{-2mm}

We begin with the conventional control setting described by the general single-qubit sensing Hamiltonian 
\begin{equation} \label{eq:sensing_ham}
    H(t) = \frac{1}{2}B(t)Z + H_c(t) \ ,
\end{equation}
where $H_c(t)$ contains all controls applied to the sensor and $B(t) = A^T f(t)$ as above. A \textit{conventional} protocol can prepare an initial state, interrogate $H(t)$ for a chosen time, and measure the sensor, possibly adapting future controls to earlier measurement outcomes. We do not assume that the signal resets between each measurement. 

\begin{figure*}
    \centering
    \includegraphics[width=\linewidth]{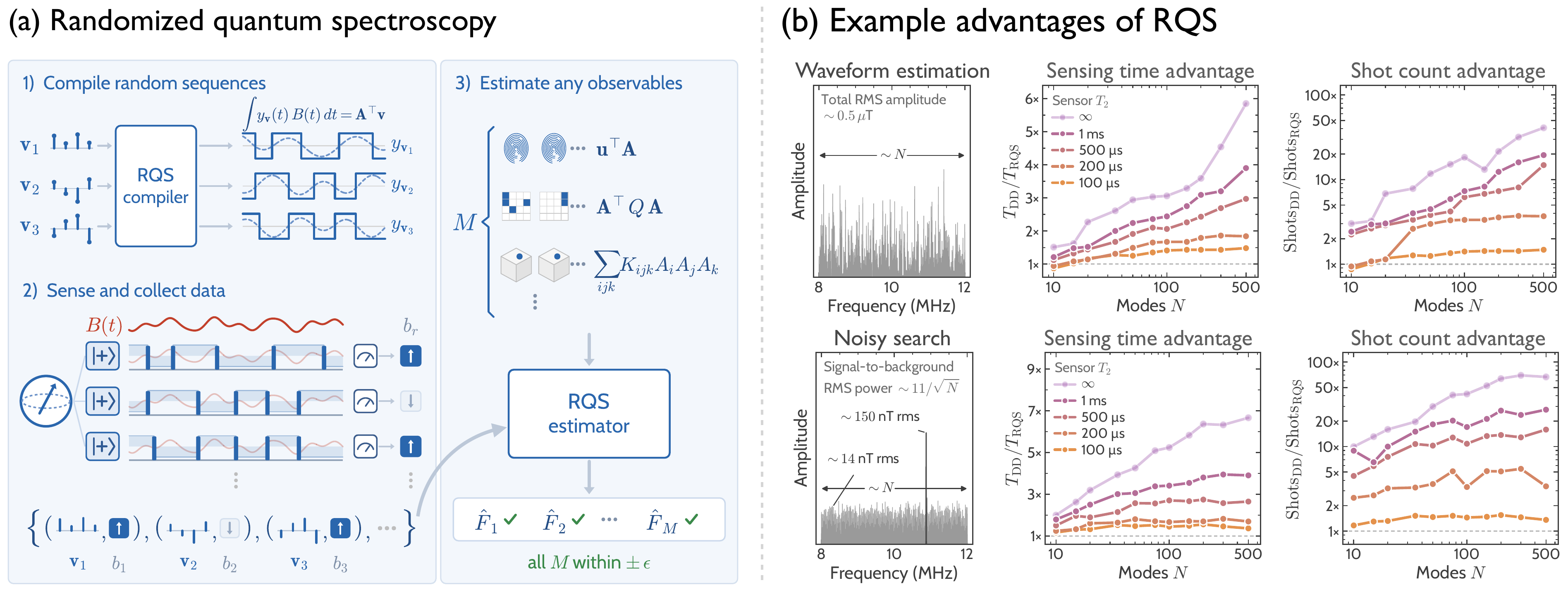}        \par\vspace{-1mm}

    \caption{
(a) \textit{Overview of RQS algorithm}. Before sensing, RQS samples a bank of random vectors and compiles sequences of $\pi$ pulses realizing them as linear responses. The sequences are executed with $X$ and $Y$-basis readout, and the signal need not restart between shots. After sensing, any set of $M$ linear or nonlinear observables can be requested, and the same dataset of response vectors and measured bits estimates all of them to precision $\epsilon$ with $\sim \log M$ overhead in the number of sensing rounds. (b) \textit{Simulations of RQS vs. DD}. We simulate (i) learning all $2N$ quadrature amplitudes of a weak signal with Gaussian-random amplitudes and uniformly random phase and (ii) identifying a weak signal with larger amplitude than its background, using RQS and CPMG-based DD, and plot the improvement factor enabled by RQS for various sensor $T_2$ times. This RQS implementation uses Rademacher randomness; see code. Signals with total RMS amplitude $<0.5\,\mu$T in the MHz regime are used, and the search simulation holds signal power constant as $N$ grows, so improvement need not be monotone in $N$. In counting sensing time, sensor readout is assumed instantaneous. RQS requires far fewer readouts than DD; the shot advantage achieved by our open-source pulse sequence compiler is depicted for both tasks. Our simulations adopt kHz-scale spacings, but applications with larger carrier frequencies and spacings can produce equivalent advantage with shorter sensor $T_2$s.} 
    \label{fig:rqs_main}
    \vspace{-4mm}
\end{figure*}
This Hamiltonian encompasses a wide variety of existing strategies, including Rabi spectroscopy and Ramsey interferometry, through suitable choices of $H_c$. Pulse sequence protocols such as dynamical decoupling (DD) spectroscopy choose $H_c(t) = \frac{\pi}{2}\sum_{j = 1}^{n} \delta(t-t_j)X$, applying $\pi$-pulses at chosen times $t_1,\ldots,t_{n}$. A longstanding approach to AC sensing is to sweep the relevant frequency band with evenly-spaced $\pi$-pulses at incrementally finer periods, resolving one frequency component at a time; ideal DD spectroscopy estimates each component with Heisenberg-limited precision $\epsilon$ in time $\sim 1/\epsilon$.

For general tasks, however, it is unclear whether a particular control sequence is optimal. For instance, DD excels at resolving specific tones, but may be suboptimal when many tones with uneven amplitudes are simultaneously involved. Moreover, pulse sequence design usually relies on pulses in the X-Y plane of the Bloch sphere, which produce an interaction-picture Hamiltonian with only a single Pauli generator whose signal sensitivity can be easily quantified. More exotic choices of $H_c$, such as multi-axis pulses, preclude such simple interaction Hamiltonians but may offer harder-to-quantify sensing gains, making it difficult to assert that a chosen DD protocol is optimal. We therefore ask: how can we design sensing algorithms within Hamiltonian~\eqref{eq:sensing_ham} that are \textit{provably optimal} for any AQS instance?

\vspace{3mm}
{\renewcommand{\addcontentsline}[3]{}%
\subsection{Randomized quantum spectroscopy}}
\vspace{-2mm}

We begin with randomized quantum spectroscopy (RQS), which uses classical randomness to collect a dataset from which many observables can be estimated downstream. The sensing procedure only needs $\pi$-pulses in the X-Y plane, requires no adaptivity, and is independent of the observables being estimated, so the collected data can be reused without repeating experiments. This protocol is suited to tasks involving many quantities, such as spectrum or waveform estimation, identifying strong tones in a broad band, and nuclear or molecular spectroscopy; such problems appear across fundamental physics and chemistry~\cite{bax_two-dimensional_1989, magesan_compressing_2013,szankowski_spectroscopy_2016, Ying2017SMILE, liang_ultrasensitive_2021, backes_quantum_2021, Anderegg_2023, Cotler_2026}. 

Before presenting the algorithm, we recall how linear responses appear from pulse sequences. If a sensor begins in $\ket{+}$ and only $X$-axis $\pi$-pulses are applied during sensing, a $Y$-measurement at time $\tau$ has expectation $\langle Y\rangle = \sin(\phi)$, where $\phi = \int_0^\tau y(t) B(t)dt$ and $y(t)\in \{\pm 1\}$ is a square-wave modulation that flips sign on each pulse. Writing $\phi = v^T A$, the vector $v$ is the protocol's \textit{linear response}, determined by $y(t)$ and $\tau$, which captures how the pulse sequence filters the signal's spectral components. Consider a DD sequence which isolates frequency $\omega_j$. Its linear response $v$ is close to one at entry $j$ and zero at every other entry. By selecting $v$ to target frequency combinations of interest, we can reverse-engineer pulse modulations $y(t)$ realizing that response. 

In each RQS shot, we sample a Gaussian-random vector $g\sim\mathcal N(0,\mathds 1_{2N})$ and classically compile a pulse sequence on a time interval $I$ whose linear response is $g$, so that its modulation satisfies $\int_I y_g(t)f(t)\,dt = g$. We reverse-engineer $y_g(t)$, prepare the sensor in $\ket{+}$ and apply the sequence, obtaining the state $\frac{1}{\sqrt{2}}\bigl(\ket{0} + e^{ig^T A}\ket{1}\bigr)$. We measure $Y$, and record the outcome bit $s\in\{\pm1\}$ together with $g$. The accumulated phase is $g^TA$, so $\langle Y\rangle= \mathbb E[s|g] = \sin(g^TA)$, and averaging over random $g$ gives $\mathbb E[s\cdot g] = e^{-\|A\|_2^2/2} A$. When rescaled, each shot therefore yields an unbiased estimate of the entire spectrum. We calibrate the rescaling prefactor $e^{-\|A\|_2^2/2}$ with $X$-basis measurements, since $\mathbb E[\cos (g^T A)] = e^{-\|A\|_2^2/2}$.

The full protocol draws $R$ Gaussian random vectors $g_1, \ldots, g_R$, compiles the corresponding pulse sequences, and records a $Y$-basis outcome for each, along with $X$-basis shots that calibrate the prefactor (Fig.~\ref{fig:rqs_main}a). For a linear observable $K^TA$, each shot satisfies $\mathbb E[(s\,K^Tg)^2]=\|K\|_2^2$. The estimate $\widehat A = \frac{1}{\widehat c\,R}\sum_{j=1}^R s_j g_j$, where $\widehat c$ is the calibrated value of $e^{-\|A\|_2^2/2}$, therefore estimates $K^TA$ with standard deviation $O(\|K\|_2/\sqrt R)$, \textit{independent} of $N$. RQS thus estimates spectral quantities without first reconstructing the full spectrum to precision $\epsilon$ (which would require coordinate-wise error shrinking with $N$), in the same way that classical shadows estimate quantum observables while avoiding full tomography~\cite{Huang_2020_classical}.

The same approach and guarantee apply to nonlinear observables. For example, to estimate the spectral power $p_1^2+q_1^2$ at frequency $\omega_1$, we collect two independent datasets, form estimates $\widehat A^{(1)}$ and $\widehat A^{(2)}$, and multiply their $p_1$ and $q_1$ components across datasets, giving $\widehat A^{(1)}_{p_1}\widehat A^{(2)}_{p_1}\allowbreak+\allowbreak\widehat A^{(1)}_{q_1}\widehat A^{(2)}_{q_1}$. Using independent datasets avoids the bias of squaring a single noisy estimate, so the number of shots stays independent of $N$. For a general rank-$L$ observable $K$, we collect $L$ independent datasets and return $K[\widehat A^{(1)},\ldots,\widehat A^{(L)}]$. In Appendix~\ref{sec:rank_red_error_cancellation} we show that the same datasets estimate any bank of $M$ observables to precision $\epsilon$, with the number of shots growing only as $\log M$ while remaining cheaper than full reconstruction. 

RQS requires controls whose linear response is a random vector over all $2N$ quadratures. A na\"{i}ve construction would need challenging, smoothly varying control and an interrogation time linear in $N$ per shot. That would leave RQS with no advantage in $N$ over an ideal DD sweep, which estimates all $N$ components in time $\sim N/\epsilon$. Our compiler (Algorithm~\ref{algo:gaussian_response_compiler}) instead synthesizes Gaussian-random responses from $\pi$-pulses alone, in time $\sim\sqrt{N}$ per shot. Since the number of shots is independent of $N$, the total interrogation time scales as $\sqrt{N}(\kappa/\epsilon)^2$, which gives the first branch of Theorem~\ref{thm:rqs_informal}.

\medskip
\begin{theorem}\label{thm:rqs_informal}
    Randomized quantum spectroscopy solves $\textnormal{AQS}[\{\omega_j\}, \epsilon, \mathcal 
    K]$, with $N$ frequencies and $M$ observables, using     total interrogation time
    \vspace{-1mm}
    \begin{equation}
        T \lesssim \widetilde O\!\left(\min\!\left\{\sqrt N\left(\frac{\kappa}{\epsilon}\right)^2\!, N\frac{\kappa}{\epsilon}\right\}\log M\right) \ .
    \end{equation}
\end{theorem}
Theorem~\ref{thm:rqs_informal} exhibits two precision regimes. At \textit{standard} precision, $\epsilon/\kappa \gtrsim 1/\sqrt{N}$, the random responses in RQS yield interrogation time scaling as $\sqrt N (\kappa/\epsilon)^2$. At \textit{high} precision, $\epsilon/\kappa \lesssim 1/\sqrt{N}$, the $\sqrt{N}$ advantage of randomized pulses is dominated by demanding precision cost, so a strategy with improved $\epsilon$-dependence is needed. Since a DD shot of duration $\tau$ accumulates a phase $\phi \propto \tau A_j$ that wraps around the unit circle, our high-precision protocol uses DD sequences of incrementally longer duration to learn the sinusoidal branch on which each $A_j$ lies before resolving its value (the ``\textit{phase unwrapping}'' strategy of Ref.~\cite{higgins_demonstrating_2009}), matching the asymptotic performance of DD with improved constant factors. Both regimes are given in Appendix~\ref{sec:randomized_spec}, and we next show that the tradeoff between $\sqrt{N}$ and $\kappa/\epsilon$ is unavoidable without quantum computation, so RQS saturates both ends of the AC Heisenberg limit depicted in Fig.~\ref{fig:Fig_1}. Our approach differs from that of Ref.~\cite{liu2026optimalquantumsensingalgorithms}, which randomizes the modulations $y(t)$ rather than the sensor response $v$, achieving optimal detection and single-tone search but incurring an $N$-dependent overhead for more general learning tasks. RQS recovers the same scaling for detection and search, but applies more broadly to any AQS instance and can be quadratically more efficient in $N$.

While we have discussed a single sensor, the same pulses can be applied globally to an ensemble of $S$ sensors. This provides a $\sim 1/\sqrt{S}$ reduction in total time, while using $S$-qubit GHZ or N00N states provides the standard metrological enhancement up to $\sim 1/S$. These basic multi-sensor enhancements apply to all subsequent upper and lower bounds and we omit them for clarity.

\vspace{-4mm}
\subsubsection{Optimality and noise}
\vspace{-2mm}

We prove two lower bounds showing that RQS is optimal among all protocols using Hamiltonian~\eqref{eq:sensing_ham}. The first establishes that no strategy that supports \textit{downstream} estimation, in which observables may be provided after sensing, can perform better than RQS. The second shows that when a large number of observables is desired, as in waveform reconstruction, broadband search, or power spectroscopy, even a strategy granted prior knowledge of the observables and adaptive acquisition cannot asymptotically outperform RQS.

\begin{theorem}[AC Heisenberg limit] \label{thm:heisenberg_informal}
    Any conventional sensing protocol for $\textnormal{AQS}[\{\omega_j\}, \epsilon, \mathcal 
    K]$ that supports downstream estimation must use interrogation time
    \begin{equation}
    T\gtrsim \widetilde \Omega\!\left(\min\!\left\{\sqrt N\left(\frac{\kappa}{\epsilon}\right)^2\!, N\frac{\kappa}{\epsilon}\right\}\right)
    \end{equation}
    with logarithmic factors matching those suppressed in Theorem~\ref{thm:rqs_informal}. When $M\gtrsim N$, any adaptive sensing protocol, even with prior knowledge of the observables, must also use at least this much time.
\end{theorem}
The general-$L$ analog of this bound (Appendix~\ref{sec:nonadaptive_lower_bounds}) precisely matches the $L$-dependent extension of Theorem~\ref{thm:rqs_informal}, so RQS is optimal for linear and nonlinear AC sensing. Moreover, optimal performance does not require Gaussian randomness. RQS needs only random responses that sample all $2N$ quadrature directions equally, so that one dataset supports estimation of any observable chosen after sensing. Our open-source code (\cite{algorithmic_sensing_toolkit}) includes compilers using Rademacher-random modulations, which achieve similar asymptotic performance and can behave better in practice (Fig.~\ref{fig:rqs_main}b).

Noise can pose a major obstacle for sensing: in DC metrology, Heisenberg scaling under signal-aligned noise is provably impossible unless the noise rate vanishes with $\epsilon$~\cite{pradenne_restrictions_2026}. We establish a similar obstruction for conventional AC sensing when the sensor evolves under the master equation 
\begin{equation} \label{eq:main_text_noise_eq}
    \dot{\rho} =  -i\left[H(t),\rho\right]+\frac{\gamma}{4}\sum_{J\in\{X,Y,Z\}}\left(J\rho J-\rho\right),
\end{equation}
where $\gamma$ is approximately proportional to the sensor's $1/T_2$. This dissipator realizes full-rank depolarizing noise (isotropic white noise), but all conclusions also apply to weaker $Z$-only dephasing.
\begin{theorem}[AC Standard Quantum Limit] \label{eq:AC_SQL_informal}
    Any conventional protocol for $\textnormal{AQS}[\{\omega_j\}, \epsilon, \mathcal K]$, learning $M = \Omega(N)$ observables under depolarizing or dephasing noise with $\gamma \gtrsim \epsilon/\kappa$, must use interrogation time
    \begin{equation}
        T \geq \Omega\!\left(N\gamma\frac{\kappa^2}{\epsilon^2}\right).
    \end{equation}
    RQS achieves this scaling.
\end{theorem}
For $\gamma \lesssim \epsilon/\kappa$, noise contributes only constant overhead to the noiseless scaling. Constant-strength noise thus forces RQS, and any other conventional protocol, to revert to the AC SQL, yet RQS continues to saturate the optimal noise-aware scaling at any $\gamma$ (Appendix~\ref{sec:rqs_noise_optimality}). When $\gamma \gtrsim \Delta_\omega$, the sensor $T_2$ is too short to coherently resolve neighboring lines. For this regime we develop a variant of Qdyne spectroscopy \cite{boss_quantum_2017}, which uses a stable local clock to achieve arbitrary precision at the SQL (Appendix~\ref{sec:randomized_spec}). Together, these protocols are optimal for large banks of observables at any $N$, $\epsilon$, and noise strength.

Beyond sensing time optimality, RQS requires a factor of $\sim\sqrt{N}$ to $N$ fewer shots than DD for equivalent tasks and noise rates, a large improvement in sensitivity per unit time in experiments where readout and relaxation times dominate, e.g.~in biological applications~\cite{Kucsko_2013, zvi2023engineeringspincoherencecoreshell, Ying2017SMILE}. Finally, mitigation strategies for pulse imperfections and quasistatic fields, including XY8 phase cycling and response centering, integrate directly with RQS without increasing the number of pulses (Appendix~\ref{sec:rqs_mods}).

Since DD and other existing approaches can first reconstruct the waveform and then estimate observables, it may seem that RQS is not unique in its ability to support efficient downstream estimation. Indeed, Walsh-based sensing~\cite{cooper_time-resolved_2014}, compressed sensing with randomized orthogonal bases~\cite{magesan_compressing_2013}, randomized pulse protocols for noise spectroscopy~\cite{huang_random_2025}, and optimized spectral filters~\cite{poggiali_optimal_2018, frey_application_2017} also acquire information about multiple signal components per measurement, and can outperform frequency-selective DD at waveform reconstruction if the signal is sparse in every basis. For general signals, however, estimating downstream observables to accuracy $\epsilon$ via waveform reconstruction can require far more accuracy in the original waveform estimate, which increases the sensing time. RQS avoids full reconstruction, providing an optimal observable-independent procedure for any signal regardless of sparsity. In Appendix~\ref{sec:rqs_advantages} we consider two examples of standard sensing tasks where RQS confers significant advantages: the ``needle in a haystack'' problem of locating a weak, single-tone signal in a broad band, and lock-in detection of sparse signals. In both cases we prove that RQS provides a speedup over DD and Walsh-based sensing~\cite{cooper_time-resolved_2014} that grows with the bandwidth; a numerical simulation of the search example is given in Figure~\ref{fig:rqs_main}. 

\vspace{-1.5em}
\subsection{Matched spectroscopy}
\vspace{-2mm}

\begin{figure*}
    \centering
    \includegraphics[width=\linewidth]{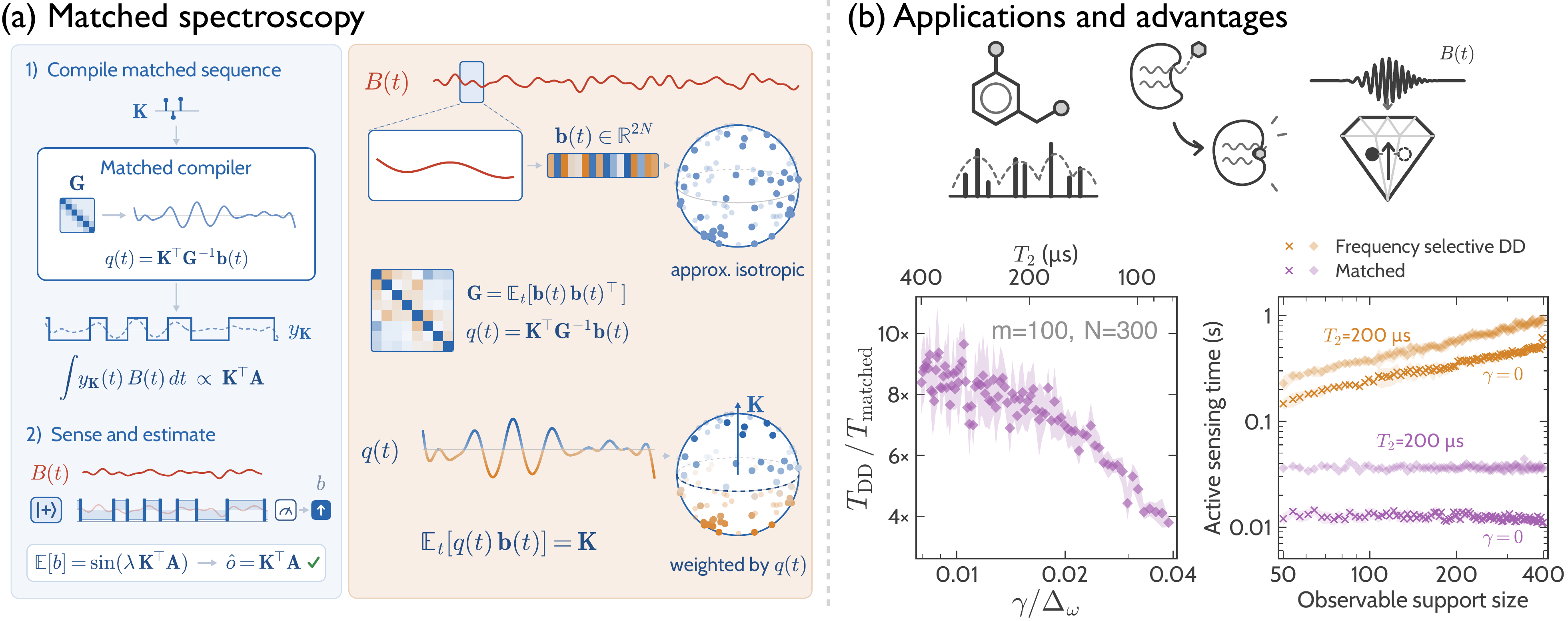}
    \par\vspace{-2mm}
    \caption{(a) \textit{Overview of matched spectroscopy}. Blue: Given an observable $K$, matched spectroscopy computes the dual filter $q(t)$ whose average response over short interrogations is $K^TA$ (or $K[A^{\otimes L}]$ more generally), and efficiently compiles it into a pulse sequence. Executing the sequence reconstructs the observable without first learning $A$. Orange: The time-dependent vector $b(t)$ captures how the quadratures evolve in a short interval. Free precession under signal $B(t)$ is filtered by $b(t)$; each shot experiences an isotropic filter on the $2N$-dimensional sphere. Constructing the dual filter by inverting the spectral frame ($G^{-1}$) and weighting by $K$ instead filters responses parallel to $K$ and suppresses responses orthogonal to it. (b) \textit{Simulations of matched filtering tasks}. Broadband matched filtering tasks appear in many settings; depicted are molecular fingerprinting, protein binding, and detection of magnetic transients with NV centers. Matched spectroscopy achieves speedups over DD across the simulated noise rates and bandwidths, and for randomly chosen observables is seen to scale independently of support size, outperforming its proven worst-case scaling. Simulations use a 10~MHz carrier frequency and 50~kHz frequency resolution.}
    \label{fig:matched}
    \vspace{-4mm}
\end{figure*}

When only a few known observables are required, acquiring the reusable dataset of RQS can cost sensing efficiency. Formally, Theorem~\ref{thm:heisenberg_informal} is no longer the fundamental optimum when $M\ll \sqrt{N}$ and the observables are known in advance. Many important tasks involve comparing weak signals to a few known waveforms and thus operate in this regime: trace-gas and molecular free-induction-decay spectroscopy, for example, detect very weak, known signal profiles among many optical frequencies \cite{thorpe_cavity-enhanced_2008, coddington_time-domain_2010,  liang_modulated_2025}, often to identify biomarkers of specific diseases or metabolic processes \cite{thorpe_cavity-enhanced_2008,  liang_ultrasensitive_2021}. Similarly, fluorescence microscopy targets frequencies associated with known fluorescent proteins to perform nanoscale imaging, and can be executed with molecular qubits~\cite{feder2025fluorescentproteinspinqubit}. We introduce matched spectroscopy to achieve the optimal sensing time for tasks where fields may be composed of many frequency modes, but we require prespecified functions of the amplitudes rather than a reusable dataset.

Consider, for simplicity, a task that requires a single linear observable $K^TA$, and let $\|K\|_\infty = 1$. While $A$ is supported on $N$ frequencies, $K$ may be supported on $m\leq N$ frequencies. A Ramsey signal interrogation of duration $\tau$, centered at an absolute time $t$, accumulates a phase $\phi = \tau b(t)^T A$, where $b(t) = \tfrac{1}{\tau}\int_{t-\tau/2}^{t+\tau/2} f(t')dt'$ captures the spectral sensitivity of that pulse-free interrogation. The Gram matrix $G = \mathbb E_t[b(t)b(t)^T]$ then quantifies how sensitive a collection of shots of duration $\tau$ is to specified waveforms. 

Matched spectroscopy observes that the \textit{dual matched filter} $q(t) = K^T G^{-1}b(t)$ satisfies 
\begin{equation} \label{eq:matched_dual_filter}
    \mathbb E_t[q(t)b(t)^T A] = K^T A \ .
\end{equation}
If $q(t)$ could be realized as a sensor modulation, the accumulated phase would be proportional to $K^TA$, with no reusable information about $A$ learned first. In Appendix~\ref{sec:small_M_optimal}, we show that any $q(t)$, tied to an observable supported on $m$ frequencies, obeys $\|q\|_\infty \leq O(\sqrt{m})$, while our Algorithm~\ref{algo:deterministic_response_compiler} compiles any response vector in time linear in its $\infty$-norm. Compiling these matched sequences and performing phase unwrapping estimate $K^TA$ in total time $\widetilde{O}(\sqrt{m}/\epsilon)$, quadratically faster in $\epsilon$ than RQS. The resulting algorithm (Fig.~\ref{fig:matched}a) is precisely the quantum realization of classical analog matched filtering, which modulates a classical receiver with a chosen dual filter to amplify sensitivity to a template waveform and suppress orthogonal waveforms.

To further improve this interrogation-time scaling, we observe that while the worst-case $\infty$-norm of $q(t)$ is $O(\sqrt{m})$, its time-averaged magnitude is $\mathbb E_t[q(t)^2] = K^T G^{-1} K = O(1)$ because $G$ is well-conditioned; conceptually, any large peaks of $q$ must occupy a small fraction of the sensing window. We therefore pick a threshold value $C$ and decompose $q(t) = a_C(t) + R_C(t)$, where $a_C(t) = \text{clip}(q(t), [-C, C])$, so that the desired observable splits into two contributions, 
\begin{equation}
    K^TA = \mathbb E_t[a_C(t)b(t)^T A] + \mathbb E_t[R_C(t) b(t)^T A].
\end{equation}
To estimate the first component we simply synthesize $a_C(t)$ as a response vector, costing time $O(C/\epsilon)$ as before. The residual satisfies $\mathbb 
E_t[|R_C(t)|] \lesssim 1/C$, so we perform importance sampling over short sensing shots at random starting times to extract the second component (Appendix~\ref{sec:small_M_optimal}), requiring time $\sim m/(C^2\epsilon^2)$. Optimizing $C$ between the two contributions gives $C \sim (m/\epsilon)^{1/3}$ and a total interrogation time of $\widetilde O(m^{1/3}/\epsilon^{4/3})$, which outperforms $\sqrt{m}$ scaling.

Matched spectroscopy extends to multiple observables, $\{K_1,...,K_M\}$, by first row-reducing the $M$ vectors to a shared subspace of rank $m\leq 2N$ and computing orthogonal basis vectors $u_1,...,u_m$. We then replace the vector $K$ in equation~\eqref{eq:matched_dual_filter} with each $u_j$, one at a time, and run matched spectroscopy with the corresponding dual filter. The resulting data is postprocessed to estimate all $M$ observables, giving the following guarantee when $M=O(1)$:

\begin{theorem}\label{thm:adaptive_small_M_lb}
    When a constant number of observables are specified before sensing, and supported on a rank-$m\leq N$ subspace, $\textnormal{AQS}[\{\omega_j\}, \epsilon, \mathcal K]$ can be solved using 
    \begin{equation}
        T_{\rm int} = \widetilde O\!\left(\min\!\left\{m^{1/3}\left(\frac{\kappa}{\epsilon}\right)^{4/3}\!, \sqrt{m}\,\frac{\kappa}{\epsilon}\right\}\right)
    \end{equation}
    total interrogation time.
\end{theorem}
\vspace{-1mm}
A similar strategy supports arbitrary nonlinear spectroscopy. Compared with Theorem~\ref{thm:heisenberg_informal}, adaptivity produces a polynomial advantage over any nonadaptive strategy when few observables are desired; in Appendix~\ref{sec:small_M_optimal} we show that this asymptotically outperforms RQS when $M=o(\sqrt{N})$. This interrogation time is optimal:

\vspace{-1mm}
\begin{theorem}[Matched Heisenberg limit]\label{thm:matched_heisenberg_informal}
    Any conventional sensing protocol for $\textnormal{AQS}[\{\omega_j\}, \epsilon, \mathcal 
    K]$ that estimates a set of observables supported on $m$ frequencies requires 
    \begin{equation}
    T_{\rm int}\gtrsim \widetilde \Omega\!\left(\min\!\left\{m^{1/3}\left(\frac{\kappa}{\epsilon}\right)^{4/3}\!, \sqrt{m}\frac{\kappa}{\epsilon}\right\}\right)
    \end{equation}
    total interrogation time.
\end{theorem}

An important distinction arises when comparing interrogation time to the total elapsed time of the protocol, including waiting between shots, which we refer to as ``wall-clock'' time. The two coincide for RQS and for the $\sqrt{m}\,\kappa/\epsilon$ branch of matched spectroscopy. The $m^{1/3}$ branch instead estimates the residual $R_C$ with short shots at random times drawn by importance sampling, and the sensor idles between them. Its wall-clock time therefore scales as $O(m(\kappa/\epsilon)^2)$, which has the form of the SQL with $m$ in place of $N$. The following theorem shows that $\sqrt{m}\,\kappa/\epsilon$ is a wall-clock limit for any protocol allowed by quantum mechanics, and that any \textit{conventional} protocol with shorter interrogation time pays SQL overhead in waiting time.
\begin{theorem}[Wall-clock Heisenberg limit and SQL]
    Any protocol for $\textnormal{AQS}[\{\omega_j\}, \epsilon, \mathcal 
    K]$ requires 
    \begin{equation}
        T_{\rm wall} \geq \widetilde \Omega\!\left(\sqrt{m}\,\frac{\kappa}{\epsilon}\right) \ ,
    \end{equation}
    when $|\mathcal K|$ is constant and all observables are supported on a rank-$m$ subspace. Any conventional protocol using interrogation time $o(\sqrt{m}\kappa/\epsilon)$ must use
    \begin{equation}
        T_{\rm wall} \geq \Omega\!\left(m\,\frac{\kappa^2}{\epsilon^2} \right) 
    \end{equation}
    total wall-clock time.
\end{theorem}
The first bound applies to any strategy, including ones with quantum computation, while the second shows that among conventional protocols with $\ll \sqrt{m}$ interrogation time, matched spectroscopy is optimal in wall-clock time. We will show later that quantum computational protocols achieve $m$-\textit{independent} interrogation time for matched tasks while keeping wall-clock time at the fundamental $\sim \sqrt m \kappa/\epsilon$ limit. 

Notably, under strong decoherence, matched spectroscopy retains a polynomial advantage over the AC SQL of Theorem~\ref{eq:AC_SQL_informal}.
\begin{theorem}[Matched SQL]
Under depolarizing or dephasing noise with $\gamma \gtrsim \epsilon/\kappa$, any conventional protocol for $\textnormal{AQS}[\{\omega_j\}, \epsilon, \mathcal K]$ with $|\mathcal{K}| = O(1)$ must use interrogation time
\begin{equation}
    T \geq \Omega\!\left(\sqrt{m\gamma\Delta_\omega}\,\frac{\kappa^2}{\epsilon^2}\right) \ .
\end{equation}
Matched spectroscopy achieves this scaling.
\end{theorem}
Even when the noise-induced linewidth matches the desired frequency resolution ($\gamma \sim \Delta_\omega$), matched spectroscopy retains a $\sim\sqrt{m}$ advantage over established protocols like DD. Matched filter design avoids resolving individual broadened lines and instead filters the desired observable coherently, enabling a meaningful robustness advantage in noisy quantum sensing. Moreover, the worst-case $m$-dependence arises only for templates whose dual filter has sharp peaks, $\|q\|_\infty = \Theta(\sqrt{m})$, which requires the frequency components in the support of $K$ to adversarially add in phase at a common time. Typical template waveforms lack such peaks, and for randomly chosen templates our simulations show interrogation time scaling independently of $m$ (Fig.~\ref{fig:matched}b). DD cannot do this, because it resolves the $m$ frequencies one at a time. 

Together, RQS and matched spectroscopy provide optimal conventional sensing strategies for any number of observables and at any noise strength. Replacing DD sequences with the pulse compilers provided in our accompanying code can thus improve existing platforms without introducing new hardware requirements. 

\vspace{-3mm}
{\renewcommand{\addcontentsline}[3]{}%
\subsection{Pooled spectroscopy}} 
\vspace{-2mm}

The previous protocols were designed for settings in which each of the $2N$ Fourier coefficients may contain relevant information. In other settings, the spectrum is populated by nuisance degrees of freedom and the sought-after signal is hidden among a broadband noise background. Rather than slowly learning each spectral line, we would like to read out coarse-grained features of the signal, such as the total power within a frequency band $[\omega_0, \omega_0+\Delta]$. However, the $X$- and $Y$-axis $\pi$-pulses used in conventional sequences create the \textit{commuting} interaction-picture Hamiltonian $H_{\rm int}(t) \propto y(t) B(t) Z$, so each shot accumulates a phase that depends linearly on $B$. Pooled spectroscopy instead designs pulse sequences around all axes of the Bloch sphere to coherently perform nonlinear signal processing before readout, estimating coarse-grained properties of a band containing $m$ frequencies polynomially faster in $m$ than \textit{any} strategy using commuting control, including DD, randomized, or matched spectroscopy.

\vspace{1em}
\subsubsection{Polynomial phase mixing mechanism}
\vspace{-2mm}

The main mechanism behind pooled spectroscopy can be understood as follows. Suppose a signal interrogation evolves the sensor by $R_Z(\theta)$, an angle $\theta$ rotation around the $Z$ axis. By applying $X$ and $Y$ pulses between such interrogations, we can create the evolution $C(\theta) \coloneqq R_X(\theta)R_Y(\theta)R_X(-\theta)R_Y(-\theta) \approx R_Z(\theta^2) + O(\theta^3)$, and the third-order dependence can be suppressed by taking $C(\theta/\sqrt{2})C(-\theta/\sqrt 2) = R_Z(\theta^2) + O(\theta^4)$. Eight noncommuting pulses have thus created a sensor response $\approx \theta^2$, rather than $\propto \theta$, without first learning $\theta$. Physically, the sensor Hamiltonian no longer commutes with itself at different times, so different time-ordered paths interfere, enclosing an area $\approx \theta^2$ in state space.  More generally, carefully chosen noncommuting pulses let us turn rotations $R_Z(x_1), ..., R_Z(x_m)$ into any polynomially transformed rotation $R_Z(cx_1^{L_1} \cdots x_m^{L_m})$ without learning the individual phases, a primitive we call \textit{polynomial phase mixing} (Appendix~\ref{sec:poly_phase_mixing}).

To estimate coarse-grained nonlinear properties, all that remains is to develop a procedure for synthesizing rotations $R_Z(\theta)$ such that $\theta$ aggregates phases accumulated by different frequency modes in the desired band. For this, we use the random commuting sequences from RQS. To estimate the total power in a band $\mathcal B$ of $m$ frequencies with spectrum vector $A_{\mathcal B}$, we draw Gaussian-random responses $v_{\mathcal B}$ supported only on the band, so that each interrogation accumulates the phase $v_{\mathcal B}^T A_{\mathcal B}$, which the squaring sequence converts into $(v_{\mathcal B}^TA_{\mathcal B})^2$. Since the entries of $v_{\mathcal B}$ are independent, cross terms average out and $\mathbb E[(v_{\mathcal B}^T A_{\mathcal B})^2] = \sum_{j\in \mathcal B} (p_j^2 + q_j^2)$, so concatenating many such squared phases rotates the sensor by an angle concentrating around a known multiple of the band power, which is read out without learning any single spectral line. The full algorithm is given in Appendix~\ref{sec:pooled_spectroscopy}. 

Unlike the usual matrix-element formulation of quantum signal processing (QSP)~\cite{low_optimal_2016}, pooled spectroscopy transforms the entire sensing unitary. In the example above, we implement $R_Z(\theta)\mapsto R_Z(c\theta^2)$, controlling the error in operator norm. Crucially, the nonlinear transformation is encoded as a phase, $e^{-i\theta/2}\mapsto e^{-ic\theta^2/2}$. In contrast, QSP usually manipulates the \textit{amplitude} of the top-left matrix element while the remaining matrix elements can transform arbitrarily. Pooled spectroscopy is therefore fundamentally distinct from QSP compilation and requires stronger guarantees.

\vspace{-1.5em}
\subsubsection{Advantages of pooled spectroscopy}
\vspace{-2mm}

Continuing with the power in a band of $m$ frequencies, we show in Appendix~\ref{sec:pooled_spectroscopy} that DD spectroscopy, which learns each quadrature amplitude $p_j, q_j$ and then sums their squares, requires time $\sim m^{3/2}/\epsilon$ and $\sim m$ terminal measurements to reach precision $\epsilon$, while RQS improves the time to $\sim m/\epsilon$ with $\sim\sqrt{m}$ total measurements. Pooled spectroscopy outperforms both, achieving up to a factor $\sim m^{1/2}$ speedup over RQS and $\sim m$ over na\"{i}ve DD, depending on the precision regime, and using only $\widetilde O(1)$ terminal measurements, which is valuable when readout time dominates, as in liquid-state spectroscopy. Beyond these benchmarks, we prove that noncommutativity is fundamentally responsible for the speedup.
\begin{theorem} \label{thm:pooled_adv}
    Let $\mathcal B$ be a frequency band containing $m$ of the $N$ signal frequencies, with total power $P_{\mathcal B} = \sum_{j\in \mathcal B}(p_j^2+q_j^2)$. Any conventional protocol whose interaction-picture Hamiltonian commutes with itself at all times, such as a sequence of $X$- and $Y$-axis $\pi$-pulses, requires at least $\widetilde{\Omega}(m^{1/8})$ times more interrogation time than pooled spectroscopy to estimate $P_{\mathcal B}$ to precision $\epsilon$.
\end{theorem}
\noindent This bound holds even against phase-insensitive DD strategies tailored for power estimation~\cite{Kotler_2013}. Our construction of pooled spectroscopy is purely analytical, so while it exhibits large asymptotic speedups, its pulse count can be demanding. In practice, substantially smaller constants in the interrogation time can be achieved by using heuristic classical optimizers and variational pulse control to synthesize pooled polynomial responses~\cite{sinanan-singh_single-shot_2024, prabhu_quantum_2026}. Experimentally realizing pooled spectroscopy speedups with such methods remains a valuable direction for future work.

\vspace{-1.5em}
{\renewcommand{\addcontentsline}[3]{}%
\section{Quantum computational spectroscopy}}
\vspace{-2mm}

\begin{figure*}
    \centering
    \includegraphics[width=\linewidth]{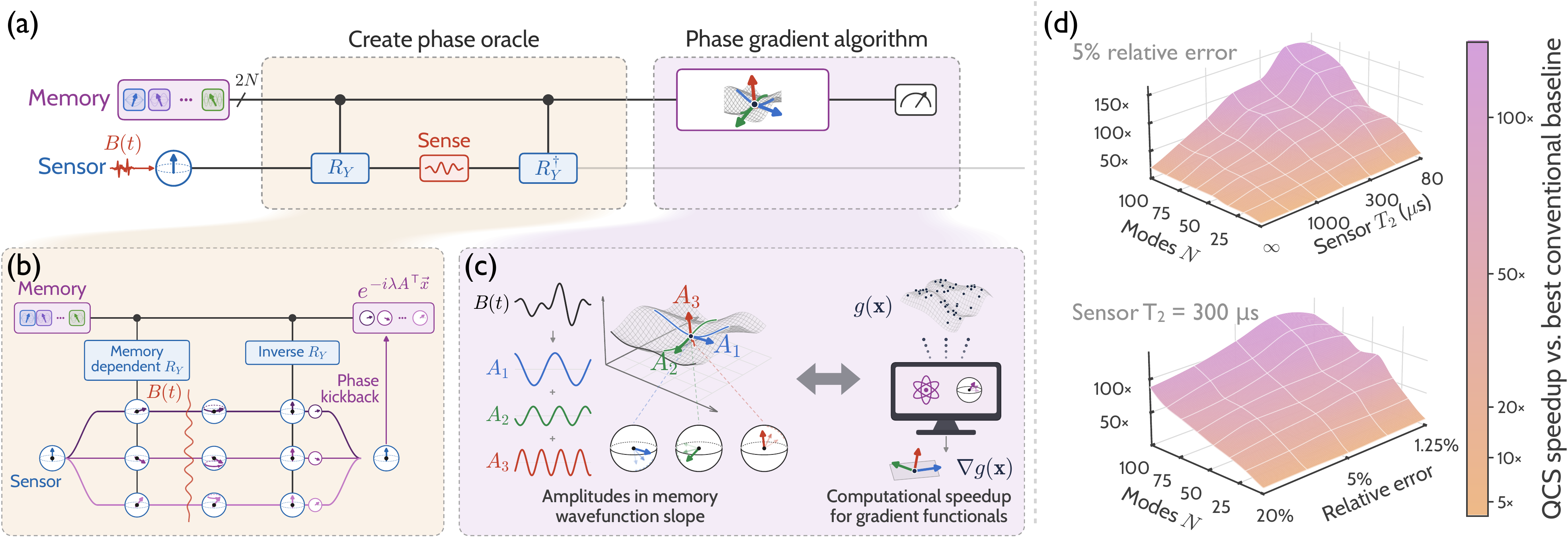}
    \caption{(a) \textit{Components of QCS}. The QCS algorithm uses a qubit sensor and a memory register of qubits, spins, or bosonic modes, and has two stages. (b) The first stage is the \textit{quantum computational frequency comb}, which applies memory-controlled rotations before and after sensing blocks to create a phase oracle out of the analog signal. Physically, the signal is written into the position-basis wavefunction (or spin angular momentum) of the memory, and the choice of rotation makes the protocol robust to noise (see below). (c) After constructing the oracle, we apply a quantum algorithm for estimating functionals of the encoded phase, providing a quantum algorithmic speedup for sensing; a special case is quantum gradient estimation \cite{jordan_fast_2005}. Physically, the memory's trajectory is read out in all directions at once by digitally measuring the conjugate momentum (or transverse angular momentum), and nonlinear signal information corresponds to nonlinear measurements such as photon counting or spin squeezing. (d) \textit{Noise-robust quantum sensing advantage}. We compare QCS to the better of DD and Qdyne spectroscopy for the same signal ensembles in Fig.~\ref{fig:rqs_main}b (top). At fixed relative error $\epsilon$, the advantage grows with $N$ and noise rate, showing the beyond-quadratic gap between the SQL performance of any protocol without quantum computation, and the noise-robust generalized Heisenberg scaling of QCS.  The ratios are large because we assume instantaneous control operations; practical advantages must be benchmarked on real devices.}
    \label{fig:qcs_main}
    \vspace{-3mm}
\end{figure*}

We now move beyond conventional pulse sequence protocols and ask whether coupling the sensor to a quantum processor enables better algorithms for AQS. We develop quantum computational spectroscopy (QCS), a family of protocols which use quantum algorithmic speedups to surpass Heisenberg-limited sensitivity and sustain it under constant-strength full-rank noise on the sensor. We first treat the ancillary system as a fault-tolerant quantum computer to present the most general form of the algorithm, then illustrate how QCS can be implemented in a simplified form on near-term platforms with as few as one ancilla qubit (see Fig.~\ref{fig:exp} below).

\vspace{1.5em}
{\renewcommand{\addcontentsline}[3]{}%
\subsection{Algorithm components}}
\vspace{-2mm}
QCS consists of two subroutines: noise-robust transduction of signal information into quantum memory, and quantum algorithms for reading out the encoded data in superposition (Fig.~\ref{fig:qcs_main}). We begin with the readout, since it is significantly simpler and determines the form in which the transduction gadget must store the signal.

\vspace{1em}
\subsubsection{Quantum algorithms for gradient functionals}
\vspace{-2mm}
QCS can be implemented with qubit or spin memories, but its algorithmic speedup is most easily understood by first treating the memory as a register of $2N$ bosonic modes with position and momentum operators $X_\mu$ and $P_\mu$ satisfying $[X_\mu, P_\nu] = i\delta_{\mu\nu}$. Writing $X = (X_1,..., X_{2N})$, suppose a black box provides access to the unitary $U_A(\lambda) = \exp(-i\lambda A^T X)$ for an unknown vector $A\in\mathbb R^{2N}$ and scalar $\lambda$. On a position eigenstate, $U_A(\lambda)\ket{x} = e^{-i\lambda A^Tx}\ket{x}$, so $U_A(\lambda)$ is a \textit{phase oracle} for the linear function $\lambda A^Tx$, whose gradient is $\lambda A$. 

First consider extracting $A$ itself. Initialize the memory in the vacuum state $\ket{0}$ and apply $U_A$. In the Heisenberg picture, the oracle leaves each position operator unchanged and shifts each momentum, $U_A^\dagger P_\mu U_A = P_\mu - \lambda A_\mu$, displacing the vacuum to a coherent state with momentum $-\lambda A_\mu$ in mode $\mu$. Measurement of all $2N$ momenta, rescaled by $-1/\lambda$, estimates every coordinate of $A$ at once, each with vacuum-limited error $\sim 1/\lambda$. Taking $\lambda\sim 1/\epsilon$, a single oracle query reconstructs every $A_\mu$ to precision $\sim\epsilon$, independent of the dimension of $A$. This is the quantum speedup for \textit{gradient estimation} of Ref.~\cite{jordan_fast_2005}. Classically, $\Omega(N)$ queries to $f_A(x) = A^Tx$ are necessary to learn its gradient $A$. In recent work, Ref.~\cite{liu2026optimalquantumsensingalgorithms} leveraged this primitive for a particular $L=1$, large-$M$ task; we solve AQS more broadly, show next that gradient estimation is a member of a larger family of quantum algorithms useful for sensing, and develop variations compatible with near-term implementation.

The same oracle enables efficient estimation of quadratic functions $A^TKA$. Define the annihilation operators $a_\mu = (X_\mu + iP_\mu)/\sqrt{2}$. Since the oracle displaces the vacuum to the $2N$-mode coherent state $\ket{\Psi_A} \coloneqq \ket{-i\lambda A/\sqrt{2}}$, each annihilation operator acts as $a_\mu\ket{\Psi_A} = -(i\lambda A_\mu/\sqrt{2})\ket{\Psi_A}$. The normally ordered operator $O_K = \frac{2}{\lambda^2}\sum_{\mu,\nu}K_{\mu\nu}a_\mu^\dagger a_\nu$ therefore has expectation
\begin{equation}
\begin{aligned}
 \bra{\Psi_A}O_K\ket{\Psi_A}
 &= \frac{2}{\lambda^2}\sum_{\mu,\nu}\frac{K_{\mu\nu}}{2} (i\lambda A_\mu)(-i\lambda A_\nu) \\
  & \qquad =A^TKA \ ,
\end{aligned}
\end{equation}
so measuring it exactly recovers the desired quadratic function. Importantly, because $O_K$ is normally ordered, the vacuum fluctuations of the $2N$ modes do not contribute to its variance, which is at most $2\kappa^2/\lambda^2$ for $\|K\|_\infty\leq\kappa$, independent of $N$. In contrast, first estimating $A$ and evaluating $A^TKA$ in postprocessing accumulates vacuum noise from every mode, adding a variance contribution that grows with $N$ (see Appendix~\ref{sec:qcs}). On a bosonic register, measuring $O_K$ amounts to photon counting after a passive optical transformation that diagonalizes $K$.

These two readouts suffice for any higher-order function: because the phase oracle produces coherent states, which are characterized by their first two moments, an order-$L$ function $K[A^{\otimes L}]$ can be estimated by applying the oracle to $\lceil L/2\rceil$ independent copies and measuring a quadratic observable on $\lfloor L/2\rfloor$ copies, together with a linear observable on the remaining copy when $L$ is odd.

These ideas translate directly to spin and qubit memories. For a large-$J$ spin memory, linear readout measures transverse angular momentum while quadratic readout measures spin squeezing. Meanwhile, in Appendix~\ref{sec:qcs} we use a qubit register to reproduce the bosonic relations with only polylogarithmic overhead in $N$ and $1/\epsilon$ in the number of qubits, and exponentially small truncation error. Readout is then realized by a local change of basis followed by computational-basis measurement. Alternatively, the $2N$-dimensional gradient can be encoded into the phases of $2N$ qubits and read out by phase unwrapping; we present this variation below. Practically, the quantum algorithmic component of QCS simply requires measurement in an appropriate conjugate basis, and does not require sophisticated quantum circuits.

\vspace{-1.5em}
\vspace{-1mm}
\subsubsection{Computational interpretation}
\vspace{-2mm}

The readout methods above define quantum algorithms for estimating functionals of a high-dimensional gradient that are of interest beyond sensing. Consider a function $h:\mathbb R^d\rightarrow\mathbb R$ encoded in the unitary $U_h = \exp(-i\lambda h(X))$. Since $[h(X), P_\mu] = i\partial_\mu h(X)$, the oracle shifts each momentum by the corresponding component of the gradient, $U_h^\dagger P_\mu U_h = P_\mu - \lambda\,\partial_\mu h(X)$. The oracle thus imprints $h$ as a position-dependent phase on the memory wavefunction, and the conjugate momentum records the local slope of that phase. Measuring all $d$ momenta after one query reveals the local gradient of $h$, whereas a classical algorithm needs $d+1$ queries to perform finite differences. Our linear readout therefore reproduces Jordan's quantum algorithm for gradient estimation~\cite{jordan_fast_2005}.

Readouts beyond linear order, however, lead to primitives fundamentally different from the quantum Fourier transform underlying Jordan's algorithm~\cite{jordan_fast_2005}. Quadratic readout measures the occupation number $a^\dagger a = \tfrac{1}{2}(X^2+P^2-1)$, which does not commute with $P$. Its eigenstates are the Hermite functions, which are the eigenfunctions of the Fourier transform itself. Quadratic readout therefore realizes the quantum Hermite transform~\cite{jain_efficient_2025}. For $L\geq 3$, the relevant observables are diagonal in neither the Fourier nor the Hermite basis, nor in any product of these bases. QCS measures them using several copies of the memory, so that each order-$L$ observable becomes a product of linear and quadratic normally ordered observables on a multi-copy Hilbert space. Whether a richer orthogonal-polynomial transform could measure these observables on a single copy, or whether the larger Hilbert space is necessary to avoid dimension-dependent penalties, remains open.

Viewed this way, QCS readout points to a hierarchy of quantum algorithmic primitives in which the quantum Fourier transform is only the first instance: momentum yields the Fourier transform, occupation number yields the Hermite transform, and higher-order functionals may require new families of spectral measurement. These primitives estimate polynomial functionals of high-dimensional gradients with far fewer queries than are classically required. Their applications beyond sensing and relation to other orthogonal polynomial and representation-theoretic transforms remain to be explored.

\vspace{-1.5em}
\subsubsection{Quantum computational frequency comb \\ for phase oracle transduction}
\vspace{-2mm}

\begin{figure}
    \centering
    \includegraphics[width=\linewidth]{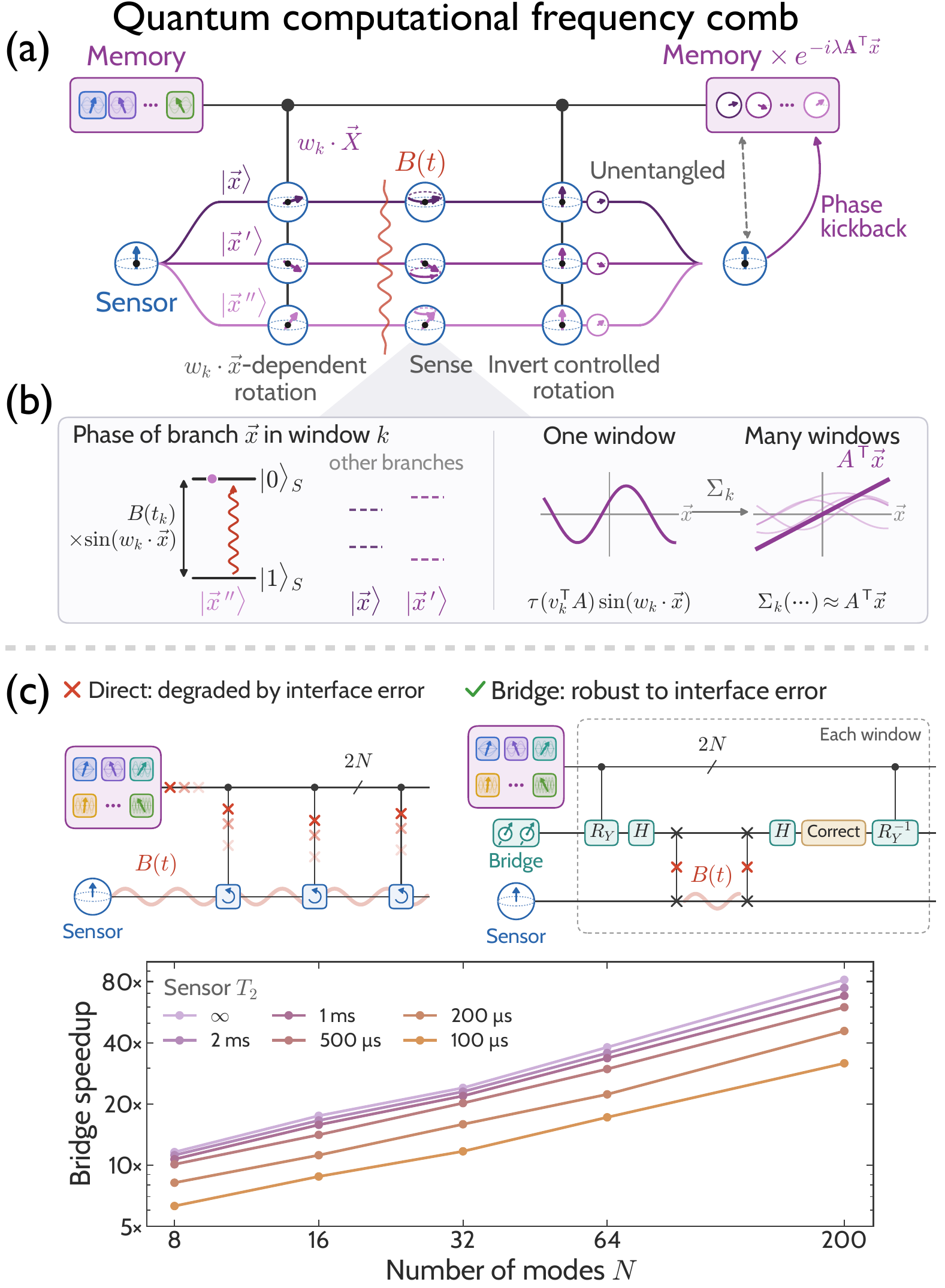}
    \caption{(a) \textit{Construction of phase oracle transducer}. The memory is prepared in a superposition of position eigenstates. In each transducer iteration, a vector $w$ is chosen and a memory-controlled rotation $R_Y(\pi/2 - w^T \Vec{x})$ is applied, for each memory branch $\ket{\Vec{x}}$. After sensing, the rotation is inverted, and the memory experiences phase kickback. (b) The vectors $w$ are chosen so that after several repetitions the accumulated memory phase is proportional to $A^Tx$, despite $A$ being unknown; this realizes a phase oracle. (c) \textit{Bridge gadget}. The bridge gadget corrects interface errors which otherwise degrade QCS. Depicted is the bridge gadget speedup for the same tasks from Fig.~\ref{fig:rqs_main}b (top) and Fig.~\ref{fig:qcs_main}d, when every interface gate causes a sensor error with probability $1\%$, and for various $N$ and sensor $T_2$ values. This gadget enables end-to-end beyond-Heisenberg-limited scaling under noise.}
    \label{fig:transducer}
    \vspace{-1em}
\end{figure}

We now develop the gadget that robustly creates the phase oracle $\exp(-i\lambda A^T X)$ (Fig.~\ref{fig:transducer}a,b). Since imprinting each coefficient $A_\mu$ requires entangling the sensor with $2N$ memory registers, sensor noise would na\"{i}vely degrade the memory coherence responsible for the speedup, as is unavoidable in DC metrology~\cite{pradenne_restrictions_2026} and has obstructed previous AC sensing constructions~\cite{allen_quantum_2025, liu2026optimalquantumsensingalgorithms}. We avoid this by using controlled rotations that rotate the sensor by nearly the same angle in every memory branch. A sensor error then reveals little to the environment about which branch the memory occupies, so the memory decoheres at a suppressed rate relative to signal accumulation.

Write the initial state of the memory and sensor as $\ket{\Psi}_{\rm M}\ket{0}_s$, and let $O$ be a Hermitian operator on the memory, specified below, with eigenstates $O\ket{x}_{\rm M} = o_x\ket{x}_{\rm M}$. Before sensing, we apply a rotation of the sensor about the $Y$ axis by an angle that depends on the memory's eigenvalue of $O$,
\begin{align}
\begin{aligned}
R_Y^{\rm M\rightarrow s}\!\left(\tfrac{\pi}{2} - O\right) &\coloneqq \exp\!\left[-\tfrac{i}{2}\left(\tfrac{\pi}{2}\mathds{1} - O\right)\otimes Y_s\right] \\
&= \sum_x \ket{x}\!\bra{x}_{\rm M}\otimes R_Y\!\left(\tfrac{\pi}{2} - o_x\right),
\end{aligned}
\end{align}
where $R_Y(\theta) = e^{-i\theta Y/2}$, and the sum becomes an integral for a continuous-variable memory. On the branch with memory state $\ket{x}_{\rm M}$, the joint state becomes
\begin{equation}
\begin{aligned}
 \ket{x}_{\rm M}\left[
 \cos\!\left(\frac{\frac{\pi}{2}-o_x}{2}\right)\ket{0}_{s} +\sin\!\left(\frac{\frac{\pi}{2}-o_x}{2}\right)
 \ket{1}_{s}\right].
\end{aligned}
\end{equation}
Writing the cosine and sine terms as $c_x, s_x$ respectively, the sensor accumulates a phase $\theta$ upon interrogating the signal, and the joint state becomes
\begin{equation}
 \ket{x}_{\rm M}\left[
 c_x
 e^{-i\theta/2}\ket{0}_{s} +s_x
 e^{i\theta/2}\ket{1}_{s}\right].
\end{equation}
While a local reference drive may be applied to demodulate large carrier frequencies, we do not require pulse sequences to actively filter the signal during interrogation. Finally, applying $R_Y^{\dagger}(\pi/2 - O)$ returns the sensor to $\ket{0}$, and the accumulated phase kicks back onto the memory, mapping $\ket{x}_M$ to $\exp(-i\theta\sin(o_x)/2)\ket{x}_M$. The accumulated phase is a linear response $\theta = v^T A$, with an interrogation-dependent $v$. Then for small $o_x$, $\sin(o_x)\approx o_x$, and the resulting memory operator is approximately $U_A = \exp(-\tfrac{i}{2} v^TA\, O)$. Now we set $O = w^T X$, a linear combination of position operators on the memory, and we choose random vectors $w$, based on the known linear response $v$, so that $\mathbb E[vw^T] \propto \mathds 1$, giving
\begin{equation}\label{eq:ideal_unitary_oracle}
    U_A \approx \exp(-\tfrac{i}{2}(v^TA)(w^T X)) \approx \exp (-i\lambda A^T X)
\end{equation}
and realizing the phase oracle for an interrogation time-dependent constant $\lambda$. Formally, we select time intervals and vectors $w$ such that the accumulated phase is proportional to $\arcsin(o_x)$ rather than $o_x$ (see Appendix~\ref{sec:qcs}). 

This procedure can be viewed as a quantum computational analog of an optical frequency comb~\cite{frequency_combs_ye,YeCundiff2005, Picqu__2019}. Frequency comb spectroscopy prepares an electromagnetic field whose frequency-domain representation approximates a Dirac comb across a wide band. In AQS language, an optical frequency comb selects a discrete set of frequencies $\{\omega_j\}$, prepares a broadband collection of bosonic modes, interrogates the signal, and measures the modified mode amplitudes to estimate many components of $A$ with narrowband precision. The set of sensitive frequencies can then be tuned for different tasks. Similarly, our transducer prepares a quantum memory and sensor such that their joint state is a receiver for a selected set of frequencies $\{\omega_j\}$ with narrowband precision. The sensitive frequencies are tuned classically by choosing the vector $w$ that defines the memory operator $O$, and after interrogation, each register of the memory contains a record of one frequency component of the signal. The architecture is flexible and allows us to apply both linear and nonlinear filters \textit{during} the transduction, which we exploit below in matched QCS. Unlike a classical frequency comb, the resulting state remains available for quantum postprocessing.

Next, we combine the subroutines and explain why QCS is noise-robust.

\vspace{-1em}
{\renewcommand{\addcontentsline}[3]{}%
\subsection{Complete QCS procedure and origin of noise-robust advantage}}

The QCS procedure begins by synthesizing the phase oracle $\exp(-i\lambda A^TX)$ described above. A total interrogation time $T$ realizes this unitary to exponentially small error with $\lambda \propto T/\sqrt{N}$ (Appendix~\ref{sec:qcs}), so we choose $T \propto \sqrt{N}/\epsilon$, capturing the $O(N)$ entries of $A$ while paying only a $\sqrt{N}$ loss of amplitude per coefficient. We then run linear readout to obtain a reusable estimate $\widehat A$, or the appropriate normally-ordered nonlinear readout when few nonlinear observables are required. Because gradient estimation confers an $O(N)$ speedup, and transduction only attenuates each amplitude by $\sqrt{N}$, we obtain a $\sqrt{N}$ speedup over all approaches without quantum computation.

\vspace{2mm}
\begin{theorem} \label{thm:qcs_main}
    Quantum computational spectroscopy solves $\textnormal{AQS}[\{\omega_j\}, \epsilon, \mathcal K]$, with $N$ frequencies and $M$ linear observables, using 
    \begin{equation}
        T \lesssim \widetilde O\!\left(\sqrt{N}\,\frac{\kappa}{\epsilon}\right) 
    \end{equation}
    total interrogation time.
\end{theorem}
Nonlinear readout solves nonlinear spectroscopy with schematic scaling $N^{1-1/L}(\kappa/\epsilon)^{2/L}$, polynomially better than the conventional lower bound on nonlinear tasks (Appendix~\ref{sec:lower_bounds}); in particular, even power spectroscopy, $L=2$, costs only $\sim\sqrt{N}\kappa/\epsilon$. QCS therefore improves on the AC SQL of Theorem~\ref{eq:AC_SQL_informal} beyond-quadratically, by both $\sqrt{N}$ and $\kappa/\epsilon$. This beyond-Heisenberg-limited sensitivity is optimal.
\begin{theorem}\label{thm:qcs_lb_main}
    Any protocol for $\textnormal{AQS}[\{\omega_j\}, \epsilon, \mathcal 
    K]$ with $M\gtrsim N$ observables must use at least
    \begin{equation}
    T\gtrsim \widetilde \Omega\!\left(\sqrt{N}\,\frac{\kappa}{\epsilon}\right) 
    \end{equation}
    total interrogation time.
\end{theorem}
This bound is a special case of the generalized Heisenberg limit of Ref.~\cite{kannan_universal_2026} and constitutes the fundamental lower bound on AC sensing with many observables. Unlike Heisenberg-limited protocols, however, QCS retains this sensitivity under the depolarizing noise of Eq.~\eqref{eq:main_text_noise_eq} acting continuously on the sensor, and under any weaker noise model, including $Z$-axis dephasing. 

To understand the effect of this noise, we examine how it propagates through the quantum computational frequency comb onto the memory. In one call to the transducer, noise replaces the ideal memory unitary of Eq.~\eqref{eq:ideal_unitary_oracle} with the master equation
\begin{equation}
    \dot \rho_M = -\frac{i}{2} \!\left[B(t)\sin(w^T X), \rho_M\right] + \frac{\gamma}{2}(\mathcal{Q}(\rho_M) - \rho_M) \ ,
\end{equation}
with dissipator (see Appendix~\ref{sec:qcs} for the derivation)
\begin{equation} \label{eq:memory_dissipator}
    \mathcal Q(\rho) = \frac{1}{2}\big(e^{iw^TX}\rho e^{-iw^TX} + e^{-iw^TX}\rho e^{iw^TX}\big) \ . 
\end{equation}
Sensor depolarization thus realizes a memory noise channel of random phase rotations $e^{\pm iw^TX}$, and since $e^{iw^TX}P_\mu e^{-iw^TX} \allowbreak=\allowbreak P_\mu - w_\mu$, an error displaces the memory momentum coordinate by a random kick $\Delta P = \pm w(t)$. Now notice that for any two memory branches $\ket{x}_{\rm M}, \ket{y}_{\rm M}$, which ideally rotate the sensor by $e^{iw^T x}, e^{i w^T y}$, the loss of coherence caused by Eq.~\eqref{eq:memory_dissipator} on the matrix element $\ketbra{x}{y}$ is \begin{equation}\label{eq:quadratic_coherence}
\frac{\gamma}{2}[1-\cos(w^Tx - w^Ty)] \simeq \frac{\gamma}{4}(w^T(x-y))^2    
\end{equation}
That is, the amount of information the environment learns from an error is suppressed \textit{quadratically} in the difference between memory branches, because the QCS oracle couples the sensor and memory so that changing one memory coordinate by a constant changes the sensor's rotation angle by only $O(1/\sqrt{N})$.

More precisely, writing $G = \mathbb E_t[v(t)v(t)^T]$ for the average sensor response Gram matrix, QCS takes $w(t) \simeq G^{-1}v(t)/\sqrt{N}$. The factor $G^{-1}$ ensures that $\mathbb E_t[w(t)v(t)^T] = \mathds{1}/\sqrt{N}$, so that the phases written in different intervals add up to a phase proportional to $A^TX$, while $\mathbb E_t[w(t)w(t)^T] = G^{-1}/N \preceq c\,\mathds{1}/N$ for a constant $c$. Consequently, sensor decoherence of rate $\gamma$ only decoheres memory correlations at a rate $O(\gamma/N)$, while the signal phase accumulated over total interrogation time $T$ is $\Phi \simeq T A^T \mathbb E[v(t) w(t)^T] X = TA^TX/\sqrt{N}$: the signal couples to the \textit{memory} quadratically more strongly than the noise does, despite both coupling to the sensor with comparable strength.

This can be contrasted with DC metrology, where a sensor accumulating phase under strength-$\gamma$ dephasing has Fisher information $\mathcal{I}(T) = T^2 e^{-2\gamma T}$, maximized at $T\sim 1/\gamma$, which is standard quantum limited for any $\gamma\gtrsim \epsilon$~\cite{pradenne_restrictions_2026}. As such, even a quantum computational protocol is standard quantum limited. In QCS, the Fisher information of one memory coordinate is instead, schematically, 
\begin{equation}
    \mathcal{I}_{\rm QCS}(T) \sim \frac{T^2}{N}\exp\!\left(\!-O\left(\frac{\gamma T}{N}\right)\right) \ ,
\end{equation}
because the signal phase grows as $T/\sqrt{N}$ while the noise grows as $T\gamma/N$. This is maximized around $T\sim N/\gamma$, far beyond the sensor's own coherence time of $\sim 1/\gamma$. More precisely, the ideal shot-limited uncertainty of QCS scales as $({\rm accumulated\ signal\ phase})^{-2}\sim N/T^2$ while the noise contributes variance of order $\gamma / T$, so under noise QCS requires time
\begin{equation}
    T\sim \max \{\sqrt{N}\kappa/\epsilon, \gamma\kappa^2/\epsilon^2\} \ .
\end{equation} Noise only dominates if $\gamma \gtrsim \sqrt{N}\epsilon/\kappa$, so outside the high-precision regime $\epsilon/\kappa \ll 1/\sqrt{N}$, the noiseless scaling persists even when $\gamma$ is constant. 

\begin{theorem}
    When the signal is broadband, so that $\sqrt{N}\epsilon/\kappa \gtrsim 1$, QCS solves $\textnormal{AQS}[\{\omega_j\}, \epsilon, \mathcal 
    K]$ with beyond-Heisenberg sensitivity $T = \Theta(\sqrt{N}\kappa/\epsilon)$, even under constant-strength sensor depolarization.
\end{theorem}

The noise threshold $\gamma \gtrsim \sqrt{N}\epsilon/\kappa$ is optimal; no protocol, even using arbitrarily sophisticated quantum error correction, can be more noise-robust. We note that the spectroscopy algorithm of Ref.~\cite{liu2026optimalquantumsensingalgorithms} is standard-quantum-limited for any noise rate beyond the vanishing threshold $\gamma \gtrsim \epsilon/(\kappa\sqrt{N})$. To our knowledge, QCS is the first quantum-enhanced protocol which achieves beyond-Heisenberg sensitivity under constant-strength noise aligned with the signal, and it establishes that the HNLS condition of Ref.~\cite{zhou_achieving_2018} is not necessary for asymptotic quantum advantage in AC sensing. Moreover, it suggests that quantum error correction is not the essential mechanism that makes sensing asymptotically more noise-robust.

\vspace{-1em}
\subsubsection{Gadget for correcting interface errors}
\vspace{-2mm}

We have assumed, as is common in error-corrected metrology~\cite{zhou_achieving_2018}, that sensor control is fast and noiseless; however, in DC sensing, signal-aligned errors during sensor-memory interfacing, which arise because gates are not instantaneous and may suffer errors during their implementation, can force SQL scaling even when the memory is fault-tolerant~\cite{cotler_noisy_2026}. We now suppose that every sensor-ancilla interaction is subject to random $Z$ errors, aligned with the signal axis, a setting where Refs.~\cite{cotler_noisy_2026, pradenne_restrictions_2026} showed that error-corrected DC metrology breaks down.

Unlike the DC case, QCS can be made fault-tolerant against such errors by adding a simple bridge of two logical qubits $a, b$ between the sensor and memory (Fig.~\ref{fig:transducer}c), with initial joint state $\ket{x}_{\rm M}\ket{00}_{ab}\ket{0}_s$ conditioned on memory branch $\ket{x}_{\rm M}$. Instead of applying a controlled rotation directly between memory and sensor, we apply $R_Y^{\rm M \rightarrow a}(\pi/2 - O)$ to bridge qubit $a$, couple $a, b$ via ${\rm CNOT}_{a\rightarrow b}$, apply a Hadamard gate to $a$, and SWAP qubit $a$ with the sensor; assuming ancilla resources are fault-tolerant, the only uncorrected error occurs during the SWAP with the sensor. At this stage, without error, the total state is
\begin{equation}
 \ket{x}_{\rm M}\left(
 c_x
 \ket{0}_a\ket{0}_b\ket{0}_s +s_x
 \ket{0}_a\ket{1}_b\ket{1}_s\right),
\end{equation}
so the joint state of qubits $s$ and $b$ has even parity. After sensing, we again SWAP $s$ and $a$, and apply a Hadamard to $a$. A $Z$ error during either SWAP gate does not destroy the accumulated phase; it simply flips the relative parity between $s$ and $b$, which after the second SWAP moves the parity relation to qubits $a$ and $b$. Measuring $Z_aZ_b$ and applying an $X$ correction to qubit $a$ if its parity is odd thus corrects the interface fault without affecting the memory channel (up to a small constant attenuation factor, see Appendix~\ref{sec:qcs}). 

\vspace{-1.5em}
{\renewcommand{\addcontentsline}[3]{}%
\subsection{QCS variations}}
\vspace{-2mm}

We now describe a variation of QCS that encodes the signal into the relative phases of memory qubits rather than in displacements, and use it to state a matched analog of QCS which outperforms reusable readout when few observables are desired and requires few or even one ancilla qubit. These variations enable near-term implementations demonstrating significant metrological gains without fault-tolerant processing. 

\vspace{4mm}
\subsubsection{QCS using relative qubit phases}
\vspace{-2mm}
Consider a memory register of $2N$ qubits with Pauli operators $Z_M = (Z_1, ..., Z_{2N})$ and one sensor. We use the same transducer, but choose $O = \arcsin(w^T Z_M)$, so that the phase generated by the transducer, proportional to $\sin O$, is linear in $w^T Z_M$, with $w$ chosen so that in expectation the transduced phase is $A^T Z_M$. With total sensing time $T$, this procedure creates the phase oracle
\begin{equation}
    U_A = \exp\!\Bigg(\!-i\lambda \sum_{j=1}^{2N}A_j Z_j\Bigg)
\end{equation}
where $\lambda \propto T/\sqrt{N}$. Given a memory initially in $\ket{+}^{\otimes 2N}$, the final state of memory qubit $j$ is $\tfrac{1}{\sqrt{2}}(\ket{0}_j + e^{2i\lambda A_j}\ket{1}_j)$, so $A_j$ is written directly into its relative phase. Standard phase unwrapping, identical to high-precision RQS, then reconstructs the vector $\widehat A$, resulting in total evolution time $T = \widetilde O(\sqrt{N}\kappa/\epsilon)$. This variation is equivalent to our displacement-based constructions up to polylogarithmic measurement overhead, and all downstream estimation and noise-robustness guarantees follow, while it is easier to implement with e.g. optically addressable two-level memory systems. 

\vspace{-2mm}
\subsubsection{Matched QCS with few ancillas}
\vspace{-2mm}

\begin{figure}
    \centering
    \includegraphics[width=\linewidth]{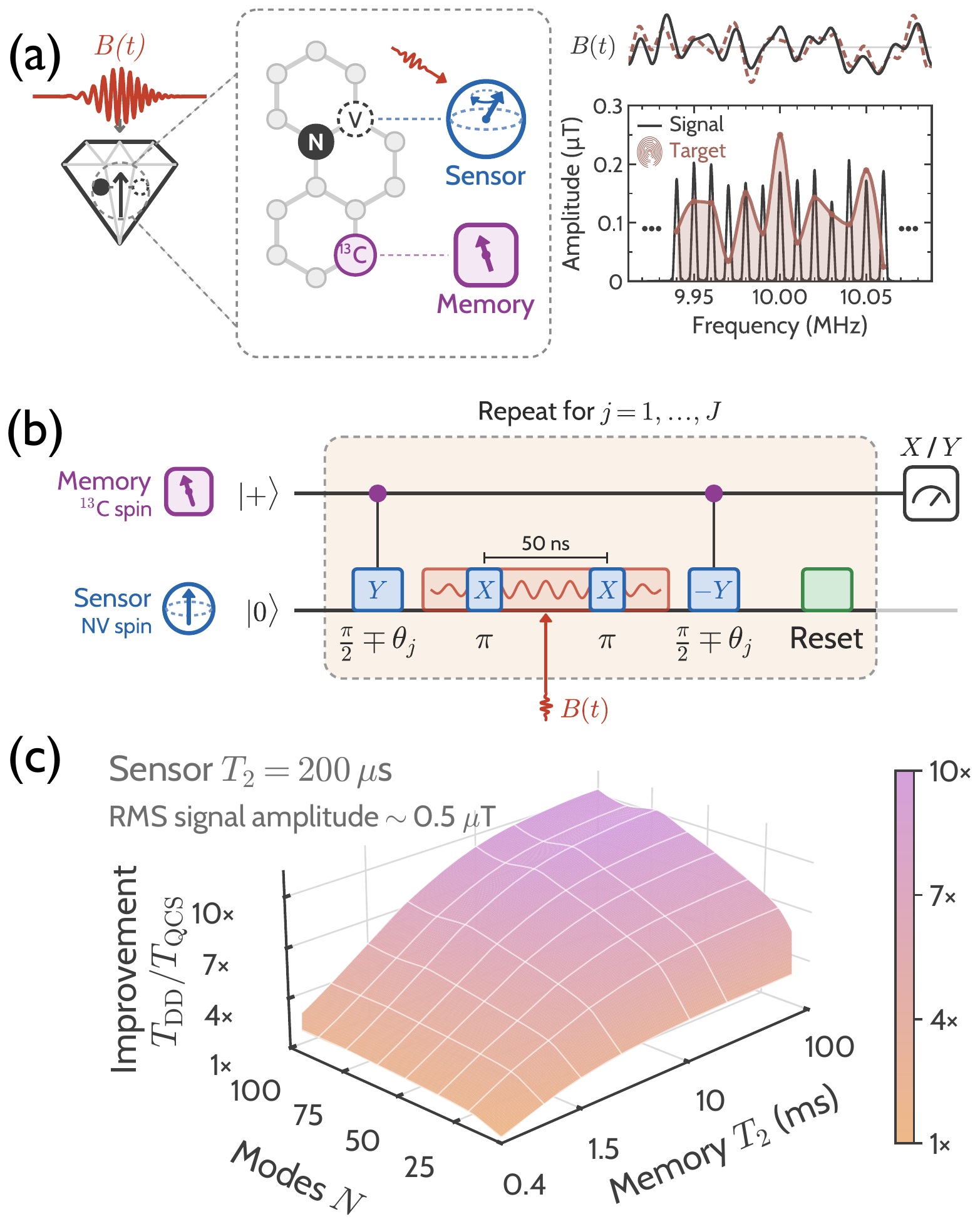}
    \caption{
    (a) Proposed near-term implementation of QCS, with a nitrogen-vacancy electron spin as the sensor and a nearby $^{13}$C nuclear spin as the memory, sensing a weak magnetic field with RMS amplitude $\sim 0.5\,\mu$T and a 10\,MHz carrier frequency. Using the $^{13}$C spin as its single memory qubit, matched QCS measures the overlap of the signal with a chosen waveform. (b) Control sequence. Memory-controlled $Y$ rotations of the electron spin, interleaved with sensing intervals, create the phase oracle. The $\pi$-pulses shown simply demodulate the 10\,MHz carrier, and a static drive can replace them. The protocol can be modified so the sensor need not be reset after every repetition, and sensing intervals of 100\,ns or longer suffice. (c) Simulations with sensor $T_2 = 200\,\mu$s show an advantage over DD that grows with the memory lifetime. Unlike conventional metrological speedups that grow only with sensor number, QCS enables an additional computational speedup growing with signal bandwidth, which can be realized without multiple sensors or ancillas.
    }
    \label{fig:exp}
    \vspace{-2mm}
\end{figure}

Thus far, the quantum computational frequency comb has written each component of $A$ into a separate memory register, so that every component can be read out simultaneously. When the quantities of interest are known in advance, however, the transducer can process the signal coherently as it is written, encoding only the desired observables. The resulting matched analog of QCS outperforms reusable QCS when few observables are desired, and requires one memory qubit per observable (independent of $N$) so a broadband QCS speedup can be obtained with a single ancilla qubit.

Consider again a single linear observable $K^TA$ supported on $m$ frequencies, with $\|K\|_\infty = 1$, and recall the dual matched filter $q(t) = K^TG^{-1}b(t)$, which satisfies $\mathbb E_t[q(t)b(t)^TA] = K^TA$. Matched spectroscopy was limited by the peaks of $q(t)$, which reach $\|q\|_\infty = O(\sqrt{m})$ even though its magnitude averages to $O(1)$. Matched QCS instead accumulates $q(t)$ directly in the phase of a single memory qubit. We divide the sensing window into short cells, and in the cell centered at $t$ we interrogate the signal for only a fraction $|q(t)|/\|q\|_\infty$ of the cell, then transduce the acquired phase onto the memory with the sign of $q(t)$. Each cell thus contributes a memory phase proportional to $q(t)\,b(t)^TA$. Between cells the sensor is reset while the memory retains its coherence, so these contributions add up to a multiple of $\mathbb E_t[q(t)b(t)^TA] = K^TA$. A wall-clock time $T_{\rm wall}$ produces
\begin{equation}
    U_M \simeq \exp\!\left(-i\lambda (K^TA)Z_M\right), \qquad \lambda \propto \frac{T_{\rm wall}}{\|q\|_\infty},
\end{equation}
writing $K^TA$ into the relative phase of the memory qubit. Since the sensor is active for a fraction $|q(t)|/\|q\|_\infty$ of each cell and $\mathbb E_t|q(t)| = O(1)$, the total interrogation time is only $T_{\rm int}\sim T_{\rm wall}/\|q\|_\infty \sim \lambda$. Phase unwrapping, as in the relative-phase variation above, requires $\lambda\sim 1/\epsilon$, so $T_{\rm int}\sim 1/\epsilon$ while $T_{\rm wall}\sim\|q\|_\infty/\epsilon = O(\sqrt{m}/\epsilon)$.
\begin{theorem}[Matched QCS]\label{thm:matched_qcs}
    In the absence of noise, matched QCS estimates a single specified linear observable supported on $m$ frequencies to precision $\epsilon$ using one memory qubit, with
    \begin{equation}
        T_{\rm int}=\widetilde O\!\left(\frac{\kappa}{\epsilon}\right), \qquad T_{\rm wall}=\widetilde O\!\left(\sqrt{m}\,\frac{\kappa}{\epsilon}\right).
    \end{equation}
\end{theorem}
Matched QCS thus achieves $m$-independent interrogation time, a $\sqrt{m}$ advantage over any conventional protocol, while saturating the wall-clock Heisenberg limit. This result shows that a QCS advantage growing with the signal bandwidth does not require a large quantum processor. Figure~\ref{fig:exp} proposes and analyzes a candidate experiment in which a single NV-center electron spin senses an oscillating magnetic field while a nearby $^{13}$C nuclear spin serves as the memory. The nuclear-spin-controlled rotations of the electron can be implemented through the native hyperfine interaction, which shifts the electron's resonance depending on the nuclear-spin state and enables frequency-selective conditional rotations~\cite{Jelezko_2004}. For weak fields with a carrier frequency near 10\,MHz and an electron coherence time $T_2\sim200\,\mu{\rm s}$, our simulations exhibit a significant advantage over DD spectroscopy for matched filtering with one nuclear-spin memory. Realizing it requires control fidelity sufficient for repeated transduction in intervals of several hundred nanoseconds.

\vspace{-1em}
{\renewcommand{\addcontentsline}[3]{}%
\section{Statistical learning vs. parameter estimation}}
\vspace{-1mm}

\begin{figure}
    \centering
    \includegraphics[width=\linewidth]
    {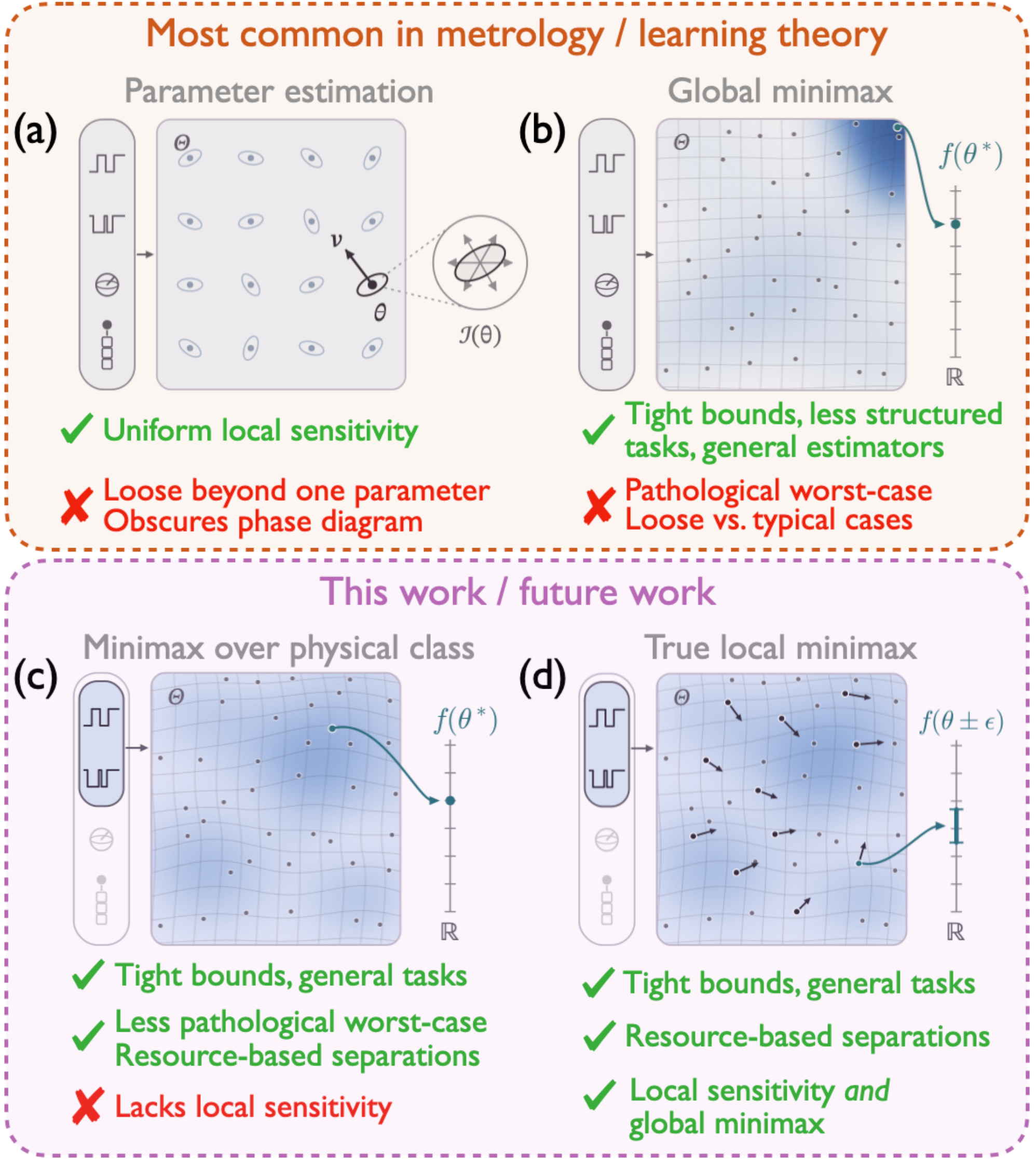}
    \par \vspace{-1mm}
    \caption{Each panel depicts a parameter space $\Theta$, with shading indicating instance hardness, together with the class of experiments that induces it. (a) Parameter estimation bounds the sensitivity at every point $\theta$ and along every direction $v$, through the quantum Fisher information $\mathcal I(\theta)$; lower bounds are uniform over the full space and all experiments. (b) Global minimax lower bounds against broad experiment classes certify quantum advantages for a fixed observable $f$, but the hardest instances are often confined to a small, highly structured region of $\Theta$. (c) Recent advances in quantum learning, and this work, restrict the class of experiments in minimax bounds, so the induced worst case $f(\theta^*)$ reflects typical instances and separations based on available experimental resources emerge. (d) A combined local and minimax theory would quantify, for a given class of experiments, the sensitivity to perturbations $f(\theta\pm\epsilon)$ at every point, recovering (a)-(c) as special cases.}
    \label{fig:estimation_vs_learning}
    \vspace{-4mm}
\end{figure}

We have studied AC sensing through the lens of statistical learning theory rather than the more familiar parameter estimation perspective centered on the quantum Fisher information. This choice is what allowed us to discover the phase diagram of Fig.~\ref{fig:Fig_1}b, and it points to the need for a theory of quantum sensing incorporating both perspectives (Fig.~\ref{fig:estimation_vs_learning}).

Parameter estimation, often through the quantum Fisher information, bounds the achievable sensitivity at every point of parameter space and along every direction (Fig.~\ref{fig:estimation_vs_learning}a). Because such lower bounds hold uniformly over the parameter space, they are unattainable when the measurements optimal for different directions are incompatible~\cite{szczykulska_multi-parameter_2016, ragy_compatibility_2016, albarelli_perspective_2020}, and it is usually unclear how loose they are, or how their attainability changes when resources such as ancillary quantum processing are added. Moreover, physical observables are often functions of the parameters rather than the parameters themselves, and the theory's focus on unbiased estimators behaves poorly for nonlinear functions~\cite{kannan_exponential_2026}. These limitations are why fundamental limits for less structured tasks, such as waveform estimation, have drawn on Bayesian and hypothesis-testing tools closer to statistical learning theory even while operating in a parameter-estimation framework~\cite{tsang_fundamental_2011, Tsang_2012, berry_quantum_2015}. Without knowing whether Fisher-information bounds are attainable, one cannot tell whether a gap between two classes of experiments, such as the gap between the AC and generalized Heisenberg limits in Fig.~\ref{fig:Fig_1}b, is real or an artifact of a loose bound.

Statistical learning theory has the complementary strength. Its minimax (worst-case) lower bounds can certify that an added quantum resource yields a provable advantage. Incorporating measurement restrictions into these bounds is technically difficult, however, so existing separations usually compare nearly unconstrained classes, such as single-copy measurements with arbitrary quantum processing versus two-copy measurements~\cite{chen2021exponentialseparationslearningquantum, Huang_2022}. Even the weaker class in such separations lies beyond any realistic experiment. The hard instances are structured ensembles built to hide information from that entire class, so the provable speedups apply only in a small corner of parameter space (Fig.~\ref{fig:estimation_vs_learning}b). It then remains unclear how much advantage survives for realistic problems and modest quantum resources~\cite{cotler_noisy_2026}.

Our formulation adopts a minimax standpoint but restricts it to physically realistic classes of signals, observables, and experiments (Fig.~\ref{fig:estimation_vs_learning}c). As a result, worst-case hardness does not diverge substantially from typical instances, and comparing the minimax limits of experiment classes with and without quantum computation reveals the phase diagram of Fig.~\ref{fig:Fig_1}b; the AC Heisenberg limit is a worst-case statement about a restricted class of experiments rather than a pointwise sensitivity bound, so its separation from the generalized Heisenberg limit only becomes visible once sensing is posed as a learning problem. However, this framework still lacks tight bounds on the hardness of estimating small changes in an observable around a prescribed point in the parameter space. The natural goal is a unified theory in which both perspectives emerge as special cases, quantifying resource-based advantage at each point of parameter space and in the worst case, for observables of the parameters rather than the parameters themselves (Fig.~\ref{fig:estimation_vs_learning}d). The quantum phase-space inference formalism of Ref.~\cite{kannan_exponential_2026}, which is minimax over learning observables at each point, provides such a framework. Developing it for AC sensing, including a local theory of the phase diagram in Fig.~\ref{fig:Fig_1}b, remains a valuable direction~\cite{kannan_universal_2026}.

\vspace{1em}
{\renewcommand{\addcontentsline}[3]{}%
\section{Outlook}}
\vspace{-2mm}
By adopting an algorithmic and learning-theoretic picture of AC sensing, we have introduced new families of sensing protocols that are provably optimal and outperform longstanding experimental methods. We have also shown that, contrary to expectations suggested by no-go theorems for single-parameter sensing, quantum computational AC sensing admits a metrological limit beyond Heisenberg scaling that is stable under noise. 

Our conventional protocols are readily implementable in platforms which use DD-based pulse control, including electronic or nuclear spins, circuit-QED systems, atomic clocks and atom interferometers, and trapped ions. Tailoring randomized, matched, and pooled spectroscopy to such platforms and realizing the resulting sensing gains is a high-value direction for future experimental work, and our code repository containing practical pulse compilers may be a useful starting point. 

More broadly, formulating sensing as a statistical learning problem opens it to techniques from randomized algorithms, yielding protocols that outperform established approaches through principled algorithm design alone, without demanding more advanced experimental control. Algorithmic formulations of other sensing tasks involving many unknown parameters, such as continuous frequency estimation, noise spectroscopy, or field imaging, may uncover similar gains~\cite{kannan_universal_2026}.

Our quantum computational protocols open several avenues at the intersection of quantum metrology and quantum algorithms. Previous asymptotic analyses of the Heisenberg limit share two parameters across applications: the number of sensors and the precision $\epsilon$. We have identified a third common axis, the amount of frequency-domain information $N$ in the signal, which quantifies \textit{computational multiplexing}: while the conventional $1/\epsilon$ Heisenberg limit still scales linearly in $N$, the fundamental quantum-mechanical limit permits a \textit{second} speedup, in $N$, due to processing information in superposition. For broadband signals this limit survives constant-strength noise, whereas Heisenberg scaling survives only when the noise rate vanishes with the target precision. These observations apply to all AQS problems, unlike the quadratic speedup of Ref.~\cite{allen_quantum_2025} for detecting a single tone. Ref.~\cite{kannan_universal_2026} reveals the nature of this new metrological limit beyond the AQS tasks studied here. Importantly, this quantum multiplexing speedup is not limited to weak-signal regimes where quantum shot noise is already the main obstruction, nor does it require growing quantum sensor arrays, so it may be applicable to classical receivers and sensing tasks where measurement uncertainty is classically limited.

The noise-robust performance of QCS demonstrates that the absence of signal-aligned noise is not a necessary condition for noise-robust multiparameter AC sensing~\cite{zhou_achieving_2018}, suggesting a more complete criterion to be identified~\cite{kannan_universal_2026}. Understanding how these limits vary across the parameter space requires an integrated theory of parameter estimation and statistical learning, like the Q$\Psi$ or Accessible Feature Information framework of Ref.~\cite{kannan_exponential_2026}. Developing the theory of AC sensing through this lens may lead to new, more fine-grained metrological gains enabled by quantum computation. 

Finally, our quantum computational frequency combs allow quantum-enhanced sensitivity to persist far beyond the sensor's own coherence time. They deliberately keep the branches of the memory wavefunction close together, so that sensor noise only slightly degrades memory coherence while signal accumulates parametrically faster, with robustness improving as the memory grows. Other multiparameter sensing protocols whose advantages degrade under noise may become similarly robust by adapting these transduction gadgets and shifting quantum processing away from exposed sensors and onto an ancillary memory, rather than correcting the sensors themselves~\cite{Humphreys_2013, Eldredge_2018, Zhuang_2020, Gorecki_2020, kannan2026faulttolerantquantumprocessingphysical}. Classical signals can now be converted into quantum computational oracles using simple, noise-robust operations. Practical, scalable quantum computational speedups for AC sensing may therefore be within near-term reach.

\vspace{2mm}
{\renewcommand{\addcontentsline}[3]{}%
\subsection*{Accompanying Code}}
\vspace{-3mm}
The code repository at the link below contains Python scripts that admit user-defined frequencies and coherence times and compile pulse sequences that can be exported for immediate use. They also handle postprocessing of imported experimental data. Users may refer to the contained Jupyter notebooks for walkthroughs and example simulations. Link:\href{https://github.com/avenkatesh20/Algorithmic-Sensing-Toolkit}{https://github.com/avenkatesh20/Algorithmic-Sensing-Toolkit}

\par
\vspace*{3mm}
{\renewcommand{\addcontentsline}[3]{}%
\subsection*{Acknowledgements}}
\vspace{-3mm}
We thank Soonwon Choi, Lysander Christakis, Adam Cohen, Luke Coffman, Siddharth Dandavate, Dhruva Ganapathy, Shimon Kolkowitz, Peter Maurer,  Peter McMahon, Rohan Mehta, Kang-Kuen Ni, Sridhar Prabhu, and Susanne Yelin for helpful discussions and valuable comments on this manuscript. IK is supported by the Nobile Research Initiative and the Air Force Office of Scientific Research under award number FA9550-25-C-B010 in the amount of \$134600 through the DoD NDSEG Fellowship. AV acknowledges support from the National Science Foundation Graduate Research Fellowship Program under Grant No. DGE-2140743. ChatGPT 6.1 Sol and Claude Opus 5.5 were used in developing the accompanying code, which was verified by the authors. ChatGPT 6 Sol and Astra aided in identifying complex analysis results used in Algorithm~\ref{algo:classical_polynomial_compiler}, proving the matched Heisenberg limit in Section~\ref{sec:adaptive_conventional_lower_bound}, extending the $L=1$ lower bound to $L\geq 2$ in Section~\ref{sec:nonadaptive_lower_bounds}, and in performing the noise computations in Section~\ref{sec:deferred_qcs_target_specific}. They were also used to check and polish various aspects of the manuscript.

\bibliographystyle{apsrev4-2}
\bibliography{references}

\clearpage
\onecolumngrid
\appendix
\makeatletter
\let\addcontentsline\AQSsavedaddcontentsline
\makeatother

\counterwithin{algocf}{section}
\renewcommand{\thealgocf}{\thesection.\arabic{algocf}}
\pagestyle{header}
\thispagestyle{header}
\begin{center}
{\Large\bfseries Appendices for ``Algorithmic Quantum Sensing''\par}
\vspace{1em}
{\normalsize
Ishaan Kannan\textsuperscript{1},
Aditi Venkatesh\textsuperscript{1},
and Jordan Cotler\textsuperscript{1}\par}
\vspace{0.5em}
{\small\itshape
\textsuperscript{1}Department of Physics, Harvard University, Cambridge, MA 02138 USA\par}
\end{center}

\begin{appendixabstract}
\quad Section~\ref{sec:app_A} formally states the AQS problem and sensing models used throughout, and provides a summary table of all protocols. Section \ref{sec:spec_toolbox} gathers the algorithmic subroutines that power our conventional protocols. Sections \ref{sec:randomized_spec}, \ref{sec:small_M_optimal}, \ref{sec:pooled_spectroscopy}, and~\ref{sec:qcs} each provide a self-contained description of one protocol and prove their formal scaling. Longer derivations are deferred to Section~\ref{sec:deferred} to improve readability. Lower bounds for each task are given in Section~\ref{sec:lower_bounds}. Clicking on the page number in the upper-right corner will return the reader to this table of contents.
\vspace{-2mm}
\end{appendixabstract}

\phantomsection
\label{apptoc}

\setcounter{tocdepth}{2}
\makeatletter
\let\l@subsubsection\@gobbletwo
\makeatother
\tableofcontents
\newpage

\IfFileExists{appendices_body.tex}{%
  \section{Algorithmic quantum sensing} \label{sec:app_A}
\subsection{Signal model and the AQS problem}

In AC sensing, a quantum sensor acquires information about a time-dependent classical field through a linear coupling. Any such field $B(t)$ can be written in a basis of $N$ frequencies $0 < \omega_1 < \cdots < \omega_N$, as
\begin{equation}
    B(t) = \sum_{j=1}^N \left(p_j \cos(\omega_j t) + q_j \sin(\omega_j t)\right),
\end{equation}
where the unknown coefficients $p_j$ and $q_j$ are the two quadrature amplitudes at frequency $\omega_j$. We define the minimum spacing $\Delta_\omega = \min\left\{\omega_1,\min_{1\leq j<N}(\omega_{j+1}-\omega_j)\right\}$, and we assume that it is a constant strictly greater than zero. The amplitude of each frequency component is $(p_j^2 + q_j^2)^{1/2}$, but different frequency components may be incident on the sensor with different, unknown relative phases $\phi_j = \text{atan2}(q_j,p_j)$. Any signal is then characterized by a quadrature vector,
\begin{equation}
    A = (p_1, \ldots, p_N, q_1, \ldots, q_N)^T \in \mathbb{R}^{2N}, \qquad \|A\|_2 \leq 1 \ .
\end{equation}
The constraint $\|A\|_2\le1$ simply normalizes the overall amplitude scale of the signal, which is absorbed by the parameter $\kappa$ below. By writing the (known) oscillating components into 
\begin{equation}
    f(t) = \big(\!\cos(\omega_1 t), \ldots, \cos(\omega_N t), \sin(\omega_1 t), \ldots, \sin(\omega_N t)\big)^T,
\end{equation}
we can represent a general signal as $B_A(t)=A^T f(t)$. This is without loss of generality: if a frequency component is absent, its entries in $A$ are zero, but the frequency component remains in $f(t)$. 

AC sensing tasks, which learn properties of the signal, can thus be formalized as learning functionals of $A$. A simple example is reconstructing the spectrum itself, which requires learning dot products of $A$ against each of the $2N$ basis vectors $e_1,... e_{2N}$, and thereby learning $A_1,..., A_{2N}$; these dot products are \textit{linear functionals}. Many tasks learn the overlap between the given signal and template waveforms consisting of many frequencies, the problem of \textit{matched filtering}; in our language, such tasks provide vectors $K \in \mathbb R^{2N}$ describing each template waveform, and the overlap is precisely the linear functional $K^T A$.  

This description of AC sensing extends beyond linear functionals. For instance, learning the power spectrum of a signal involves nonlinear quantities $p_j^2+q_j^2$; this example can be written as a \textit{bilinear} functional $K_j\in\mathbb R^{2N\times 2N}$, such that $A^T K_j A$ is precisely $p_j^2 + q_j^2$. In this manner, we can write the most general observable by fixing a real-valued order-$L$ tensor $K$ and taking
\begin{equation}
F_K(A) \coloneqq K[A^{\otimes L}] \coloneqq K[\underbrace{A, \ldots, A}_{L\ {\rm copies}}] = \sum_{\mu_1, \ldots, \mu_L = 1}^{2N} K_{\mu_1 \cdots \mu_L} A_{\mu_1} \cdots A_{\mu_L}.
\label{eq:general_spectroscopic_observable}
\end{equation}
Only the symmetric part of $K$ contributes to $F_K(A)$, so without loss of generality we may assume $K$ to be symmetric. We measure the size of $K$ with the full multilinear norm
\begin{equation}
\|K\|_\infty \coloneqq \sup_{\|x_1\|_2, \ldots, \|x_L\|_2 \leq 1} |K[x_1, \ldots, x_L]|.
\label{eq:multilinear_norm}
\end{equation}
For $L=1$ this is the Euclidean norm of a vector, and for $L=2$ it is the operator norm of a symmetric matrix. For a list $\mathcal K=(K_1,\ldots,K_M)$, we define $\kappa\coloneqq\max_{j\le M}\|K_j\|_\infty$. Since $\|A\|_2\le1$, every observable in the list satisfies $|F_{K_j}(A)|\le\kappa$; the overall scale of the signal may be absorbed into $\kappa$.

The order $L$ determines which quantities can be expressed. $L=1$ captures linear AC sensing, the most common setting that includes waveform reconstruction, matched filtering, signal search, and lock-in detection. $L=2$ describes powers and correlations between frequency components, while higher-order tensors capture spectral resonances, non-Gaussianity, and polyspectra which can be used to study the noise environment of a spin system or patterns in a dense signal profile. We now formalize these concepts by defining the AQS problem central to this work.

\begin{graydefinitionbox}
\begin{definition}[Algorithmic quantum sensing problem]
    Fix the candidate frequencies $\omega_1,\ldots,\omega_N$, an integer $L\ge1$, a finite list $\mathcal K=(K_1,\ldots,K_M)$ of symmetric order-$L$ tensors, an accuracy $\epsilon>0$, and a failure probability $\delta\in(0,1)$. The Algorithmic Quantum Sensing problem $\textnormal{AQS}[\{\omega_j\}, \epsilon, \delta, \mathcal K]$ is to output classical estimates $\widehat F_1,\ldots,\widehat F_M$ such that, for every $A\in\mathbb R^{2N}$ with $\|A\|_2\le1$,
\begin{equation}
\Pr\!\left[ \max_{j \leq M} \bigl|\widehat F_j - F_{K_j}(A)\bigr| \leq \epsilon \right] \geq 1 -\delta.
\label{eq:algorithmic_spectroscopy_task}
\end{equation}
\end{definition}
\end{graydefinitionbox}

The probability in Eq.~\eqref{eq:algorithmic_spectroscopy_task} is over measurement outcomes and protocol randomness, while the vector $A$ and chosen observables are fixed in each AQS instance. The task is trivial if $\max_j \|K_j\|_\infty = \kappa\le\epsilon$, since returning zero is already $\epsilon$-accurate for each estimate, so we always take $\kappa>\epsilon$. Since $\kappa$ is a rescaling of the problem, one can equivalently take $\kappa = 1$ and replace $\epsilon$ with $\epsilon/\kappa$; this is why $\kappa/\epsilon$ appears jointly throughout our work. 

\subsection{Conventional and quantum computational sensing}
Throughout this work we consider two physical access models for AC sensing. In the \textit{conventional} model, the sensor can be initialized in any single-qubit state and evolved under a single-qubit Hamiltonian of the form 
\begin{equation}
H_A(t) = \frac{1}{2} B_A(t) Z + H_c(t),
\label{eq:general_sensing_hamiltonian}
\end{equation}
where $B_A(t) = A^T f(t)$ is a general $N$-mode signal, and $H_c(t)$ is a known control Hamiltonian chosen by the experimentalist. The sensor is then measured in any single-qubit basis, and subsequent shots can be tailored using this measurement information. Decoherence is incorporated by introducing a dissipative term describing the noise, such that the sensor's evolution is described by a Lindblad master equation whose Hamiltonian term is $H_A$. Importantly, no ancillary quantum memory resources are persistent between shots.

This model includes a wide range of conventional sensing approaches. For example, $H_c(t)$ may describe the pulses in Rabi spectroscopy, Ramsey interferometry, or dynamical-decoupling experiment. Conventional pulse design aims to choose $H_c$ to maximize measurement sensitivity to chosen properties of $B_A(t)$. We assume pulses to be instantaneous, which for many platforms is a good approximation; for instance, pulse durations in solid-state spins can be three or more orders of magnitude shorter than coherence times, allowing hundreds or thousands of pulses per interrogation. Many platforms also admit global control of many-spin ensembles; this can be thought of as parallel interrogations under $H_A(t)$. 

Our second model allows \textit{quantum computational} sensing, where the controls of Hamiltonian~\ref{eq:general_sensing_hamiltonian} remain available, but an ancillary quantum memory can be coupled to the sensor and may remain coherent between resets of the sensor. Our lower bounds do not place any restriction on the size, coherence time, or control capabilities of the memory system, and assume all memory operations to be instantaneous. However, our quantum computational sensing protocols use memories whose size is at most linear (up to polylogarithmic factors) in the bandwidth of the given signal, and can provide metrological gain even when the memory has a lifetime that is not infinite but is sufficiently longer than the sensor's own coherence time (see e.g. Fig.~\ref{fig:exp}). 

In both settings the goal is to design, given an AC sensing problem $\textnormal{AQS}[\{\omega_j\}, \epsilon, \delta, \mathcal K]$, a protocol within the admissible class that solves the problem in the minimum possible interrogation time. More precisely, if a sensing protocol makes a total of $S$ measurements, each interrogating the signal for a time $\tau_s, s\in \{1,2,...,S\}$, we wish to minimize the total time $T_{\rm int} =\sum_{s=1}^S \tau_s$. The signal $B(t)$ continues to oscillate between shots, i.e. we do not assume that each shot begins with automatic access to $B(0)$. This raises a distinction that will be important when we discuss matched spectroscopy: under the above definitions, a protocol can choose to pause between measurements without interrogating the signal and wait for a favorable interrogation start time. Neglecting such waiting time can result in misleading definitions of optimality, since a protocol may spend much longer time waiting than it spends sensing, achieving a small interrogation time on paper but a large total time in practice. We thus distinguish between the total interrogation time, $T_{\rm int}$, and the total \textit{wall-clock} time, $T_{\rm wall}$, which charges both waiting and sensing time. Readout is assumed instantaneous in both, but we will discuss measurement cost case-by-case. For randomized, pooled, and standard quantum computational spectroscopy, $T_{\rm int}\simeq T_{\rm wall}$ up to small constants, so we will not explicitly discriminate between the two. For conventional and quantum computational \textit{matched} spectroscopy, we can have $T_{\rm int} \leq o(T_{\rm wall})$, and in these cases we will explicitly state both timescales. 

We note that some experimental applications may indeed provide the ability to restart the signal at an effective time $t=0$ between shots, such as when performing spectroscopy of a repeatedly driven solution or quantum system. In such settings interrogation time and wall clock time will generally coincide, even for matched spectroscopy.

Finally, under both conventional and quantum computational models, we develop protocols for two broad kinds of experimental tasks. In the first, we consider tasks which require estimating rich information about the entire signal and reconstructing a large number of observables; this includes problems like waveform reconstruction, multifrequency lock-in detection, power spectroscopy, and broadband search. For such tasks, we develop protocols that require no advance knowledge of the observables to be estimated, instead producing a \textit{reusable} dataset from which many quantities can be estimated downstream in postprocessing. This is possible without first reconstructing the full waveform to high accuracy, achieving a gain analogous to how classical shadow tomography estimates many quantum observables without performing full tomography~\cite{Huang_2020_classical}; similarly, reusable strategies have the practical benefit that the one dataset can be stored and used later without repeating the experiment. In the second setting, we consider matched filtering problems in which we are given advance knowledge of a few observables we wish to estimate, which is the setting of many detection tasks. Here we develop \textit{adaptive} protocols which tailor their measurements to the desired observables while deliberately suppressing sensitivity to the rest of the signal, trading reusability for increased efficiency. 

The classification, reusability, asymptotic scaling, and lower bounds for each protocol in our work are summarized in the Table~\ref{tab:aqs_protocol_comparison}. 

\begin{table*}[h!]
\centering
\small
\setlength{\tabcolsep}{3pt}
\renewcommand{\arraystretch}{1.3}
\setlength{\aboverulesep}{0pt}
\setlength{\belowrulesep}{0pt}
\renewcommand{\tabularxcolumn}[1]{m{#1}}

\begin{tabularx}{\textwidth}{@{}
 >{\raggedright\arraybackslash\vspace{3pt}}m{.115\textwidth}<{\endgraf\vspace{3pt}}|
 >{\centering\arraybackslash\vspace{3pt}}m{.17\textwidth}<{\endgraf\vspace{3pt}}
 >{\centering\arraybackslash\vspace{3pt}}m{.09\textwidth}<{\endgraf\vspace{3pt}}
 >{\centering\arraybackslash\vspace{3pt}}m{.18\textwidth}<{\endgraf\vspace{3pt}}
 >{\centering\arraybackslash\vspace{3pt}}m{.20\textwidth}<{\endgraf\vspace{3pt}}
 >{\centering\arraybackslash\vspace{3pt}}X<{\endgraf\vspace{3pt}}@{}}
\toprule
\textbf{Protocol} & \makecell*[c]{\textbf{Access}\\\textbf{model}} & \makecell*[c]{\textbf{Reusable}\\\textbf{vs.}\\\textbf{Adaptive}} & \makecell*[c]{\textbf{Interrogation}\\\textbf{time}} & \makecell*[c]{\textbf{Wall-clock}\\\textbf{time}} & \makecell*[c]{\textbf{Optimal}\\\textbf{Scaling}} \\
\midrule
Randomized spectroscopy & conventional (commuting $H_c$) & reusable &
\makecell*[c]{$\widetilde O\bigl(\min\{\sqrt N\,z^2\log M,$\\$N z\sqrt{\log M}\}\bigr)$} &
same as $T_{\rm int}$ & ACHL (Thm.~\ref{thm:nonadaptive_posthoc_lower_bound}) \\
\midrule
Matched spectroscopy & conventional (commuting $H_c$) & adaptive &
$\widetilde O\bigl(\min\{N^{1/3}r^{2/3}z^{4/3},$\newline$\sqrt{Nr}\,z\}\bigr)$ &
First branch $\widetilde O(Nz^2)$,\newline second $\widetilde O(\sqrt{Nr}\,z)$ &
Matched HL (Thm.~\ref{thm:adaptive_conventional_linear_lb}), \ Wall-clock HL and SQL (Thm.~\ref{thm:matched_wall_clock})\\
\midrule
Pooled spectroscopy & conventional \mbox{(noncommuting $H_c$)} & adaptive &
Precision-dependent; given in Thm.~\ref{thm:compressed_power_complexity} &
same as $T_{\rm int}$ & Bound given against commuting $H_c$ (Thm.~\ref{thm:power_discrimination_lower_bound}) \\
\midrule
Standard QCS & Quantum computational & reusable &
$\widetilde O\bigl(\sqrt N\,z\log M\bigr)$ &
same as $T_{\rm int}$ & Generalized HL (Thm.~\ref{thm:all_memory_lower_bound}) \\
\midrule
Matched QCS & Quantum computational & adaptive & $\widetilde O(\sqrt{r}z)$ & $\widetilde O(\sqrt{N}z)$ & Wall-clock HL (Thm.~\ref{thm:matched_wall_clock})\\
\midrule
Qubit phase QCS & Quantum computational & reusable &
$\widetilde O\bigl(\sqrt N\,z\log M\bigr)$ &
same as $T_{\rm int}$ & Generalized HL (Thm.~\ref{thm:all_memory_lower_bound}) \\
\bottomrule
\end{tabularx}
\caption{\textit{Overview of all protocols}. We write $z=\kappa/\epsilon$. $M$ denotes the total number of observables, while $r$ (in matched spectroscopy and matched QCS) is the dimension of the subspace spanned by the $M$ observables, so $r\leq M$. Displayed upper bounds are all for $L=1$ (except $L=2$ for pooled spectroscopy); tight $L$-dependent bounds are proven in subsequent appendices. Dependence on $\delta$ is suppressed and is always logarithmic in $\delta^{-1}$.}
\label{tab:aqs_protocol_comparison}
\end{table*}

\newpage
\section{Algorithmic toolbox}\label{sec:spec_toolbox} 
In this section, we develop a toolbox of algorithmic subroutines for quantum AC sensing. These protocols will be used repeatedly in our algorithms, and in subsequent appendices we will put them together and prove time complexity bounds for the composite protocols.

A core component of single-spin spectroscopy is the engineering of \textit{pulse sequences}, which are specially timed pulses inducing rotations of the sensor Bloch vector around chosen Pauli axes. These pulses  filter an electromagnetic field in the frequency domain to enable coherent accumulation of phase, preventing phase cancellation that otherwise occurs due to the signal's periodicity. The basic principle can be seen as follows. Given the Hamiltonian $H(t) = \frac{B(t)}{2}Z + H_c(t)$, select $H_c(t)$ to be a sequence of instantaneous $X$ operators at $m$ chosen times, producing the pulse sequence Hamiltonian
\begin{equation}\label{eq:pulse_ham}
    H(t) = \frac{B(t)}{2}Z + \frac{\pi}{2}\sum_{j = 1}^m \delta(t-t_j)X \ . 
\end{equation}
The factor of $\pi/2$ is chosen so that $e^{iX\pi/2 }\propto X$. We then move into the \textit{toggling frame}, the time-dependent Pauli frame obtained by conjugating the sign of a $Z$ operator according to $X^{n(t)}ZX^{n(t)}$, where $n(t)$ denotes the number of pulses that have been applied before time $t$. This can also be understood as the interaction picture relative to the Hamiltonian $H_0(t) = \frac{\pi}{2}\sum_{j = 1}^m \delta(t-t_j)X$. In this frame, the Hamiltonian \eqref{eq:pulse_ham} becomes
\begin{equation} \label{eq:Commuting_interaction_pic_ham}
    \Tilde{H}(t) = \frac{1}{2}B(t)y(t)Z \ , \qquad y(t) = (-1)^{n(t)} \ .
\end{equation}
More generally, the interaction picture Hamiltonian will always take this form if $H_c(t)$ is supported on a single Pauli axis, with different modulations $y(t)$. Notably, the interaction Hamiltonian commutes with itself at all times, since it now consists of only a $Z$-term, so its propagator is
\begin{equation}
    U_H(t, t') = \exp\left(-\frac{iZ}{2}\int_t^{t'} y(s)B(s)\,ds\right) \ .\label{eq:commuting_propagator}
\end{equation}
Substituting the mode expansion $B(t) = \sum_{j=1}^N p_j \cos(\omega_j t) + q_j\sin(\omega_j t)$, and letting $A = (p_1, ..., p_N, q_1,...,q_N)^T$, we can write
\begin{equation}
    \int_{t_0}^{t_1} y(t)B(t)\,dt = v^TA, \qquad
    v_j =
    \begin{cases}
\int_{t_0}^{t_1}\cos(\omega_j t)y(t)\,dt,
        & j\in\{1,\ldots,N\},\\[0.5em]
        \int_{t_0}^{t_1}\sin(\omega_{j-N}t)y(t)\,dt,
        & j\in\{N+1,\ldots,2N\}.
    \end{cases}
\end{equation}
If the sensor is initially prepared in $\ket{+}$ and evolved for total time $T$ through the pulse sequence, the final expectation value of an $X$-basis measurement is $\langle X\rangle = \cos(v^T A)$. The goal of linear spectroscopy is to estimate entries or linear combinations of entries of $A$, so the goal is to design $y(t)$ that realizes vectors $v$ aligned with the entries of interest; then $X$ (or $Y$) basis measurements confer information about the desired spectroscopic observables.  

\subsection{Efficient pulse sequence compilers}\label{subsec:pulse_compile}

The response-vector formalism above reduces pulse-sequence design to an inverse problem: given a desired response $v$, construct a bounded modulation $y(t)$ that realizes it with the minimum sensing time. For many standard spectroscopy protocols this inverse problem is largely hidden, because the desired response is concentrated near one or a few spectral modes. For example, conventional Ramsey spectroscopy uses $y(t) = 1$, targeting the static or near-zero-frequency component of the field. The algorithms we develop below require more general choices of $v$, including Gaussian-random response vectors, for which the inverse problem of constructing a suitable $y(t)$ is less immediate. We note that these more complex choices of $v$ need not imply more complicated pulse control; a complicated choice of $v$ may be realizable by very simple pulse sequences. 

This synthesis problem is analogous to the inverse problem of finite-impulse-response (FIR) filter design~\cite{rabiner_fir_1975}. In FIR design, one prescribes the desired frequency response at selected frequencies and solves for a finite-duration set of filter coefficients that realizes or approximates it. Here $v$ specifies the desired in-phase and quadrature components of the frequency response at a known, possibly nonuniform set of coherent spectral modes. We provide an efficient classical algorithm to construct a bounded, finite-duration modulation $y(t)$, and subsequently a binary modulation $y_{\rm p}(t)$ whose frequency response matches $v$, with explicit guarantees on interrogation time and synthesis error.  

The algorithm will first compute a continuous function $y(t)$ that can take on any value in $[-1/2, 1/2]$ at each point in time. We then use a subroutine to convert this continuous modulation into an experimentally implementable sign modulation composed of $\pi$-pulses by using a pulse-width-modulation construction~\cite{holmes_pulse_2003}. In each time cell, the duty cycle of $y_{\rm p}(t) \in \{-1,+1\}$ is chosen to reproduce the cell average of $y(t)$ exactly. Related pulse-width-modulation approximations converting continuous controls into these bang--bang sequences have also been considered in quantum control~\cite{chen_pulse_2017}. 

\begin{algorithm}
    \DontPrintSemicolon
    \caption{Discrete compilation of a smooth modulation}
    \label{algo:discretize_response}
    \KwIn{A known function $y(t)$ on $I=[t_0,t_0+T]$ satisfying $|y(t)|\leq 1/2$, and response accuracy $\epsilon_{\rm dis}>0$.}
    \KwOut{A sign modulation $y_{\rm p}(t)\in\{-1,+1\}$ satisfying
    \begin{equation}
        \left\|
        \int_I \bigl(y_{\rm p}(t)-y(t)\bigr)f(t)dt
        \right\|_2
        \leq \epsilon_{\rm dis}.
    \end{equation}}

    Divide $I$ into
    \begin{equation}
        Q=\max\left\{1,\left\lceil \frac{3\omega_N\sqrt{N}T^2}{8\epsilon_{\rm dis}}\right\rceil\right\}
    \end{equation}
    equal cells of width $h=T/Q$.\;

    In each cell $C=[a,a+h]$, calculate the average modulation
    \begin{equation}
        \overline y_C=\frac{1}{h}\int_C y(t)dt
    \end{equation}
    and define $u_C=a+h(1+\overline y_C)/2$. Set $y_{\rm p}(t)=+1$ on $[a,u_C)$ and $y_{\rm p}(t)=-1$ on $[u_C,a+h)$.\;

    Return the resulting sign modulation, implemented by an instantaneous $X$ pulse at every sign change and, if necessary, a final $X$ pulse at $t_0+T$ to restore the original pulse frame.\;
\end{algorithm}
The proof of the algorithmic guarantee for this subroutine is as follows.
\begin{proof}
    In each cell $C=[a,a+h]$, observe that the time at which $y_p(t)$ flips signs is chosen such that $\int_C \bigl(y_{\rm p}(t)-y(t)\bigr)dt=0$. Let $t_C = a+h/2$ denote the center of the cell. It follows that
    \begin{equation}
        \int_C \bigl(y_{\rm p}(t)-y(t)\bigr)f(t)dt
        =
        \int_C \bigl(y_{\rm p}(t)-y(t)\bigr)\bigl(f(t)-f(t_C)\bigr)dt\ ,
    \end{equation}
    because subtracting any constant from $f(t)$ will leave the integral unchanged. Since $|y(t)| \leq 1/2$ and $|y_p(t)| = 1$, it follows that $|y_{\rm p}(t)-y(t)|\leq 3/2$. Moreover, $\|f'(t)\|_2^2 = \sum_j \omega_j^2 (\sin(\omega_jt)^2 + \cos(\omega_jt)^2) \leq N\omega_N^2$, so $\sup_t\|f'(t)\|_2\leq \omega_N\sqrt{N}$. Thus by triangle inequality,
    \begin{equation}
         \left\|
        \int_C \bigl(y_{\rm p}(t)-y(t)\bigr)f(t)dt
        \right\|_2
        \leq 
        \int_C \left|y_{\rm p}(t)-y(t)\right|\|f(t) - f(t_C)\|_2 \ dt
         \leq \frac{3\sqrt{N}\omega_N}{2}\int_C |t-t_C| dt
    \end{equation}
    Computing the integral gives $\int_C |t-t_C| dt = 2\int_0^{h/2} t \ dt = h^2/4$, so the approximation error in one time bin is at most $\frac{3}{8}\omega_N\sqrt{N}h^2$.
    Summing over all $Q$ time bins gives
    \begin{equation}
        \left\|
        \int_I \bigl(y_{\rm p}(t)-y(t)\bigr)f(t)dt
        \right\|_2
        \leq \frac{3\omega_N\sqrt{N}T^2}{8Q}
        \leq \epsilon_{\rm dis}.
    \end{equation}
\end{proof}
We now present the pulse sequence compiler and proof of its guarantee. 
\begin{algorithm}
    \DontPrintSemicolon
    \caption{Compile a pulse sequence with a specified response}
    \label{algo:deterministic_response_compiler}
    \KwIn{Known frequencies $\omega_1<\cdots<\omega_N$ satisfying $\omega_1>\Delta_\omega$ and $\omega_{j+1}-\omega_j\geq\Delta_\omega$, a response vector $v\in\mathbb{R}^{2N}$, accuracy $\epsilon_{\rm syn}>0$, and a starting time $t_0$.}
    \KwOut{A pulse sequence defined by $y_{\rm p}(t)\in\{-1,+1\}$ on an interval $I=[t_0,t_0+T]$ satisfying
    \begin{equation}
        \left\|
        \int_I y_{\rm p}(t)f(t)dt-v
        \right\|_2
        \leq \epsilon_{\rm syn}.
    \end{equation}}

    Choose
    \begin{equation}
        T=C_{\rm comp}\left(\|v\|_1+\Delta_\omega^{-1}\right),
    \end{equation}
    where $C_{\rm comp}$ is a sufficiently large numerical constant. Let $t_c=t_0+T/2$ and define $w(t)=1-\frac{2|t-t_c|}{T}$ for all $t\in I$.
    
    Form the weighted Gram matrix
    \begin{equation}
        G=\int_I w(t)f(t)f(t)^Tdt
    \end{equation}
    and solve $Gc=v$ for $c$. Define
    \begin{equation}
        y(t)=w(t)f(t)^Tc.
    \end{equation}

    Run Algorithm~\ref{algo:discretize_response} on $y(t)$ over $I$ with precision $\epsilon_{\rm syn}$, and return the resulting sign modulation $y_{\rm p}$.\;
\end{algorithm}
\begin{proof}
    The goal is to show that with the specified interrogation time, the weighted Gram matrix $G$ is always well-conditioned and invertible; it is then simple to show that the resulting $y(t)$ satisfies the desiderata. Write $t=t_c+s$, where $t_c=t_0+T/2$ and $s\in[-T/2,T/2]$. Then we can rewrite the triangular weight function $w(t)$ as $ w_T(s)=1-\frac{2|s|}{T}$, which is the standard Bartlett window \cite{harris_use_1978}. Its Fourier transform is
    \begin{equation}
        W_T(\nu)=\int_{-T/2}^{T/2} w_T(s)e^{i\nu s}ds
        =\frac{T}{2}\operatorname{sinc}^2\left(\frac{\nu T}{4}\right),
    \end{equation}
    and therefore $|W_T(\nu)|\leq CT/(1+(\nu T)^2)$.

    We now bound $G=\int_I w(t)f(t)f(t)^Tdt$.  The entries of $G$ are integrals of products of sines and cosines against $w(t)$. By the standard product-to-sum identities, these are linear combinations of Fourier coefficients $W_T(\omega_j-\omega_k)$ and $W_T(\omega_j+\omega_k)$. Each diagonal entry picks up an additive $T/4$ factor due to the cancelling frequencies. All remaining terms correspond to nonzero differences between frequencies in the set $        \{-\omega_N,\ldots,-\omega_1,\omega_1,\ldots,\omega_N\}$. By assumption, neighboring positive frequencies differ by at least $\Delta_\omega$, while $\omega_1>\Delta_\omega$, so this entire frequency set is separated by at least $\Delta_\omega$.   We may therefore write
    \begin{equation}
        G=\frac{T}{4}\mathds{1}+E,
    \end{equation}
    where $E$ contains all of the nonzero-frequency terms. For any fixed row of $E$, the $r$th closest nonzero frequency difference therefore has magnitude at least $r\Delta_\omega$. Using the bound on $W_T$ gives
    \begin{equation}
        \|E\|_{1\rightarrow 1}
        \leq CT\sum_{r=1}^\infty \frac{1}{1+(r\Delta_\omega T)^2}
        \leq \frac{C}{\Delta_\omega^2T}.
    \end{equation}
    Hence, when $T\geq C\Delta_\omega^{-1}$, we obtain $\|G^{-1}\|_{1\rightarrow 1}\leq C/T$. Since $c=G^{-1}v$, we obtain $\|c\|_1\leq C\|v\|_1/T$. Moreover, $0\leq w(t)\leq1$ and every entry of $f(t)$ has magnitude at most one, so
    \begin{equation}
        |y(t)|=|w(t)f(t)^Tc|
        \leq \|c\|_1
        \leq \frac{C\|v\|_1}{T}.
    \end{equation}
    Choosing $C_{\rm comp}$ sufficiently large in $T=C_{\rm comp}(\|v\|_1+\Delta_\omega^{-1})$ therefore guarantees both that $G$ is well-conditioned and invertible, and that $|y(t)|\leq1/2$. Finally, the definition of $G$ gives
    \begin{equation}
        \int_I y(t)f(t)dt
        =
        \int_I w(t)f(t)f(t)^Tc\,dt
        =
        Gc
        =
        v.
    \end{equation}
    Thus the continuous modulation has exactly the desired response. Algorithm~\ref{algo:discretize_response} changes this response by at most $\epsilon_{\rm syn}$, which proves
    \begin{equation}
        \left\|
        \int_I y_{\rm p}(t)f(t)dt-v
        \right\|_2
        \leq \epsilon_{\rm syn}.
    \end{equation}
\end{proof}
In practice, we generally require vectors $v$ whose entries have $O(1)$ magnitude, as opposed to magnitude growing with $N$, since we want to extract the mode amplitudes or basic linear combinations. Our compilation algorithm therefore allows us to synthesize linear response vectors with interrogation time no worse than $O(N)$, and as small as $O(1)$ when we want to target specific frequencies.

\subsection{High-gain phase readout}\label{sec:high_gain_readout}
The pulse sequence compiler above in Section~\ref{subsec:pulse_compile} allows us to engineer a prescribed linear phase $v^TA$. Next we describe how to extract this quantity efficiently from the sensor. Preparing the qubit in $\ket{+}$ and measuring in the $X$ or $Y$ basis gives us $\langle X \rangle  = \cos(v^TA)$ or $\langle Y \rangle  = \sin(v^TA)$, so the accumulated phase is only observed modulo $2 \pi$. This periodicity is unimportant when $v^TA$ is known a priori to lie in a sufficiently small interval, but becomes essential when we deliberately amplify a desired response direction $u$ by taking $v=Hu$. We refer to $H$ as the \emph{gain}: increasing $H$ causes the same linear observable $u^TA$ to accumulate a proportionally larger phase $Hu^TA$, thereby increasing the sensitivity of the measurement.

The overall strategy is simple. We first use measurements at successively increasing gains to obtain a coarse estimate of the phase, $a_0$, that determines the sinusoidal branch on which the desired phase lies. We then use additional $X$ and $Y$ measurements at a large gain to construct bounded random variables whose mean is the sine of the residual phase. Finally, because the residual is known to lie on an invertible branch of the sine function, we approximately invert this response and add the resulting correction back to $a_0$. These three readout steps resolve the $2 \pi$ ambiguity, isolate the small residual error, and estimate that residual with accuracy set by the largest gain, recovering Heisenberg-limited estimation of the desired phase up to polylogarithmic overhead. 

The first step is closely related to standard classical phase-unwrapping methods which reconstruct a continuous phase by selecting consistent $2 \pi$-lifts of wrapped phase measurements~\cite{tribolet_new_1977}, and multiscale quantum phase-estimation protocols which similarly use measurements at successively increasing phase gains to resolve the branch ambiguity while retaining Heisenberg-limited sensitivity~\cite{higgins_demonstrating_2009, kimmel_robust_2015}. Algorithm~\ref{algo:phase_unwrapping} adapts this multiscale phase-estimation to our spectroscopy setting, where the amplified phases are generated by compiled linear responses of the unknown spectral vector.

\begin{algorithm}
    \DontPrintSemicolon
    \caption{Phase unwrapping}
    \label{algo:phase_unwrapping}
    \KwIn{Known frequencies $\{\omega_j\}_{j=1}^N$, integer $J$, pulse compilation accuracy $\epsilon_{\rm syn}$ and failure probability $\delta$.}
    \KwOut{A vector $a_0\in\mathbb{R}^{2N}$ satisfying $\|a_0-A\|_\infty\leq 1/(2^{J+1})$ except with probability at most $\delta/4$.}

    Let $h_j=2^j/4$ for $j=0,\ldots,J$, and set
    $r_{\rm p}=\left\lceil 512\log\!\left(32N(J+1)/\delta\right)\right\rceil$.\;

    For every coordinate $\mu=1,\ldots,2N$ and every $j=0,\ldots,J$, run Algorithm \ref{algo:deterministic_response_compiler} with precision $\epsilon_{\rm syn}$ and $v = h_je_\mu$. Take $r_{\rm p}$ $X$ measurements and $r_{\rm p}$ $Y$ measurements. Let $\overline X_{\mu,j}$ and $\overline Y_{\mu,j}$ denote their empirical means, and set
    $\widetilde\phi_{\mu,j}=\mathrm{Arg}(\overline X_{\mu,j}+i\overline Y_{\mu,j})\in(-\pi,\pi]$.\;

    For each $\mu$, set $\widehat\phi_{\mu,0}=\widetilde\phi_{\mu,0}$. For $j=1,\ldots,J$, choose the integer $k_{\mu,j}$ for which
    $\widetilde\phi_{\mu,j}+2\pi k_{\mu,j}$ is closest to $2\widehat\phi_{\mu,j-1}$, and set
    $\widehat\phi_{\mu,j}=\widetilde\phi_{\mu,j}+2\pi k_{\mu,j}$.\;

    Return $(a_0)_\mu=4\widehat\phi_{\mu,J}/2^J$ for $\mu=1,\ldots,2N$.\;
\end{algorithm}
Algorithm \ref{algo:high_gain_residual_sampling} then estimates the residual error in the coarse estimate $a_0$ of $A$. The $X$ and $Y$ measurement outcomes provide the sine and cosine quadratures of the accumulated phase, which are combined with the known reference phase to produce bounded scores. These scores isolate the small residual error and will be used to remove the bias of the final estimator.
\begin{algorithm}
    \DontPrintSemicolon
    \caption{Residual sampling}
    \label{algo:high_gain_residual_sampling}
    \KwIn{Known frequencies $\{\omega_j\}_{j=1}^N$, gain $H$, reference vector $a_0$, number of samples $R$, and pulse compilation accuracy $\epsilon_{\rm syn}$.}
    \KwOut{For each coordinate $\mu=1,\ldots,2N$, independent random variables $W_{\mu,1},\ldots,W_{\mu,R}\in[-2,2]$ satisfying
    \begin{equation}
        \E[W_{\mu,r}\mid a_0]=\sin\!\left(H(A_\mu-(a_0)_\mu)\right).
    \end{equation}}

    For every $\mu=1,\ldots,2N$ and $r=1,\ldots,R$, run Algorithm \ref{algo:deterministic_response_compiler} with precision $\epsilon_{\rm syn}$ and $v = He_\mu$, run the resulting pulse sequence while interrogating the signal, choose the measurement independently and uniformly from $\{X,Y\}$, and record the outcome $B_{\mu,r}\in\{-1,+1\}$.\;

    After $a_0$ has been obtained, set
    \[
        W_{\mu,r}
        =
        \begin{cases}
            -2B_{\mu,r}\sin\left(H(a_0)_\mu\right), & X\text{ measured},\\
            2B_{\mu,r}\cos\left(H(a_0)_\mu\right), & Y\text{ measured}.
        \end{cases}
    \]

    Return the variables $W_{\mu,r}$.\;
\end{algorithm}

Finally, we must recover the residual phase from the sinusoidal expectation values produced by Algorithm~\ref{algo:high_gain_residual_sampling}. We do so using the bounded sine-inversion procedure in Algorithm~\ref{algo:sine_inversion}. The preceding phase-unwrapping step guarantees that the residual phase lies within a known interval on which the sine function is one-to-one. Thus, estimating the inverse sine of the residual score recovers the desired correction without any remaining phase-wrap ambiguity.
\begin{algorithm}[h!]
    \DontPrintSemicolon
    \caption{Bounded sine inversion}
    \label{algo:sine_inversion}
    \KwIn{Recorded values $W_1,\ldots,W_n\in[-2,2]$ from independent, identically distributed scores with unknown mean $s$ satisfying $|s|\leq1/8$, and accuracy $\epsilon_{\rm inv}\in(0,1/2]$. Assume $n\geq C\log^2(2/\epsilon_{\rm inv})$ for a sufficiently large numerical constant $C$.}
    \KwOut{A scalar $S$ satisfying $|S|\leq\pi/6$ for every input realization and $|\E[S]-\arcsin(s)|\leq\epsilon_{\rm inv}$, where the expectation is over the input data.}

    Set $P=\left\lfloor(\sqrt n-1)/2\right\rfloor$ and $q=\left\lfloor n/(2P+1)\right\rfloor$. Divide the first $q(2P+1)$ values into disjoint groups $\mathcal I_1,\ldots,\mathcal I_{2P+1}$ of size $q$.\;

    For each group $r$, calculate
    \begin{equation}
        Q_r=\operatorname{clip}\!\left(\frac{1}{q}\sum_{j\in\mathcal I_r}W_j,\left[-\frac12,\frac12\right]\right).
    \end{equation}

    Return
    \begin{equation}
        S=\sum_{k=0}^P\frac{\binom{2k}{k}}{4^k(2k+1)}\prod_{r=1}^{2k+1}Q_r.
    \end{equation}
\end{algorithm}

\subsection{Improved Gaussian random pulse compilers}
While the pulse sequence compilers in Section~\ref{subsec:pulse_compile} are useful for deterministic linear responses, we now present a key refinement that will allow us to design an optimal single-qubit AC sensing algorithm for reconstructing many observables with the same dataset. Naively, if we drew $v\sim \mathcal{N}(0, \mathds{1}_{2N})$, a typical vector would have $\ell_1$-norm $N$. Despite this $\ell_1$ scaling, we now show that we can realize Gaussian-random pulse sequence responses using only $O(\sqrt{N})$ interrogation time. The origin of the improvement lies in replacing the worst-case $\ell_1$ bound on a deterministic vector with a probabilistic bound that leverages Gaussianity and is thereby controlled by typical, rather than worst-case, magnitude. This improvement propagates into our randomized and pooled spectroscopy protocols, enabling substantial improvements over conventional frequency-selective approaches.

\begin{algorithm}
    \DontPrintSemicolon
    \caption{Compile a pulse sequence with a Gaussian random response}
    \label{algo:gaussian_response_compiler}
    \KwIn{Known frequencies $\omega_1<\cdots<\omega_N$, accuracy $\epsilon_{\rm syn}>0$, failure probability $\delta_{\rm syn}>0$, and a starting time $t_0$.}
    \KwOut{A sampled vector $g\sim\mathcal{N}(0,\mathds{1}_{2N})$ and a pulse sequence $y_{\rm p}(t)\in\{-1,+1\}$ satisfying
    \begin{equation}
        \left\|
        \int_I y_{\rm p}(t)f(t)dt-g
        \right\|_2
        \leq \epsilon_{\rm syn},
    \end{equation}
    or FAIL with probability at most $\delta_{\rm syn}$.}

    For a sufficiently large numerical constant $C_G$, choose
    \begin{equation}
        T=C_{\rm G}\left(\Delta_\omega^{-1}+\sqrt{N\log\left(\frac{CN}{\delta_{\rm syn}}\right)}\right),
    \end{equation}
    let $I=[t_0,t_0+T]$, and define the $w(t)$ as in Algorithm~\ref{algo:deterministic_response_compiler}. Form $G=\int_I w(t)f(t)f(t)^Tdt$.\;

    Sample $g\sim\mathcal{N}(0,\mathds{1}_{2N})$ and define $a_0(t)=f(t)^TG^{-1}g$. Let $a(t)$ be $a_0(t)$ clipped to the interval $[-1/4,1/4]$, and calculate the residual
    \begin{equation}
        r=\int_I w(t)f(t)\bigl(a_0(t)-a(t)\bigr)dt.
    \end{equation}

    Solve $Gc_r=r$. If $\sqrt{N}\|c_r\|_2>1/4$, return FAIL. Otherwise define
    \begin{equation}
        y(t)=w(t)\left(a(t)+f(t)^Tc_r\right).
    \end{equation}

    Run Algorithm~\ref{algo:discretize_response} on $y(t)$ with precision $\epsilon_{\rm syn}$, and return $g$ and the resulting pulse sequence $y_{\rm p}$.\;
\end{algorithm}
\begin{proof}[Proof of compilation guarantee]
    By the same argument as in Algorithm~\ref{algo:deterministic_response_compiler}, for $T\geq C\Delta_\omega^{-1}$ the weighted Gram matrix satisfies $\|G^{-1}\|_{\rm op}\leq C/T$ and $\|G\|_{\rm op}\leq CT$. Now note that at any fixed time $t$, the random variable
    \begin{equation}
        a_0(t)=f(t)^TG^{-1}g
    \end{equation}
    is Gaussian with variance
    \begin{equation}
        \mathbb{E}[a_0(t)^2]
        =
        f(t)^TG^{-2}f(t)
        \leq \frac{CN}{T^2},
    \end{equation}
    where we used $\|f(t)\|_2^2=N$. A standard Gaussian tail bound therefore gives
    \begin{equation}
        \mathbb{E}\left[\bigl(a_0(t)-a(t)\bigr)^2\right]
        \leq \frac{CN}{T^2}\exp\!\left(-\frac{cT^2}{N}\right).
    \end{equation}
    To bound the residual $r$ that remains after clipping $a_0(t)$, let $\xi(t)=a_0(t)-a(t)$. For any unit vector $u$, Cauchy--Schwarz gives
    \begin{align}
        |u^Tr|^2
        &=
        \left|\int_I w(t)u^Tf(t)\xi(t)dt\right|^2 \\
        &\leq
        \left(\int_I w(t)(u^Tf(t))^2dt\right)
        \left(\int_I w(t)\xi(t)^2dt\right) \\
        &\leq
        CT\int_I w(t)\xi(t)^2dt.
    \end{align}
    Taking the supremum over $u$ to turn the initial expression into $\|r\|_2^2$ and then taking the expectation gives
    \begin{equation}
        \mathbb{E}\|r\|_2^2
        \leq
        CN\exp\left(-\frac{cT^2}{N}\right)
    \end{equation}
    by the previous tail bound. Since $c_r=G^{-1}r$, a failure can occur only if $\|r\|_2\geq cT/\sqrt{N}$. Markov's inequality therefore gives
    \begin{equation}
        \Pr[\mathrm{FAIL}]
        \leq
        \frac{CN^2}{T^2}\exp\left(-\frac{cT^2}{N}\right)
        \leq \delta_{\rm syn}
    \end{equation}
    for a sufficiently large choice of $C_{\rm G}$. When the compiler succeeds (with probability $\geq 1-\delta_{\rm syn}$), $|a(t)|\leq1/4$ and $|f(t)^Tc_r|\leq\sqrt{N}\|c_r\|_2\leq1/4$, so $|y(t)|\leq1/2$. Moreover,
    \begin{align}
        \int_I y(t)f(t)dt
        &=
        \int_I w(t)f(t)a(t)dt+Gc_r \\
        &=
        (g-r)+r=g.
    \end{align}
    Thus the continuous modulation realizes the sampled Gaussian response exactly, and Algorithm~\ref{algo:discretize_response} changes the response by at most $\epsilon_{\rm syn}$.
\end{proof}

We also generalize the Gaussian-random pulse sequence compiler of Algorithm~\ref{algo:gaussian_response_compiler}. Whereas Algorithm~\ref{algo:gaussian_response_compiler} treats a full standard Gaussian response $g \sim \mathcal N(0, \mathds{1}_{2N})$, Lemma~\ref{lem:random_response_compiler} applies to a broader class of random response vectors supported on a $d$-dimensional coordinate subspace, assuming only spherical symmetry conditional on the norm and sub-Gaussian one-dimensional marginals. This will be relevant to pooled spectroscopy, described in Section~\ref{sec:pooled_spectroscopy}, where we are interested in quantities within a specified frequency bin and want to suppress frequencies outside it.

\begin{lemma}
\label{lem:random_response_compiler}
Let $v\in\mathbb R^{2N}$ be a random response vector supported on a known $d$-dimensional coordinate subspace of $\mathbb R^{2N}$, which we identify with $\mathbb R^d$. Supposing $\|v\|_2$ is sub-Gaussian,
\begin{equation}
    \E e^{t u^Tv}
    \leq \exp \left(\frac{\sigma^2 t^2}{2}\|u\|_2^2\right)
    \label{eq:subgaussian_vector}
\end{equation}
for every $u\in\mathbb R^{d}$ and $t\in\mathbb R$. Then given the vector $v$, a value $h>0$, synthesis accuracy $\epsilon_{\rm syn}>0$, and failure probability $\delta_{\rm syn}>0$, there is a classical pulse compiler which implements the response $hv$ to accuracy $\epsilon_{\rm syn}$, except with probability at most $\delta_{\rm syn}$ over the random choice of $v$, in interrogation time
\begin{equation}
    T = \widetilde O\!\left(\Delta_\omega^{-1} + h\sigma\sqrt{d\log\frac{CN}{\delta_{\rm syn}}}
    \right).
    \label{eq:random_response_compiler}
\end{equation}
\end{lemma}

\begin{proof}
    The procedure follows the Gaussian random-response compiler of Algorithm~\ref{algo:gaussian_response_compiler}, replacing Gaussian concentration by the assumed sub-Gaussian bound~\eqref{eq:subgaussian_vector}. Relabeling coordinates if necessary, we assume $v_j=0$ for $j > d$. As above, for $T\geq C\Delta_\omega^{-1}$ the weighted Gram matrix satisfies
    \begin{equation}
        \|G^{-1}\|_{\rm op}\leq \frac{C}{T},
        \qquad
        \|G^{-1}\|_{\infty\rightarrow\infty}\leq \frac{C}{T},
        \qquad
        \|G\|_{\rm op}\leq CT.
    \end{equation}
    The algorithm solves $Gc=hv$ and defines the unconstrained modulation $a_0(t)=f(t)^Tc = h\,f(t)^TG^{-1}v.$
    Then equation \eqref{eq:subgaussian_vector} implies
    \begin{equation}
        \E e^{\lambda a_0(t)}
        \leq
        \exp\left[
            \frac{\lambda^2h^2\sigma^2}{2}
            \left\|G^{-1}f(t)\right\|_2^2
        \right].
    \end{equation}
    Since every entry of $f(t)$ has magnitude at most one,
    \begin{equation}
        \left\|G^{-1}f(t)\right\|_2^2
        \leq
        d\,\|G^{-1}f(t)\|_\infty^2
        \leq
        \frac{Cd}{T^2}.
    \end{equation}
    Hence, \begin{equation}\label{eq:random-a0-tail}
        \Pr \bigl(|a_0(t)|>u\bigr) \leq 2 \exp \left(-\frac{cT^2u^2}{h^2\sigma^2d}\right).
    \end{equation}
    As above, we clip $a_0(t)$ to $[-1/4,1/4]$, and write $\xi(t)=a_0(t)-a(t)$. Integrating the sub-Gaussian tail gives
    \begin{equation}
        \E[\xi(t)^2]
        \leq
        \frac{Ch^2\sigma^2d}{T^2}
        \exp\left(
            -\frac{cT^2}{h^2\sigma^2d}
        \right).
    \end{equation}
    Repeating the residual calculation from the Gaussian compiler therefore gives
    \begin{equation}
        \E\|r\|_2^2
        \leq
        Ch^2\sigma^2d
        \exp\left(
            -\frac{cT^2}{h^2\sigma^2d}
        \right).
    \end{equation}
    The remainder of the proof then proceeds unchanged, and we obtain both precision $\epsilon_{\rm syn}$ and failure probability $\delta_{\rm syn}$ when $T$ is chosen as in the lemma statement.
\end{proof}

\subsection{Rank reduction and error cancellation for nonlinear spectroscopy} \label{sec:rank_red_error_cancellation} 

In addition to linear spectroscopy, we are interested in estimating nonlinear functionals of the signal. Recall that in the most general formulation of nonlinear spectroscopy of order $L$, our goal is to estimate the quantity $K[A^{\otimes L}]\coloneqq \sum_{j_1,...,j_L} K_{j_1,...,j_L}A_{j_1}...A_{j_L}$, where $K$ is a symmetric rank-$L$ tensor. Since the sequences of $X$-pulses considered in Section~\ref{subsec:pulse_compile} provide access to linear functionals of the form $v^T A$, a natural first approach is to estimate $A$ via linear spectroscopy, then contract $L$ copies of the resulting estimate for $A$ against the tensor $K$. 

At first glance, this naive approach seems to fail. As an illustrative example, consider bilinear spectroscopy, where $K$ is a $2N\times2N$ real matrix with operator norm $\|K\|_{\rm op} \leq \kappa$. Suppose we obtain a noisy estimate $\hat{A} = A + e$, and we wish to estimate $A^T K A$ via the naive strategy. It then follows that
\begin{equation}
    |\hat{A}^TK\hat A - A^TK A| = |2e^TKA + e^TKe| \leq 2\kappa\|e\|_2  + \kappa \|e\|^2_2 \ ,
\end{equation}
a worst-case bound on the resulting error. For this error to drop below $\epsilon$, we require $\|e\|_2, \|e\|_2^2\lesssim \epsilon/\kappa$. If we model $e$ as mean-zero noise with $\mathbb{E}[ee^T] = \sigma^2 \mathds{1}_{2n}$, so that a typical vector $e$ satisfies $\|e\|_2^2 \lesssim 2N\sigma^2$, we require $\sigma\lesssim \epsilon/(\kappa\sqrt{N})$: the error in our estimate of $A$ must now fall with $\sqrt{N}$, incurring much more severe overhead than basic linear spectroscopy. As the tensor rank increases, this worst-case overhead compounds. 

However, a simple modification removes this obstruction. Continuing with the quadratic example, consider running two \textit{independent} rounds of linear spectroscopy to obtain estimates $\hat{A}_1 = A + e_1, \hat{A}_2 = A+e_2$. Then, under the same mean-zero, diagonal-covariance model for $e_1, e_2$, the error in $\hat{A}_1^T K \hat{A}_2$ is
\begin{equation}
    \hat{A}_1^T K A_2 - A^T K A = e_1^T K A + A^T K e_2 + e_1^T K e_2 \ .
\end{equation}
Now the \textit{mean squared error} in this estimate is $2\kappa^2\sigma^2 + 2N\kappa^2 \sigma^4$; where the initial approach had factors of $2\kappa^2 N \sigma^2$ we have removed the $N$-dependence. Then in the high-precision regime $\epsilon/\kappa \lesssim 1/\sqrt{N}$, we find that we only require $\sigma \lesssim \epsilon/\kappa$ to achieve mean-squared precision $\epsilon$. In this regime, there is no overhead beyond the linear case. Moreover, we need only estimate $A$ a fixed number of times $L$, which is small compared with the number of spectral lines $N$. The core strategy is thus to use $L$ independent estimates from a linear spectroscopy algorithm followed by controlling the mean squared error in the final nonlinear observable. We now make the above argument quantitative and prove an error-cancellation lemma that we use repeatedly to lift our optimal linear spectroscopy algorithms to optimal nonlinear ones. This argument will also allow for bias in each estimate of $A$, which will be useful later.

\begin{lemma}[Error cancellation for nonlinear estimation]
\label{lem:independent_nonlinear_estimation}
Let $K$ be a symmetric rank-$L$ tensor satisfying $\|K\|_\infty \leq \kappa$, and let $A \in \mathbb R^{2N}$ satisfy $\|A\|_2 \le 1$. Suppose that $X_1,\ldots, X_L \in \mathbb R^{2N}$ are independent random vectors satisfying, 
\begin{equation}
    \left\|\mathbb E[X_i] - A\right\|_2 \leq b, \qquad \operatorname{Cov}(X_i)
\preceq
\sigma^2\mathds{1}_{2N}
\end{equation}
for every $i = 1,\ldots,L$. Then the estimator $\widehat{F}_K=K[X_1,\ldots,X_L]$ satisfies, 
\begin{equation}
    \left|
\mathbb E[\widehat{F}_K]-K[A^{{L}}]
\right|
\leq
L\kappa b(1+b)^{L-1},\qquad \operatorname{Var}[\widehat{F}_K]
\leq
\frac{\kappa^2}{2N}
\left[
\left((1+b)^2+2N\sigma^2\right)^L
-(1+b)^{2L}
\right].
\end{equation}
Consequently if each $X_i$ is unbiased for $A$, then $\widehat{F}_K$ is unbiased for $K[A^{{L}}]$, and hence, 
\begin{equation}\label{eq:nonlinear_est_MSE}
    \mathbb E\left[
\left|
\widehat{F}_K-K[A^{{L}}]
\right|^2
\right]
\leq
\frac{\kappa^2}{2N}
\left[
\left(1+2N\sigma^2\right)^L-1
\right].
\end{equation}
\end{lemma}

\begin{proof}
    Let $\mu_i=\mathbb E[X_i]$,  $b_i=\mu_i-A$, and $\xi_i=X_i-\mu_i$. Then $X_i=\mu_i+\xi_i=A+b_i+\xi_i$ with $\|b_i\|_2\leq b$, $\E[\xi_i]=0$, and $\E[\xi_i\xi_i^T] = \operatorname{Cov}(X_i)
\preceq
\sigma^2\mathds{1}_{2N}$. Independence implies $\E[\widehat{F}_K]=K[\mu_1,\ldots,\mu_L]$. To compare the estimator with the true $K[A^{\otimes L}]$, replace the $\mu_i$ by $A$ one at a time. This gives the exact telescoping identity
    \begin{equation}
        K[\mu_1,\ldots,\mu_L]-K[A^{\otimes L}]
        =
        \sum_{i=1}^L
        K[A^{\{i-1\}},b_i,\mu_{i+1},\ldots,\mu_L].
    \end{equation}
    Since $\|A\|_2\leq1$, $\|b_i\|_2\leq b$, and $\|\mu_i\|_2\leq1+b$, each term has magnitude at most $\kappa b(1+b)^{L-i}$. Therefore
    \begin{equation}
        \left|\E[\widehat{F}_K]-K[A^{\otimes L}]\right|
        \leq
        \kappa b\sum_{i=1}^L(1+b)^{L-i}
        \leq
        L\kappa b(1+b)^{L-1}.
    \end{equation}
    We next bound the variance. Expanding each $X_i=\mu_i+\xi_i$ gives
    \begin{equation}
        \widehat{F}_K
        =
        \sum_{S\subseteq\{1,\ldots,L\}} Z_S,
    \end{equation}
    where $Z_S$ denotes the contraction in which the $i$th slot contains $\xi_i$ if $i\in S$ and $\mu_i$ otherwise. The term $S=\emptyset$ is $K[\mu_1,\ldots,\mu_L]=\E[\widehat{F}_K]$, and every other $Z_S$ has zero mean. Moreover, if $S\neq S'$, choose an index belonging to one of the two subsets but not the other. The corresponding mean-zero random vector $\xi_i$ appears in only one of $Z_S$ and $Z_{S'}$, so independence and $\E[\xi_i]=0$ imply $\E[Z_SZ_{S'}]=0$. It follows that
    \begin{equation}
        \Var[\widehat{F}_K]
        =
        \sum_{S\neq\emptyset}\E[Z_S^2].
    \end{equation}
    Fix a subset $S$ containing $q$ indices. Contracting the other $L-q$ slots of $K$ against the corresponding $\mu_i$ defines an order-$q$ tensor $T_S$. Since $\|\mu_i\|_2\leq1+b$,
    \begin{equation}
        \|T_S\|_\infty
        \leq
        \kappa(1+b)^{L-q}.
    \end{equation}
    Writing the entries of $T_S$ in the standard basis, independence of the $q$ noise vectors and the covariance bound $\E[\xi_i\xi_i^T]\preceq\sigma^2\mathds{1}_{2N}$ give
    \begin{equation}
        \E[Z_S^2]
        =
        \E\left[T_S[\xi_{i_1},\ldots,\xi_{i_q}]^2\right]
        \leq
        \sigma^{2q}\|T_S\|_F^2.
    \end{equation}
    Finally,
    \begin{align}
        \|T_S\|_F^2
        &=
        \sum_{j_1,\ldots,j_{q-1}=1}^{2N}
        \left\|
        T_S[e_{j_1},\ldots,e_{j_{q-1}},\cdot]
        \right\|_2^2 \\
        &\leq
        (2N)^{q-1}\|T_S\|_\infty^2,
    \end{align}
    because each vector $T_S[e_{j_1},\ldots,e_{j_{q-1}},\cdot]$ has Euclidean norm at most $\|T_S\|_\infty$. Thus
    \begin{equation}
        \E[Z_S^2]
        \leq
        \kappa^2(1+b)^{2(L-q)}(2N)^{q-1}\sigma^{2q}.
    \end{equation}
    There are $\binom{L}{q}$ subsets $S$ of size $q$, so summing over $q=1,\ldots,L$ gives
    \begin{align}
        \Var[\widehat{F}_K]
        &\leq
        \kappa^2
        \sum_{q=1}^L
        \binom{L}{q}
        (1+b)^{2(L-q)}
        (2N)^{q-1}\sigma^{2q}\\
        &=
    \frac{\kappa^2}{2N}
    \left[
        \left((1+b)^2+2N\sigma^2\right)^L
        -(1+b)^{2L}
    \right],
    \end{align}
    where the final identity follows from the binomial theorem. If each $X_i$ is unbiased for $A$, then $b_i=0$ for every $i$. Hence $\E[\widehat{F}_K]=K[A^{\otimes L}]$, so the variance is also the mean squared error and the final statement in the lemma follows.
\end{proof}

The utility of this error-cancellation lemma is captured by the term~\eqref{eq:nonlinear_est_MSE}. In particular, the leading contribution is independent of $N$. Bounding each noise vector through its Euclidean norm would instead introduce an additional factor of $2N$ even in the leading term. This improvement lets us take
\begin{equation}
    \sigma^2
    \lesssim_L
    \min\left\{
    \left(\frac{\epsilon}{\kappa}\right)^2,
    (2N)^{-(L-1)/L}
    \left(\frac{\epsilon}{\kappa}\right)^{2/L}
    \right\}.
\end{equation}
Thus when $\epsilon/\kappa\lesssim N^{-1/2}$, the first term dominates and it is sufficient to have $\sigma\lesssim\epsilon/\kappa$. The required linear estimates need not be any more precise than for the linear spectroscopy problem. Our algorithms will handle the more common $\epsilon/\kappa \gtrsim N^{-1/2}$ regime separately, where we will see that the problem simplifies and sophisticated error cancellation becomes unnecessary.

A second useful strategy is to use noisy estimates of $A$ to reduce the tensor rank of the spectroscopy problem itself. After contracting $L-1$ slots of $K$ against previously obtained estimates of $A$, the remaining slot defines a known linear observable that can be estimated using any of our linear spectroscopy algorithms. Writing these preliminary estimates as $X_i = A + e_i$, however, a direct contraction contains error terms of the form $K[e_i, A^{(L-1)}]$ which are first order in the estimation errors. We instead construct an affine function $Q_X(a)$ whose coefficients are determined by the $X_i$, so that evaluating $Q_X(A)$ still requires only a linear spectroscopy measurement. The combination is chosen so that every term containing exactly one error $e_i$ cancels algebraically, leaving only terms involving at least two errors. This second-order error dependence asymptotically reduces the overhead of rank reduction and makes the construction an essential primitive in some of our nonlinear spectroscopy algorithms.

\begin{lemma}[First-order error cancellation]
\label{lem:first_order_error_cancellation}
Let $K$ be a symmetric rank-$L$ tensor satisfying $\|K\|_\infty\leq\kappa$, and let $X_i=A+e_i$, with $i=1,\ldots,L$.
Define the affine function
\begin{equation}
    Q_X(a)
    =
    \sum_{i=1}^L
    K[X_1,\ldots,X_{i-1},a,X_{i+1},\ldots,X_L]
    -(L-1)K[X_1,\ldots,X_L].
\end{equation}
Then
\begin{equation}
    Q_X(A)-K[A^{\otimes L}]
    =
    \sum_{q=2}^L
    (1-q)
    \sum_{\substack{S\subseteq\{1,\ldots,L\}\\|S|=q}}
    K[e_S,A_{\bar S}],
\end{equation}
where $K[e_S,A_{\bar S}]$ denotes the contraction obtained by inserting $e_i$ in the slots indexed by $S$ and $A$ in the remaining slots. In particular, every contribution that is first order in the errors $e_i$ cancels exactly. If the $e_i$ are independent, $\E[e_i]=0$, and $\E[e_ie_i^T]\preceq\sigma^2\mathds{1}_{2N}$, then
\begin{equation}
    \E\left[
    \left|Q_X(A)-K[A^{\otimes L}]\right|^2
    \right]
    \leq
    \kappa^2
    \sum_{q=2}^L
    (q-1)^2
    \binom{L}{q}
    (2N)^{q-1}\sigma^{2q}.
\end{equation}
\end{lemma}

\begin{proof}
    Expand each $X_i=A+e_i$. Consider a fixed subset $S\subseteq\{1,\ldots,L\}$ with $|S|=q$, corresponding to the term $K[e_S,A_{\bar S}]$ in the expansion. We first determine its coefficient in $Q_X(A)$. In the term
    \begin{equation}
        K[X_1,\ldots,X_{i-1},A,X_{i+1},\ldots,X_L],
    \end{equation}
    the $i^{\rm th}$ slot is fixed to $A$, so $K[e_S,A_{\bar S}]$ appears if and only if $i\notin S$. There are $L-q$ such values of $i$. The final term $-(L-1)K[X_1,\ldots,X_L]$ contributes the same contraction with coefficient $-(L-1)$. Hence the total coefficient of $K[e_S,A_{\bar S}]$ is
    \begin{equation}
        (L-q)-(L-1)=1-q.
    \end{equation}
    For $q=0$, this coefficient is $1$, giving $K[A^{\otimes L}]$. For $q=1$, the coefficient vanishes. Therefore
    \begin{equation}
        Q_X(A)
        =
        K[A^{\otimes L}]
        +
        \sum_{q=2}^L
        (1-q)
        \sum_{\substack{S\subseteq\{1,\ldots,L\}\\|S|=q}}
        K[e_S,A_{\bar S}],
    \end{equation}
    proving the first statement. Next, we consider the setting where all errors are independent and mean-zero. Terms corresponding to distinct subsets $S\neq S'$ are orthogonal in expectation, since some $e_i$ appears in exactly one of the two contractions. Thus
    \begin{equation}
        \E\left[
        \left|Q_X(A)-K[A^{\otimes L}]\right|^2
        \right]
        =
        \sum_{q=2}^L
        (q-1)^2
        \sum_{|S|=q}
        \E\left[K[e_S,A_{\bar S}]^2\right].
    \end{equation}
    For a fixed subset $S$ of size $q$, contracting the remaining $L-q$ slots against $A$ produces an order-$q$ tensor $T_S$ satisfying $\|T_S\|_\infty\leq\kappa$. As in Lemma~\ref{lem:independent_nonlinear_estimation},
    \begin{equation}
        \E\left[K[e_S,A_{\bar S}]^2\right]
        \leq
        \sigma^{2q}\|T_S\|_F^2
        \leq
        \kappa^2(2N)^{q-1}\sigma^{2q}.
    \end{equation}
    Summing over the $\binom{L}{q}$ subsets of size $q$ proves the result.
\end{proof}

\subsection{Polynomial phase mixing} \label{sec:poly_phase_mixing}
In this and the next section we develop the algorithmic pulse sequence primitives central to pooled spectroscopy, which is detailed in Section~\ref{sec:pooled_spectroscopy}. The main goal of pooled spectroscopy is to access macroscopic spectral features of a signal with a very dense spectrum by compressing a large number of spectral data in a desired band $[\omega_0, \omega_0 +\Delta]$ into a single collective amplitude via pulse-sequence preprocessing, removing noise before measurement.

The first core primitive is polynomial phase mixing. When we want to access coarse-grained spectral information such as the total power in a frequency bin, we want to estimate nonlinear observables such as
\begin{equation}
\sum_{\omega_j\in[\omega_0, \omega_0+\Delta]} (p_j^2 + q_j^2)
\end{equation}
without needing to learn each $p_j$ and $q_j$ separately. Polynomial phase mixing is a pulse sequence protocol that gives us coherent access to these squared (or more generally, nonlinear polynomial) amplitudes. Consider the following illustrative example.

\begin{observation}[Coherent squaring of a spectral response]
\label{obs:coherent_squaring}
Let $x=v^TA$ be any linear spectral response that can be generated by a commuting pulse sequence, and define $R_\mu(x)=e^{-ix\sigma_\mu/2}$ for $\mu\in\{X,Y,Z\}$. Consider the following evolution operators obtained by applying $\pi/2$ pulses along $X$ and $Y$ axes while interrogating the signal:
\begin{equation}
    C(x)=R_X(x)R_Y(x)R_X(-x)R_Y(-x), \qquad V(x)=C(x/\sqrt{2})C(-x/\sqrt{2}) \ .
\end{equation}
Then,
\begin{equation}
    V(x) = R_Z(x^2) + O(|x|^4).
\end{equation}
Thus an unknown linear response $x=v^TA$ can be coherently converted into a phase proportional to $x^2$ without first estimating $x$. The sequence uses eight signal-dependent rotations.
\end{observation}

\begin{proof}
    The original sensing interaction resulting from a commuting pulse sequence produces $R_Z(x)$; the following simple $\pi/2$ pulses let us realize $R_X(x), R_Y(x)$: 
    \begin{equation}
        R_X(x) = R_Y(\pi/2)R_Z(x)R_Y(-\pi/2), \qquad  R_Y(x) = R_X(-\pi/2)R_Z(x)R_X(\pi/2) \ .
    \end{equation}
    The underlying commuting pulse sequence can be used to obtain $R_Z(-x)$ rather than $R_Z(x)$ (without needing to know $x$), so applying the above transformations similarly produces $R_X(-x), R_Y(-x)$.

    Now let $A_x=-ixX/2$ and $B_x=-ixY/2$. Using the Baker-Campbell-Hausdorff formula and taking logs gives
    \begin{equation}
        \log C(x)
        =
        [A_x,B_x]
        +
        \frac{1}{2}[A_x+B_x,[A_x,B_x]]
        +
        O(x^4).
    \end{equation}
    Using $[X,Y]=2iZ$, $[X,Z]=-2iY$, and $[Y,Z]=2iX$, this becomes
    \begin{equation}
        \log C(x)
        =
        -\frac{i}{2}x^2Z
        -\frac{i}{4}x^3(X-Y)
        +O(x^4).
    \end{equation}
    Replacing $x$ by $-x$ leaves the quadratic term unchanged while reversing the cubic term. Consequently,
    \begin{align}
        \log V(x)
        &=
        \log C(x/\sqrt{2})+\log C(-x/\sqrt{2})+O(x^5) \\
        &=
        -\frac{i}{2}x^2Z+O(x^4),
    \end{align}
    which proves the stated result.
\end{proof}
Observation~\ref{obs:coherent_squaring} shows us that if we have access to an unknown phase $x=v^TA$, noncommuting single-qubit control can transform this phase into nonlinear polynomial functions of $x$ without ever learning the value of $x$ classically. For spectroscopy we will instead need a systematic procedure that, for any fixed $L \in \{2, 3, 4\}$ implements $R_Z(x) \longmapsto R_Z(ax^L)$ to a prescribed accuracy uniformly over the full interval $|x| \leq \pi/2$. 

We first solve the problem of decomposing the desired nonlinear response into single qubit rotations as, 
\begin{equation}
    W_0R_Z(s_1 x)W_1R_Z(s_2x)\cdots
W_{d_L-1}R_Z(s_{d_L}x)W_{d_L}
\approx
R_Z(ax^L)
\end{equation}
where $W_j$ are single qubit rotations and  $s_j \in \{-1,+1\}$ for every $|x|\leq \pi/2$. Determining this decomposition can be done classically before the algorithm is run. The structure of this synthesis problem becomes easier to see by introducing $z = e^{ix/2}$. Since, 
\begin{equation}
    R_Z(x)=
\begin{pmatrix}
z^{-1}&0\\
0&z
\end{pmatrix}
\end{equation}
any finite sequence of $R_Z(\pm x)$ interspersed with single qubit rotations has matrix entries that are Laurent polynomials in $z$. We can therefore construct the desired sequence by approximating the target phase $e^{-iax^L/2}$ with an appropriate Laurent polynomial, completing it to a unitary $2 \times 2$ Laurent-polynomial matrix, and then factoring that matrix into elementary factors corresponding to $R_Z(\pm x)$ and known rotations. Algorithm~\ref{algo:classical_polynomial_compiler} implements this classical compilation procedure. 

\begin{algorithm}
    \DontPrintSemicolon
    \caption{Classical compilation of a polynomial phase sequence}
    \label{algo:classical_polynomial_compiler}
    \KwIn{Polynomial order $L\in\{2,3,4\}$, coefficient $a\in\mathbb{R}$, and target accuracy $\eta>0$.}
    \KwOut{A sequence length $d_L$, signs $s_1,\ldots,s_{d_L}\in\{-1,+1\}$, and known single-qubit rotations $W_0,\ldots,W_{d_L}$.}

    Construct a smooth $2\pi$-periodic function $\phi_L(x)$ that agrees with $x^L$ for every $|x|\leq\pi/2$, using the cutoff construction in the proof of Lemma~\ref{lem:polynomial_phase_mixing}.\;

    Define
    \begin{equation}
        f(x)=e^{-ia\phi_L(x)/2},
    \end{equation}
    and choose
    \begin{equation}
        d=
        O_L\left(
        (1+|a|)
        \log^2\!\left(\frac{2+|a|}{\eta}\right)
        \right).
    \end{equation}
    Compute the Fourier coefficients of $f$ for $|k|\leq d$ and form the corresponding truncated Fourier polynomial $p_0(x)$.\;

    Multiply $p_0$ by a factor $1-O(\eta^2)$, chosen so that the resulting polynomial $p(x)$ satisfies $|p(x)|<1$ for all $x$ while remaining within $O(\eta^2)$ of $f(x)$.\;

    Set $z=e^{ix/2}$ and write $p$ as a Laurent polynomial in $z$. Define
    \begin{equation}
        p^*(z)=\overline{p(1/\bar z)}.
    \end{equation}
    Apply the Fej\'er--Riesz factorization to the nonnegative Laurent polynomial $1-p(z)p^*(z)$ to obtain $q(z)$ satisfying
    \begin{equation}
        1-p(z)p^*(z)=q(z)q^*(z).
    \end{equation}

    Form the unitary Laurent-polynomial matrix
    \begin{equation}
        F(z)=
        \begin{pmatrix}
            p(z) & q(z) \\
            -q^*(z) & p^*(z)
        \end{pmatrix}.
    \end{equation}

    Factor $F(z)$ (as a matrix-valued polynomial) recursively into degree-one factors. At each step, identify the outermost Laurent coefficient of the current matrix, choose a known single-qubit rotation that places this coefficient in one computational-basis component, and choose $s_j\in\{-1,+1\}$ so that multiplication by $R_Z(-s_jx)$ removes that outer Laurent power. Record the corresponding $s_j$ and rotation, and repeat until the remaining matrix is independent of $x$.\;

    Write the degree reduction procedure in reverse to obtain
    \begin{equation}
        F(e^{ix/2})
        =
        W_0R_Z(s_1x)W_1R_Z(s_2x)\cdots
        W_{d_L-1}R_Z(s_{d_L}x)W_{d_L}.
    \end{equation}

    Return $d_L$, the signs $\{s_j\}$, and the rotations $\{W_j\}$.\;
\end{algorithm}

The mathematical construction is closely related to Laurent-polynomial factorizations used in quantum signal processing~\cite{haah_product_2019}. Here we require the stronger task of approximating the entire single-qubit unitary $R_Z(ax^L)$, rather than only a selected matrix element. We therefore choose the target Laurent polynomial to have modulus $1-O(\eta^2)$, which forces the complementary polynomial obtained from the unitary completion to have magnitude $O(\eta)$. 

 The following lemma proves that the sequence produced by Algorithm~\ref{algo:classical_polynomial_compiler} approximates the desired polynomial phase with the claimed sequence length.

\begin{lemma}[Polynomial phase synthesis]
\label{lem:polynomial_phase_mixing}
Let $L\in\{2,3,4\}$ be the fixed order of the desired nonlinear spectra, and suppose a commuting pulse sequence implements $R_Z(x)$ for $x=v^TA$ satisfying $|x|<\pi/2$. For any $a\in\mathbb{R}$ and $\eta>0$, we can realize a length-$d_L$ pulse sequence
\begin{equation}
    U_L(x)
    =
    W_0R_Z(s_1x)W_1R_Z(s_2x)\cdots W_{d_L-1}R_Z(s_{d_L}x)W_{d_L},
\end{equation}
where each $s_j\in\{-1,+1\}$ and each $W_j$ is a known single-qubit rotation independent of $A$, such that
\begin{equation}
    \left\|U_L(x)-R_Z(ax^L)\right\|_{\rm op}\leq\eta
\end{equation}
for every $|x|\leq \pi/2$. The required number of signal interrogation blocks is
\begin{equation}
    d_L = O_L\!\left( (1+|a|) \log^2\left(\frac{2+|a|}{\eta}\right) \right).
\end{equation}
Thus, for the constant amplitudes $a$ required in our algorithms, the total number of pulses is constant. 
\end{lemma}

\begin{proof}
    We construct a pulse sequence that approximates the desired polynomial $ax^L$ on the interval $[-\pi/2, \pi/2]$. Note that because we assume $|x|\leq\pi/2$, we are free to first construct functions that match the desired polynomial only on this interval, which are then easier to approximate with an appropriate pulse sequence whose response is a periodic trigonometric polynomial. 
    
    To do this, choose a smooth even function $\chi(x)$ on $[-\pi,\pi]$ satisfying
    \begin{equation}
        \chi(x)=1 \text{ for } |x|\leq\pi/2,
        \qquad
        \chi(x)=0 \text{ for } |x|\geq3\pi/4.
    \end{equation}
    An appropriate function can be constructed explicitly from, for example, the function
    \begin{equation}
        \rho(t)=
        \begin{cases}
            e^{-1/t}, & t>0,\\
            0, & t\leq0,
        \end{cases}
    \end{equation}
    by defining
    \begin{equation}
        s(t)=\frac{\rho(t)}{\rho(t)+\rho(1-t)}.
    \end{equation}
    Then $s(t)=0$ for $t\leq0$, $s(t)=1$ for $t\geq1$, and $s$ is smooth. Taking
    \begin{equation}
        \chi(x)
        =
        s\left(
        \frac{3\pi/4-|x|}{\pi/4}
        \right)
    \end{equation}
    on the region $\pi/2<|x|<3\pi/4$ gives us the desired cutoff. Now define
    \begin{equation}
        \phi_L(x)=\chi(x)x^L
    \end{equation}
    on $[-\pi,\pi]$, and then extend $\phi_L$ periodically. Since $\chi(x)$ vanishes on a neighborhood of $\pm\pi$, $\phi_L$ and all of its derivatives agree at $\pm \pi$, so the periodic extension is smooth. Moreover, $\phi_L(x)=x^L$ for every $|x|\leq\pi/2$ as desired.     We next bound the derivatives of $\phi_L$. The smooth cutoff constructed above satisfies
    \begin{equation}
        \|\chi^{(r)}\|_\infty\leq C^{r+1}(r!)^2
    \end{equation}
    for some numerical constant $C$. To see this, repeated differentiation of $\rho(t)=e^{-1/t}$ produces $e^{-1/t}$ multiplied by a polynomial in $1/t$ of degree $r$. The maximum of terms of the form $t^{-k}e^{-1/t}$ is obtained by setting $1/t$ of order $k$, which gives a bound of order $C^r(r!)^2$.

    Since $\phi_L(x)=\chi(x)x^L$ and $L$ is fixed, Leibniz's rule gives
    \begin{equation}
        \phi_L^{(r)}(x)
        =
        \sum_{j=0}^{\min\{r,L\}}
        \binom{r}{j}
        \chi^{(r-j)}(x)
        \frac{L!}{(L-j)!}x^{L-j}.
    \end{equation}
    On $[-\pi,\pi]$, the factors involving $x$ and $L$ are bounded by constants depending only on $L$. Using the derivative bound on $\chi$, and absorbing the finite sum into the constant, we obtain
    \begin{equation}
        \|\phi_L^{(r)}\|_\infty
        \leq C_L^{r+1}(r!)^2.
    \end{equation}
    Next, consider the periodic function
    \begin{equation}
        f(x)=e^{-ia\phi_L(x)/2}.
    \end{equation}
    By the Fa\`a di Bruno formula, its $r^{\rm th}$ derivative $f^{(r)}(x)$ is a sum of terms of the form
    \begin{equation}
        f(x)a^k\prod_{\ell=1}^k \phi_L^{(j_\ell)}(x),
        \qquad
        j_1+\cdots+j_k=r,
        \qquad
        1\leq k\leq r,
    \end{equation}
    multiplied by combinatorial coefficients. Since $|f(x)|=1$, the above bound on the derivatives of $\phi_L$ implies
    \begin{equation}
        \|f^{(r)}\|_\infty
        \leq
        \left[C_L(1+|a|)\right]^r(r!)^2.
    \end{equation}
    Here we have used that $\prod_{\ell=1}^k j_\ell!\leq r!$ whenever $j_1+\cdots+j_k=r$, and absorbed the number of partitions of $r$ and the remaining numerical factors into $C_L^r$. To implement the pulse sequence, we will approximate $f$ by its Fourier series. If $\widehat f_k$ denotes its $k$th Fourier coefficient, integrating by parts $r$ times gives
    \begin{equation}
        |\widehat f_k|
        \leq
        \frac{\|f^{(r)}\|_\infty}{|k|^r}.
    \end{equation}
    Combining our derivative bounds gives $\|f^{(r)}\|_\infty\leq[C_L(1+|a|)]^r(r!)^2$. Choosing $r$ of order $\sqrt{|k|/(1+|a|)}$ therefore gives
    \begin{equation}
        |\widehat f_k|
        \leq
        C_L(1+|a|)
        \exp\!\left(-c_L\sqrt{\frac{|k|}{1+|a|}}\right).
    \end{equation}
    Hence, truncating the Fourier series at degree
    \begin{equation}
        d=
        O_L\!\left(
        (1+|a|)
        \log^2\left(\frac{2+|a|}{\eta}\right)
        \right)
    \end{equation}
    gives a trigonometric polynomial $p_0(x)$ satisfying $|p_0(x)-f(x)|\leq c\eta^2$ for all $x$.

    We now show how this phase function can be realized in the diagonal entries of a unitary single-qubit matrix. First, multiply $p_0$ by $1-c'\eta^2$, for a sufficiently large numerical constant $c'$, and call the resulting polynomial $p(x)$. The constant is chosen to shrink $p_0$ such that $|p(x)|<1$ for every $x$, while $p(x)$ still differs from $f(x)$ by only $O(\eta^2)$. 
    
    Now change to complex coordinates $z=e^{ix/2}$, so that $p$ can be written as a Laurent polynomial in $z$. Define $p^*(z)=\overline{p(1/\bar z)}$. On the complex unit circle $|z| = 1$, $1/\overline{z} = z$, so $p^*(z) = \overline{p(z)}$. Therefore the function $1-p(z)p^*(z) = 1-|p(z)|^2$ is real and nonnegative on the unit circle. The Fejer-Riesz theorem states that if a polynomial $r(z)$ is real and nonnegative on the unit circle, there is a polynomial $q(z)$ such that $r(z) = q^*(z)q(z)$. As such, we have $1-p(z)p^*(z)=q(z)q^*(z)$. With these constructed polynomials, consider the two by two complex matrix 
    \begin{equation}
        F(z)=
        \begin{pmatrix}
            p(z) & q(z) \\
            -q^*(z) & p^*(z)
        \end{pmatrix}.
    \end{equation}
    By our construction, it follows that for all $|z| = 1$,
    \begin{equation}
        F(z)^\dagger F(z) = \begin{pmatrix}
            |p(z)|^2 + |q(z)|^2 & 0 \\
            0 & |p(z)|^2 + |q(z)|^2
        \end{pmatrix}
        = \mathds{1} \ ,
    \end{equation}
    so $F$ is exactly unitary on the unit circle. Moreover, since $|p(x)|=1-O(\eta^2)$, we have $|q(x)|=O(\eta)$, and therefore
    \begin{equation}
        \left\|
        F(e^{ix/2})-R_Z(a\phi_L(x))
        \right\|_{\rm op}
        \leq C\eta.
    \end{equation}
    Rescaling $\eta$ by a numerical constant gives the desired error tolerance. Finally, we convert $F(z)$ into a physical pulse sequence. Starting from the largest power of $z$ appearing in $F$, choose a known single-qubit rotation that maps its coefficient onto one computational-basis component. Multiplying by either
    \begin{equation}
        D_+(z)=
        \begin{pmatrix}
            z^{-1} & 0 \\
            0 & z
        \end{pmatrix}
        =R_Z(x)
    \end{equation}
    or $D_-(z)=D_+(z)^{-1}=R_Z(-x)$ then cancels that Laurent coefficient and reduces the degree by one. Repeating this procedure until the degree reaches zero gives
    \begin{equation}
        F(e^{ix/2})
        =
        W_0R_Z(s_1x)W_1R_Z(s_2x)\cdots W_{d_L-1}R_Z(s_{d_L}x)W_{d_L},
    \end{equation}
    where every $W_j$ is a known rotation and every $s_j\in\{-1,+1\}$. The number of degree-reduction steps is proportional to the Laurent degree, so
    \begin{equation}
        d_L=
        O_L\!\left(
        (1+|a|)
        \log^2\left(\frac{2+|a|}{\eta}\right)
        \right).
    \end{equation}
    Finally, on the physical interval $|x|\leq\pi/2$ we have $\phi_L(x)=x^L$, so the resulting sequence approximates $R_Z(ax^L)$ as claimed.
\end{proof}

Algorithm~\ref{algo:classical_polynomial_compiler} determines how repeated applications of the unknown linear phase $x= v^TA$ with interspersed single qubit rotations have combined evolution that approximates $R_Z(ax^L)$.  We now implement this sequence physically in Algorithm~\ref{algo:polynomial_phase_mixing}. Each required factor $R_Z(s_jx)$ is generated by interrogating the signal with the compiled response $s_jv$, while the known rotations $W_j$ are applied between successive sensing intervals.

\begin{algorithm}
    \DontPrintSemicolon
    \caption{Polynomial phase mixing}
    \label{algo:polynomial_phase_mixing}
    \KwIn{Known frequencies $\{\omega_j\}_{j=1}^N$, a response vector $v\in\mathbb R^{2N}$, order $L\in\{2,3,4\}$, coefficient $a\in\mathbb R$, accuracy $0<\eta<1$, and starting time $t_0$. If $v$ is random, a failure probability $\delta_{\rm syn}$}
    \KwOut{A pulse sequence $U_L$ and its ending time $t_1$, satisfying $\|U_L-R_Z(a(v^TA)^L)\|_{\rm op}\leq\eta$ whenever $|v^TA|\leq\pi/2$.}

    Run Algorithm~\ref{algo:classical_polynomial_compiler} with inputs $L,a,\eta/2$. Denote its outputs by $d_L$, signs $s_1,\ldots,s_{d_L}$, and rotations $W_0,\ldots,W_{d_L}$.\;

    Initialize the sequence $U_L$ to begin with $W_{d_L}$ and set $t=t_0$.\;

    \For{$i=d_L,\ldots,1$}{
        Compile the response $s_iv$ to accuracy
        $\epsilon_{\rm syn}=\eta/\max\{1,d_L\}$ at starting time $t$.
        If $v$ is a random response satisfying Lemma~\ref{lem:random_response_compiler}, use the random response compiler with failure probability $\delta_{\rm syn}/\max\{1,d_L\}$. Otherwise use Algorithm~\ref{algo:deterministic_response_compiler}. Denote the resulting pulse sequence by $y_i$, on $[t,t+T_i]$\;

        Append $y_i$ and then $W_{i-1}$ to $U_L$, and set $t=t+T_i$.\;
    }

    Return $U_L$ and $t_1=t$; the resulting unitary is $U_L(x) = W_0R_Z(s_1x)W_1R_Z(s_2x)\cdots W_{d_L-1}R_Z(s_{d_L}x)W_{d_L}$.\;
\end{algorithm}

Thus polynomial phase mixing coherently converts a compiled linear response $x = v^TA$ into the nonlinear phase $ax^L$, without first estimating $x$. For fixed $L$ and $a=O(1)$, this requires only $d_L=O_L\left(\log^2(1/\eta)\right)$ applications of the compiled linear phase $R_Z(\pm v^TA)$, independent of the spectral dimension $N$.

\subsection{Random convolution}
The second primitive for pooled spectroscopy is random convolution. Its purpose is to coherently compress the enormous spectroscopic data present in a dense spectrum into a single response $x$ within a chosen bin $[\omega_0, \omega_0 + \Delta]$ or capture collective responses across different spectral bins. 

It is convenient here to combine the two real quadratures at each frequency into a single complex amplitude, $\alpha_j=p_j-iq_j$, so that \begin{equation}
    B_A(t)=\operatorname{Re}\left(\sum_{j=1}^N \alpha_j e^{i\omega_j t}\right).
\end{equation}
$\alpha_b$ denotes the vector of amplitudes whose frequencies lie in bin $b$. The simplest example is the total power in one bin, 
\begin{equation}
    P_b=\sum_{j\in\mathcal B_b}|\alpha_j|^2=\|\alpha_b\|_2^2
\end{equation}

If $g$ is a standard complex Gaussian vector supported on that bin, and $z=g^\dagger\alpha_b$ then $\mathbb E[gg^\dagger]=I$, 
\begin{equation}
    \mathbb E|z|^2
    =
    \alpha_b^\dagger\mathbb E[gg^\dagger]\alpha_b
    =
    \|\alpha_b\|_2^2
    =
    P_b.
\end{equation}
Thus a single random projection compresses all of the microscopic amplitudes in the bin into one complex number, while its squared magnitude recovers the total bin power in expectation. 

Random convolution extends to an arbitrary fixed-order coarse observable, 
\begin{equation}
    F_K(\alpha)
=
K[\widetilde\alpha_{b_1},\ldots,\widetilde\alpha_{b_L}],
\end{equation}
where each $\widetilde\alpha_{b_\ell}$ denotes either $\alpha_{b_\ell}$ or its complex conjugate depending on the desired observable.  The construction uses $L-1$ random Gaussian projections together with one final $K$-dependent projection so that the product of the resulting $L$ complex responses has expectation $F_K(\alpha)$. Because the pulse-sequence compiler acts on real response vectors, the Algorithm~\ref{algo:random_convolution} finally separates each complex response into its real and imaginary parts. We provide Algorithm~\ref{algo:random_convolution} below.

\begin{algorithm}
    \DontPrintSemicolon
    \caption{Random convolution pooling}
    \label{algo:random_convolution}
    \KwIn{Frequency bins $b_1,\ldots,b_L$; an order-$L$ observable $F_K(\alpha)=K[\widetilde \alpha_{b_1},\ldots,\widetilde \alpha_{b_L}]$
    where each $\widetilde \alpha_{b_l}$ is specified to be either $\alpha_{b_l}$ or $\bar \alpha_{b_l}$; precision $\epsilon_{\rm syn}$.}
    \KwOut{$2L$ real response vectors $v_{\ell,R},v_{\ell,I}\in\mathbb R^{2N}$ such that defining $x_\ell\coloneqq v_{\ell,R}^TA+i\,v_{\ell,I}^TA$, it holds that
    \begin{equation}\label{eq:convolution_identity}
        \mathbb E\left[\prod_{\ell=1}^L x_\ell\right]
        =
        F_K(\alpha).
    \end{equation} 

    }

    Draw $L-1$ independent complex Gaussian vectors $g_1,\ldots,g_{L-1}$, where $g_\ell$ is supported only on bin $b_\ell$ and every nonzero coordinate is independently distributed as $(x+iy)/{\sqrt2}, \ x,y\sim\mathcal N(0,1)$

    Contract these $L-1$ vectors through the first $L-1$ slots of $K$ and compute the unique vector $h$, supported on bin $b_L$, satisfying
    \begin{equation}
        h^\dagger u
        =
        K[g_1,\ldots,g_{L-1},u]
    \end{equation}
    for every complex vector $u$ supported on $b_L$.\;

    \For{\(\ell=1,\ldots,L\)}{
        Set \(c=g_\ell\) if \(\ell<L\), and \(c=h\) if \(\ell=L\). Write
        \(c=a+ib\), with \(a,b\in\mathbb R^N\).\;

        If the \(\ell\)th slot contains \(\alpha_{b_\ell}\), set
        \[
            v_{\ell,R}
            =
            \begin{pmatrix}a\\-b\end{pmatrix},
            \qquad
            v_{\ell,I}
            =
            \begin{pmatrix}-b\\-a\end{pmatrix}.
        \]
        If instead it contains \(\overline{\alpha}_{b_\ell}\), set
        \[
            v_{\ell,R}
            =
            \begin{pmatrix}a\\b\end{pmatrix},
            \qquad
            v_{\ell,I}
            =
            \begin{pmatrix}-b\\a\end{pmatrix}.
        \]
        In either case, set all coordinates outside bin \(b_\ell\) to zero.\;
    }
    Return the $2L$ response vectors.
\end{algorithm}

\begin{proof}
    Recall the original real spectral vector $A = (p_1,...,p_N,q_1,...,q_N)^T$ and the notation $\alpha = p - iq$. We first verify that the real response vectors passed to the pulse-sequence compiler of Algorithm~\ref{algo:deterministic_response_compiler} reproduce the desired complex projections. For $c = a +ib$,
    \begin{align}
        c^\dagger\alpha
    &=
    a^Tp-b^Tq
    -i(b^Tp+a^Tq)\\
    c^\dagger\overline\alpha
    &=
    a^Tp+b^Tq
    +i(-b^Tp+a^Tq).
    \end{align}
    The two cases in the algorithm therefore give $z_\ell = c^\dagger \widetilde \alpha_{b_\ell}$. For $\ell < L$, $z_\ell = g_l^\dagger \widetilde \alpha_{b_\ell}$ while the definition of $h$ gives, 
    \begin{equation}
        z_L = h^\dagger \widetilde \alpha_{b_L} = K[g_1,...,g_{L-1},\widetilde \alpha_{b_L}].
    \end{equation}
Hence, 
    \begin{equation}
        \prod_{\ell=1}^L z_\ell
    =
    \left(
    \prod_{\ell=1}^{L-1}
    g_\ell^\dagger\widetilde\alpha_{b_\ell}
    \right)
    K[g_1,\ldots,g_{L-1},\widetilde\alpha_{b_L}].
    \end{equation}
A standard complex Gaussian vector satisfies $\E[gg^\dagger] = I$, and therefore, for any fixed vector $u$, $\E[(g^\dagger u)g] = u$. Using this identity successively for the independent vectors $g_1,\ldots, g_{L-1}$, and using multilinearity of $K$ we get
\begin{align}
     \mathbb E\!\left[\prod_{\ell=1}^L z_\ell\right]
    &=
    K[
    \widetilde\alpha_{b_1},
    \ldots,
    \widetilde\alpha_{b_L}]\\
    &=
    F_K(\alpha),
\end{align}
as claimed. 
\end{proof}

\newpage
\section{Randomized Quantum Spectroscopy}\label{sec:randomized_spec}
In this section, we introduce the randomized quantum spectroscopy algorithm and prove that it is optimal among all protocols without persistent quantum memory for estimating spectroscopic observables of any rank, including in the presence of decoherence. The algorithm is independent of the observables of interest, and the collected dataset can be reused to estimate many observables after data collection has concluded; remarkably, the algorithm remains optimal even relative to any adaptive single-qubit protocol with prior knowledge of the chosen observables, when the number of observables or the noise is large.

\subsection{Algorithm for general spectroscopic rank}
The main result we will prove in this section is the existence of an algorithm using single-spin pulse sequences to perform linear or nonlinear spectroscopy with the following guarantee.
\begin{graydefinitionbox}
\begin{theorem}
\label{thm:randomized_spectroscopy_upper_bound}
Fix parameters $\kappa>0$,  $0<\epsilon<\kappa$, and $0<\delta<1/4$. There exists an algorithm using a single spin and a control Hamiltonian of the form \eqref{eq:pulse_ham} satisfying the following guarantee. The algorithm interrogates the unknown signal for time at most
\begin{equation}
T\leq
\begin{cases}
\widetilde O\left(
\min\left\{
\sqrt N\left(\frac{\kappa}{\epsilon}\right)^2\log\left(\frac{M}{\delta}\right),
N\frac{\kappa}{\epsilon}\sqrt{\log\left(\frac{M}{\delta}\right)}
\right\}\right), & L=1,\\[2mm]
\widetilde O_L\left(
\max\left\{
N^{3/2-1/L}\left(\frac{\kappa}{\epsilon}\right)^{2/L}\log\left(\frac{M}{\delta}\right)^{1/L}
,
N\frac{\kappa}{\epsilon}\sqrt{\log\left(\frac{M}{\delta}\right)}
\right\}\right), & L\geq2 \ ,
\end{cases}
\label{eq:randomized_spectroscopy_combined_upper_compact}
\end{equation}
and given any set of $M$ rank-$L$ tensors $K_1,...,K_M$ all satisfying $\|K_j\|_\infty\leq \kappa$ after interrogation has completed, produces estimates $\widehat K_1, ..., \widehat K_M$ such that
\begin{equation}
\max_{j=1,\ldots,M}\left|\widehat K_j-K_j[A^{\otimes L}]\right|\leq\epsilon
\end{equation}
with probability at least $1-\delta$. 
\end{theorem}
\end{graydefinitionbox}
\begin{proof}
The general algorithm is composed of two protocols (Algorithms~\ref{algo:randomized_spectroscopy_coarse} and \ref{algo:randomized_spectroscopy_fine}) whose guarantees are proved below; we reference those results here. Let $p=\max\{2, \log(4M/\delta)\}$. If $\epsilon/\kappa\geq N^{-1/2}\sqrt{p/2}$, then $p\leq2N(\epsilon/\kappa)^2$, and Proposition~\ref{prop:gaussian_randomized_spectroscopy} gives the first terms in each case of \eqref{eq:randomized_spectroscopy_combined_upper_compact}. If $\epsilon/\kappa\leq N^{-1/2}\sqrt{p/2}$, then $2N(\epsilon/\kappa)^2\leq p$, and Proposition~\ref{prop:high_precision_randomized_spectroscopy} gives the second terms. At the crossover both scalings are $\widetilde O_L(N^{3/2})$. Comparing the two expressions on either side of this point gives \eqref{eq:randomized_spectroscopy_combined_upper_compact}.
\end{proof}

The algorithm is composed of two protocols which respectively dominate in the regimes $\epsilon/
\kappa\gtrsim 1/\sqrt{N}$ and $\epsilon/\kappa \lesssim 1/\sqrt{N}$. The goal of these protocols is to avoid reconstructing \(A\) to high precision before performing downstream estimation, which would incur costly $N$-dependent overheads. Instead, we take carefully-designed noisy sketches of $A$ whose raw noise is large, but is arranged so that it cancels when contracted against the multilinear forms of interest. More precisely, every returned vector may have a large Euclidean error, but any fixed tensor estimated later sees only small directional projections of that error, and independence across tensor slots removes the nonlinear bias.

\subsubsection{Gaussian randomized spectroscopy}
The central idea of Gaussian randomized spectroscopy is to probe random linear combinations of the signal's spectral amplitudes by engineering pulse sequences with Gaussian-distributed spectral responses. In each sensing round, we sample a Gaussian response vector $g \sim \mathcal{N}(0,\mathds{1}_{2N})$ and compile a pulse sequence whose accumulated phase is $g^TA$. Measuring the sensor in the $Y$ basis produces a bit $B\in \{\pm 1\}$ with conditional expectation $\mathbb E[B \mid g]= \sin(g^TA)$. Then, Gaussian integration by parts gives the exact identity, 
\begin{equation}
    \mathbb E_{g,B}[Bg] = \mathbb E_g[g \sin(g^TA)] = e^{-\|A\|_2^2/2}A.
\end{equation}
Thus, averaging the measurement outcomes weighted by their Gaussian responses gives an estimate of the entire spectral vector, up to a normalization factor that can be independently calibrated using $X$-basis measurements. Crucially, the fluctuations of $Bg$ along any fixed direction $k$ have variance at most $\|k\|_2^2$, independent of $N$. This allows us to estimate spectral observables directly, without first reconstructing every spectral amplitude to high precision. 

First we state the protocol for the $\epsilon/\kappa \gtrsim 1/\sqrt{N}$ regime. Below, we take $C_L=(12L)^2$, while in practice the necessary constant may be substantially smaller.

\begin{algorithm}
    \DontPrintSemicolon
    \caption{Randomized spectroscopy}
    \label{algo:randomized_spectroscopy_coarse}
    \KwIn{Known frequencies $\{\omega_j\}_{j=1}^N$, tensor rank $L$, list size $M$, accuracy $\epsilon$, norm bound $\kappa$, failure probability $\delta$.}
    \KwOut{Vectors $U_1,\ldots,U_L\in\mathbb{R}^{2N}$.}

    Let $p$ be the smallest even integer satisfying $p\geq \max\left\{2,\log\frac{4M}{\delta}\right\}$, and choose
    \begin{equation}
    \label{eq:gauss_rand_m_chosen}
        m
        =
        \left\lceil
        C_L\max\left\{
        p\left(\frac{\kappa}{\epsilon}\right)^2,
        (2N)^{1-1/L}p^{1/L}
        \left(\frac{\kappa}{\epsilon}\right)^{2/L}
        \right\}
        \right\rceil ,
        \qquad
        r_0
        =
        \left\lceil
        C_L\left(\frac{\kappa}{\epsilon}\right)^2
        \log\frac{8}{\delta}
        \right\rceil .
    \end{equation}
    with $C_L$ a constant depending only on $L$. 

    Run Algorithm~\ref{algo:gaussian_response_compiler} with $\epsilon_{\rm syn} = \delta/2R, \delta_{\rm syn} = \delta/{4R}$, $R=r_0+Lm$ times to obtain $R$ Gaussian-random pulse sequences and their random vectors $g_r\sim\mathcal{N}(0,\mathds{1}_{2N})$.
    
    Initialize in $\ket{+}$ and run all $R$ pulse sequences while interrogating the signal; measure $X$ for the first $r_0$ responses and $Y$ for the remaining $Lm$ responses, recording outcomes $B_r\in\{-1,+1\}$.\;

    Set
    \begin{equation}
        \widehat\mu
        =
        \operatorname{clip}\left(
        \frac{1}{r_0}\sum_{r=1}^{r_0}B_r,
        [e^{-1/2},1]
        \right),
    \end{equation}
    divide the remaining $Lm$ shots into $L$ consecutive groups $\mathcal I_1,\ldots,\mathcal I_L$ of size $m$, and return
    \begin{equation}
        U_i
        =
        \frac{1}{\widehat\mu m}
        \sum_{r\in\mathcal I_i}B_rg_r,
        \qquad
        i=1,\ldots,L.
    \end{equation}
\end{algorithm}

The sensing rounds in Algorithm~\ref{algo:randomized_spectroscopy_coarse} are independent of the observables to be estimated. After the transcript $U_1,\ldots,U_L$ has been obtained, any fixed list of rank-$L$ tensors $K_1,\ldots, K_M$, chosen independently of the acquisition randomness and outcomes, can be evaluated by computing $\widehat F_j = K_j [U_1,\ldots, U_L]$ with $j = 1,\ldots,M$. Thus, the same experimental record can be reused to answer many spectroscopic queries supplied only after the sensing experiment has completed. The interrogation time cost of the protocol is as follows.

\begin{prop}[Time complexity of Algorithm \ref{algo:randomized_spectroscopy_coarse}]
\label{prop:gaussian_randomized_spectroscopy} For every fixed list of rank-$L$ tensors $K_1,\ldots,K_M$ satisfying $\|K_j\|_\infty\leq\kappa$, the vectors returned by Algorithm~\ref{algo:randomized_spectroscopy_coarse} satisfy
\begin{equation}
    \Pr\left[\max_{j=1,\ldots,M}\left|K_j[U_1,\ldots,U_L]-K_j[A^{\otimes L}]\right|\leq\epsilon\right]\geq1-\delta.
\end{equation}
The total interrogation time is bounded as
\begin{equation}
    T\leq O_L \left(\max\left\{p\left(\frac{\kappa}{\epsilon}\right)^2,(2N)^{1-1/L}p^{1/L}\left(\frac{\kappa}{\epsilon}\right)^{2/L}\right\}\left(\Delta_\omega^{-1}+\sqrt{N\log\left(\frac{4CNR}{\delta}\right)}\right)\right),
\label{eq:gaussian_spectroscopy_time_cap}
\end{equation}
where $C$ is the numerical constant from Algorithm \ref{algo:gaussian_response_compiler}. Suppressing constants,
\begin{equation}
     T\leq\begin{cases}
         O \left(\sqrt{N}\left(\frac{\kappa}{\epsilon}\right)^2\log\left(\frac{M}{\delta}\right)\right), \qquad &\textnormal{linear spectroscopy }(L=1) \\
         O\left(N^{3/2-1/L}\left(\frac{\kappa}{\epsilon}\right)^{2/L}\log^{1/L}\left(\frac{M}{\delta}\right)\right), &\textnormal{nonlinear spectroscopy } (L\geq 2)
     \end{cases} \label{eq:gaussian_spectroscopy_sample_count}
\end{equation}
\end{prop}
\noindent For presentational clarity, we defer the formal proofs of all RQS performance guarantees to Section~\ref{sec:deferred_rqs_perf_guarantees}.

\subsubsection{High-precision randomized spectroscopy}

Gaussian randomized spectroscopy gains efficiency by probing many spectral amplitudes simultaneously, but improving precision requires averaging over increasingly many measurements. At high precision, $\epsilon/\kappa \lesssim 1/\sqrt{N}$, it becomes more efficient to exploit coherent phase accumulation, improving the precision dependence at the expense of worse scaling with $N$. As we show, this tradeoff is fundamental to conventional memory-free spectroscopy.

By amplifying a coordinate response by a factor $h$, an order-one uncertainty in the measured phase translates into an $O(1/h)$ uncertainty in the corresponding spectral amplitude, achieving Heisenberg-limited precision scaling. The resulting phase ambiguity modulo $2\pi$ is resolved through phase unwrapping (Algorithm~\ref{algo:phase_unwrapping}), which yields a preliminary estimate of the spectrum. To estimate general nonlinear observables, we must additionally control the bias and fluctuations of the spectral estimates. We accomplish this through high-gain residual sampling (Algorithm~\ref{algo:high_gain_residual_sampling}) and bounded sine inversion (Algorithm~\ref{algo:sine_inversion}), producing noisy vector estimates suitable for subsequent tensor contractions. Importantly, all sensing measurements can be specified in advance, and consist of different dynamical decoupling sequences.

We now state the complete high-precision protocol, which combines these subroutines to construct $L$ vector estimates $U_1,\ldots,U_L$ for downstream estimation of spectroscopic observables. In the following algorithm, $\sigma$ controls the size of the fluctuations in each coordinate, while $m$ sets the number of residual measurements per coordinate used to construct each vector.

\begin{algorithm}[h!]
    \DontPrintSemicolon
    \caption{High-precision randomized spectroscopy}
    \label{algo:randomized_spectroscopy_fine}
    \KwIn{Known frequencies $\{\omega_j\}_{j=1}^N$, tensor rank $L$, list size $M$, accuracy $0<\epsilon<\kappa$, norm bound $\kappa$, and failure probability $\delta$.}
    \KwOut{Vectors $U_1,\ldots,U_L\in\mathbb R^{2N}$.}

    Let $p$ be the smallest even integer satisfying $p\geq\max\{2,\log(4M/\delta)\}$, and choose
    \begin{equation}
        \sigma=\frac{1}{16L}\min\left\{
        \frac{\epsilon}{\kappa\sqrt p},
        (2N)^{-(L-1)/(2L)}p^{-1/(2L)}
        \left(\frac{\epsilon}{\kappa}\right)^{1/L}
        \right\}.
    \end{equation}

    Set $\epsilon_{\rm inv}=\min\{1/2,\sqrt{p/(2N)}\}$ and $m=\left\lceil C\log^2(2/\epsilon_{\rm inv})\right\rceil$, with $C$ a sufficiently large numerical constant.\;

    Set $J=\left\lceil\log_2(4/\sigma)\right\rceil$, $H=2^J/4$, and
$r_{\rm p}=\left\lceil512\log\left(\frac{32N(J+1)}{\delta}\right)\right\rceil$.
Set $R=2N[2(J+1)r_{\rm p}+Lm]$ and $\epsilon_{\rm syn}=\delta/R$.\;

    Run Algorithm~\ref{algo:phase_unwrapping} with inputs $\{\omega_j\}_{j=1}^N,J,\epsilon_{\rm syn},\delta$ to obtain $a_0$.\;

    For each $i=1,\ldots,L$, run Algorithm~\ref{algo:high_gain_residual_sampling} with inputs $\{\omega_j\}_{j=1}^N$, gain $H$, $a_0$, shot count $m$, and compilation accuracy $\epsilon_{\rm syn}$, using independent data in each call. Denote the returned values by $W_{i,\mu,r}$ for $\mu=1,\ldots,2N$ and $r=1,\ldots,m$.\;

    For every $i$ and $\mu$, pass the $m$ recorded values $W_{i,\mu,1},\ldots,W_{i,\mu,m}$ to Algorithm~\ref{algo:sine_inversion} with accuracy $\epsilon_{\rm inv}$, and denote the returned scalar by $S_{i,\mu}$.\;

Return $U_1,\ldots,U_L$ with entries $(U_i)_\mu=(a_0)_\mu+S_{i,\mu}/H$.\;
\end{algorithm}

\begin{prop}[Time complexity of Algorithm \ref{algo:randomized_spectroscopy_fine}]
\label{prop:high_precision_randomized_spectroscopy}
For every fixed list of rank-$L$ tensors $K_1,\ldots,K_M$ satisfying $\|K_j\|_\infty\leq\kappa$ and chosen independently of the acquisition randomness and outcomes, the vectors returned by Algorithm~\ref{algo:randomized_spectroscopy_fine} satisfy
\begin{equation}
    \Pr\left[\max_{j=1,\ldots,M}\left|K_j[U_1,\ldots,U_L]-K_j[A^{\otimes L}]\right|\leq\epsilon\right]\geq1-\delta.
\end{equation}
The total interrogation time is bounded as
\begin{equation}
    T\leq O_L\left(
    N\left[
    \Delta_\omega^{-1}
    +
    \max\left\{
    \frac{\kappa\sqrt p}{\epsilon},
    (2N)^{(L-1)/(2L)}p^{1/(2L)}
    \left(\frac{\kappa}{\epsilon}\right)^{1/L}
    \right\}
    \right]
    \log^2\left(\frac{CNp\kappa}{\epsilon\delta}\right)
    \right),
    \label{eq:high_precision_spectroscopy_time_cap}
\end{equation}
where $C$ is a numerical constant. In the high-precision regime $2N(\epsilon/\kappa)^2\leq p\leq2N$, suppressing constants and the remaining logarithmic factors gives
\begin{equation}
    T\leq\widetilde O_L\left(
    N\frac{\kappa}{\epsilon}\sqrt{\log\left(\frac{4M}{\delta}\right)}
    \right).
    \label{eq:high_precision_spectroscopy_scaling}
\end{equation}
\end{prop}
\noindent The proof of this performance guarantee is found in Section~\ref{sec:deferred_rqs_perf_guarantees}.

\subsubsection{Reusable Qdyne algorithm for large noise}

The preceding algorithms rely on coherently synthesizing spectrally selective responses. This becomes inefficient once the sensor coherence time is shorter than the inverse spectral spacing. Indeed, distinguishing two neighboring frequencies separated by $\Delta_\omega$ using a single coherent pulse sequence requires an interrogation of duration at least $O(\Delta_\omega^{-1})$, since elementary linear-response theory illustrates that a sequence of duration $\tau$ cannot produce spectral features appreciably narrower than $1/\tau$. When $\gamma\gtrsim\Delta_\omega$, however, such an interrogation has depolarizing contrast at most $e^{-O(\gamma/\Delta_\omega)}$. Thus the sensor loses coherence before a conventional DD or RQS pulse sequence can become spectrally selective, and attempting to run such protocols leads to an exponential degradation in performance. This noise regime is a well-studied obstruction for amplitude estimation as well as frequency estimation.

The workaround we present in this section is closely related to Qdyne spectroscopy, which was introduced precisely for such settings~\cite{schmitt_submillihertz_2017, boss_quantum_2017, glenn_high-resolution_2018}. Rather than preserving a quantum sensor for time $1/\Delta_\omega$, Qdyne performs many short measurements using freshly initialized sensors and tracks the absolute wall-clock time at which each measurement and initialization is made. By tracking these times closely, Qdyne spectroscopy forgoes spectral filtering and instead fits the measured responses to the wall-clock times to reconstruct the waveform. Effectively, a classical clock retains the phase coherence that a very noisy quantum probe cannot, at the cost of standard quantum limited performance. As a simple example for how Qdyne spectroscopy works, note that signals oscillating at two frequencies $\omega_j$ and $\omega_k$ separated by $\Delta_\omega$ produce a relative phase $(\omega_j-\omega_k)t$ after time $t$, so timestamped measurements taken over a total interval of order $1/\Delta_\omega$ can distinguish them even when no individual quantum state remains coherent for that long.

We use the same idea here. In the following protocol, each shot is deliberately too short to resolve individual frequencies and instead gives a broadband measurement of $A^Tf(t)$. Since the frequencies are known, the absolute measurement times specify a known classical sensing matrix. We then reconstruct the spectral amplitudes by comparing the measurement outcomes and collected timestamps to this matrix and performing an appropriate closed-form regression. 

\begin{algorithm}
    \DontPrintSemicolon
    \caption{Reusable Qdyne spectroscopy}
    \label{algo:large_noise_rqs}
    \KwIn{Known frequencies $\{\omega_j\}_{j=1}^N$, noise strength $\gamma\geq\Delta_\omega$, accuracy $0<\epsilon<\kappa$, norm bound $\kappa$, list size $M$, and failure probability $\delta$.}
    \KwOut{A reusable vector $\widehat A\in\mathbb R^{2N}$ from which any later-supplied linear observable can be estimated.}

    Set $\tau=c/(N\Delta_\omega+\gamma)$ for a sufficiently small numerical constant $c$, and set
    \[
        R=
        \left\lceil
        C\frac{1}{\tau^2}
        \left(\frac{\kappa}{\epsilon}\right)^2
        \log\left(\frac{4M}{\delta}\right)
        \right\rceil
    \]
    for a sufficiently large numerical constant $C$.\;

    \For{$r=1,\ldots,R$}{
        Set $I_r=[(r-1)\tau,r\tau]$ and $t_r=(r-\frac12)\tau$.\;

        Prepare a fresh sensor in $\ket{+}$, interrogate the signal throughout $I_r$, and measure $Y$, recording the outcome $B_r\in\{-1,+1\}$ and its known center time $t_r$.\;
    }

    Compute
    \[
        b_r=\frac{1}{\tau}\int_{I_r}f(t)\,dt,
        \qquad
        G=\frac{1}{R}\sum_{r=1}^R b_rb_r^T
    \]
    and form
    \[
        \widetilde A
        =
        \frac{e^{\gamma\tau}}{R\tau}
        G^{-1}
        \sum_{r=1}^R B_rb_r.
    \]

    \eIf{$\epsilon/\kappa\geq1/\sqrt N$}{
        Return $\widehat A=\widetilde A$.\;
    }{
        Define the known map
        \[
            \Phi(x)
            =
            \frac{1}{R\tau}
            G^{-1}
            \sum_{r=1}^R
            b_r\sin(\tau b_r^Tx).
        \]
        Starting from $A^{(0)}=\widetilde A$, iterate
        \[
            A^{(s+1)}
            =
            A^{(s)}
            +
            \widetilde A-\Phi(A^{(s)})
        \]
        until $\|A^{(s+1)}-A^{(s)}\|_2\leq\epsilon/(10\kappa)$, and return $\widehat A=A^{(s+1)}$.\;
    }
\end{algorithm}

Although each $b_r$ in Algorithm~\ref{algo:large_noise_rqs} is a broadband response, its entries carry the phases $\cos(\omega_jt_r)$ and $\sin(\omega_jt_r)$. As $r$ increases, these known phases rotate at different rates for different frequencies. The classical matrix $G$ therefore becomes well-conditioned over a long sequence of short measurements without requiring any individual sensor to remain coherent over the corresponding time.

\begin{prop}[Time complexity of Algorithm~\ref{algo:large_noise_rqs}]
\label{prop:large_noise_rqs}
For $\gamma\geq\Delta_\omega$ (sensor coherence time shorter than neighboring frequency resolution), Algorithm~\ref{algo:large_noise_rqs} produces a reusable vector such that, for any later-supplied fixed list $K_1,\ldots,K_M$ with $\|K_j\|_2\leq\kappa$,
\begin{equation}
    \Pr\left[
    \max_{j=1,\ldots,M}
    |K_j^T\widehat A-K_j^TA|
    \leq\epsilon
    \right]
    \geq1-\delta.
\end{equation}
Its total interrogation time is
\begin{equation}
    T
    =
    \widetilde O\left(
    (N\Delta_\omega+\gamma)
    \left(\frac{\kappa}{\epsilon}\right)^2
    \right).
    \label{eq:large_noise_rqs_time}
\end{equation}
\end{prop}
\noindent The proof is given in Section~\ref{sec:deferred_rqs_perf_guarantees}.

\subsection{Optimality and signal-aligned noise}\label{sec:rqs_noise_optimality}
Randomized quantum spectroscopy has a particularly attractive practical property: it is a data collection procedure that does not require specifying observables of interest beforehand. As a result, the collected classical dataset can be reused downstream for various estimation tasks. In Section~\ref{sec:nonadaptive_lower_bounds}, we prove that our algorithm is exactly optimal among all possible strategies that enable this kind of post-hoc learning. In particular, Theorem~\ref{thm:nonadaptive_posthoc_lower_bound} states that any conventional spectroscopy protocol for AQS compatible with post-hoc estimation must use a total interrogation time 
\begin{equation}
T\geq
\begin{cases}
\Omega\left(
\min\left\{
\sqrt N\left(\frac{\kappa}{\epsilon}\right)^2\log(M),
N\frac{\kappa}{\epsilon}\sqrt{\log M}
\right\}\right), & L=1,\\[2mm]
\Omega_L\left(
\max\left\{
N^{3/2-1/L}
\left(\frac{\kappa}{\epsilon}\right)^{2/L}(\log M)^{1/L},
N\,\frac{\kappa}{\epsilon}\sqrt{\log M}
\right\}\right), & L\geq2.
\end{cases}
\end{equation}
Notice that this exactly matches the time complexity of RQS given in Theorem~\ref{thm:randomized_spectroscopy_upper_bound}, up to logarithmic factors. RQS is therefore the optimal post-hoc conventional protocol for the standard formulation of AQS. 

A natural question now arises: if the protocol were allowed prior knowledge of the observables of interest (as is often the case in practice), could better performance be achieved? Remarkably, we find that RQS remains optimal for large, practically important parameter regimes, even when adaptive spectroscopy protocols are allowed. In the parameter regimes where RQS is suboptimal, we develop simple adaptive pulse sequence protocols that saturate the remaining bounds; these are given in Section~\ref{sec:small_M_optimal}. 

In particular, we show in Section~\ref{sec:adaptive_conventional_lower_bound} that any conventional protocol with prior knowledge of the observables still requires total sensing time
\begin{equation}
    T \geq \Omega_L\left(\min\left\{ N^{1/3} M^{2/3} \left( \frac{\kappa}{\epsilon} \right)^{4/3}, \sqrt{N M} \, \frac{\kappa}{\epsilon} \right\}\right) \ .
\end{equation}
In the $M\sim N$ regime relevant to e.g. waveform reconstruction, power spectrum learning, and other tasks in which the entire spectrum is of interest, this scales as $N\kappa/\epsilon$. Thus for the most relevant observables, $L=1,2$, RQS is optimal even relative to adaptive protocols. 

Next we ask how the optimality of RQS changes when signal-aligned noise is incorporated. We now show that when noise is incorporated, RQS returns to standard quantum limited performance, but remains optimal relative to all adaptive strategies over even broader parameter regimes than in the noiseless case. Before stating the result, let us understand how noise is parameterized, and connect our understanding to DC quantum metrology. We work with fully isotropic noise, to illustrate that our results are not the choice of a particular noise axis, while still incorporating constant-strength noise along the signal direction. For noise of strength $\gamma >0$, our sensor state $\rho$ evolves according to the Lindblad master equation
\begin{equation}
    \dot \rho = -\frac{i}{2}[A^T f(t) Z, \rho] + \frac{\gamma}{4}\sum_{J\in{X, Y, Z}} (J\rho J - \rho) \ .
\end{equation}
This evolution describes interrogation under sensor depolarization with coherence time $1/\gamma$. Let us first recall what such noise does in the well-studied setting of single-parameter DC metrology, which will greatly inform our intuition for AC sensing. Consider the signal Hamiltonian $H = bZ$, where $b>0$ is scalar. Standard Ramsey interrogation for time $\tau$, beginning in the state $\ket{+}$, creates a state $\propto e^{-ib\tau}\ket{0} + e^{ib\tau}\ket{1}$ with relative phase $2b\tau$, so that measuring $X$ has outcome probabilities $p_{\pm} = (1\pm \cos(2b\tau))/2$. Under depolarizing noise, the contrast is modified to $\pm e^{-\gamma\tau}\cos(2b\tau)$, resulting in a Fisher information scaling as $F \sim \tau^2 e^{-2\gamma \tau}$. Simply optimizing this Fisher information shows that it stops increasing around $\tau \sim 1/\gamma$.

When the desired estimation accuracy is $\epsilon$, a noiseless experiment would ideally take $\tau\sim 1/\epsilon$. This argument qualitatively illustrates that noise is asymptotically negligible up until $\gamma\sim \epsilon$, because the optimal scaling does not change. In this simple, single-sensor example, the crossover from the Heisenberg limit ($\tau\sim 1/\epsilon$) to the SQL ($\tau\sim 1/\epsilon^2$) occurs at $\gamma\sim \epsilon$. 

A second way to obtain Heisenberg-limited scaling is to leverage entanglement. Suppose $S$ sensors are entangled and prepared in $(\ket{0}^{\otimes S} + \ket{1}^{\otimes S})/\sqrt 2$. It follows that the noiseless Fisher information scales as $S^2\tau^2$, while the coherence decay now modifies the Fisher information to $\sim S^2\tau^2 e^{-S\gamma\tau}$. Note that \textit{both} $S$ and $\epsilon$ now couple to the noise, $S$ literally, and $\epsilon$ implicitly through the final desired precision. A second crossover point now follows: the entangled Fisher information is only approximately preserved up to $\gamma \sim 1/(S\tau)$. Fundamentally, we note that the two mechanisms that can produce Heisenberg scaling (longer coherent interrogation and shorter, entangled sensing) are not independent, because the key coherence resource is the product $S\tau$, which determines the crossover point for when noise becomes non-negligible. Together, we obtain the crossover points $\gamma \sim \epsilon$ and $\gamma \sim 1/\sqrt{S}$, beyond which SQL scaling is unavoidable. 

Now let us lay out the same picture for AC sensing and RQS. In our setting, we are concerned with two independent parameters: $N$ (the number of amplitudes) and $\epsilon/\kappa$ (the effective normalized precision); sensor count $S$ can also be introduced by a similar argument as in the DC case, but we neglect it here. RQS, like canonical DC metrology, can leverage two different approaches to achieve optimal scaling. In the high-precision regime $\epsilon/\kappa \lesssim 1/\sqrt{N}$, recall that our Algorithm~\ref{algo:randomized_spectroscopy_fine} makes high-gain phase-unwrapping measurements, which require interrogation time at most $\sim \kappa/\epsilon$. Like DC metrology, this mechanism for optimal ``Heisenberg-limited" scaling relies on long, coherent evolution. Because the mechanism is the same, noise changes the scaling identically: high-gain RQS under depolarizing noise scales as
\begin{equation}
    T\sim N\max\left\{\frac{\kappa}{\epsilon}, \gamma \frac{\kappa^2}{\epsilon^2}\right\} \ ,
\end{equation}
with the crossover point $\gamma \sim \epsilon/\kappa$ denoting when Heisenberg scaling is lost. The second mechanism behind RQS appears in the standard-precision regime, where Algorithm~\ref{algo:randomized_spectroscopy_coarse} uses Gaussian-random pulse sequences. Here, the use of the quadratically more efficient Gaussian pulse compiler allows each measurement to occur in time $\sim \sqrt{N}$, rather than the naive $\sim N$ scaling required for arbitrary deterministic sequences, resulting in the scaling $\sqrt{N}\kappa^2/\epsilon^2$. Using the same argument from the DC case, coherence is roughly unaffected only when $\gamma\sqrt{N}\lesssim 1$, giving us a second noise crossover point of $\gamma \sim 1/\sqrt{N}$. Thus, in the standard-precision regime, RQS scales as
\begin{equation}
    T\sim \frac{\kappa^2}{\epsilon^2}\max\!\left\{\sqrt{N}, N\gamma\right\} \ .
\end{equation}
Thus, despite the fact that $N$, as a measure of spectral complexity, is distinct from $n_S$ as a measure of available entanglement, the qualitative behavior is the same: there are two mechanisms for Heisenberg scaling, and two crossover points beyond which noise forces SQL scaling in the respective parameters.

Having understood the core intuition, let us formally state the noise-aware time complexity of RQS in the linear spectroscopy setting. We take $\gamma \geq \epsilon/\kappa$, because smaller noise can essentially be ignored and noiseless scaling persists up to constant overhead. 

\begin{prop}
    For $\gamma > \epsilon/\kappa$, running the best of the three RQS subroutines above yields a time complexity of 
    \begin{equation}
        T \leq \widetilde O\left(\begin{cases}
            \min\left\{\frac{N\kappa}{\epsilon}\left(1+\gamma\frac{\kappa}{\epsilon}\right), (\sqrt{N} + N\gamma)\frac{\kappa^2}{\epsilon^2}\right\}, \ &\gamma\lesssim \Delta_\omega \\
            (N\Delta_\omega + \gamma)\frac{\kappa^2}{\epsilon^2}, &\gamma \gtrsim \Delta_\omega
        \end{cases}\right)
    \end{equation}
\end{prop}
We give the formal proof in Section~\ref{sec:deferred_rqs_perf_guarantees}, but the main idea comes from the discussion above. In the large-noise regime $\gamma \gtrsim \Delta_\omega$ the scaling simply comes from running reusable Qdyne spectroscopy (Algorithm~\ref{algo:large_noise_rqs}). When noise is smaller and precision is standard, we obtain the $(\sqrt{N} + N\gamma)\frac{\kappa^2}{\epsilon^2}$ scaling from running Gaussian randomized spectroscopy. Finally, when precision is high, we run the high-gain subroutine to obtain $\frac{N\kappa}{\epsilon}\left(1+\gamma\frac{\kappa}{\epsilon}\right)$. The addition of constant-strength noise, as in canonical metrology, degrades Heisenberg-limited protocols to the standard quantum limit, which in this case corresponds to linear-in-$N$ and quadratic-in-$1/\epsilon$ scaling. Still, we now show that RQS saturates the optimal noise-aware scaling for conventional metrology by stating the analog of the Standard Quantum Limit for spectroscopy. This fundamental lower bound applies to conventional \textit{adaptive} protocols in the same noise model, and allows us to understand why noise actually broadens the regime in which RQS achieves optimal performance.
\begin{theorem}
    For $\gamma \geq \epsilon/\kappa$, any adaptive conventional spectroscopy protocol that estimates $\Omega(N)$ observables with norm bounded by $\kappa$ to within precision $\epsilon$ requires total time
    \begin{equation}
        T \geq \Omega\!\left(\max\left\{\gamma, \frac{N\gamma\Delta_\omega}{\gamma + \Delta_\omega}\right\}\cdot \frac{\kappa^2}{\epsilon^2}\right) \ .
    \end{equation}
    This applies to protocols with arbitrary single-qubit control.
\end{theorem}
The full statement and proof are given in Section~\ref{sec:noisy_conventional_lower_bound}. Putting together upper and lower bounds, we have the following clean picture. For any number of observables and any noise strength $\gamma>\epsilon/\kappa$, the time complexity of linear spectroscopy is
\begin{equation}
    T = \widetilde \Theta\left(\begin{cases}
        N\gamma\frac{\kappa^2}{\epsilon^2}, \qquad &\epsilon/\kappa \lesssim \gamma \lesssim \Delta_\omega \\
        \max\{N\Delta_\omega, \gamma\}\frac{\kappa^2}{\epsilon^2}, \qquad &\gamma \gtrsim \Delta_\omega
    \end{cases}\right) \ .
\end{equation}
The lower bound holds against adaptive strategies with prior knowledge of all observables and arbitrary, possibly noncommuting controls, but it is saturated by the subroutines of RQS, which each only require commuting pulse sequences and support post-hoc estimation. 

This lower bound pertains to tasks with many observables. However, when noise and the number of observables are small, an adaptive protocol can take advantage of known observable structure and longer coherent evolution to design more efficient measurement strategies than RQS. We address this regime with matched spectroscopy in Section~\ref{sec:small_M_optimal}.

\subsection{Advantages over existing approaches} \label{sec:rqs_advantages}

We have rigorously established RQS as the optimal strategy for conventional AC sensing, across linear and nonlinear tasks. The breadth of the AQS framework suggests that this optimality guarantee carries over to a number of concrete, well-studied sensing tasks, such as waveform reconstruction, matched filtering, lock-in detection, and frequency search. Here, we compare RQS to well-studied and often utilized approaches for such tasks, to demonstrate how its optimality translates into practical gains.

Conventional sensing techniques largely fall into two categories: frequency-selective and multiplexed. Frequency-selective sensing protocols tailor their linear response to accurately characterize a particular frequency mode within a potentially broader band; the most commonly-used family of protocols is dynamical decoupling spectroscopy, with additional modifications made for decoherence mitigation. In contrast, broadband tasks such as waveform reconstruction can benefit from protocols that are simultaneously sensitive to different frequency modes. In multiplexed settings, the standard strategy is to design a set of sensor response vectors that are supported on many frequencies, but are mutually orthogonal in a chosen basis. By doing this, each shot can contain useful information about various frequencies in the band, while different shots contain strictly complementary information. In classical signal processing and conventional quantum spectroscopy, a common multiplexed control strategy uses Walsh-Hadamard matrices to define the filter function basis because of their natural compatibility with the bang-bang control available in digital electronics or quantum pulse sequences~\cite{Hayes_2011, de_lange_single-spin_2011, magesan_compressing_2013, cooper_time-resolved_2014, degen_quantum_2017}. We briefly review how Walsh-based waveform reconstruction works in the examples below.

Let us now see how RQS enables practical advantages over these common frequency-selective and multiplexed approaches with concrete examples. At the end of the section, we will offer quantitative intuition for this universally optimal (quantum computation-free) performance.

\subsubsection{RQS speedups for search}
Locating a single-tone signal within a broad band is a fundamental sensing task in many experimental settings, such as searches for axionic dark matter or NMR spectroscopy. The most basic formulation of the task in AQS language is as follows.

Consider a frequency band $W$ populated by many frequencies $\omega_1, ..., \omega_N$, each corresponding to some potential signal of interest. In the search for axionic dark matter, for instance, such a band could span dozens or hundreds of MHz, and be populated by millions of potential lines, and the goal is to test each line for whether a nonzero amplitude is present or absent. The full amplitude vector $A = (p_1,...p_N, q_1,...,q_N)^T$ therefore has $p_j, q_j = 0$ for every $j$ except a particular marked index $J$, and the AQS task is to estimate all linear functionals $A^T(e_j + e_{j+N})$ (where $e_j$ is the unit vector which has entry $1$ at index $j$ and $0$ everywhere else). If the signal has amplitude at least $2\epsilon$, solving this AQS task to precision $\epsilon$ and locating the bin based on which of the $N$ functionals has value $>\epsilon$ successfully solves the search task with high success probability.

For search tasks of this form, a simple strategy is to apply a dynamical decoupling pulse sequence tailored to each frequency $\omega_j$ (repeating for in-phase and out-of-phase quadratures), and to learn each amplitude to precision $\epsilon$. Since an $\epsilon$-accurate DD sequence takes time $T\sim 1/\epsilon$, and a sweep must estimate every frequency equally accurately, DD takes time $O(N/\epsilon)$.

By instead using Gaussian-random pulse sequences, we can do better. Let $u_J = \cos(\phi) e_J + \sin(\phi)e_{J+N}$ for some phase $\phi$, and suppose that the true signal has a spectrum vector $A = 2\epsilon u_J$. We now run a simplified form of Gaussian RQS that does not require calibration. First sample $g\sim\mathcal{N}(0,\mathds{1}_{2N})$, and use Algorithm~\ref{algo:gaussian_response_compiler} to compile the response vector $v = g/(2\epsilon)$. This causes the sensor to accumulate phase $v^TA = g^Tu_J$, and $Y$-basis measurements, which result in outcomes $B\in \{\pm1 \}$, have distribution $\mathbb E[B|g] = \sin(g^T u_J)$. As usual, store the product $Bg$. Repeating this $R$ times, we collect $R$ vectors $Z_r = B_r g_r, r\in\{1,2,...,R\}$.

The mean of the collected vectors then selects $u_J$. Since $\mathbb E[Z_r|g_r] = g_r\sin(g_r^T u_J)$, we can use Stein's identity, which says that for a function $f(x)$, 
\begin{equation}
    \mathbb E_g[g f(g^T u)] = u \,\mathbb E[f'(G)], \qquad G\sim \mathcal{N}(0, 1),
\end{equation}
to obtain that $\mathbb E[Z] = u_J \mathbb E[\cos G] = e^{-1/2}u_J$. This follows from taking $f(x) = \sin(x)$. Up to a fixed multiplicative factor, the mean of our collected outcomes is an unbiased estimator for $u_J$. Moreover, the empirical mean $\overline Z = \frac{1}{R}\sum_{r=1}^R Z_r$ is a sum of signed Gaussian random vectors, so each coordinate of each $Z_r$ has constant sub-Gaussian magnitude. Gaussian concentration and a union bound then give
\begin{equation}
    \textnormal{Pr}\left[\|\overline Z - \mathbb E[Z]\| > t\right]\leq 2CNe^{-cRt^2},
\end{equation}
for some numerical constants $c, C > 0$. We only need our estimate to be within a constant factor of the mean with high probability, so letting $t = e^{-1/2}/4$ gives that $R = O(\log N/\delta)$ is a sufficient number of shots to achieve success probability at least $1-\delta$. From the Gaussian compiler performance guarantee, each shot takes time $\widetilde{O}(\sqrt{N}/\epsilon)$. As a result, this simple protocol solves the search task in time $O(\sqrt{N}/\epsilon)$. 

\begin{observation}[RQS vs. dynamical decoupling speedup for conventional search]
    Randomized quantum spectroscopy detects a single-tone signal of amplitude $\sim \epsilon$ in a band containing $N$ frequencies using total time $\sim \sqrt{N}/\epsilon$. Meanwhile, sweeping the band with any frequency-selective dynamical decoupling protocol requires time $\sim N/\epsilon$. RQS thus enables a practical quadratic speedup for conventional search.
\end{observation}

Next, we consider a generalization of the conventional search task that is common in digital signal processing. Rather than analyzing a signal by sinusoidal Fourier frequency components, Walsh analysis reconstructs time-varying fields in a square-wave basis compatible with digital filters and has been used for magnetic field sensing with spin systems \cite{cooper_time-resolved_2014}. Rather than writing $B(t)$ as a sum over sine and cosine components, we can write it as a sum of Walsh functions, which also form a complete basis if the function is discrete. A Walsh function $W_k(x)$ is specified by a finite integer $m$, which can always be written using the binary expansion $\sum_{k=0}^{\infty} m_k 2^k$. The resulting function is $W_m(x) = \prod_{k\geq 0} (-1)^{g_k(m) \lfloor 2^{k+1}x\rfloor}$; this realizes a square wave which alternates signs $m$ times between $x\in[0, 1]$. In signal processing, one fixes a maximum $K$ for the basis of Walsh functions used, which sets the temporal resolution. For a time window $[0, T]$, this choice defines the basis of Walsh functions
\begin{equation}
    f_{\rm Walsh}(t) = (W_0(t/T), ..., W_{K-1}(t/T)) \ .
\end{equation}
Given appropriate time resolution, we can write our usual signal in two different representations: $B(t) = A^Tf_{\rm Fourier}(t) \approx \beta^T f_{\rm Walsh} (t)$. Walsh analysis can be understood as AQS in this basis.

A simplified Walsh-analysis search task can be stated as follows: given a signal whose Walsh-basis vector is $\beta = 2\epsilon e_J$, determine the index $J$. This is precisely the Walsh-basis equivalent of the previous Fourier-basis search task, and we use it to compare RQS to well-studied digital multiplexing protocols. 

The standard Walsh-Hadamard spectroscopy protocol measures these coefficients separately by applying a digital filter corresponding to each Walsh function. Specifically, applying a sequence of $\pi$ pulses at the sign changes of $W_m(t/T)$ causes the sensor to accumulate phase
\begin{equation}
    \varphi_m = \int_0^T W_m(t/T) B_A(t)dt = T\beta_m,
\end{equation}
where $\beta_m$ is the $m$-th Walsh coefficient. To see how these measurements reconstruct the waveform, divide $[0,T]$ into $K$ equal time bins, and denote the average field in bin $\ell$ by $b_\ell$. Since each Walsh function is constant on these bins, we have
\begin{equation}
    \beta_m = \frac{1}{K}\sum_{\ell=0}^{K-1} H_{m\ell}b_\ell,\qquad H_{m\ell}=W_m((\ell+1/2)/K).
\end{equation}
The matrix $H$ is a Walsh-Hadamard matrix satisfying $HH^T=K\mathds{1}$, so measuring all $K$ Walsh coefficients allows us to reconstruct the time-bin averages by applying the inverse Hadamard transform. This is the conventional Walsh reconstruction approach used in quantum sensing \cite{cooper_time-resolved_2014}.

For our search problem where $\beta=2\epsilon e_J$, a Walsh pulse sequence corresponding to $W_m$ accumulates zero phase unless $m=J$. Thus, as in frequency-selective DD spectroscopy, the conventional Walsh approach must search through the possible channels individually. Even allowing Heisenberg-limited sensing of each coefficient, distinguishing a zero coefficient from one of magnitude $2\epsilon$ requires interrogation time $\Omega(1/\epsilon)$. Since there are $K=\Theta(N)$ possible channels, a Walsh-Hadamard search requires total interrogation time $\Omega(N/\epsilon)$. 

We can again do better using Gaussian-random pulse sequences. Let $V$ denote the known linear transformation from Fourier amplitudes to Walsh coefficients, so that $\beta=VA$. We sample $g\sim\mathcal N(0,\mathds{1}_K)$, and use Algorithm~\ref{algo:gaussian_response_compiler} to compile the response vector
\begin{equation}
    v=\frac{V^Tg}{2\epsilon}.
\end{equation}
This protocol can be interpreted as RQS in the Walsh basis. Under our promise $\beta=2\epsilon e_J$, RQS causes the sensor to accumulate phase
\begin{equation}
    v^TA=\frac{g^TVA}{2\epsilon}=g^Te_J=g_J.
\end{equation}
Measuring in the $Y$ basis gives an outcome $B\in\{\pm1\}$ satisfying $\mathbb E[B|g]=\sin(g_J)$. As in the Fourier-search protocol, we retain the classical vector $Z=Bg$. Using Stein's identity, we obtain
\begin{equation}
    \mathbb E[Z]=\mathbb E[g\sin(g_J)]=e^{-1/2}e_J.
\end{equation}
Thus the expected Gaussian score points directly toward the occupied Walsh coefficient; the effect of RQS is therefore independent of the orthonormal basis chosen. Identical analysis then leads us to interrogation time $\widetilde O(\sqrt N/\epsilon)$.

\begin{observation}[RQS advantage for search is basis-independent]
    For locating a single Walsh sequence, RQS achieves a quadratic speedup over Walsh-Hadamard search. More generally, RQS achieves a search speedup in any orthonormal basis of signal modes over protocols that are mode-selective. It is the optimal random frame in which to search any signal basis.
\end{observation}

\subsubsection{RQS speedups for lock-in detection and phase matching}

A second important task in sensing and signal processing is lock-in detection. In general, the phase acquired during a sensing protocol may be difficult to characterize because of a relative phase difference between the two frequency quadratures. The standard method for lock-in detection involves applying DD sequences at incrementally shifted phases until the acquired signal is maximized, indicating that the pulse sequence is locked to one signal quadrature. More generally, AC signals can contain many modes whose respective quadrature components are out of phase by different amounts. Matching joint patterns in such phases $(\phi_1,...\phi_N)$ is useful for radar and sonar, multitone spectroscopy, magnetic resonance, and gravitational wave detection~\cite{bax_two-dimensional_1989,coddington_time-domain_2010,kotler_single-ion_2011, abbott_observation_2016, boss_quantum_2017}.

To illustrate this task, suppose that only $k$ out of our $N$ candidate frequencies have nonzero amplitudes, but we do not know which frequencies are occupied or their relative phases. We take the occupied frequencies to have equal amplitudes $1/\sqrt{k}$, so that $\|A\|_2 = 1$. More precisely, for some unknown set $S \subseteq [N]$ with $|S| = k$, we have
\begin{equation}
    A = \frac{1}{\sqrt{k}}\sum_{j\in S}\left(\cos(\phi_j)e_j+\sin(\phi_j)e_{j+N}\right),
\end{equation}
where $\phi_j$ is the phase of the $j$-th frequency relative to a common reference clock. Our task is to identify the occupied frequencies and estimate each phase $\phi_j$ to some angular precision $\vartheta$. This is a natural broadband generalization of conventional lock-in detection, since we want to determine the relative phase of every appreciable frequency constituent, rather than just one.

A conventional frequency-selective DD approach would scan through the $N$ frequencies, estimating both quadratures at each frequency. Since the occupied modes have amplitude $1/\sqrt{k}$, resolving whether a particular frequency is occupied requires quadrature accuracy of order $1/\sqrt{k}$. Even allowing Heisenberg-limited estimation of each quadrature, this initial scan takes time $\widetilde O(N\sqrt{k})$. Once the occupied frequencies have been identified, we can refine the estimates of their quadratures. To estimate a phase to angular precision $\vartheta$, we need absolute quadrature accuracy $O(\vartheta/\sqrt{k})$, since the phase uncertainty is the quadrature uncertainty divided by the signal amplitude. Refining each of the $k$ occupied modes therefore costs $\widetilde O(\sqrt{k}/\vartheta)$, giving total interrogation time
\begin{equation}
    T_{\rm DD} = \widetilde O\left(N\sqrt{k}+\frac{k^{3/2}}{\vartheta}\right).
\end{equation}
This is already better than simply estimating all $N$ frequency quadratures to the final desired precision, since the second stage only interrogates frequencies identified as occupied.

Now consider applying RQS instead. We can use our Gaussian-random acquisition to estimate all $2N$ quadratures simultaneously, without knowing which frequencies are occupied. As in our previous examples, sample $g\sim\mathcal N(0,\mathds 1_{2N})$, compile the response vector $v=g$, and measure in the $Y$ basis to obtain an outcome $B\in\{\pm1\}$ satisfying $\mathbb E[B|g]=\sin(g^TA)$. We again store the classical score $Z=Bg$. By Stein's identity,
\begin{equation}
    \mathbb E[Z]=e^{-\|A\|_2^2/2}A=e^{-1/2}A,
\end{equation}
where we used $\|A\|_2=1$. Thus the Gaussian measurement record gives an unbiased estimate of $A$ up to a known constant factor, regardless of which frequencies are occupied or their relative phases.

To determine the interrogation cost, repeat the experiment $R$ times and form $\overline Z=R^{-1}\sum_{r=1}^R Z_r$. Since each coordinate of $Z_r$ has constant sub-Gaussian magnitude, concentration and a union bound over the $2N$ coordinates give
\begin{equation}
    \Pr\left[\|\overline Z-e^{-1/2}A\|_\infty>t\right]\leq 4N e^{-cRt^2}
\end{equation}
for some numerical constant $c>0$ and $0<t<1$. We want to recover the quadratures of each occupied frequency to accuracy $O(\vartheta/\sqrt{k})$, so it suffices to take $t=c'\vartheta/\sqrt{k}$ for a sufficiently small constant $c'>0$. Consequently, choosing
\begin{equation}
    R=O\left(\frac{k}{\vartheta^2}\log\frac{N}{\delta}\right)
\end{equation}
gives the required coordinatewise accuracy with probability at least $1-\delta$.

To see that this is sufficient for phase recovery, define $\widehat A=e^{1/2}\overline Z$. Every unoccupied frequency has estimated quadrature norm $O(\vartheta/\sqrt{k})$, whereas every occupied frequency has estimated quadrature norm $1/\sqrt{k}+O(\vartheta/\sqrt{k})$. For sufficiently small $\vartheta$, we can therefore identify the occupied set simply by retaining those frequency pairs whose estimated norm exceeds $1/(2\sqrt{k})$. Moreover, for every occupied frequency, the estimated phase
\begin{equation}
    \widehat\phi_j=\operatorname{atan2}(\widehat q_j,\widehat p_j)
\end{equation}
differs from $\phi_j$ by at most $O(\vartheta)$, since the estimated quadrature vector is within distance $O(\vartheta/\sqrt{k})$ of a vector with norm $1/\sqrt{k}$. From the Gaussian compiler, each randomized interrogation takes time $\widetilde O(\sqrt N)$, so the total interrogation time is
\begin{equation}
    T_{\rm RQS}=\widetilde O\!\left(\frac{\sqrt N k}{\vartheta^2}\right).
\end{equation}
For constant angular precision $\vartheta=\Theta(1)$, sufficiently small for the preceding phase estimates, and $k\ll N$, the conventional DD scan takes time $\widetilde O(N\sqrt{k})$, whereas RQS takes time $\widetilde O(\sqrt N k)$. Thus RQS offers an improvement by a factor of $\sqrt{N/k}$, up to logarithms. In particular, when only a constant number of frequencies are occupied, we recover the $\sqrt N$ advantage of the single-tone search problem. Conversely, when $k=\Theta(N)$, the advantage disappears, since resolving all $N$ independent phases requires correspondingly more information. In practice, signals of interest are generally sparse in the frequency band of interest, suggesting a meaningful practical speedup for broadband lock-in detection. 

\begin{observation}[RQS advantage for broadband lock-in detection]
    Randomized spectroscopy enables a speedup up to quadratic in the number of frequencies for broadband lock-in detection compared to frequency-selective sensing.
\end{observation}

In such settings, multiplexed pulse sequences may also be used. While conventional Walsh-Hadamard multiplexing suffers the same obstructions as frequency-selective dynamical decoupling, compressed sensing techniques can improve performance \cite{magesan_compressing_2013}. However, particular choices of compressed-sensing bases (like the Walsh functions) can overlap poorly with Fourier modes of interest, and vice versa; RQS remains the optimal approach since Gaussian random pulse sequence control enables optimal estimation without favoring any one mode basis. 

\subsection{Modifications for pulse-imperfection and static noise robustness} \label{sec:rqs_mods}

We have studied the performance of RQS in noiseless settings as well as in the presence of continuous depolarizing noise. We have not studied its integration with pulse sequence protocols designed for other forms of error mitigation such as spin-echo protocols. Here we show that the basic modifications used to improve the robustness of conventional dynamical decoupling protocols \cite{meiboom_modified_1958} can be ported to RQS. 

Error-mitigating pulse modifications for single-spin sensors generally aim to correct the leading-order effects of pulse imperfections, loss of transverse coherence from slowly varying longitudinal fields, and transverse anisotropy. We now discuss two simple modifications that improve robustness to some of these error sources.

\subsubsection{Pulse rotation errors}

First consider systematic errors in the rotation angles of each pulse. Specifically, consider a sensing pulse sequence containing $P$ $\pi$ pulses, each with a small overrotation $\eta$. A pulse about the axis $sX$, for $s=\pm1$, is then implemented as $\widetilde P_s = e^{-is(\pi+\eta)X/2} = P_s e^{-is\eta X/2}.$ When every pulse uses the same rotation axis, these errors can accumulate coherently, producing an accumulated error of order $P|\eta|$ in the overall evolution. Consider a sensor with density matrix $\rho=(\mathds{1}+xX+yY)/2$ (lying in the XY plane) immediately before each pulse. An overrotation by $s\eta$ about $X$, where $s=\pm1$, changes this state according to
\begin{equation}
    \rho \mapsto \rho+\frac{s\eta y}{2}Z-\frac{\eta^2 y}{4}Y+O(\eta^3).
\end{equation}
Describe $\rho$ by a vector on the Bloch sphere. Then, we went from a vector initially pointing along $(x, y, 0)$ to $(x, y, O(\eta))$ (to first order in $\eta$). The next pulse, with overrotation, then creates the coordinate correction $y\rightarrow -y + \eta z$. Since $z = O(\eta)$ from the previous error, the $y$ component is now affected by $O(\eta^2)$. With $P$ pulses, there are $O(P^2)$ opportunities for such second-order errors to appear, leading to $O(P^2\eta^2)$ error in each measurement outcome:
\begin{equation}
    \mathbb E[B|g]=\sin(g^TA)+O(P^2\eta^2).
\end{equation}
Note that this quadratic scaling in $\eta$ occurs due to the alignment of the initial state with the transverse axis; this is precisely the Meiboom-Gill modification \cite{meiboom_modified_1958}. However, since $P$ scales with $N$, this can still be damaging. We can suppress this quadratic dependence on $P$ with a simple modification. Upon running the pulse compiler Algorithm~\ref{algo:gaussian_response_compiler}, simply choose a random sign $s = \pm 1$ for each pulse, and apply the $\pi$-pulse around the Bloch axis $sX$ instead of $X$. Note that a $\pi$-pulse around any axis in the XY plane maps $Z\rightarrow-Z$ exactly, so this modification leads to identical measurement outcomes. However, averaging over these random choices gives
\begin{equation}
    \frac12\sum_{s=\pm1}e^{-is\eta X/2}\rho e^{is\eta X/2}
    =\rho+\sin^2(\eta/2)(X\rho X-\rho).
\end{equation}
Thus the coherent rotation error is converted into a stochastic Pauli error of probability $O(\eta^2)$. These errors can no longer combine across pulses, so the dependence on $P$ is linear. This reduces the total deviation to $O(P\eta^2)$, a quadratic improvement~\cite{wallman_noise_2016, wang_randomization_2019}. This is standard Pauli-twirling error suppression, and applying deterministic, alternating pulses to our randomized pulse spacings confers some of the advantages associated with randomizing pulse phases in ordinary DD pulse trains~\cite{wang_randomization_2019}.

Because of the property that any XY-plane pulse leaves the response unchanged, more sophisticated phase-cycling schemes can be used to suppress error further or to protect against transverse perturbations. We do not perform a rigorous asymptotic analysis here, because the advantages of combining sequences like XY4 or XY8 with RQS are more problem-dependent; however, this combination is possible without any other modification due to the symmetries of RQS, and is a valuable direction for future experimental and theoretical work.

\subsubsection{Robustness against quasistatic longitudinal fields}
Another common source of error is random, quasistatic longitudinal fields acting on the sensor. These fields modify the effective Hamiltonian to
\begin{equation}
    H(t) = (B(t) + \xi)Z + H_c(t),
\end{equation}
where $\xi$ is a mean-zero unknown variable that remains constant across each shot. This modifies the accumulated phase of a length-$\tau$ interrogation by $\xi\int_0^\tau y(t) dt$, where the interaction-frame modulation from $H_c(t)$ is denoted by $y(t)$. If $\xi$ is Gaussian with variance $\sigma^2$, averaging over the noise attenuates the coherence of our measurement by $\exp\left[-\frac{\sigma^2}{2}\left(\int_0^\tau y(t)dt\right)^2\right]$, which can be exponentially severe in $\sigma$.

To mitigate this error, we simply require that for every shot, $\int_0^\tau y(t) dt = 0$. While the modulations that realize Gaussian-random response need not initially satisfy this property, we can impose it as follows. In the Gaussian-random compiler, Algorithm~\ref{algo:gaussian_response_compiler}, we may pad $f(t)$ with a single constant entry of value $1$, to obtain $\overline f = (1, \cos(\omega_1 t),...,\cos(\omega_Nt), \sin(\omega_1 t),...,\sin(\omega_Nt))$. We also append a zero to the sampled Gaussian vector, to give $\overline g = (0, g)$. Constructing the Gram matrix and then running the algorithm as usual results in a pulse sequence whose response on the coordinates $1...N$ is still Gaussian-random, but it also satisfies $\int_0^\tau 1\cdot y(t) dt = 0$. All guarantees of RQS follow, but we exactly cancel the quasistatic error~\cite{hahn_spin_1950, biercuk_dynamical_2011}.

\newpage
\section{Matched Spectroscopy} \label{sec:small_M_optimal}
In this section, we introduce \textit{matched spectroscopy}, which exploits
prior knowledge of the observables to reduce interrogation time by tailoring
the sensing pulse sequences to the specified targets.

\begin{graydefinitionbox}
\begin{theorem}[Matched Spectroscopy]
\label{thm:specified_spectroscopy_upper_bound}
Fix $\kappa>0$, $0<\epsilon<\kappa$, and $0<\delta<1/4$, and suppose the observables are specified before interrogation. There exists a single-spin protocol producing estimates with additive error at most $\epsilon$ and success probability at least $1-\delta$, using interrogation time
\begin{equation}
T_{\rm int}
\leq
\begin{cases}
\widetilde O\!\left(
\min\left\{
N^{1/3}r^{2/3}\left(\frac{\kappa}{\epsilon}\right)^{4/3},
\sqrt{Nr}\frac{\kappa}{\epsilon}
\right\}
\right),
& L=1,\\[2mm]
\widetilde O_L\!\left(
N^{3/2-1/(L-1)}
\left(\frac{\kappa}{\epsilon}\right)^{2/(L-1)}
\right),
& L\geq2,\ M=1,
\end{cases}
\label{eq:specified_spectroscopy_upper_bound}
\end{equation}
where for $L=1$, $r=\dim\operatorname{span}\{K_1,\ldots,K_M\}$ and $\kappa=\max_j\|K_j\|_2$, while for $L\geq2$, $F_K(A)=K[A,\ldots,A]$ and $\|K\|_\infty\leq\kappa$.
\end{theorem}
\end{graydefinitionbox}

\begin{proof}
For $L=1$, Algorithms~\ref{algo:specified_linear_spectroscopy_coarse}
and~\ref{algo:specified_linear_spectroscopy_fine}, together with
Propositions~\ref{prop:specified-linear-coarse}
and~\ref{prop:specified-linear-fine}, give the two branches in
Eq.~\eqref{eq:specified_spectroscopy_upper_bound}. For $L=2$, the result follows from
Lemma~\ref{lem:quadratic_covariance_estimation}. For fixed $L\geq3$ and
$M=1$, it follows from
Algorithm~\ref{algo:specified_nonlinear_spectroscopy} and
Proposition~\ref{prop:specified-nonlinear-upper-bound}.
\end{proof}

As in classical matched filtering, we tailor the sensor's temporal response so that the desired combination of spectral amplitudes accumulates while unwanted contributions cancel. This allows us to measure a spectral overlap directly, without reconstructing its individual components. For multiple linear observables, the same idea extends to their common $r$-dimensional span. The resulting matched filter has a small averaged square magnitude, but can exhibit large peaks that are costly to implement coherently. At standard precision, $\kappa/\epsilon \lesssim \sqrt{N/r}$, we therefore coherently accumulate its bounded component and estimate the remaining contributions using short Ramsey measurements. At high precision, $\kappa/\epsilon \gtrsim \sqrt{N/r}$, it becomes more favorable to accumulate the entire response coherently. These approaches yield the two branches of Eq.~\eqref{eq:specified_spectroscopy_upper_bound} at $L=1$, whose crossover at $\kappa/\epsilon\sim\sqrt{N/r}$ is fundamental.

For nonlinear observables, we follow a different approach that exploits the freedom to choose the distribution of random spectral responses. The key observation is that averaging the sensor's response over Gaussian-random phases produces a signal whose contrast depends on the variance of those phases. By choosing two Gaussian ensembles with covariances tailored to a desired quadratic form, we can isolate that observable by comparing their average signals, without learning the individual spectral amplitudes. This yields an interrogation time of Eq.~\eqref{eq:specified_spectroscopy_upper_bound} for quadratic observables, $L=2$ . Higher-order observables can then be reduced to quadratic estimation by contracting the tensor against independent noisy estimates of the spectrum, giving the scaling in Eq.~\eqref{eq:specified_spectroscopy_upper_bound} for fixed $L\geq 3$. Unlike the linear case, however, the general minimax complexity of single-target higher-order spectroscopy remains open.

Noise limits coherent interrogation time, but does not eliminate the advantage of measuring a known observable directly. By avoiding unnecessary spectral reconstruction, linear matched spectroscopy retains a polynomial advantage even under substantial depolarizing noise, reaching a factor of $\widetilde\Theta(\sqrt N)$ near $\gamma\sim\Delta_\omega$. This advantage comes from targeting the desired spectral information, rather than suppressing the noise itself. The corresponding strong-noise complexity for nonlinear observables remains open.

\subsection{Linear matched filtering}
When the observables of interest are not known during acquisition, Algorithm~\ref{algo:randomized_spectroscopy_coarse} uses Gaussian-random responses drawn isotropically from the full spectral space, producing a reusable record that can be contracted against observables supplied only after sensing. When the observables are known beforehand, this universality is unnecessary. Instead, we can tailor the pulse sequences to probe only the subspace relevant to the desired queries.

Let $r=\dim\operatorname{span}\{K_1,\ldots,K_M\}$, and choose an orthonormal
basis $U=(u_1,\ldots,u_r)\in\mathbb R^{2N\times r}$ for this query subspace.
Writing
\begin{equation}
    \frac{K_j}{\kappa}=Uv_j,
    \qquad
    \|v_j\|_2\leq1,
\end{equation}
gives $K_j^TA=\kappa v_j^TU^TA$. Thus every requested observable depends only
on the $r$-dimensional projection $U^TA$, and the sensing protocol need not
acquire information in directions orthogonal to the query subspace.

We first describe the protocol for the standard-precision regime, which combines coherent
sensing within this subspace with shared short-time measurements of the
remaining response.

\begin{algorithm}[h!]
    \DontPrintSemicolon
    \caption{Matched spectroscopy}
    \label{algo:specified_linear_spectroscopy_coarse}
    \KwIn{Observables $K_1,\ldots,K_M$, accuracy $\epsilon$, failure probability $\delta$.}
    \KwOut{Estimates $\widehat F_1,\ldots,\widehat F_M$.}

    Let $z=\kappa/\epsilon$, choose an orthonormal basis
    $U\in\mathbb R^{2N\times r}$ for $\operatorname{span}\{K_j\}$, and write
    $K_j/\kappa=Uv_j$.\;

    Form
    \begin{equation}
        q(t)=U^TG^{-1}b(t),
        \qquad
        a=\left(\frac{Nz}{r}\right)^{1/3},
    \end{equation}
    and decompose
    \begin{equation}
        q(t)=c(t)+R(t),
        \qquad
        c(t)=\min\left\{1,\frac{a\sqrt r}{\|q(t)\|_2}\right\}q(t).
    \end{equation}

    Estimate
    \begin{equation}
        C(A)=\int_I c(t)b(t)^TA\,d\nu(t)
    \end{equation}
    using the pulse compiler and high-gain readout procedures of
    Section~\ref{sec:spec_toolbox}.\;

    With $S=\int_I\|R(t)\|_2d\nu(t)$, sample
    $t\sim\|R(t)\|_2d\nu(t)/S$ and use short Ramsey measurements and
    Algorithm~\ref{algo:sine_inversion} to estimate
    \begin{equation}
        R(A)=\int_I R(t)b(t)^TA\,d\nu(t).
    \end{equation}

    Set $\widehat A_U=\widehat C+\widehat R$. Repeat independently
    $O(\log(M/\delta))$ times and return
    \begin{equation}
        \widehat F_j
        =
        \kappa\operatorname{median}_s
        v_j^T\widehat A_U^{(s)}.
    \end{equation}
\end{algorithm}

\begin{prop}[Time complexity of Algorithm~\ref{algo:specified_linear_spectroscopy_coarse}]
\label{prop:specified-linear-coarse}
Let $K_1,\ldots,K_M\in\mathbb R^{2N}$ satisfy
$\kappa=\max_j\|K_j\|_2>0$, and let
$r=\dim\operatorname{span}\{K_1,\ldots,K_M\}$. Suppose $8\leq \frac{\kappa}{\epsilon}\leq\sqrt{\frac{N}{r}}$. Then Algorithm~\ref{algo:specified_linear_spectroscopy_coarse} returns estimates satisfying
\begin{equation}
    \Pr\left[
        \max_{j\leq M}
        |\widehat F_j-K_j^TA|
        \leq\epsilon
    \right]
    \geq1-\delta
\end{equation}
for every $\|A\|_2\leq1$, using total interrogation time
\begin{equation}
    T_{\rm int}
    \leq
    C\ell^4
    N^{1/3}r^{2/3}
    \left(\frac{\kappa}{\epsilon}\right)^{4/3},
    \qquad
    \ell=\log\left(e+\frac{NM\kappa}{\epsilon\delta}\right).
\end{equation}
\end{prop}

\begin{proof}
Write $z=\kappa/\epsilon$. For $\tau_0=c/N$, the short Ramsey response
$b(t)$ is uniformly well conditioned, and the weighted-Gram construction gives
$G\asymp\mathds1$. Hence, for
$q(t)=U^TG^{-1}b(t)$,
\begin{equation}
    \int_I q(t)b(t)^T\,d\nu(t)=U^T,
    \qquad
    \int_I q(t)q(t)^T\,d\nu(t)\preceq C\mathds1_r.
\end{equation}
In particular,
\begin{equation}
    \int_I\|q(t)\|_2^2\,d\nu(t)\leq Cr,
    \qquad
    \int_I(v^Tq(t))^2\,d\nu(t)\leq C
\end{equation}
for every $\|v\|_2\leq1$.

Radially clip $q=c+R$ at radius $a\sqrt r$, and let
$S=\int_I\|R(t)\|_2d\nu(t)$. The preceding bounds imply
\begin{equation}
    S\leq\frac{C\sqrt r}{a},
    \qquad
    S\int_I
    \frac{(v^TR(t))^2}{\|R(t)\|_2}\,d\nu(t)
    \leq\frac{C}{a^2}.
    \label{eq:specified-coarse-residual}
\end{equation}
Moreover, the query-space basis can be chosen so that
$\max_i\sup_t|c_i(t)|\leq Ca\sqrt\ell$.

Choose final readout gain $H=\Theta(z)$ and response duration
$W_0=\Theta(az\ell)$. The high-gain readout procedures of
Section~\ref{sec:high_gain_readout} then estimate the $r$ coherent
components with directional bias $O(z^{-1})$ and variance $O(z^{-2})$
using $O(\ell^2)$ shots per component. Their total interrogation cost is
\begin{equation}
    T_C\leq Cra z\,\ell^3.
\end{equation}

For the residual, sample centers with density
$\|R(t)\|_2d\nu(t)/S$. Equation~\eqref{eq:specified-coarse-residual}
and Algorithm~\ref{algo:sine_inversion} imply that
\begin{equation}
    n_R
    =
    O\left(
        \frac{z^2}{a^2\tau_0^2}
    \right)
\end{equation}
sampled centers suffice to obtain the same $O(z^{-1})$ directional bias and $O(z^{-2})$ variance for every requested contraction. Since each center uses $O(\ell^2)$ Ramsey shots of duration $\tau_0=\Theta(N^{-1})$,
\begin{equation}
    T_R
    \leq
    C\frac{Nz^2}{a^2}\ell^2.
\end{equation}

Repeating the common acquisition $O(\log(M/\delta))$ times and taking
the median after each requested contraction gives simultaneous success
probability at least $1-\delta$. Thus
\begin{equation}
    T_{\rm int}
    \leq
    C\ell^4
    \left(
        raz+\frac{Nz^2}{a^2}
    \right).
\end{equation}
Choosing $a=\left(\frac{Nz}{r}\right)^{1/3}$ balances the two terms and gives
\begin{equation}
    T_{\rm int}
    \leq
    C\ell^4
    N^{1/3}r^{2/3}z^{4/3},
\end{equation}
as claimed.
\end{proof}

For larger $\kappa/\epsilon$, the optimal clipping threshold in the preceding construction approaches the largest value allowed by the query-space response. Beyond the crossover $\kappa/\epsilon\sim\sqrt{N/r}$, it is therefore more efficient to just implement the full query-space response coherently. This gives the high-precision protocol.
\begin{algorithm}[h!]
    \DontPrintSemicolon
    \caption{High-precision matched spectroscopy}
    \label{algo:specified_linear_spectroscopy_fine}
    \KwIn{Observables $K_1,\ldots,K_M$, accuracy $\epsilon$, failure probability $\delta$.}
    \KwOut{Estimates $\widehat F_1,\ldots,\widehat F_M$.}

    Let $z=\kappa/\epsilon$, choose an orthonormal basis
    $U\in\mathbb R^{2N\times r}$ for $\operatorname{span}\{K_j\}$, and write
    $K_j/\kappa=Uv_j$.\;

    Form
    \begin{equation}
        q(t)=U^TG^{-1}b(t),
        \qquad
        U^TA=\int_I q(t)b(t)^TA\,d\nu(t).
    \end{equation}

    Choose the query-space basis as in the proof and coherently compile the
    $r$ responses defined by $q_i(t)$. Estimate each component of $U^TA$
    using the high-gain readout procedures of Section~\ref{sec:spec_toolbox},
    obtaining $\widehat A_U$.\;

    Repeat independently $O(\log(M/\delta))$ times and return
    \begin{equation}
        \widehat F_j
        =
        \kappa\operatorname{median}_s
        v_j^T\widehat A_U^{(s)}.
    \end{equation}
\end{algorithm}

\begin{prop}[Time complexity of Algorithm~\ref{algo:specified_linear_spectroscopy_fine}]
\label{prop:specified-linear-fine}
Let $K_1,\ldots,K_M\in\mathbb R^{2N}$ satisfy $\kappa=\max_j\|K_j\|_2>0$, and let $r=\dim\operatorname{span}\{K_1,\ldots,K_M\}$. Suppose $\frac{\kappa}{\epsilon}\geq\sqrt{\frac{N}{r}}$. Then Algorithm~\ref{algo:specified_linear_spectroscopy_fine} returns estimates satisfying
\begin{equation}
    \Pr\left[
        \max_{j\leq M}
        |\widehat F_j-K_j^TA|
        \leq\epsilon
    \right]
    \geq1-\delta
\end{equation}
for every $\|A\|_2\leq1$, using total interrogation time
\begin{equation}
    T_{\rm int}
    \leq
    C\ell^4
    \sqrt{Nr}\frac{\kappa}{\epsilon},
    \qquad
    \ell=\log\left(e+\frac{NM\kappa}{\epsilon\delta}\right).
\end{equation}
\end{prop}

\begin{proof}
Write $z=\kappa/\epsilon$. The weighted-Gram estimates established in the
proof of Proposition~\ref{prop:specified-linear-coarse} give
\begin{equation}
    U^TA
    =
    \int_I q(t)b(t)^TA\,d\nu(t),
    \qquad
    \sup_{t\in I}\|q(t)\|_2\leq C\sqrt N.
\end{equation}
Choose $a=C\sqrt{\frac{N}{r}}$. Then $a\sqrt r\geq\sup_t\|q(t)\|_2$, so the radial clipping is inactive, so $c(t)=q(t)$ and $R(t)=0$. The entire query-space response is estimated coherently.

As in the preceding proof, the query-space basis can be chosen so that
\begin{equation}
    \max_{i\leq r}\sup_{t\in I}|q_i(t)|
    \leq
    Ca\sqrt\ell.
\end{equation}
Taking final readout gain $H=\Theta(z)$ therefore allows each query-space coordinate to be compiled on an interval of length
\begin{equation}
    W_0 =  O(az\ell)  =  O\!\left(  \sqrt{\frac Nr}\,z\ell \right).
\end{equation}
The high-gain readout procedures of Section~\ref{sec:high_gain_readout} estimate each of the $r$ coordinates with directional bias $O(z^{-1})$ and variance $O(z^{-2})$ using $O(\ell^2)$ shots. Hence one common acquisition costs
\begin{equation}
    T_{\rm int}
    \leq
    CrW_0\ell^2
    \leq
    C\ell^3\sqrt{Nr}\,z.
\end{equation}
Repeating independently $O(\log(M/\delta))$ times and taking the median after each requested contraction gives simultaneous success probability at least $1-\delta$, yielding
\begin{equation}
    T_{\rm int}
    \leq
    C\ell^4\sqrt{Nr}\,z.
\end{equation}
\end{proof}
\subsection{Nonlinear matched filtering}

We now consider a single specified nonlinear observable $F_K(A)=K[A,\ldots,A]$ of fixed order $L\geq2$, with $\|K\|_\infty\leq\kappa$. Knowing $K$ before sensing allows the acquisition procedure itself to depend on the desired nonlinear observable. Rather than first constructing a reusable estimate of $A$ and then evaluating $K$ classically, we can choose random pulse-sequence responses whose distribution is tailored to the specified target.

The basic idea is particularly transparent for a quadratic observable $F_K(A)=A^TKA$. In randomized spectroscopy, the response vector is sampled isotropically as $g\sim\mathcal N(0,\mathds1_{2N})$. When $K$ is known in advance, we instead sample Gaussian response vectors with covariances that depend on $K$, and compile each sampled vector into a pulse sequence. For $g\sim\mathcal N(0,\Sigma)$, averaging the terminal $X$ measurement over the random response gives
\begin{equation}
    \mathbb E_g\cos(g^TA)
    =
    \exp\left(-\frac12 A^T\Sigma A\right).
\end{equation}
By comparing two suitably chosen covariance matrices, the common $\|A\|_2^2$ contribution cancels and the desired quadratic form $A^TKA$ can be extracted directly. We first make this construction precise.

\begin{lemma}[Covariance-based quadratic estimation]
\label{lem:quadratic_covariance_estimation}
Let $K\in\mathbb R^{2N\times 2N}$ be a known symmetric matrix satisfying $\|K\|_{\rm op}\leq\kappa$. There is a single-spin protocol that estimates $A^TKA$ to additive error $\epsilon$ with failure probability at most $\delta$ using interrogation time
\begin{equation}
    T_{\rm int}  =  \widetilde O\!\left( \sqrt N \left(\frac{\kappa}{\epsilon}\right)^2 \right).
\end{equation}
\end{lemma}

\begin{proof}
Because $K$ is known before sensing, we can choose the distribution of the random response vectors used to generate the pulse sequences. Define
\begin{equation}
    \Sigma_\pm
    =
    \mathds1_{2N}\pm\frac{K}{\kappa}.
\end{equation}
Since $\|K\|_{\rm op}\leq\kappa$, both $\Sigma_\pm$ are positive semidefinite and satisfy $\Sigma_\pm\preceq2\mathds1_{2N}$, so they are valid Gaussian covariance matrices.

For each experiment, sample $g_\pm\sim\mathcal N(0,\Sigma_\pm)$, compile $g_\pm$ as the response vector of a pulse sequence, and measure $X$ at the end of the interrogation. For a fixed response $g_\pm$, the $X$ expectation is $\cos(g_\pm^TA)$. Averaging also over the Gaussian response and using the Gaussian characteristic function gives
\begin{align}
    \mu_\pm
    &=
    \mathbb E_{g_\pm}\cos(g_\pm^TA) \\
    &=
    \exp\left(-\frac12 A^T\Sigma_\pm A\right) \\
    &=
    \exp\left[
        -\frac12\left(
        \|A\|_2^2
        \pm\frac{A^TKA}{\kappa}
        \right)
    \right].
\end{align}
Taking the difference of the logarithms cancels the common $\|A\|_2^2$ contribution exactly:
\begin{equation}
    A^TKA
    =
    \kappa\left(\log\mu_--\log\mu_+\right).
    \label{eq:quadratic-covariance-identity}
\end{equation}

Since $\|A\|_2\leq1$ and $\Sigma_\pm\preceq2\mathds1_{2N}$, $e^{-1}\leq\mu_\pm\leq1$. Thus the logarithm is uniformly Lipschitz on the relevant interval, and estimating each $\mu_\pm$ to accuracy
$O(\epsilon/\kappa)$ suffices to estimate Eq.~\eqref{eq:quadratic-covariance-identity} to accuracy $O(\epsilon)$. This requires $O((\kappa/\epsilon)^2\log(1/\delta))$ bounded $X$ measurements. Because
$\Sigma_\pm\preceq2\mathds1_{2N}$, each Gaussian response can be compiled in interrogation time $\widetilde O(\sqrt N)$ by Lemma~\ref{lem:random_response_compiler}, giving the stated bound.
\end{proof} 

We now extend Lemma~\ref{lem:quadratic_covariance_estimation} to $L\geq3$. Averaging $k$ Gaussian-response measurements gives a random vector $X$ with
\begin{equation}
    \mathbb E[X]=\mu A,
    \qquad
    \operatorname{Cov}(X)\preceq k^{-1}\mathds1_{2N},
    \qquad
    \mu=e^{-\|A\|_2^2/2}.
\end{equation}
For independent $X_1,\ldots,X_{L-2}$, define
\begin{equation}
    K_X
    =
    K[X_1,\ldots,X_{L-2},\cdot,\cdot].
\end{equation}
Then
\begin{equation}
    \mathbb E[A^TK_XA]
    =
    \mu^{L-2}K[A,\ldots,A],
\end{equation}
reducing the problem to quadratic estimation together with an estimate of $\mu$.

\begin{algorithm}[h!]
    \DontPrintSemicolon
    \caption{Matched nonlinear spectroscopy}
    \label{algo:specified_nonlinear_spectroscopy}
    \KwIn{A specified rank-$L$ tensor $K$, $L\geq3$, accuracy $\epsilon$, and failure probability $\delta$.}

    Set
    \begin{equation}
        k
        =
        \left\lceil
        C_L(2N)^{(L-2)/(L-1)}
        \left(\frac{\kappa}{\epsilon}\right)^{2/(L-1)}
        \right\rceil .
    \end{equation}

    Construct independent $X_1,\ldots,X_{L-2}$, each by averaging $k$ Gaussian-response measurements.\;

    Form
    \begin{equation}
        K_X
        =
        K[X_1,\ldots,X_{L-2},\cdot,\cdot].
    \end{equation}

    Estimate $A^TK_XA$ using Lemma~\ref{lem:quadratic_covariance_estimation}, and independently estimate $\mu=e^{-\|A\|_2^2/2}$.\;

    Divide the quadratic estimate by $\widehat\mu^{\,L-2}$.\;

    Repeat $O(\log(1/\delta))$ times and return the median.
\end{algorithm}

\begin{prop}[Time complexity of Algorithm~\ref{algo:specified_nonlinear_spectroscopy}]
\label{prop:specified-nonlinear-upper-bound}
Fix $L\geq2$ and $\|K\|_\infty\leq\kappa$. If $C_L\leq\kappa/\epsilon\leq c\sqrt N$, there is a single-spin protocol returning $\widehat F_K$ such that
\begin{equation}
    \Pr\!\left[
        |\widehat F_K-K[A,\ldots,A]|\leq\epsilon
    \right]
    \geq1-\delta
\end{equation}
for every $\|A\|_2\leq1$, using interrogation time
\begin{equation}
    T_{\rm int}
    =
    \widetilde O_L\left(
        N^{3/2-1/(L-1)}
        \left(\frac{\kappa}{\epsilon}\right)^{2/(L-1)}
    \right).
    \label{eq:specified-nonlinear-upper-bound}
\end{equation}
For $L=2$ this is Lemma~\ref{lem:quadratic_covariance_estimation}, while for $L\geq3$ the protocol is Algorithm~\ref{algo:specified_nonlinear_spectroscopy}.
\end{prop}

\begin{proof}
The case $L=2$ follows from Lemma~\ref{lem:quadratic_covariance_estimation}. Suppose $L\geq3$ and write $z=\kappa/\epsilon$. For the independent vectors $X_1,\ldots,X_{L-2}$ constructed by the algorithm,
\begin{equation}
    \mathbb E[A^TK_XA]
    =
    \mu^{L-2}K[A,\ldots,A],
\end{equation}
and Lemma~\ref{lem:independent_nonlinear_estimation} gives
\begin{equation}
    \mathbb E\left|
        A^TK_XA-\mu^{L-2}K[A,\ldots,A]
    \right|^2
    \leq
    C_L\kappa^2
    \sum_{j=1}^{L-2}
    \frac{(2N)^{j-1}}{k^j}.
    \label{eq:specified-nonlinear-contraction-error}
\end{equation}
For the choice of $k$ in Algorithm~\ref{algo:specified_nonlinear_spectroscopy} and $z^2\lesssim N$, this is $O_L(\epsilon^2)$. Moreover, $\mathbb E\|X_i\|_2^2\leq1+2N/k$. Discarding the repetition if any $\|X_i\|_2>C_L\sqrt{1+2N/k}$ changes the success probability only by a constant, and otherwise
\begin{equation}
    \|K_X\|_{\rm op}
    \leq
    C_L\kappa
    \left(1+\frac{2N}{k}\right)^{(L-2)/2}.
\end{equation}
The algorithm of Lemma~\ref{lem:quadratic_covariance_estimation} therefore estimates $A^TK_XA$ in time
\begin{equation}
    \widetilde O_L\left[
        \sqrt N\,z^2
        \left(1+\frac{2N}{k}\right)^{L-2}
    \right].
\end{equation}
Constructing the $X_i$ costs $\widetilde O_L(\sqrt N\,k)$, and estimating $\mu$ costs $\widetilde O_L(\sqrt N\,z^2)$. Hence
\begin{equation}
    T_{\rm int}
    =
    \widetilde O_L\left[
        \sqrt N\left(
        k+z^2+
        z^2\left(1+\frac{2N}{k}\right)^{L-2}
        \right)
    \right].
\end{equation}
For the chosen $k$ and $z^2\lesssim N$, both $z^2$ and $z^2(1+2N/k)^{L-2}$ are $O_L(k)$. Thus
\begin{equation}
    T_{\rm int}
    =
    \widetilde O_L(\sqrt N\,k)
    =
    \widetilde O_L\left(
        N^{3/2-1/(L-1)}z^{2/(L-1)}
    \right).
\end{equation}
Finally, $\mu\geq e^{-1/2}$, so estimating $\mu$ to $O_L(1/z)$ contributes only $O_L(\epsilon)$ error after normalization. Standard median concentration achieves any failure probability $0<\delta<1/3$ with logarithmic overhead in $1/\delta$.
\end{proof}
\subsection{Performance under noise}

We consider Markovian depolarizing noise,
\begin{equation}
    \frac{d\rho}{dt}
    =
    -i[H_A(t),\rho]
    +
    \frac{\gamma}{4}\sum_{P\in\{X,Y,Z\}}(P\rho P-\rho).
\end{equation}
Depolarization reduces the measurement contrast by $e^{-\gamma\tau}$, limiting the benefit of long coherent interrogations. Nevertheless, matched spectroscopy can retain an advantage over reusable sensing protocols like DD or RQS by measuring the desired spectral overlap without resolving its individual frequency components.

\begin{prop}[Matched linear spectroscopy under depolarizing noise]
\label{prop:specified-linear-noisy}
For a single specified linear observable $F_K(A)=K^TA$ with $\|K\|_2\leq\kappa$, the worst-case minimax interrogation time over the fixed-density frequency class satisfies, at fixed confidence and sufficiently small relative error,
\begin{equation}
\label{eq:noisy_matched_total}
T_{\rm int}^*
=
\widetilde\Theta\!\left(
    \min\left\{
        N^{1/3}z^{4/3},\,\sqrt N\,z
    \right\}
    +z^2\max\left\{
        \gamma,\,
        \min\left\{
            N\gamma,\sqrt{N\gamma\Delta_\omega}
        \right\}
    \right\}
\right),
\end{equation}
where $z=\kappa/\epsilon$. The upper bound is uniform over the supplied targets and allowed frequency sets, while the lower bound holds for a worst-case target and an allowed frequency spectrum.
\end{prop}

\begin{proof}
The lower bound follows from Theorem~\ref{thm:noisy_matched_lower_bound}, proved in Appendix~\ref{sec:noisy_matched_lower_bound}.

For the upper bound, we adapt the clipped-response construction of Algorithm~\ref{algo:specified_linear_spectroscopy_coarse}. Recall that the matched filter $q(t)$ satisfies $\int q(t)^2\,d\nu(t)=O(1)$ and $\|q\|_\infty=O(\sqrt N)$. Clipping at threshold $a$ gives $q=c_a+r_a$, with $\|c_a\|_\infty\leq a$ and $\int|r_a(t)|\,d\nu(t)=O(1/a)$.

For $\gamma\lesssim N\Delta_\omega$, the residual can be estimated as in Proposition~\ref{prop:specified-linear-coarse}, using short Ramsey interrogations of duration $\Theta((N\Delta_\omega)^{-1})$. These retain constant contrast under depolarization, giving a residual cost $\widetilde O(N\Delta_\omega z^2/a^2)$.

For the bounded component, coherent interrogations are limited to duration $O(1/\gamma)$. When $\gamma\lesssim\Delta_\omega$, the matched response can still be compiled over a well-conditioned spectral window, and the phase-readout procedures of Algorithms~\ref{algo:phase_unwrapping}--\ref{algo:sine_inversion} give a cost $\widetilde O(az+\gamma a^2z^2)$.

For $\Delta_\omega\lesssim\gamma\lesssim N\Delta_\omega$, we instead divide the matched response into intervals of duration $\tau=\Theta(1/\gamma)$ and sample them independently. The spectral Gram bound gives a mean squared accumulated phase $O(\tau^2)$ across these intervals. Using the same phase-readout procedures, the rescaled estimates have variance $O(a^2/\tau^2)$, yielding a total cost $\widetilde O(\gamma a^2z^2)$ for the bounded component.

Combining these estimates, for $\gamma\lesssim N\Delta_\omega$ we obtain
\begin{equation}
T_{\rm int}
=
\widetilde O\left(
az+\gamma a^2z^2+
\frac{N\Delta_\omega z^2}{a^2}
\right),
\end{equation}
where the residual term vanishes when $a\geq\|q\|_\infty$. At weak noise, $\gamma\lesssim\Delta_\omega/N$, the noise contribution is bounded by $N\gamma z^2$, while the noiseless clipping choice remains available. In the intermediate regime, balancing the two noise-dependent terms gives $a\asymp(N\Delta_\omega/\gamma)^{1/4}$ and a cost $\widetilde O(\sqrt{N\gamma\Delta_\omega}\,z^2)$. Optimizing against the noiseless construction gives the first term in Eq.~\eqref{eq:noisy_matched_total}.

Finally, when $\gamma\gtrsim N\Delta_\omega$, Ramsey interrogations of duration $\Theta(1/\gamma)$ are short enough to preserve a well-conditioned spectral frame. Directly estimating the matched overlap using these measurements and bounded sine inversion costs $\widetilde O(\gamma z^2)$. Together, these regimes establish the upper bound.
\end{proof}

Consider the illustrative regime $\gamma\sim\Delta_\omega$, where noise-induced linewidth matches the desired spectral resolution. Here, 
\begin{equation}
T_{\rm matched}
=\widetilde\Theta(\sqrt N\,\Delta_\omega z^2),
\qquad
T_{\rm reusable}
=\widetilde\Theta(N\Delta_\omega z^2),
\end{equation}
where $T_{\rm reusable}$ denotes the optimal sensing time achievable by any strategy capable of producing a reusable dataset, such as RQS or DD. Notably, matched spectroscopy retains a $\widetilde\Theta(\sqrt N)$ advantage even when depolarization enforces quadratic precision dependence. The gain comes from estimating the desired observable directly, rather than reconstructing broadened spectral amplitudes, illustrating that matched filter design is very valuable in realistic, noisy quantum sensing applications.

\newpage
\section{Pooled Spectroscopy} \label{sec:pooled_spectroscopy}
While extensive literature has studied optimized pulse sequence protocols composed of $X, Y$ and $Z$ pulses, the goal of these pulse sequences is generally to minimize losses in contrast due to disorder, spin-spin interactions, and decoherence. On the other hand, the pulses used for sensing generally lie in a single basis (for instance, $X$-basis $\pi$ pulses used in Ramsey or dynamical decoupling spectroscopy) and are timed to filter specific frequencies while suppressing the rest of the spectrum. The filter-function behavior of a single-axis pulse sequence can be studied via Equation \eqref{eq:commuting_propagator}; however, the utility of noncommuting multi-axis sequences for sensing is comparatively poorly understood.

There is reason to believe that noncommuting sequences may confer a sensing advantage. Note that when both $X$ and $Y$ pulses are used, the toggling-frame Hamiltonian will cease to lie in the span of a single-axis operator like $Z$. In turn, the resulting propagator will have nested integrals in time that are not present in equation \eqref{eq:commuting_propagator}, causing terms that are multilinear in entries of $A$ to appear in the final measurement probability. Carefully-designed multi-axis sequences may be able to coherently access nonlinear observables like the power spectrum of a signal while suffering comparatively smaller shot noise than single-axis approaches. We introduce pooled spectroscopy to enable precisely this kind of advantage.

Pooled spectroscopy is designed for experimental settings in which the frequency band is very broad or very dense. This is natural, for instance, in biological imaging applications, where a massive number of environmental nuisance factors populate the signal spectrum, and the sensing application requires coarse-grained resolution of spectral features rather than a precise description of every spectral line. In such settings, the goal is often to estimate the total spectral power contained within frequency bands which may individually contain a large number of nuisance spectral lines and few genuine signal peaks. Instead of estimating each line in such a bin, pooled spectroscopy uses noncommuting pulse sequences, interleaved with randomized commuting pulse sequences, to aggregate the total signal phase applied by all frequencies in the bin and coherently apply a polynomial transformation to the phase using quantum processing. The pulse sequence can be tailored to engineer specific polynomial sequences of interest, but a natural and pragmatically useful example is to map a phase $x\mapsto x^2$, which allows the sensor phase to directly encode spectral power. By deliberately discarding information about the nuisance degrees of freedom, pooled spectroscopy allows us to estimate these cumulative "binned" observables much more efficiently than is possible using conventional commuting pulse sequences.

\subsection{Algorithms for power and correlation}
Here we give the pooled spectroscopy algorithm for power and correlation. For any specified $m$-dimensional spectrum vector $u$, the power is given by
\begin{equation}
    P(u)\coloneqq \|u\|_2^2=\sum_{r=1}^m |u_r|^2.
\end{equation}
Thus $P(A_b)=P_b$, while choosing
$u=(A_b\pm A_c)/\sqrt2$ or $u=(A_b\pm iA_c)/\sqrt2$ allows us to measure correlations between amplitudes in different frequency bins. Since
$\|A\|_2\leq1$, each of these normalized choices satisfies $P(u)\leq1$.

\begin{algorithm}
    \DontPrintSemicolon
    \caption{Pooled spectroscopy for quadratic observables}
    \label{algo:compressed_power}
    \KwIn{Known frequencies $\{\omega_j\}_{j=1}^N$;  a choice
    $u=A_b$, $(A_b\pm A_c)/\sqrt2$, or $(A_b\pm iA_c)/\sqrt2$,
    for bins $b,c$ of size $m$; scale $\kappa>0$; accuracy $0<\epsilon<\kappa/2$; and failure probability $0<\delta<1/4$.}
    \KwOut{An estimate $\widehat P$ satisfying $|\widehat P-\kappa P(u)|\leq\epsilon$ with probability at least $1-\delta$; otherwise FAIL.}

    Set $J=\lceil\log_2(23\kappa/(24\epsilon))\rceil$, $\ell=\log(9(J+1)/\delta)$, and $R=\lceil44\ell\rceil$.\;

    \For{$j=0,\ldots,J$}{
        Set $s=2^j$ and $k=\lceil22s^2\ell\rceil$.\;

        \eIf{$k\leq m$}{
            Draw $k$ independent standard complex Gaussian vectors $g_1,\ldots,g_k$, and set $h=\sqrt{s/k}$ and $a=1$.\;
        }{
            Draw a uniformly random orthonormal basis $g_1,\ldots,g_m$, set $k=m$,
            \[
                h=\min\left\{\sqrt{s},\frac{\pi}{2}\sqrt{\frac{m-1}{\ell+\log m}}\right\},
                \qquad
                a=\frac{s}{h^2},
            \]
            with $h=\min\{\sqrt{s},\pi/2\}$ when $m=1$.\;
        }

        \For{$r=1,\ldots,k$}{
            Form the complex response vector $ c_r^\dagger A=g_r^\dagger u$. Writing $c_r=a_r+ib_r$, define
            \[
            v_{r,R}=
            \begin{pmatrix}a_r\\-b_r\end{pmatrix},
            \qquad
            v_{r,I}=
            \begin{pmatrix}-b_r\\-a_r\end{pmatrix}.
            \]

        }
        \For{each of $R$ $X$ measurements and $R$ $Y$ measurements}{
        Prepare a fresh qubit in $\ket{+}$.\;

        \For{$r=1,\ldots,k$}{
            Run Algorithm~\ref{algo:polynomial_phase_mixing} on
            $hv_{r,R}$ and $hv_{r,I}$ with $L=2$, coefficient $a$,
            accuracy $\eta=1/(20k)$, failure probability
            $\delta=e^{-\ell}/(4Rk)$, and their starting time, and store the pulse sequence. If the algorithm fails, return FAIL.
            \;
        }

        Concatenate all $k$ pulse sequences and perform the $X$ or $Y$ basis measurement, storing $X_r$ or $Y_r$.\;
    }

    Let $\overline X_j,\overline Y_j$ be the empirical means of all $X_r$ and $Y_r$ at this $j$, and set
    $\widetilde\phi_j=\operatorname{Arg}(\overline X_j+i\overline Y_j)$.\;
    }

    Set $\widehat\phi_0=\widetilde\phi_0$ and, for $j=1,\ldots,J$, let $\widehat\phi_j$ be the $2\pi$ lift of $\widetilde\phi_j$ closest to $2\widehat\phi_{j-1}$.\;

    Return $\widehat P=\kappa\widehat\phi_J/2^J$.\;
\end{algorithm}
\begin{algorithm}
    \DontPrintSemicolon
    \caption{Power and correlation spectroscopy}
    \label{algo:power_correlation_spectroscopy}
    \KwIn{Known frequencies; paired bins $\mathcal B_b,\mathcal B_c$ of size $m$; a target $\kappa P_b$, $\kappa\operatorname{Re}C_{bc}$, or $\kappa\operatorname{Im}C_{bc}$; accuracy $\epsilon$; and failure probability $\delta$.}
    \KwOut{An estimate of the specified target with error at most $\epsilon$ and failure probability at most $\delta$.}

    \If{the target is $\kappa P_b$}{
        Run Algorithm~\ref{algo:compressed_power} on $u=A_b$ with parameters $\kappa,\epsilon,\delta$, and return its output.\;
    }

    Set
    \[
        u_\pm=
        \begin{cases}
            (A_b\pm A_c)/\sqrt2, & \operatorname{Re}C_{bc},\\
            (A_b\pm iA_c)/\sqrt2, & \operatorname{Im}C_{bc}.
        \end{cases}
    \]

    Run Algorithm~\ref{algo:compressed_power} on $u_+$ and $u_-$ with parameters $\kappa,\epsilon,\delta/2$, obtaining $\widehat P_+$ and $\widehat P_-$.\;

    Return $(\widehat P_+-\widehat P_-)/2$.\;
\end{algorithm}

\begin{theorem}
\label{thm:compressed_power_complexity}
Fix $\kappa>0$, $0<\epsilon<\kappa/2$, and $0<\delta<1/4$. Then Algorithm~\ref{algo:power_correlation_spectroscopy}, estimates one bin power or one correlation component between distinct bins containing $m$ frequencies each to absolute precision $\epsilon$, using total interrogation time
\begin{equation}
    T \leq
    \begin{cases}
        \widetilde O\!\left(
            \sqrt m\left(\frac{\kappa}{\epsilon}\right)^{3/2}
            +\Delta_\omega^{-1}\left(\frac{\kappa}{\epsilon}\right)^2
        \right),
        & \displaystyle \frac{\kappa}{\epsilon}\leq\sqrt m,\\
        \widetilde O\!\left(
            m\sqrt{\frac{\kappa}{\epsilon}}
            +m\Delta_\omega^{-1}
        \right),
        & \displaystyle \sqrt m<\frac{\kappa}{\epsilon}\leq m,\\
        \widetilde O\!\left(
            (\sqrt m+\Delta_\omega^{-1})\frac{\kappa}{\epsilon}
        \right),
        & \displaystyle \frac{\kappa}{\epsilon}>m,
    \end{cases}
    \label{eq:compressed_power_complexity}
\end{equation}
and succeeds with probability at least $1-\delta$. It uses $O\left(
\log\frac{\kappa}{\epsilon}       \log\left(\frac{\log(\kappa/\epsilon)}{\delta}\right)\right)$
terminal measurements.
\end{theorem}

\begin{proof}
We first prove the correctness guarantee. Consider Algorithm~\ref{algo:compressed_power} for a fixed choice of $u$, with $P(u)\leq1$. At each value of $s$, any $g_r$ is associated with two vectors $v_{r, R}, v_{r, I}$ associated with $g_r$ such that $v_{r, R}^T A = {\rm Re}(g_r^\dagger u), v_{r, I}^T A = {\rm Im}(g_r^\dagger u)$. The two pulse sequences returned by Algorithm \ref{algo:polynomial_phase_mixing} therefore realize $R_Z(ah^2{\rm Re}(g_r^\dagger u)^2)$, $R_Z(ah^2{\rm Im}(g_r^\dagger u)^2)$. Since both are $Z$-rotations, concatenating the sequences simply adds their angles. Concatenating the resulting $2k$ sequences therefore (assuming ideal pulse sequence compilation) creates the phase
\begin{equation}
    \Phi_j
    =
    ah^2\sum_{r=1}^k
    \left[
        \operatorname{Re}(g_r^\dagger u)^2
        +\operatorname{Im}(g_r^\dagger u)^2
    \right]
    =
    ah^2\sum_{r=1}^k|g_r^\dagger u|^2.
    \label{eq:compressed_power_phase}
\end{equation}
The purpose of the choices of $k,h,a$ is to make this phase close to $sP(u)$ while minimizing the time needed to produce it. For independent complex Gaussian vectors $g_r$, $g_r^\dagger u \sim \mathcal{N}_\mathbb{C}(0, P(u))$, so we can write $g_r^\dagger u = \sqrt{P/2}(X_r + i Y_r), X_r, Y_r \sim \mathcal{N}(0, 1)$. Therefore, $Z_r \coloneqq |g^\dagger u|^2/P = (X_r^2 + Y_r^2)/2$. Noting that the numerator is a $\chi^2_2$ distribution, we use the standard fact that a $\chi^2_2$ distribution is equivalent to an exponential distribution with parameter $1/2$ to conclude that $Z_r$ has exponential distribution with parameter $1$. Applying exponential Markov's inequality to $\sum_r Z_r$ gives
\begin{equation}
    \Pr\left[|\Phi_j-sP(u)|>\frac13\right]
    \leq
    2\exp\left[
        -\frac{k}{s^2}\left(\frac13-\log\frac43\right)
    \right]
    \leq2e^{-\ell},
    \label{eq:compressed_power_phase_error}
\end{equation}
where the last inequality uses $k\geq22s^2\ell$ and $22(1/3-\log(4/3))>1$. When the algorithm instead uses an orthonormal basis, Parseval's identity gives $\Phi_j=s\sum_{r=1}^m|g_r^\dagger u|^2=sP(u)$, so there is no phase approximation error.

We next prove that the arguments passed to polynomial phase mixing lie in the allowed $[\pi/2, \pi/2]$ range. For Gaussian vectors, the exponential tail gives
\begin{equation}
    \Pr\left[\max_{r\leq k}h|g_r^\dagger u|>\frac{\pi}{2}\right]
    \leq k\exp\left(-\frac{\pi^2k}{4s}\right)
    \leq e^{-\ell}.
\end{equation}
For a uniform complex unit vector and $m>1$, $\Pr[|g_r^\dagger u|^2>q]\leq e^{-(m-1)q}$ for $0\leq q\leq1$. A union bound and the choice of $h$ consequently give the same failure bound for an orthonormal basis. When $m=1$, the argument is bounded deterministically. These bounds control both real quadratures. On the good event that all $h|g_r^\dagger u| \leq \pi/2$, each polynomial sequence has operator error at most $\eta=1/(20k)$ from compilation. There are $2k$ such sequences in one shot, so telescoping their product gives
\begin{equation}
    \|U_j-R_Z(\Phi_j)\|_{\rm op}\leq2k\eta=\frac1{10}.
\end{equation}
Starting from $\ket{+}$, this changes either terminal Pauli expectation by at most $1/5$. If the pulse compilation were perfect, Hoeffding's inequality and $R\geq44\ell$ imply that both empirical means differ from their physical means by at most $1/\sqrt{22}$, except with probability $4e^{-\ell}$. Therefore, combining errors,
\begin{equation}
    |\overline X_j+i\overline Y_j-e^{i\Phi_j}|
    \leq
    \sqrt2\left(\frac15+\frac1{\sqrt{22}}\right)
    <\sin(5/8).
\end{equation}
The corresponding angular error is less than $5/8$. Combining this with Equation~\eqref{eq:compressed_power_phase_error}, the measured phase differs from $sP(u)$, modulo $2\pi$, by less than $1/3+5/8=23/24$.

We now show inductively that phase unwrapping selects the correct $2\pi$ branch at every gain. At $j=0$, the true phase is $P(u)\in[0,1]$, while the measured phase is within $23/24<\pi-1$ of it, so the principal value is already the correct branch.

Now suppose that for a fixed $j>0$, $\left|\widehat\phi_{j-1}-2^{j-1}P(u)\right|<\frac{23}{24}$.  Then doubling the previous estimate gives $\left|2\widehat\phi_{j-1}-2^jP(u)\right| < \frac{23}{12}$. The new measurement is within $23/24$ of the true phase modulo $2\pi$, so the correct branch lies within $\frac{23}{12}+\frac{23}{24} = \frac{23}{8} <\pi$ of $2\widehat\phi_{j-1}$. Every other branch differs by at least $2\pi$, so the algorithm uniquely selects the correct one. The fresh measurement then gives $\left|\widehat\phi_j-2^jP(u)\right| <\frac{23}{24}$. Thus the bound holds for every $j$. At the final gain,
\begin{equation}
    |\widehat P-\kappa P(u)|
    \leq\frac{23\kappa}{24\,2^J}
    \leq\epsilon
\end{equation}
so the algorithmic guarantee holds. 

Next, we bound the interrogation time. First consider the case where $k\leq m$, so the chosen response vectors are Gaussian. For each $j$, we call Algorithm \ref{algo:polynomial_phase_mixing} $2k\cdot 2R$ times. Since the failure probability given to Algorithm \ref{algo:polynomial_phase_mixing} is $e^{-\ell}/4Rk$, a union bound tells us that total failure probability is at most $e^{-\ell}$.

Recall that Algorithm \ref{algo:polynomial_phase_mixing} uses the efficient compiler from Lemma \ref{lem:random_response_compiler}, because our Gaussian-random responses satisfy equation \eqref{eq:subgaussian_vector}. Lemma \ref{lem:random_response_compiler} gave the interrogation time per compilation of $O(\Delta_\omega^{-1} + h\sigma\sqrt{d\log(CN/\delta_{\rm syn})}$; for a Gaussian vector of $\sigma = O(1)$, restricted to a $d=O(m)$-dimensional subspace, this becomes $T = \widetilde O(h\sqrt{m} + \Delta_\omega^{-1})$. To bound the total time with high probability, the same exponential concentration used above gives
\begin{equation}
    \Pr\left[\sum_{r=1}^k\|g_r\|_2^2>\frac43mk\right]
    \leq
    \exp\left[-mk\left(\frac13-\log\frac43\right)\right]
    \leq e^{-\ell}.
\end{equation}
Outside this exponentially unlikely event, $\sum_r\|g_r\|_2\leq \sqrt{k}(\sum_{r=1}^k \|g_r\|_2^2)^{1/2}\leq O(k\sqrt m)$ by Cauchy-Schwarz. Lemma \ref{lem:polynomial_phase_mixing} gives us that each call to polynomial phase mixing uses $O(\log^2 k)$ blocks of interleaved pulse sequences. Substituting $k=O(s^2\ell)$ and $h=\sqrt{s/k}$, the interrogation time is
\begin{equation}
    \widetilde O\left(
        kh\sqrt m+k\Delta_\omega^{-1}
    \right)
    =
    \widetilde O\left(
        \sqrt m\,s^{3/2}+\Delta_\omega^{-1}s^2
    \right).
\end{equation}
The number of repetitions is $O(RJ)$, which only scales as $O(\log\frac{\kappa}{\epsilon}\log\log\frac{\kappa/\epsilon}{\delta})$, so it is subleading.

Now consider the case $k>m$, where we use a random orthonormal basis instead of Gaussian random vectors, and set $k=m$. The failure-probability accounting and the $2R$ repetitions are unchanged. Since each $g_r$ is a uniformly random unit vector in an $O(m)$-dimensional real subspace, Equation~\eqref{eq:subgaussian_vector} holds with $\sigma=O(m^{-1/2})$. Lemma~\ref{lem:random_response_compiler} therefore compiles a response amplified by $h$ in time $\widetilde O\left(h+\Delta_\omega^{-1}\right)$. Unlike the Gaussian case, $\|g_r\|_2=1$ deterministically, so no additional concentration bound is required. Here we run polynomial phase mixing with coefficient $a=s/h^2$ rather than $a=1$, which from Lemma \ref{lem:polynomial_phase_mixing} incurs $O(1+a)$ interleaved pulse sequences. Repeating over $m$ blocks and once again suppressing the subleading $O(RJ) = O(\log\frac{\kappa}{\epsilon}\log\log\frac{\kappa/\epsilon}{\delta})$ overhead,
\begin{equation}
    T
    =
    \widetilde O\left[
        m\left(1+\frac{s}{h^2}\right)
        \left(h+\Delta_\omega^{-1}\right)
    \right].
\end{equation}
Up to logarithmic factors, the choice of $h$ in Algorithm~\ref{algo:compressed_power} is
$h=\sqrt{\min\{s,m\}}$. Hence
\begin{equation}
    T
    =
    \begin{cases}
        \widetilde O\left(m\sqrt{s}+m\Delta_\omega^{-1}\right),
        & s\leq m,\\
        \widetilde O\left(\sqrt m\,s+\Delta_\omega^{-1}s\right),
        & s\geq m.
    \end{cases}
\end{equation}
Comparing the three scalings above yields the interrogation time upper bound.
\end{proof}

\subsection{Advantage over commuting spectroscopy}\label{sec:nc_advantage_pooled}
We now rigorously quantify the spectroscopic advantage conferred by noncommutative quantum control. In Section~\ref{sec:pooled_power_lower_bound}, we prove a lower bound on the total time required by any adaptive spectroscopy protocol which uses only commuting (single-axis) control of the sensor qubit for estimating the power within a frequency bin containing $m$ spectral lines. Optimizing the choice of parameters in this bound gives us the following result.
 
\begin{observation}
\label{cor:compressed_power_commuting_separation}
For fixed $\Delta_\omega$ and sufficiently large $m$, optimizing the choice of $\rho$ in Theorem~\ref{thm:power_discrimination_lower_bound} gives the commuting lower bound
\begin{equation}
    T_{\rm int}^{\rm comm}
    =
    \begin{cases}
        \widetilde\Omega\!\left(
            \sqrt m\left(\frac{\kappa}{\epsilon}\right)^2
        \right),
        & \displaystyle 1\lesssim\frac{\kappa}{\epsilon}\lesssim m^{1/4},\\[2mm]
        \widetilde\Omega\!\left(
            m^{4/5}\left(\frac{\kappa}{\epsilon}\right)^{4/5}
        \right),
        & \displaystyle m^{1/4}\lesssim\frac{\kappa}{\epsilon}\lesssim m^{3/2},\\[2mm]
        \widetilde\Omega\!\left(
            m^{5/4}\sqrt{\frac{\kappa}{\epsilon}}
        \right),
        & \displaystyle \frac{\kappa}{\epsilon}\gtrsim m^{3/2}.
    \end{cases}
    \label{eq:optimized_commuting_power_lower_bound}
\end{equation}
Therefore, there exist values of $m, \kappa, \epsilon$ such that pooled spectroscopy (Algorithm \ref{algo:compressed_power}) achieves an interrogation-time advantage of at least $\Tilde{\Omega}(m^{1/8})$ over any conventional commuting approach to spectroscopy.
\end{observation}

\begin{proof}
Writing $z=\kappa/\epsilon$, the two terms in Theorem~\ref{thm:power_discrimination_lower_bound} balance when $\rho=m^{1/5}z^{-4/5}$. For $z\lesssim m^{1/4}$ this exceeds the allowed constant upper bound on $\rho$, so taking $\rho=\Theta(1)$ gives $\widetilde\Omega(\sqrt m z^2)$. For $m^{1/4}\lesssim z\lesssim m^{3/2}$ the balancing choice is admissible and gives $\widetilde\Omega(m^{4/5}z^{4/5})$. For $z\gtrsim m^{3/2}$, the constraint $z\rho\lesssim\sqrt m$ becomes active, and taking $\rho=\Theta(\sqrt m/z)$ gives $\widetilde\Omega(m^{5/4}\sqrt z)$.

At $z=m^{1/4}$ the first two lower-bound branches both give $\widetilde\Omega(m)$. Equation~\eqref{eq:compressed_power_complexity} gives $\widetilde O(\sqrt m z^{3/2}+z^2)=\widetilde O(m^{7/8})$ at the same precision. 
\end{proof}

We have therefore established that noncommutative single-qubit pulse sequences enable spectroscopy of coarse-grained observables like bin power possibly polynomially more efficiently than \textit{any}, arbitrarily clever conventional strategy using commuting control.

This lower bound applies to all single-qubit protocols and is therefore rigorous evidence of advantage due to noncommuting control. More operationally, however, it is interesting to understand the advantage gained relative to practical commuting alternatives, which must be at least as large as the $m^{1/8}$ advantage. In fact, the advantage is considerably larger. We first consider the conventional benchmark of dynamical decoupling (DD) spectroscopy by sweeping the frequency range and estimating each amplitude to sufficient accuracy that the collective power estimate is $\epsilon$-accurate. The interrogation time cost for this approach is as follows. 

Let $\{\omega_j\}_{j=1}^{m}$ denote the set of frequencies of interest, whose collective power we wish to estimate. A conventional DD sweep chooses one $\omega_j$ in each shot and applies evenly spaced $\pi$-pulses along $X$ resonant with $\omega_j$; we assume perfect timing so that upon preparing the $\ket{+}$ state, running the DD sequence, and measuring $\langle X\rangle$, the expectation value depends only on the cosine-quadrature amplitude $p_j$. Then shifting all pulses by a quarter period estimates the sine-quadrature amplitude $q_j$. Performing this procedure a number of times to be specified per frequency gives an estimate $\widehat A_b$ of the true amplitude vector $A_b$ up to some error which we call $e$. 

Writing $\widehat A_b = A_b + e$ and recalling that the power $P_b$ is given by $\|A_b\|_2^2$, it follows that the error obtained by setting $\widehat P_b = \|\widehat A_b\|_2^2$ is
\begin{equation}
    |\widehat P_b - P_b| = |2A_b^T e + \|e\|_2^2| \leq 2\|e\|_2 + \|e\|^2.
\end{equation}
For this to be $\epsilon$-accurate with scale $\kappa$ we require $\|e\|_2 = O(\epsilon/\kappa)$. Suppose the DD sweep at each frequency estimates the corresponding $p_j, q_j$ to accuracy $\eta$; then $\|e\|_2 \leq \sqrt{2m}\eta$. Therefore, we must choose $\eta = \Theta(\epsilon/(\kappa\sqrt m))$. Let us assume the most favorable implementation of each DD sequence, achieving Heisenberg-limited scaling at each frequency so that each frequency occupies time $\widetilde O(1/\eta) = \widetilde O(\sqrt m \kappa / \epsilon)$ (where suppressed logarithmic overhead would in practice come from phase unwrapping overhead). Since there are two quadratures, we obtain the final scaling 
\begin{equation}
T_{DD} = \widetilde O(m^{3/2}\kappa/\epsilon)    
\end{equation}
A more detailed analysis yields the same scaling. Combining this scaling with each case in Theorem \ref{thm:compressed_power_complexity}, pooled spectroscopy outperforms the conventional DD approach by at least a factor of $\widetilde{\Omega}(m^{3/4})$, and up to a factor of $\widetilde{\Omega}(m)$ depending on the precision regime. For noisy settings populated by dense spectral lines, this may confer a dramatic speedup. Moreover, in many realistic settings the total time required to read out the sensor qubit may dominate the total experimental time, so minimizing the number of shots is essential. Pooled spectroscopy only requires $\widetilde{O}(1)$ terminal measurements in contrast to the $\widetilde{O}(m)$ required by DD.

A stronger commuting benchmark is the randomized quantum spectroscopy algorithm from Section~\ref{sec:randomized_spec}, which is optimal for estimating non-binned conventional spectroscopic observables. Here, write $z = \kappa/\epsilon$.

First consider the Gaussian randomized spectroscopy protocol, which dominates for $z\lesssim\sqrt m$. Each shot applies a Gaussian-random response $g\in\mathbb R^{2m}$ across the entire bin and produces a binary measurement outcome. Averaging $r$ such measurements gives a noisy estimate $U=A_b+e$ whose error is approximately isotropic, with coordinate variance $O(1/r)$. For the quadratic observable we use two independent estimates $U_1=A_b+e_1$ and $U_2=A_b+e_2$ and return $\widehat P_b=\kappa U_1^TU_2$. The estimation error is
\begin{equation}
    \widehat P_b-\kappa P_b
    =
    \kappa\left(
        A_b^Te_1+A_b^Te_2+e_1^Te_2
    \right).
\end{equation}
Since $\|A_b\|_2\leq1$ and the errors are independent and isotropic, its mean-squared size is, up to constants,
\begin{equation}
    \E\left[
        |\widehat P_b-\kappa P_b|^2
    \right]
    \lesssim
    \kappa^2\left(
        \frac1r+\frac{m}{r^2}
    \right).
\end{equation}
Thus accuracy $\epsilon$ is obtained provided $r=\widetilde O\left(\max\left\{ z^2,\sqrt m\,z\right\}\right)$. In the regime $z\lesssim\sqrt m$, the second term dominates, so only $\widetilde O(\sqrt m\,z)$ Gaussian measurements are required. A Gaussian response across a $2m$-dimensional bin can be synthesized in interrogation time $\widetilde O(\sqrt m)$ by Lemma~\ref{lem:random_response_compiler}. Multiplying the cost per measurement by the number of measurements gives
\begin{equation}
    T_{\rm rand}
    =
    \widetilde O\left(
        m\frac{\kappa}{\epsilon}
    \right).
\end{equation}

For $z\gtrsim\sqrt m$, the high-precision randomized spectroscopy protocol instead estimates each of the $2m$ quadratures using long coherent interrogations. Phase unwrapping allows a response of gain $H=\widetilde\Theta(z)$ to estimate a coordinate to precision $O(1/z)$ using only polylogarithmically many terminal measurements. Each coordinate therefore requires total interrogation time $\widetilde O(z)$, and sweeping the $2m$ quadratures again gives
\begin{equation}
    T_{\rm rand}
    =
    \widetilde O\left(
        m\frac{\kappa}{\epsilon}
    \right).
\end{equation}
Thus, while the randomized quantum spectroscopy algorithm is a factor of $\sim\sqrt{m}$ more efficient than DD, pooled spectroscopy outperforms it by at least a factor ranging between $\Omega(m^{1/4})$ and $\Omega(m^{1/2})$ in total time and measurements.

\newpage
\section{Quantum Computational Spectroscopy}
\label{sec:qcs}

In this section, we present the quantum computational spectroscopy (QCS) algorithm. Whereas our other algorithms utilize standard pulse-sequence control, here we envision quantum sensors enhanced by coupling to an ancillary quantum processor that can leverage quantum algorithms to significantly improve sensing precision. In future devices the processor may be a fault-tolerant quantum computer, but near-term platforms with elementary quantum memory capabilities can already realize some of the benefits of QCS.

Recall that in Section~\ref{sec:randomized_spec}, we proved lower bounds that applied to any protocol using single-qubit control. Schematically, the two complementary lower bounds scaled as $N\kappa/\epsilon$ and $\sqrt{N}(\kappa/\epsilon)^2$. In this section, the ancillary quantum memory used by the protocol places it outside the purview of these lower bounds, allowing us to develop a protocol whose interrogation time cost for fixed precision $\epsilon$ scales like $\sqrt{N}\kappa/\epsilon$, achieving a quadratic improvement over \textit{any} conventional spectroscopy protocol in either the mode number or the precision. 

First, we lay out the sensing model. As usual, we take the sensor to be a single qubit evolving under the Hamiltonian $H(t) = \frac{1}{2}B(t) Z + H_c(t) = \frac{1}{2}A^T f(t)Z + H_c(t)$, where $A^T = (p_1,...,p_N,q_1,...,q_N)$ is the unknown vector of the signal's mode amplitudes, while $f(t) = (\cos(\omega_1 t), ..., \cos(\omega_Nt), \sin(\omega_1 t), ...,\sin(\omega_Nt))^T$ is the known time-dependent vector of signal modes. As before, we allow a known single-qubit control term $H_c(t)$. In fact, the only Hamiltonian control we will require \textit{during} sensing is a static transverse term, analogous to the constant drive used to generate Rabi oscillations. In addition, we allow a circuit model where ancillary quantum memory registers can be processed through quantum circuits and repeatedly coupled to the qubit. Specifically, we take the joint system to be
\begin{equation}
    \ket{\Psi}_{\rm mem}\otimes \ket{\phi}_{\rm int} \otimes \ket{\xi}_{\rm sensor}
\end{equation}
where $\ket{\Psi}_{\rm mem}$ represents a quantum memory register which never interacts directly with the qubit, $\ket{\phi}_{\rm int}$ denotes a small register of ``interface" qubits which transport quantum information between the memory and the sensor throughout the protocol, and $\ket{\xi}_{\rm sensor}$ is our usual single-qubit sensor. Because $\ket{\Psi}_{\rm mem}\otimes \ket{\phi}_{\rm int}$ is never exposed to the signal, we treat it as a register of logical qubits on which fault-tolerant processing is allowed. Importantly, however, the sensor is always treated as an unprotected physical qubit; recent work has made rigorous that a sensor cannot generically be embedded into an error-correcting code while remaining sensitive to the kinds of signals we study \cite{pradenne_restrictions_2026}. Therefore, when we incorporate noise into our analysis, we will subject the qubit to continuous depolarizing noise while assuming that every controlled gate acting on the qubit creates an error with constant probability. 

Note that if gate errors are ignored, the construction below can omit the interface qubits and apply the memory-controlled operation directly to the sensor; robustness to sensor depolarization is still retained. In this section we will present the full analysis including the interface qubits, introduced as the ``bridge'' gadget in the main text, and we simply note that removing these qubits entirely and directly applying the controlled rotation between memory and sensor applies the same channel on both the sensor and memory in the absence of control errors.

Quantum metrology has established a concrete criterion for when a particular noise channel rules out asymptotic speedups for canonical parameter estimation. This condition, often referred to as Hamiltonian-not-in-Lindblad-Span (HNLS) \cite{zhou_achieving_2018}, states that for single-parameter DC metrology, if the signal Hamiltonian lies in the span of the noise Lindbladian, the maximum achievable quantum Fisher information for estimating a desired parameter degrades from the Heisenberg limit to the quadratically worse standard quantum limit. Recently, \cite{pradenne_restrictions_2026} has strengthened this obstruction to Fisher information-independent arguments, certifying that unless the noise strength per sensor decays polynomially with the number of sensors (an unphysical condition), HNLS will always degrade the asymptotic time complexity of a sensing protocol to the classical limit irrespective of the kinds of quantum error correction that are available. Unfortunately, in practical settings noise often acts through the same generators as the signal (for instance, a Ramsey or DD sensor dephases along the signal axis, so its coherence time is finite). A major motivating question in metrology is therefore to understand whether there are practically important settings in which HNLS does not apply; thus far, a concrete and broadly applicable setting in which this is the case has not been found.

We show that spectroscopy is precisely such a setting. Motivated by HNLS and the fact that our signal Hamiltonian is generated by a Pauli Z operator, we model the density matrix $\rho$ of our sensor as evolving under the master equation
\begin{equation}\label{eq:noisy_signal_evol}
    \frac{d}{dt}\rho = -i\left[\frac{1}{2}A^Tf(t)Z+H_c(t), \rho\right] + \frac{\gamma}{4}\sum_{J\in\{X,Y,Z\}}\left(J\rho J-\rho\right) \ .
\end{equation}
This is a strong noise model physically equivalent to isotropic white noise of equal strength at every frequency. Moreover, whenever our algorithm couples the sensor to an interface qubit, we assume the corresponding two-qubit control operation has a constant probability $0\leq p<1/2$ of producing a $Z$ fault on each participating qubit. We take the fault events in distinct calls to be independent. Despite the signal Hamiltonian clearly lying in the span of the noise, we find that whenever $\epsilon$ is not in the extremely fine regime $\epsilon<1/\sqrt{N}$, even if $\gamma$ is a positive constant, our algorithm retains an advantage of $\sqrt{N}$ or $\kappa/\epsilon$ over the corresponding Heisenberg limit on conventional single-qubit AC sensing. To our knowledge, this constitutes the first quantum metrological advantage for AC sensing retaining an asymptotic quantum advantage under constant signal-aligned noise.

\subsection{Quantum computational frequency comb for phase oracle transduction}
The first key piece of our algorithm is a procedure for transforming an ambient classical field sensed by an unprotected single-spin sensor into a phase oracle acting on encoded quantum memory. Importantly, the protocol transfers all relevant quantum information onto the encoded memory and away from the sensor, so that even if the sensor decoheres immediately afterwards the relevant information is retained. 

\subsubsection{Elementary transduction}
We begin by isolating the basic operation through which a short interval of physical sensing writes information into the quantum memory; this is not yet the complete oracle transduction gadget. This operation requires a known Hermitian operator $O$ acting on the memory Hilbert space and a sensing interval $I$; for now it is agnostic to the actual quantum state of the memory. 

In the following algorithm, we sense over a time interval $I=[t_I-\tau/2,t_I+\tau/2]$ and take noise parameters $\gamma, p \geq 0$ as above. We set $\omega_c=(\omega_1+\omega_N)/2$, $\alpha=1-2p$, and define the signal phase
\begin{equation}
    \theta_{c,I}=\int_{-\tau/2}^{\tau/2} B_A(t_I+u)\cos(\omega_cu)\,du.
\end{equation}
We also define
\begin{equation}
    \mathcal{Q}_O(\rho)=\sin(O)\rho\sin(O)+\cos(O)\rho\cos(O).
\end{equation}
The goal of this elementary procedure is to act on a memory state $\rho_{\rm mem}$ with a channel $\mathcal{T}_{O,I}$ satisfying, whenever $(\sqrt{N}+\gamma)\tau\leq 1$,
\begin{equation}
    \left\|\mathcal{T}_{O,I}
    -
    \exp\left[
    -\frac{i\alpha\theta_{c,I}}{2}[\sin(O),\cdot]
    +
    \frac{\gamma\tau}{2}(\mathcal{Q}_O-\mathcal{I})
    \right]\right\|_\diamond
    \leq c(N+\gamma^2)\tau^2.
\end{equation}
The algorithm itself requires access to the unitary 
\begin{equation}
    U_O = \exp\left[-\frac{i}{2}\left(\frac{\pi}{2}\mathds{1}_{\rm mem} - O\right)\otimes Y_a  \ \right]\ ,
\end{equation}
where $a$ labels one of the interface qubits below. Without the bridge gadget, we simply replace $Y_a$ with $Y_s$. Later, when we choose the operator $O$ for the sensing algorithm, we will give an efficient circuit that realizes this unitary. Below we also use $H$ to denote a Hadamard gate; if $H$ is instead a Hamiltonian it will be made explicit.

\begin{algorithm}[h]
\caption{Elementary signal-to-memory transduction}
\label{alg:qcs_transduction}
\KwIn{A memory register in state $\rho_{\rm mem}$, three interface qubits labeled by $a, b, c$, query access to the unitary $U_O$, and a sensing interval $I=[t_I-\tau/2,t_I+\tau/2]$ satisfying $\tau \leq c/(\sqrt{N} + \gamma)$.}
\KwOut{The memory state updated by the channel $\mathcal{T}_{O,I}$ defined above. All interface qubits and the sensor are disentangled from $M$ and reset.}

Initialize logical interface qubits $a, b, c$ and physical sensor qubit $s$ in $\ket{0}$.

Apply $U_O$ jointly to the memory and qubit $a$, followed by $\operatorname{CNOT}_{a\rightarrow b}$. On a memory eigenstate $O\ket{o}=o\ket{o}$, this prepares
\begin{equation}
    \ket{o}\ket{00}_{ab}
    \longmapsto
    \ket{o}\left[
    \cos\left(\frac{\pi/2-o}{2}\right)\ket{00}_{ab}
    +
    \sin\left(\frac{\pi/2-o}{2}\right)\ket{11}_{ab}
    \right].
\end{equation}

Apply $H_a$, then $\operatorname{SWAP}_{as}$, then $H_s$.

Apply $R_x(-\omega_c\tau/2)$ to $s$, allow the sensor to evolve under \eqref{eq:noisy_signal_evol} over interval $I$ with the static control $H_c=\omega_c X_s/2$, and then apply $R_x(-\omega_c\tau/2)$ again.

Apply $H_s$, then $\operatorname{SWAP}_{sa}$, then $H_a$.

Measure the parity $Z_aZ_b$ using the syndrome qubit $c$. If the measured parity is odd, apply $X_a$.

Apply $\operatorname{CNOT}_{a\rightarrow b}$ and apply $U_O^\dagger$ to the memory and $a$.

Measure and reset qubits $a, b, c, s$.
\end{algorithm}
One can see that Algorithm \ref{alg:qcs_transduction} satisfies the claimed channel as follows. We first work with a pure state of the memory
\begin{equation}
    \ket{\Psi}_{\rm mem}=\sum_o c_o\ket{o},
    \qquad
    O\ket{o}=o\ket{o},
\end{equation}
since the transformation extends to arbitrary density matrices by linearity. We now explicitly track the joint state of the memory, interface, and sensor.

After applying the initial $U_O$ and $\operatorname{CNOT}_{a\rightarrow b}$, the joint state is
\begin{align}
    \ket{\Psi_1}
    =
    \sum_o c_o\ket{o}
    \bigg[
    &\cos\left(\frac{\pi/2-o}{2}\right)\ket{00}_{ab}
    +
    \sin\left(\frac{\pi/2-o}{2}\right)\ket{11}_{ab}
    \bigg]
    \ket{0}_s.
\end{align}
Thus different-eigenvalue branches of the memory coherently prepare different states of the interface qubits. Next upon applying $H_a$, $\operatorname{SWAP}_{as}$, and $H_s$, the branch-dependence gets transferred to the sensor,
\begin{align}
    \ket{\Psi_2}
    =
    \sum_o c_o\ket{o}
    \bigg[
    &\cos\left(\frac{\pi/2-o}{2}\right)\ket{0_a0_b0_s}
    +
    \sin\left(\frac{\pi/2-o}{2}\right)\ket{0_a1_b1_s}
    \bigg].
\end{align}
Next consider the sensing period. The two rotations and the static control $H_c=\omega_cX_s/2$ multiply to the identity when both the signal and depolarization are absent. When the signal is present, write $u = t - t_I$ where $t_I$ denotes the midpoint of the interval $I$, and move into the interaction frame of the control. In this frame, the sensor operator is
\begin{equation}
    n(u)
    =
    \cos(\omega_cu)Z_s+\sin(\omega_cu)Y_s.
\end{equation}
The signal Hamiltonian is transformed in this way, while the depolarizing dissipator is invariant under the control rotation. Thus in the interaction frame the master equation is
\begin{equation}
    \frac{d}{du}\rho = -\frac{i}{2}A^Tf(t_I + u)[n(u), \rho] + \frac{\gamma}{4}\sum_{J\in\{X_s,Y_s,Z_s\}}(J\rho J - \rho).
\end{equation}
To first order, the state after sensing is
\begin{equation}
    \rho_f = \rho_i -\frac{i}{2}\int_{-\tau/2}^{\tau/2} B(u + t_I)[n(u), \rho_i] \ du + \frac{\gamma\tau}{4}\sum_{J\in\{X_s,Y_s,Z_s\}}(J\rho_iJ-\rho_i) + O((N+\gamma^2)\tau^2) \ .
\end{equation}
Now consider how this changes under the parity measurement and correction. Immediately before sensing, the state of qubits $s$ and $b$ lies in the even-parity subspace where the eigenvalue of $Z_sZ_b$ is one. Writing
\begin{equation}
    P_0=\frac{\mathds{1}+Z_sZ_b}{2},
    \qquad
    P_1=\mathds{1}-P_0,
\end{equation}
the parity measurement and correction are described, before swapping the sensor back into qubit $a$, by
\begin{equation}
    \mathcal{R}(\sigma)
    =
    P_0\sigma P_0
    +
    X_sP_1\sigma P_1X_s.
\end{equation}
Since $Z_s$ preserves parity while $X_s$ and $Y_s$ change it, the three Pauli errors transform differently under this recovery. For every state $\sigma$ supported on the even-parity subspace,
\begin{align}
    &\mathcal{R}([n(u),\sigma])
    =
    \cos(\omega_cu)[Z_s,\sigma],\\
    &\mathcal{R}(X_s\sigma X_s)=\sigma,
    \qquad
    \mathcal{R}(Y_s\sigma Y_s)=\mathcal{R}(Z_s\sigma Z_s)=Z_s\sigma Z_s.
\end{align}
Combining the two nontrivial Pauli contributions, we have
\begin{equation}
    \rho_{f,\mathrm{rec}} = \rho_i - \frac{i\theta_{c, I}}{2}[Z_s, \rho_i] + \frac{\gamma\tau}{2}(Z_s\rho_i Z_s - \rho_i) + O((N+\gamma^2)\tau^2) \ .
\end{equation}
Now, restore the $Z$ faults from the two SWAP gates. Let $E,F\in\{0,1\}$ indicate whether a $Z$ fault occurs on the qubit carrying the encoded state immediately after the first and second SWAP, respectively. Faults on the other SWAP output act on a qubit in $\ket{0}$ and have no effect. After the surrounding Hadamards the faults $E,F$ appear as $X$ faults. The $X^E$ occurring before sensing commutes with the static $X_s$ control and the two endpoint rotations, while
\begin{equation}
    X_s^E n(u)X_s^E=(-1)^E n(u).
\end{equation}
Thus the first fault reverses the sign of the recovered signal term but leaves the depolarizing term unchanged. The parity recovery removes the second fault. Averaging over $E$ therefore multiplies the signal term by
\begin{equation}
    \alpha=\mathbb{E}[(-1)^E]=1-2p,
\end{equation}
without changing the depolarizing term. Finally, after applying $\operatorname{CNOT}_{a\rightarrow b}$ and $U_O^\dagger$, the $[Z_s, \cdot]$ is replaced with $[\sin(O), \cdot]$ acting on the memory, while a $Z_s$ error term $Z_s (\cdot)Z_s$  is replaced with the channel $\mathcal{Q}_O$ on the memory. Consequently the first-order term generating the evolution of the memory state is
\begin{equation}
    -\frac{i\alpha\theta_{c,I}}{2}[\sin O,\cdot]
    +
    \frac{\gamma\tau}{2}(\mathcal{Q}_O-\mathcal{I}).
\end{equation}
Since $\|n(u)\|=1$ and $|B_A(t)|\leq\sqrt{N}$, the interaction-frame generator has diamond norm at most $\sqrt{N}+3\gamma/2$. The remaining Dyson terms have diamond norm at most $c(N+\gamma^2)\tau^2$ whenever $(\sqrt{N}+\gamma)\tau\leq1$; this timescale upper bound will always be satisfied when we use this subroutine. This gives the channel claimed above. The core point can be seen by taking the limit of short evolution and no noise, where the memory undergoes the transformation
\begin{equation}
    \rho
    \longmapsto
    \rho
    -
    \frac{i\alpha\theta_{c,I}}{2}[\sin O,\rho]
    +
    O(N\tau^2).
\end{equation}
The signal has been transferred into a branch-dependent phase generated by $\sin O$, while the physical sensor is reset after the transduction step.

\subsubsection{Memory preparation and complete transduction}

Thus far we have black-boxed the role of the quantum memory register. Now we describe how the state of the memory is prepared and how the operator $O$ is selected. The core intuition of the algorithm, to be made explicit shortly, is to treat the memory register as if it were a system of $2N$ continuous-variable modes, each with their own canonical position and momentum operators, and to select $O$ so that it acts as a position operator $\widehat{x}_1+ ... + \widehat{x}_{2N}$ would on the resulting Hilbert space. One can also interpret the memory as a system of $2N$ large-$J$ spins and select $O$ as an appropriate angular momentum operator. Then, the phase oracle engineered by the elementary transduction step acts as a generator of translations in momentum or angular momentum space. By applying a phase-space rotation operator to each memory register, we switch bases: in the bosonic interpretation, we swap position for momentum, and in the spin interpretation, we rotate the spin vector by $90^\circ$ around its polarization axis to swap $J_x$ for $J_y$. This creates a displaced ``coherent state'' of the memory, which we can measure with homodyne/angular momentum measurements. The qubit-native circuits that implement these operations are simple, only requiring logarithmic memory and circuit overhead per constituent spin or mode.

In this subsection we will make precise how this physical interpretation can be implemented on an $\Tilde{O}(N)$-sized qubit register, and use this to construct the phase oracle transduction algorithm. Henceforth let $M$ refer to the memory register. For each of the $2N$ signal quadratures, introduce one $D$-dimensional logical memory register $M_\mu$, encoded into $\log_2 D$ qubits. We will show later that $D$ need only be polynomial in $N, \kappa, 1/\epsilon$, so the qubit overhead per effective ``mode'' is logarithmic. On each register define the truncated ladder and position operators
\begin{equation}
    a_D=\sum_{n=1}^{D-1}\sqrt{n}\ket{n-1}\!\bra{n},
    \qquad
    X_D=\frac{a_D+a_D^\dagger}{\sqrt{2}}.
\end{equation}
The complete data memory is initialized in the vacuum state
\begin{equation}
    \ket{\Psi_0}_{\rm mem}
    =
    \ket{0}_{M_1}\otimes\cdots\otimes\ket{0}_{M_{2N}}.
\end{equation}
where each $\ket{0}_{M_j}$ refers to a $D$-dimensional system, and this effective vacuum state corresponds to the all$-\ket{0}$ state on $2N\log_2 D$ qubits. Before sensing, we classically diagonalize the known $D\times D$ matrix $X_D$ and apply the resulting basis change independently to each register; now we are in the ``position'' eigenbasis, with a known correspondence between physical computational basis states and position eigenstates.

We will track the state of the memory using this complete position basis. Writing $X=(X_1,\ldots,X_{2N})^T$ for the collection of position operators, the memory state may be written as
\begin{equation}
    \ket{\Psi_0}_{\rm mem}
    =
    \sum_x \Psi_0(x)\ket{x},
    \qquad
    X\ket{x}=x\ket{x}.
\end{equation}
If $D$ were infinite, $|\Psi_0(x)|^2$ would be a continuous function describing the position distribution of a vacuum field in $2N$ spatial dimensions. For finite $D$, the position space is discretized, but the only property we will need below is that every linear form $w^TX$ has sub-Gaussian tails with variance proportional to $\|w\|_2^2$.

To see how this memory is used by Algorithm~\ref{alg:qcs_transduction}, consider first the simplest choice $O=X_1$, where $X_1$ acts on the first logical memory register. On a position-basis state $\ket{x}$ we then have $O\ket{x}=x_1\ket{x}$. In the noiseless short-interval limit, Algorithm~\ref{alg:qcs_transduction} therefore acts as
\begin{equation}
    \sum_x\Psi_0(x)\ket{x}
    \longmapsto
    \sum_x\Psi_0(x)e^{-i\theta_{c,I}\sin(x_1)/2}\ket{x}
    +O(N\tau^2).
\end{equation}
Thus the signal phase acquired by the physical sensor has become a position-dependent phase on the quantum memory: different branches $\ket{x}$ acquire different phases according to their value of $x_1$, even though the sensor is reset after the transduction step. More generally, choosing any known linear combination $O=\sum_{\mu=1}^{2N}c_\mu X_\mu$ produces a phase depending on $\sin(\sum_\mu c_\mu x_\mu)$. The phase-oracle construction below will choose these coefficients and combine several such transduction steps so that the sine dependence is converted into the desired linear phase in the memory coordinates.

The remaining problem is that Algorithm~\ref{alg:qcs_transduction} generates $\sin O$, whereas the phase oracle we require is linear in $O$. To remove this nonlinearity, we make multiple carefully-designed calls to the elementary transduction step in sequence so that the resulting phase is approximately linear. Recall that the arcsin function has a power series decomposition:
\begin{equation}
    \arcsin(x)
    =
    \sum_{m=0}^{\infty}
    \frac{\binom{2m}{m}}{4^m(2m+1)}
    x^{2m+1}.
\end{equation}
Then consider
\begin{equation}
    O
    =
    \arcsin(\sin O)
    =
    \sum_{m=0}^{\infty}
    \frac{\binom{2m}{m}}{4^m(2m+1)}
    \sin^{2m+1}O,
\end{equation}
which is valid whenever the spectrum of $O$ lies entirely within the principal branch of arcsin, which happens whenever $\|O\|\leq 1/2$. We truncate the power series, use trigonometric identities, and rewrite as 
\begin{equation}
    q_b(u)
    =
    \sum_{m=0}^{b}
    \frac{\binom{2m}{m}}{4^m(2m+1)}
    \sin^{2m+1}(u)
    =
    \sum_{\ell}a_\ell\sin(\ell u).
\end{equation}
For $|u|\le 1/2$, $q_b(u)$ converges exponentially quickly to $u$. We also define $s_0=\sum_\ell |a_\ell|$, for which $1\le s_0\le \pi/2$. Importantly, all $a_\ell$ are known exactly; we can then use the elementary transduction step to realize the desired weighted combination by sampling harmonic $\ell$ with probability $|a_\ell|/s_0$ and using the sign of $a_\ell$ in the memory operator. We make this idea concrete in Algorithm \ref{alg:qcs_phase_oracle}.

\begin{algorithm}[t]
\caption{Phase-oracle transduction}
\label{alg:qcs_phase_oracle}

\KwIn{The $2N$ $D$-dimensional memory registers initialized in $\ket{\Psi_0}_{\rm mem}$, total interrogation time $T$, and a target implementation error $\eta>0$.}

\KwOut{A memory state $\rho_{\rm out}$ satisfying
\begin{equation}
    \frac{1}{2}
    \left\|
    \rho_{\rm out}
    -
    \mathbb{E}_{\xi}\left[
    e^{-i\lambda(A+\xi)^TX}
    \ket{\Psi_0}\!\bra{\Psi_0}
    e^{i\lambda(A+\xi)^TX}
    \right]
    \right\|_1
    \leq \eta,
\end{equation}
where $\lambda=\alpha T/(2s_0C)$ and $\xi$ is a mean-zero random vector satisfying $ \|\operatorname{Cov}(\xi)\|_{\rm op}
    =
    \widetilde O\left(\frac{\gamma}{\alpha^2T}\right)$.}

Set $\tau = T/J$ and
\begin{equation}
    W_0=\Theta\left(\frac{\log(2N)}{\Delta_\omega}\right), \qquad  C=\Theta\left(\sqrt{N\log\frac{1+T}{\eta}}\right) \ .
\end{equation}

\For{$k=1,\ldots,J$}{
    Set
    \begin{equation}
        u_k=t_0+\frac{\tau}{2}+(k-1)(W_0+\tau),
    \end{equation}
    and compute the Gram matrix
    \begin{equation}
        G_k
        =
        \frac{1}{W_0}
        \int_{u_k}^{u_k+W_0}
        v_\tau(t)v_\tau(t)^T\,dt.
    \end{equation}

    Independently sample $t_k$ uniformly from $[u_k,u_k+W_0]$ and sample an odd integer harmonic $\ell_k$ with probability $|a_{\ell_k}|/s_0$.

    Define
    \begin{equation}
        w_k=G_k^{-1}v_\tau(t_k),
        \qquad
        O_k
        =
        \operatorname{sgn}(a_{\ell_k})
        \frac{\ell_k}{C}w_k^T X,
    \end{equation}
    and let $I_k=[t_k-\tau/2,t_k+\tau/2]$.

    Apply Algorithm~\ref{alg:qcs_transduction} with memory operator $O_k$ and sensing interval $I_k$.}

Undo the local $X_D$-eigenbasis change on every memory register and return the memory.
\end{algorithm}

The algorithm uses the following variables. We let $T$ denote the total interrogation time allowed for the phase oracle transduction. Then, $J$ is the number of calls the algorithm needs to make to the elementary transduction step; for a phase oracle implementation error $\eta>0$, we will show that taking $J = \Theta(1+NT\Delta_\omega + (N+\gamma^2)T^2/\eta)$ is sufficient. We also choose the power series truncation cutoff $b = \Theta(\log((1+T)/\eta))$, and as above let $\omega_c = (\omega_1 + \omega_N)/2$. Finally, we denote the response vector under this static driven control by
\begin{equation}\label{eq:modulated_v_r}
    v_\tau(t)
    =
    \frac{1}{\tau}\int_{-\tau/2}^{\tau/2}
    \cos(\omega_cu)f(t+u)\,du.
\end{equation}

Let us take stock of what Algorithm \ref{alg:qcs_phase_oracle} accomplishes. Ignore noise for a moment, so $\alpha = 1$ and $\gamma=0$. Upon evolving for a time $T$, the output of the algorithm is a memory state close in trace distance to $e^{-iT(A^TX)/\sqrt{N}}\ket{\Psi_0}$. Since $X = (X_D^{(1)}, ..., X_D^{2N})$, we have stored information about all $2N$ entries of $A$, with a phase damping factor of only $\sqrt{N}$. Having used only a single sensor qubit, we have transduced an analog signal into a coherent quantum phase oracle, while \textit{demultiplexing} the large amount of spectroscopic data into a distributed operator imprinting all $O(N)$ bits of that information onto a quantum memory. Our protocol can thus be considered as an efficient analog-to-digital transducer of classical information into quantum memory, with a single qubit serving as the bridge; moreover the quantum advantage that arises from our algorithm can be attributed to the relative gain we achieve by encoding $O(N)$ amplitudes while suffering only $\sqrt{N}$ damping. Importantly, sensor depolarization only causes a random shift of the phase whose distribution is known, while constant-probability interface errors only damp the realized phase by a constant independent of $N$; neither asymptotically damages the efficiency with which the transduced data can then be read out.  We defer the rigorous proof of these algorithmic guarantees to Section~\ref{sec:deferred_oracle_transduction}.

\subsection{Quantum algorithms for leveraging phase oracles}

Algorithm~\ref{alg:qcs_phase_oracle} leaves the spectral information encoded as a displacement generated by the position operators $X$. We now give two ways to read out this displacement efficiently and capture the quantum computational advantage. The first measures the displacement coordinate by coordinate, producing a reusable classical estimate of $A$. In this case, our readout protocol generalizes the algorithm given in Ref.~\cite{jordan_fast_2005} for efficient estimation of the gradient of a smooth function given phase-oracle query access. Here, one can think of the operator-valued phase $A^TX$ as a map from $\mathbb R^{2N} \rightarrow \mathbb R$, and our readout method accesses all $2N$ gradients, or entries of $A$, in one phase-oracle query. This independently suggests that modifications of our transduction step that may encode signal data in richer, nonlinear ways could be combined with this gradient-estimation primitive to extract different kinds of data from the signal beyond its spectral amplitudes.

The second readout method instead retains the coherence of the memory and directly measures a nonlinear observable, which will give improved performance for specified nonlinear targets. Computationally, this readout method enables a hierarchy of quantum algorithms for estimating polynomial functionals of a phase oracle gradient, strictly generalizing the protocol of Ref.~\cite{jordan_fast_2005}. Our algorithms further give physical intuition for this canonical query-complexity quantum advantage — efficient gradient estimation corresponds to measuring conjugate angular momenta of a large-$J$ spin system, while higher-order polynomials of the gradient equate to estimation of nonlinear observables such as spin squeezing or one-axis twisting measurements.

Recall that the phase-oracle gain is $\lambda=\alpha T/(2s_0C)$, and define $h=\lambda/\sqrt{2}$. On each $D$-dimensional memory register let
\begin{equation}
    n_D=\sum_{n=0}^{D-1}n\ket{n}\!\bra{n},\qquad R_D=\exp\left(\frac{i\pi}{2}n_D\right).
\end{equation}
The unitary $R_D$ is a quarter-turn in the virtual oscillator phase space. In the ideal oscillator model, applying it after the phase oracle transforms
\begin{equation}
    e^{-i\lambda(A+\xi)^TX}\ket{\Psi_0}\longmapsto\ket{h(A+\xi)},
\end{equation}
where the state on the right is a product of spin coherent states with real amplitudes $h(A+\xi)$. In the finite register, $R_D$ is diagonal in the occupation basis and can be implemented directly from the binary representation of $n$. This operation thus allows us to move into a basis where the spectroscopic information is directly accessible; we use it as the starting point for both readout methods below.

\subsubsection{Linear spectroscopy readout}

For linear spectroscopy we simply measure the displaced quadrature of every memory mode. Since all of these measurements act on distinct memory registers, they commute and can be performed in parallel.

\begin{algorithm}[h]
\caption{Linear phase-oracle readout}
\label{alg:qcs_linear_readout}

\KwIn{The $2N$ memory registers returned by Algorithm~\ref{alg:qcs_phase_oracle}, with known gain $h=\lambda/\sqrt{2}$.}

\KwOut{A classical vector $U\in\mathbb R^{2N}$ estimating the complete spectral vector $A$.}

Apply the quarter-turn $R_D$ to every memory register.

\For{$\mu=1,\ldots,2N$}{
    Measure the operator $X_D=(a_D+a_D^\dagger)/\sqrt{2}$ on register $M_\mu$ by applying its known local eigenbasis change followed by computational-basis measurement. Let $x_\mu$ denote the measured eigenvalue.

    Set $U_\mu=x_\mu/(\sqrt{2}h)$.
}

Return $U=(U_1,\ldots,U_{2N})^T$.
\end{algorithm}
In the ideal oscillator model, Algorithm~\ref{alg:qcs_linear_readout} has the particularly simple representation
\begin{equation} \label{eq:ideal_oscillator_dist}
    U=A+\xi+G,\qquad G\sim\mathcal N\left(0,\frac{1}{4h^2}\mathds{1}_{2N}\right),
\end{equation}
where $G$ is independent of the noise-induced displacement $\xi$. In particular, $\E[U]=A$ and
\begin{equation}
    \operatorname{Cov}(U)=\operatorname{Cov}(\xi)+\frac{1}{4h^2}\mathds{1}_{2N}.
\end{equation}
Thus a single execution returns a classical record from which any desired linear observable $K^TA$ can subsequently be estimated as $K^TU$. We will account for the finite cutoff $D$ and the phase-oracle implementation error when proving the complete algorithmic guarantee below. These approximation errors can be thought of as a realization of the Holstein-Primakoff approximation when truncating a bosonic Hilbert space to a spin system.

Physically, this algorithm requires only homodyne readout with bosonic memory, transverse angular momentum measurement of a spin memory, or a shallow change of basis and computational basis measurement with qubits. Later we also show how to create an oracle that writes the coefficients into qubit phases instead of displacements so that linear readout becomes ordinary phase unwrapping.

\subsubsection{Nonlinear spectroscopy readout}

For nonlinear targets, measuring every coordinate and evaluating a polynomial classically discards the coherence that remains in the memory. Instead, we directly measure a normally ordered polynomial observable. Let $K$ be a real symmetric rank-$L$ tensor, let $m=\lfloor L/2\rfloor$, and let $r=\lceil L/2\rceil$. We prepare $r$ independent memory states using Algorithm~\ref{alg:qcs_phase_oracle}, using the same gain $h$ but otherwise independent acquisition randomness and sensor depolarization. Write $\Sigma^{(j)}=\operatorname{Cov}(\xi^{(j)})$ for the known displacement-noise covariance of copy $j$.

After the quarter-turn, let $c^{(j)}_\mu$ denote the truncated ladder operator $a_D$ acting on mode $\mu$ of memory copy $j$. For the first $m$ copies define
\begin{equation}
    Q^{(j)}_{\mu\nu}=\frac{(c^{(j)}_\mu)^\dagger c^{(j)}_\nu}{h^2}-\Sigma^{(j)}_{\mu\nu}\mathds{1}.
\end{equation}
If $L$ is odd, define on the final copy
\begin{equation}
    W_\mu=\frac{c^{(r)}_\mu+(c^{(r)}_\mu)^\dagger}{2h}.
\end{equation}
The Hermitian operator to be measured is then
\begin{equation}
    O_{K,D}=\sum_{\mu_1,\ldots,\mu_L=1}^{2N}K_{\mu_1\ldots\mu_L}\prod_{j=1}^{m}Q^{(j)}_{\mu_{2j-1}\mu_{2j}}\times\begin{cases}1,&L\text{ even},\\W_{\mu_L},&L\text{ odd}.\end{cases}
\end{equation}
In the ideal effective model, the normal-ordering subtraction makes its expectation exactly $\E[O_{K,D}]=K[A^{\otimes L}]$, up to the finite-$D$ approximation that we control below. In essence, $O_{K,D}$ captures the desired observable by building it from expectation values of quadratic observables, appending an additional linear term if the rank is odd. This is possible because only the first and second moments of each independent memory copy enter the expectation.

The operator $O_{K,D}$ is also efficiently measurable for fixed $L$. Expanding it gives $O_L((2N)^L)$ known Hermitian terms, each acting on at most $L$ of the $D$-dimensional mode registers. Let $\Lambda$ denote the sum of the operator norms of these terms, so $\|O_{K,D}\|_{\rm op}\leq\Lambda$. Each term can be classically diagonalized and its controlled exponential implemented with polynomially many gates because $L$ is fixed and $D$ is polynomial. We therefore measure $O_{K,D}$ by iterative phase estimation of the known unitary
\begin{equation}
    U_K=\exp\left[i\frac{O_{K,D}+2\Lambda\mathds{1}}{8\Lambda}\right].
\end{equation}
Its eigenphases lie in the fixed interval $[1/8,3/8]$, so there is no phase-wrap ambiguity. Taking $O(\log(\Lambda/\eta_{\rm out}))$ phase-estimation bits gives additive readout resolution $\eta_{\rm out}$, with controlled powers synthesized by standard product-formula simulation using polynomially many memory-only gates.

\begin{algorithm}[ht]
\caption{Nonlinear phase-oracle readout}
\label{alg:qcs_nonlinear_readout}

\KwIn{$r=\lceil L/2\rceil$ independent memory states returned by Algorithm~\ref{alg:qcs_phase_oracle}, a known real symmetric rank-$L$ tensor $K$, the corresponding known covariance matrices $\Sigma^{(1)},\ldots,\Sigma^{(r)}$, and readout precision $\eta_{\rm out}>0$.}

\KwOut{A scalar outcome $\widehat F_K$ obtained from an additive-$\eta_{\rm out}$ spectral measurement of $O_{K,D}$.}

Apply the quarter-turn $R_D$ to every mode of every memory copy.

Construct $O_{K,D}$ defined above.

Compute the known bound $\Lambda$ from its local-term decomposition and form $U_K=\exp[i(O_{K,D}+2\Lambda\mathds{1})/(8\Lambda)]$.

Perform iterative phase estimation of $U_K$ to angular precision $\eta_{\rm out}/(8\Lambda)$, using one control qubit, controlled powers of $U_K$, and adaptive measurements of the control qubit~\cite{dobsicek_iterative_2007}.  

Decode the measured phase into the corresponding eigenvalue of $O_{K,D}$ and return it as $\widehat F_K$.
\end{algorithm}

Unlike Algorithm~\ref{alg:qcs_linear_readout}, Algorithm~\ref{alg:qcs_nonlinear_readout} is a matched protocol, since different tensors $K$ define different, potentially noncommuting observables $O_{K,D}$. Thus one coherent memory preparation cannot in general be destructively reused for arbitrarily many nonlinear targets. The benefit is that the normally ordered measurement removes vacuum fluctuations before measurement rather than estimating and subtracting them from a classical record. We will thus leverage this readout method to develop a matched analog of QCS below.

Full digital implementation of the controlled unitary is not generally necessary. With a bosonic memory, nonlinear readout is equivalent to passive linear optics and photon-number readout, while spin memories require spin squeezing measurements. The matched analog of QCS presented below requires yet simpler control capabilities, for which nonlinear readout can be implemented using few memory qubits.

\subsection{Reusable QCS algorithm}

We now assemble the phase-oracle transduction and readout subroutines into complete sensing algorithms. Here we state the version of QCS analogous to RQS, which collects a classical dataset supporting post-hoc estimation of many linear or nonlinear observables downstream. The resulting algorithm is significantly more efficient and noise-robust than RQS as it uses quantum computation. 

Before stating the algorithm, we must fix the dimension $D$ of each memory register. We will always take
\begin{equation}\label{eq:qcs_combined_cutoff}
    D = \Tilde{O}_L\left(1+\frac{\alpha^2 T^2}{N}\right)
\end{equation}
where $T$ is the time used per call to Algorithm \ref{alg:qcs_phase_oracle}. This time scales sub-linearly in $N$ for any choice of $L$, so the qubit overhead for each memory register will always be at most logarithmic in $N, 1/\epsilon$.

We now present \textit{reusable} quantum computational spectroscopy. We use the following time per batch,
\begin{equation}
    T_{\rm batch}
    =
    \widetilde O_L
    \begin{cases}
    \displaystyle
    \max\left\{
    \frac{\sqrt{2N}\kappa}{\alpha\epsilon},
    \frac{\gamma\kappa^2}{\alpha^2\epsilon^2}
    \right\},
    &
    \displaystyle
    \frac{\epsilon}{\kappa}\leq\frac{1}{\sqrt{2N}},
    \\[4mm]
    \displaystyle
    \max\left\{
    \frac{(2N)^{1-\frac{1}{2L}}}{\alpha}
    \left(\frac{\kappa}{\epsilon}\right)^{1/L},
    \frac{\gamma(2N)^{1-\frac1L}}{\alpha^2}
    \left(\frac{\kappa}{\epsilon}\right)^{2/L}
    \right\},
    &
    \displaystyle
    \frac{\epsilon}{\kappa}\geq\frac{1}{\sqrt{2N}}.
    \end{cases}
\end{equation}

\begin{algorithm}[h]
\DontPrintSemicolon
\caption{Reusable quantum computational spectroscopy}
\label{alg:qcs_reusable}
\KwIn{Known frequencies $\{\omega_j\}_{j=1}^N$, tensor rank $L$, accuracy $\epsilon$, norm bound $\kappa$, failure probability $\delta$, and an upper bound $M$ on the number of observables to be estimated.}
\KwOut{Estimates for any rank-$L$ tensors $K_1,\ldots,K_M$ satisfying $\|K_j\|_\infty\leq\kappa$, all to within accuracy $\epsilon$ with failure probability at most $\delta$.}

Set $R=\lceil C_L\log(2M/\delta)\rceil$.

\For{$r=1,\ldots,R$}{
    \For{$i=1,\ldots,L$}{
        Prepare a fresh memory register and run Algorithm~\ref{alg:qcs_phase_oracle} using interrogation time $T_{\rm batch}/L$ and precision $\eta = \Tilde{O}(1/L)$.

        Apply Algorithm~\ref{alg:qcs_linear_readout} to obtain a classical vector $U_i^{(r)}\in\mathbb R^{2N}$.
    }
}

After acquisition is complete, for every supplied tensor $K_j$ and every $r$, calculate
$$Z_j^{(r)}=K_j[U_1^{(r)},\ldots,U_L^{(r)}].$$

Return
$$\widehat F_j=\operatorname{median}_{r=1,\ldots,R}Z_j^{(r)}$$
for every $j=1,\ldots,M$.
\end{algorithm}
The full proof of correctness of Algorithm \ref{alg:qcs_reusable} that treats sensor depolarization and interface errors is presented in Section~\ref{sec:qcs_reusable_proof_guarantee}. Here we sketch the core intuition for how these subroutines come together and achieve beyond Heisenberg-limited scaling. 

First consider the phase oracle transducer. In one short sensing interval of duration $\tau$, the sensor acquires a phase $\theta(t)\simeq\tau v(t)^TA$, which the transducer couples to the memory operator $w(t)^TX/C$, where $w(t)=G^{-1}v(t)$ and $G=\mathbb E_t[v(t)v(t)^T]$. Up to fixed numerical factors, the resulting memory phase is therefore
\begin{equation}
\delta\Phi(t)\simeq
\frac{\tau}{C}(v(t)^TA)(w(t)^TX).
\end{equation}
The key observation is that our choice of $w(t)$ satisfies $\mathbb E_t[v(t)w(t)^T]=\mathds{1}$. Consequently,
\begin{equation}
\mathbb E_t[\delta\Phi(t)]
\simeq
\frac{\tau}{C}A^TX
=
\frac{\tau}{C}\sum_{\mu=1}^{2N}A_\mu X_\mu.
\end{equation}
Thus, averaged over the chosen sensing times, each interval writes a phase proportional to every spectral coefficient $A_\mu$, with the same gain $\tau/C$ in every direction. The cross terms between different spectral coordinates cancel, while the desired contributions accumulate coherently.

Now consider the normalization factor $C$. The memory is initially a product of vacuum states, whose position fluctuations are independent, so
\begin{equation}
\operatorname{Var}_0(w(t)^TX)=\frac12\|w(t)\|_2^2,
\qquad
\mathbb E_t\|w(t)\|_2^2
=\operatorname{tr}(G^{-1})=\Theta(N).
\end{equation}
Although the controlled rotation involves $2N$ memory coordinates, their \textit{variances} add, so its typical magnitude grows only as $\sqrt N$, rather than $N$. We can therefore take $C=\widetilde\Theta(\sqrt N)$, keeping the controlled rotations within the range required by the transducer. As such, we accumulate a signal phase of order $1/\sqrt N$ per unit interrogation time in each of the $2N$ spectral directions. After total interrogation time $T$, these contributions produce
\begin{equation}
U_A=\exp(-i\lambda A^TX),
\qquad
\lambda=\widetilde\Theta(T/\sqrt N).
\end{equation}
Equivalently, every memory momentum coordinate is displaced by an amount $\lambda A_\mu$, while vacuum fluctuations have constant variance. Only $\widetilde O(\sqrt N)$ interrogation time is needed to obtain unit-scale signal-to-noise ratio simultaneously in all $2N$ directions, rather than $O(N)$ time to address them individually.

Once this oracle has been constructed, the remaining task is computational. The unitary $U_A$ is a phase oracle for the linear function $x\mapsto\lambda A^Tx$, whose gradient is $\lambda A$. Jordan's quantum gradient-estimation algorithm~\cite{jordan_fast_2005} shows that this entire gradient can be extracted using only $O(1)$ phase-oracle queries, whereas classical access to scalar function evaluations requires $\Omega(N)$ queries to determine the gradient. Thus, after paying the physical cost of synthesizing the oracle, the quantum algorithm can recover all spectral coordinates in constant shots. Taking $\lambda=\Theta(\kappa/\epsilon)$ gives the interrogation time $\widetilde O(\sqrt{N}\,\kappa/\epsilon)$. From this computational perspective, QCS combines an $O(N)$ quantum oracle-query speedup with a $1/\sqrt{N}$ transduction gain per unit interrogation time, leaving a net $\sqrt{N}$ advantage over conventional Heisenberg-limited spectral acquisition in the high-precision regime.

It is valuable to understand the same advantage from the Fisher-information perspective; importantly, we must look at Fisher information of the \textit{memory} state, rather than the sensor itself. Acting on the memory vacuum, the ideal phase oracle produces the state
\begin{equation}
|\Psi_A\rangle
=e^{-i\lambda A^TX}|0\rangle^{\otimes 2N},
\end{equation}
which shifts every conjugate momentum by $-\lambda A_\mu$ without changing its vacuum covariance. In particular, the memory momentum vector has mean $-\lambda A$ and covariance $\mathds{1}/2$, while its quantum Fisher information matrix for $A$ is exactly
\begin{equation}
\mathcal I_{\mu\nu}
=4\lambda^2\operatorname{Cov}_0(X_\mu,X_\nu)
=2\lambda^2\delta_{\mu\nu}
=\widetilde\Theta(T^2/N)\delta_{\mu\nu}.
\end{equation}
Importantly, this information is simultaneously attainable in every direction: the momentum operators commute, so measuring all of them in parallel produces an estimator with covariance
\begin{equation}
\operatorname{Cov}(\widehat A)
=\frac{\mathds{1}}{2\lambda^2}
=\widetilde O\left(\frac{N}{T^2}\right)\mathds{1}.
\end{equation}
Consequently, any linear observable with norm at most $\kappa$ has estimation variance $O(N\kappa^2/T^2)$, immediately recovering the claimed interrogation-time scaling. The same readout produces a reusable classical record, and contracting independent records against a rank-$L$ tensor gives the corresponding nonlinear guarantees. In the sufficiently fine-precision regime, the leading variance is again $O_L(N\kappa^2/T^2)$, while higher-order noise terms account for the additional coarse-precision cost. The Fisher picture therefore tells the same story as the computational one: QCS does not increase the intrinsic sensitivity of the physical sensing qubit, but instead constructs a memory whose full-rank Fisher information can be accessed simultaneously, rather than acquiring spectral directions one at a time.

This reveals a rather different source of metrological advantage from the familiar optimization of quantum probe states. Conventional quantum metrology primarily asks how sensitively a physical probe can respond to an unknown parameter, using coherence, entanglement, or squeezing to improve the precision of that response. QCS instead acts as a quantum receiver for classical signals: the sensor transduces a broadband waveform into a coherent computational representation, and a quantum algorithm processes its many spectral degrees of freedom in superposition before measurement. The $1/\epsilon$ improvement retains the familiar character of coherent phase estimation, but the additional improvement in $N$ is fundamentally computational. QCS's computational multiplexing advantage is not intrinsically tied to a vanishing signal amplitude or to a vanishing target precision. Its transduction operates through short, controlled interactions even for finite-strength signals, and its ability to process spectral information jointly survives in regimes where sensor noise would eliminate conventional Heisenberg scaling. Quantum sensing advantages are not limited to making a probe more sensitive to weak signals; they can also arise from using quantum computation to extract and process much more information from the same physical sensing interaction.

Formally, the reusable QCS algorithm gives us the following performance guarantee.

\begin{graydefinitionbox}
\begin{theorem}[Reusable quantum computational spectroscopy]
\label{thm:qcs_rich_queries}
Fix an integer $L\geq1$, $\kappa>0$, $0<\epsilon<\kappa$, $0<\delta<1/4$, and $\gamma\geq0$. Suppose further that each sensor-memory interface gate causes a $Z$ error with constant probability $<1/2$. Then there exists a reusable quantum computational spectroscopy protocol which, for any fixed list of
$M$ rank-$L$ tensors $K_1,\ldots,K_M$ satisfying
$\|K_j\|_\infty\leq\kappa$, returns estimates obeying
\begin{equation}
\Pr\left[
\max_{j\leq M}
\left|\widehat F_j-K_j[A^{\otimes L}]\right|
\leq\epsilon
\right]\geq1-\delta
\end{equation}
for every $\|A\|_2\leq1$, using total interrogation time
\begin{equation}
T_{\rm int}\leq
\displaystyle
\widetilde O_L\left(\max\left\{\frac{\sqrt N\,\kappa}{\epsilon},
\frac{\gamma \kappa^2}{\epsilon^2}
, N^{1-\tfrac{1}{2L}}\left(\frac{\kappa}{\epsilon}\right)^{1/L}, \gamma N^{1-\tfrac{1}{L}}\left(\frac{\kappa}{\epsilon}\right)^{2/L}\right\}\right)
\label{eq:qcs_reusable_upper_bound}
\end{equation}
when sensor interrogation is described by equation~\ref{eq:noisy_signal_evol}. The acquisition is independent of the tensors, so the same classical dataset can estimate any list of observables supplied after sensing. Logarithmic dependence on $M, 1/\delta$ is suppressed.
\end{theorem}
\end{graydefinitionbox}

When $L=1$, we achieve the scaling $\sqrt{N}\left(\frac{\kappa}{\epsilon}\right)$ whenever $\gamma \leq \sqrt{N}\epsilon/\kappa$. In particular, outside the high-precision regime where $\epsilon$ vanishes with $1/\sqrt{N}$, the noiseless scaling dominates even if the noise $\gamma$ is constant. The same fact holds at different $L$, for different precision regimes. Note that $\alpha$ is suppressed since it is a constant for any constant interface error rate.

Algorithm \ref{alg:qcs_reusable} is optimal, even compared to quantum computational sensing algorithms with prior knowledge of the observables, when the number of observables $M$ is comparable to $N$ or noise is substantial. We establish this for the $L=1$ case in Section~\ref{sec:all_memory_lower_bound} where we prove the generalized Heisenberg limit, which shows that any protocol for AQS estimating $\Omega(N)$ observables, even with arbitrarily large and fast ancillary quantum memory and adaptivity, must interrogate the signal for total time $\widetilde \Omega(\sqrt{N}\kappa/\epsilon)$.

\subsection{Matched QCS algorithm}
\label{sec:qcs_target_aware}

The reusable QCS protocols above encode all $2N$ coordinates of $A$ into the memory, which is necessary when many observables are requested or when they are only supplied after sensing. When a few observables are specified in advance, as in the matched spectroscopy protocols of Section~\ref{sec:small_M_optimal}, it is sufficient to encode only the components of $A$ on which they depend. We first modify the phase oracle transducer to encode an $r<N$-dimensional projection of $A$, then show how it can be combined with either linear readout or coherent nonlinear readout to estimate the observables directly without reconstructing the $2N$ coefficients.

\subsubsection{Matched phase-oracle transduction}

Consider linear observables $K_1,\ldots,K_M$, with $\kappa=\max_j\|K_j\|_2$, and let $r$ be the dimension of their span. Choose a matrix $U\in\mathbb R^{2N\times r}$ with orthonormal columns spanning this $r$-dimensional subspace. Then
\begin{equation}
K_j^TA=(U^TK_j)^T(U^TA),
\end{equation}
with $\|U^TA\|_2\leq1$ and $\|U^TK_j\|_2=\|K_j\|_2$, so every requested observable is a linear function of the $r$-dimensional vector $U^TA$. We would therefore like to replace the phase oracle $e^{-i\lambda A^TX}$ on $2N$ memory registers by the oracle $e^{-i\lambda(U^TA)^TX}$ on $r$ registers, whose position operators we again collect into $X=(X_1,\ldots,X_r)^T$. Linear readout of this oracle returns $U^TA$ with readout noise of covariance $\mathds{1}_r/(2\lambda^2)$, so the interrogation time needed to reach a given precision is determined by how the gain $\lambda$ grows with $T$. For the full spectrum, Algorithm~\ref{alg:qcs_phase_oracle} gives $\lambda=\widetilde\Theta(\alpha T/\sqrt N)$. The goal of this section is to obtain $\lambda=\widetilde\Theta(\alpha T/\sqrt r)$, replacing the spectral dimension by the number of directions that the observables actually probe.

The construction is based on the dual matched filter from matched spectroscopy (Section~\ref{sec:small_M_optimal}). Write $\E_k$ for uniform averaging of $t$ over the $k$th window $[u_k,u_k+W_0]$ of Algorithm~\ref{alg:qcs_phase_oracle}, and recall that the coupling vector $w_k(t)=G_k^{-1}v_\tau(t)$ used there satisfies $\E_k[v_\tau w_k^T]=\mathds{1}$, which is what made the accumulated memory phase proportional to $A^TX$. Projecting this vector onto the query subspace defines the matched filter
\begin{equation}
q_k(t)=U^Tw_k(t)=U^TG_k^{-1}v_\tau(t)\in\mathbb R^r.
\end{equation}
This is the dual filter from matched spectroscopy, with the short Ramsey response $b(t)$ replaced by the driven response $v_\tau(t)$ of Equation~\eqref{eq:modulated_v_r}, due to the static demodulation drive used in QCS. This filter satisfies $\E_k[v_\tau q_k^T]=U$, so averaging the signal phases $\tau v_\tau(t)^TA$ against it extracts exactly $U^TA$. Moreover, since $G_k$ is well-conditioned for the intervals we take (Section~\ref{sec:deferred_oracle_transduction}), the filter is small on average, with
\begin{equation}
\E_k[q_kq_k^T]=U^TG_k^{-1}U\preceq \textnormal{const}\cdot\mathds{1}_r.
\end{equation}
Replacing $w_k(t_k)$ by $q_k(t_k)$ in Algorithm~\ref{alg:qcs_phase_oracle} would therefore write a phase proportional to $(U^TA)^TX$ into $r$ registers. Note that unlike matched spectroscopy, where $q_k$ was used to synthesize a pulse sequence, here $q_k$ is used to define the memory operator while the physical sensing control remains the same.

The nontrivial challenge in adapting this dual filter to QCS lies in normalizing the memory operator. In Algorithm~\ref{alg:qcs_phase_oracle}, the argument $w_k(t_k)^TX/C$ of the truncated arcsine series must lie in $[-1/2,1/2]$ except on a small tail of the vacuum wavefunction. This requires $C$ to exceed $\|w_k(t_k)\|_2$ up to logarithmic factors, and the gain $\lambda=\alpha T/(2s_0C)$ decreases accordingly. For the full spectrum this costs nothing beyond the number of encoded coordinates: $\|w_k(t)\|_2$ is typically of order $\sqrt N$, since $\E_k\|w_k\|_2^2=\operatorname{tr}G_k^{-1}\geq2N$, and it never exceeds a constant times $\sqrt N$. The matched filter behaves differently. Its average size is only $\E_k\|q_k(t)\|_2^2\leq O(r)$, but at isolated times where the frequency components of the requested waveforms add in phase, $\|q_k(t)\|_2$ can be as large as order $\sqrt N$. These are the same peaks of the dual matched filter that limited matched spectroscopy in the worst case, and normalizing by them would leave the gain at $\lambda=\widetilde\Theta(\alpha T/\sqrt N)$.

We remove this obstacle by leveraging the fact that large filter peaks can occupy only small time intervals. We therefore sample the sensing times in proportion to the size of the filter. On window $k$, define the sampling weight
\begin{equation}
\beta_k(t)=\sqrt{r+\|q_k(t)\|_2^2}
\end{equation}
with average $S_k=\E_k\beta_k$. We draw $t_k$ from the density $\beta_k(t)/(S_kW_0)$ on the window instead of the previous uniform draw, and use the normalized filter
\begin{equation}
\widetilde q_k(t)=\frac{S_k}{\sqrt{65r}}\frac{q_k(t)}{\beta_k(t)}.
\end{equation}
The denominator normalizes by $O(\sqrt r)$; the constant is chosen because our earlier analysis of the Gram matrix condition number gave $G_k^{-1}\preceq64\mathds{1}$, but this constant is not fundamental in practice.

Write $\widetilde{\mathbb E}_k$ for averaging over $t_k$ drawn from $\beta_k(t)/(S_kW_0)$. The weight $\beta_k(t)$ in the density cancels the one dividing the coupling, so $\widetilde{\mathbb E}_k[v_\tau\widetilde q_k^T]\propto\E_k[v_\tau q_k^T]=U$, and the mean signal is preserved. Together with $\sqrt r\leq S_k\leq\sqrt{65r}$, which follows from Jensen's inequality and $\E_k\|q_k\|_2^2\leq64r$, this gives the three properties
\begin{equation}
\|\widetilde q_k(t)\|_2\leq1,\qquad
\widetilde{\mathbb E}_k[v_\tau\widetilde q_k^T]
=\frac{U}{\sqrt{65r}},\qquad
\widetilde{\mathbb E}_k[\widetilde q_k\widetilde q_k^T]
\preceq\frac8r\mathds{1}_r,
\label{eq:qcs_matched_filter_properties}
\end{equation}
which we prove formally in Section~\ref{sec:deferred_matched_qcs}. The first bounds the coupling at every sampled time independently of $N$ and $r$, so the normalization of the memory operator only needs to be logarithmic. The second shows that the gain is set by the average size $\sim\sqrt r$ of the filter rather than by its peaks. The third controls the noise, as we discuss below, and is the reason we incorporated $r$ into the definition of $\beta_k$.

This is the same mechanism that matched spectroscopy used for the residual of its clipped filter in Algorithm~\ref{algo:specified_linear_spectroscopy_coarse}, where short interrogations at random times, weighted by the magnitude of the filter, estimated the part of the observable carried by its peaks. There each short interrogation was measured separately, so its small phase contributed a single noisy bit, and the residual was estimated with standard-quantum-limited precision. Here each short interrogation instead writes its phase into the memory, where the contributions of all sampled intervals add coherently before any measurement is made. The entire filter can therefore be implemented by importance sampling while retaining the coherent $1/\epsilon$ dependence on precision.

\begin{algorithm}[h]
\DontPrintSemicolon
\caption{Matched phase-oracle transduction}
\label{alg:qcs_matched_oracle}
\KwIn{$r$ memory registers of dimension $D$ initialized in $\ket{\Psi_0}_{\rm mem}$, a matrix $U\in\mathbb R^{2N\times r}$ with orthonormal columns, total interrogation time $T$, and a target implementation error $\eta>0$.}
\KwOut{A memory state $\rho_{\rm out}$ satisfying
\begin{equation}
\frac12\left\|
\rho_{\rm out}-
\mathbb E_\xi\left[
e^{-i\lambda(U^TA+\xi)^TX}
\ket{\Psi_0}\!\bra{\Psi_0}
e^{i\lambda(U^TA+\xi)^TX}
\right]\right\|_1\leq\eta,
\end{equation}
where $\lambda=\alpha T/(2s_0S_*C_0)=\widetilde\Theta(\alpha T/\sqrt r)$, $\xi\in\mathbb R^r$ is mean-zero with $\|\operatorname{Cov}(\xi)\|_{\rm op}=\widetilde O(\gamma/(\alpha^2T))$, and $S_*=\sqrt{65r}$.}

Choose $J$, $\tau=T/J$, $W_0$, and the window starts $u_k$ as in Algorithm~\ref{alg:qcs_phase_oracle}. Set $S_*=\sqrt{65r}$ and $C_0=\Theta\bigl(\sqrt{\log(N(1+T)/\eta)}\bigr)$.\;

\For{$k=1,\ldots,J$}{
Compute $G_k$, the matched filter $q_k(t)=U^TG_k^{-1}v_\tau(t)$, the weight $\beta_k(t)=\sqrt{r+\|q_k(t)\|_2^2}$, and its average $S_k=\E_k\beta_k$.\;

Independently sample $t_k$ from the density $\beta_k(t)/(S_kW_0)$ on $[u_k,u_k+W_0]$ and an odd harmonic $\ell_k$ with probability $|a_{\ell_k}|/s_0$.\;

Set $\widetilde q_k(t_k)=\frac{S_k}{S_*}\frac{q_k(t_k)}{\beta_k(t_k)}$, $O_k=\operatorname{sgn}(a_{\ell_k})\frac{\ell_k}{C_0}\widetilde q_k(t_k)^TX$, and $I_k=[t_k-\tau/2,t_k+\tau/2]$. Apply Algorithm~\ref{alg:qcs_transduction} with memory operator $O_k$ and sensing interval $I_k$.\;
}

Undo the local $X_D$-eigenbasis change on every memory register and return the memory.\;
\end{algorithm}

Compared with Algorithm~\ref{alg:qcs_phase_oracle}, the only changes made to Algorithm~\ref{alg:qcs_matched_oracle} are the projection onto the query subspace, the sampling density, and the normalization, which is now $\sqrt{65r}C_0=\widetilde\Theta(\sqrt r)$ rather than $C=\widetilde\Theta(\sqrt N)$. These modifications involve classical preprocessing and reprogramming the controlled rotations and sensing intervals; no additional experimental capability is required. The matched oracle uses $r\leq2N$ memory registers rather than the full $2N$ registers.

The proof of the guarantee, given in Section~\ref{sec:deferred_matched_qcs}, follows that of Algorithm~\ref{alg:qcs_phase_oracle} with the properties~\eqref{eq:qcs_matched_filter_properties} in place of the corresponding properties of $w_k$. The noise is handled as before, analyzing the memory momentum kicks induced by sensor depolarization along the directions $\pm\ell_k\widetilde q_k(t_k)/C_0$, and the third property in Equation~\eqref{eq:qcs_matched_filter_properties} bounds the covariance of those kicks as $1/r$. Memory correlations therefore decohere at a rate of order $\gamma/r$ while the signal phase grows as $T/\sqrt r$, which is the relation between signal and noise of Equation~\eqref{eq:quadratic_coherence} with $N$ replaced by $r$.

We now combine this subspace transducer with our algorithmic readout methods to give end-to-end matched QCS algorithms.

\subsubsection{Linear matched QCS}

We first consider matched linear observables. Applying the linear readout of Algorithm~\ref{alg:qcs_linear_readout} to the $r$ registers returned by Algorithm~\ref{alg:qcs_matched_oracle} gives a vector $V=U^TA+\xi+G$, where $G$ is independent Gaussian noise with covariance $\mathds{1}_r/(2\lambda^2)$. Each specified observable can then be estimated as $(U^TK_j)^TV$. Since $\lambda=\widetilde\Theta(\alpha T/\sqrt r)$, the covariance of the estimation error in every direction is bounded by $\widetilde O(r/(\alpha^2T^2)+\gamma/(\alpha^2T))$. Thus, a single acquisition serves all $M$ linear observables, and the required interrogation time depends on the dimension of their span rather than the dimension of the full spectral space.

\begin{prop}
\label{prop:qcs_target_linear}
Let $K_1,\ldots,K_M\in\mathbb R^{2N}$ be specified before sensing, with $\kappa=\max_j\|K_j\|_2$ and $r=\dim\operatorname{span}\{K_1,\ldots,K_M\}$. Running Algorithm~\ref{alg:qcs_matched_oracle} followed by Algorithm~\ref{alg:qcs_linear_readout}, returning $(U^TK_j)^TV$ for each $j$, and taking the median over $O(\log(2M/\delta))$ independent repetitions estimates every $K_j^TA$ to accuracy $\epsilon$ with probability at least $1-\delta$, using total interrogation time
\begin{equation}
T_{\rm int}
=
\widetilde O\left[
\max\left\{
\frac{\sqrt r\,\kappa}{\alpha\epsilon},
\frac{\gamma\kappa^2}{\alpha^2\epsilon^2}
\right\}
\log\frac{2M}{\delta}
\right].
\end{equation}
The protocol uses $r$ memory registers, each of dimension $D=\widetilde O(1+\alpha^2T^2/r)$, where $T$ is the interrogation time of one repetition.
\end{prop}

This improves on the noiseless interrogation time $\widetilde O(\sqrt N\kappa/\epsilon)$ of reusable QCS by a factor of $\sqrt{N/r}$. Whenever $M \ll \sqrt{N}$ observables are desired, this produces an advantage over reusable QCS (because $r$ is always smaller than $M$, as it is the dimension of the subspace capturing all $M$ observables). Unlike conventional matched spectroscopy, whose worst case is set by the peaks of the dual filter and thus depends on $N$ even if only one observable is desired, the interrogation time depends on the observables only through the dimension $r$ of their span. The time overhead for estimating a single observable is thus bandwidth-independent, and we give a simplified protocol for this setting in Section~\ref{sec:qcs_one_qubit_matched}.

\subsubsection{Nonlinear matched QCS}

We now extend the same construction to matched nonlinear observables. The only difference in the transduction step is how the relevant subspace is chosen, since our observables are now higher-order tensors. For rank-$L$ tensors $K_1,\ldots,K_M$, let $U\in\mathbb R^{2N\times r}$ have orthonormal columns spanning a common support of all tensor indices, so that
\begin{equation}
K_j[x_1,\ldots,x_L]
=
K_j[UU^Tx_1,\ldots,UU^Tx_L].
\end{equation}
Define the reduced tensors $\widetilde K_j$ on $\mathbb R^r$ by
\begin{equation}
\widetilde K_j[y_1,\ldots,y_L]
=
K_j[Uy_1,\ldots,Uy_L].
\end{equation}
Then
\begin{equation}
K_j[A^{\otimes L}]
=
\widetilde K_j[(U^TA)^{\otimes L}],
\qquad
\|\widetilde K_j\|_\infty=\|K_j\|_\infty.
\end{equation}
The dimension $r$ must support every index of the supplied tensors, and need not equal the dimension of their linear span. For instance, the powers $p_j^2+q_j^2$ at $m$ specified frequencies require $r=2m$ quadrature coordinates.

We can now combine matched phase-oracle transduction with the nonlinear readout of Algorithm~\ref{alg:qcs_nonlinear_readout}. For each specified tensor, we prepare $\lceil L/2\rceil$ independent copies of the matched oracle state on $r$ memory coordinates, and apply the previously defined normally ordered measurement with $K_j$ replaced by $\widetilde K_j$. The nonlinear readout is otherwise unchanged: its expectation is precisely $\widetilde K_j[(U^TA)^{\otimes L}]=K_j[A^{\otimes L}]$, while the normal-ordering subtraction removes the contribution of the known displacement-noise covariance. Since the matched oracle has gain $\lambda=\widetilde\Theta(\alpha T/\sqrt r)$ and displacement-noise covariance $\widetilde O(\gamma/(\alpha^2T))$, the full-spectrum nonlinear readout guarantees carry over with $2N$ replaced by $r$.

We use the following interrogation time in each batch below:
\begin{equation}
T_{\rm batch}
=
\widetilde O_L
\begin{cases}
\displaystyle
\max\left\{
\frac{\sqrt r\kappa}{\alpha\epsilon},
\frac{\gamma \kappa^2}{\alpha^2\epsilon^2}
\right\},
& \frac{\epsilon}{\kappa}\geq\frac{1}{\sqrt{2N}},\\[4mm]
\displaystyle
\max\left\{
\frac{r^{\frac{L}{L+1}}}{\alpha}\left(\frac{\kappa}{\epsilon}\right)^{\frac{2}{L+1}},
\frac{\gamma^{1/3}r^{1-\frac{2}{3(L+1)}}
}{\alpha^{4/3}}\left(\frac{\kappa}{\epsilon}\right)^{\frac{4}{3(L+1)}}
\right\},
& \frac{\epsilon}{\kappa}\leq\frac{1}{\sqrt{2N}}.
\end{cases}
\label{eq:qcs_coherent_batch_time}
\end{equation}
As in the original nonlinear readout analysis, we assume $\gamma/\alpha\lesssim1$.

\begin{algorithm}[h]
\DontPrintSemicolon
\caption{Matched nonlinear quantum computational spectroscopy}
\label{alg:qcs_coherent}
\KwIn{Known frequencies $\{\omega_j\}_{j=1}^N$, rank $L\geq2$, specified tensors $K_1,\ldots,K_M$ satisfying $\|K_j\|_\infty\leq\kappa$, accuracy $\epsilon$, and failure probability $\delta$.}
\KwOut{Estimates $\widehat F_1,\ldots,\widehat F_M$ satisfying $\max_j|\widehat F_j-K_j[A^{\otimes L}]|\leq\epsilon$ with probability at least $1-\delta$.}

Choose an orthonormal basis $U\in\mathbb R^{2N\times r}$ for a common support of the tensors, and compute the reduced tensors $\widetilde K_j=K_j[U,\ldots,U]$.\;

Set $R=\lceil C_L\log(2M/\delta)\rceil$, and choose $T_{\rm batch}$ according to Equation~\eqref{eq:qcs_coherent_batch_time}.\;

\For{$j=1,\ldots,M$}{
    \For{$s=1,\ldots,R$}{
        Prepare $\lceil L/2\rceil$ independent $r$-coordinate memory registers. On each register run Algorithm~\ref{alg:qcs_matched_oracle} with subspace matrix $U$, using interrogation time $T_{\rm batch}/\lceil L/2\rceil$.\;

        Apply Algorithm~\ref{alg:qcs_nonlinear_readout} to these memories, with the ambient dimension replaced by $r$, target tensor $\widetilde K_j$, readout precision $\eta_{\rm out}=\epsilon/10$, and the displacement-noise covariance matrices of each phase-oracle state. Let the resulting scalar outcome be $Z_j^{(s)}$.\;
    }

    Return
    \[
    \widehat F_j
    =
    \operatorname{median}_{s=1,\ldots,R} Z_j^{(s)}.
    \]
}
\end{algorithm}

We defer the proof of correctness to Sections~\ref{sec:deferred_qcs_target_specific} and~\ref{sec:deferred_matched_qcs}. Because different tensors generally require different coherent measurements, we perform separate acquisitions for each supplied $K_j$, giving total interrogation time $\widetilde O_L(MT_{\rm batch})$. Unlike reusable QCS, these acquisitions require prior knowledge of the desired observables.

Combining the linear Algorithm~\ref{alg:qcs_matched_oracle} and nonlinear Algorithm~\ref{alg:qcs_coherent} gives our general matched QCS guarantee.

\begin{graydefinitionbox}
\begin{theorem}[Matched quantum computational spectroscopy]
\label{thm:qcs_few_queries}
Fix an integer $L\geq1$, $\kappa>0$, $0<\epsilon<\kappa$, and $0<\delta<1/4$. Suppose $M=O(1)$ observables $K_1,\ldots,K_M$, satisfying $\|K_j\|_\infty\leq\kappa$, are specified before interrogation. Write $z=\kappa/\epsilon$ and $\alpha=1-2p>0$. For $L=1$, let $r$ be the dimension of the span of the supplied observables; for $L\geq2$, let $r$ be the dimension of their common tensor support, as defined above. Under the noise and interface assumptions of Section~\ref{sec:qcs}, there exists a target-aware quantum computational spectroscopy protocol which returns estimates satisfying
\begin{equation}
\Pr\left[
\max_{j\leq M}
\left|\widehat F_j-K_j[A^{\otimes L}]\right|
\leq\epsilon
\right]\geq1-\delta
\end{equation}
for every $\|A\|_2\leq1$, using total interrogation time

\begin{equation}
T_{\rm int}\leq
\begin{cases}
\displaystyle \widetilde O\left(\max\left\{\frac{\sqrt r\,\kappa}{\alpha\epsilon},\frac{\gamma\kappa^2}{\alpha^2\epsilon^2}\right\}\right), & L=1,\\[3mm]
\displaystyle \widetilde O_L\left(\max\left\{\frac{\sqrt r\,\kappa}{\alpha\epsilon},\frac{\gamma\kappa^2}{\alpha^2\epsilon^2},
\frac{r^{\frac{L}{L+1}}}{\alpha}\left(\frac{\kappa}{\epsilon}\right)^{\frac{2}{L+1}},
\frac{\gamma^{1/3}r^{1-\frac{2}{3(L+1)}}}{\alpha^{4/3}}\left(\frac{\kappa}{\epsilon}\right)^{\frac{4}{3(L+1)}}\right\}\right), & L\geq2.
\end{cases}
\label{eq:qcs_matched_upper_bound}
\end{equation}
For $L\geq2$, the nonlinear coherent-readout guarantee assumes $\gamma/\alpha\lesssim1$; the linear bound holds for arbitrary $\gamma\geq0$. Logarithmic factors in $N$, $M$, $\kappa/\epsilon$, and $\delta^{-1}$ are suppressed.
\end{theorem}
\end{graydefinitionbox}

\vspace{-4mm}
\subsection{QCS with qubit phases}
\label{sec:qcs_qubit_memory}

Thus far, the memory stores each coordinate of the encoded vector as a displacement, which is why it requires $D$-level registers. Alternatively, we can store the coordinates in the relative phases of individual qubits. The transducer is unchanged except for the memory observable it couples to: the position operators $X_\mu$ are replaced by Pauli operators $Z_\mu$, and the mechanism that makes the transducer robust to sensor noise is the same. Because a phase is only defined modulo $2\pi$, readout requires an additional phase unwrapping and sine inversion step from Section~\ref{sec:high_gain_readout}, but the core principle for the algorithmic speedup remains the same. We describe the full-spectrum version, which uses $2N$ memory qubits. For matched QCS as in Section~\ref{sec:qcs_target_aware}, but with qubit phases, replace $2N$ by $r$ and $A$ by $U^TA$ throughout. We present this variant as it may be easier to experimentally implement on near-term qubit or spin array platforms.

\subsubsection{Qubit phase-oracle transduction}

Initialize $2N$ memory qubits in $\ket{+}^{\otimes2N}$, write $Z_M=(Z_1,\ldots,Z_{2N})^T$, and run Algorithm~\ref{alg:qcs_phase_oracle} with $X$ replaced by $Z_M$ and without the local eigenbasis changes. The memory operators become $O_k=\operatorname{sgn}(a_{\ell_k})\frac{\ell_k}{C}w_k^TZ_M$, so the controlled rotation required by Algorithm~\ref{alg:qcs_transduction} factorizes as
\begin{equation}
    U_{O_k}=e^{-i\pi Y_a/4}\prod_{\mu=1}^{2N}\exp\left[\frac{i\operatorname{sgn}(a_{\ell_k})\ell_k(w_k)_\mu}{2C}Z_\mu\otimes Y_a\right],
\end{equation}
a product of commuting two-qubit rotations between the interface qubit and each memory qubit, which requires no arithmetic circuit.

The proof of Algorithm~\ref{alg:qcs_phase_oracle} used only two properties of the $D$-level vacuum. Every phase and noise operation was diagonal in the position basis, so the position distribution of the memory never changed, and in that distribution every linear form $w^TX$ had sub-Gaussian tails, which kept the truncated arcsine series accurate. The computational basis now plays the role of the position basis, and $\ket{+}^{\otimes2N}$ has the same tails, since 
\begin{equation}
\bra{+}^{\otimes2N}e^{s\,w^TZ_M}\ket{+}^{\otimes2N}=\prod_\mu\cosh(sw_\mu)\leq e^{s^2\|w\|_2^2/2}.    
\end{equation}
The algorithm therefore returns, within the same error, the state
\begin{equation}
\rho_A=\E_\xi\left[\bigotimes_{\mu=1}^{2N}\ket{\psi_{A_\mu+\xi_\mu}}\!\bra{\psi_{A_\mu+\xi_\mu}}\right],\qquad \ket{\psi_x}=\frac{\ket0+e^{2i\lambda x}\ket1}{\sqrt2},
    \label{eq:qubit_memory_oracle_state}
\end{equation}
where we used $e^{-i\lambda xZ}\ket{+}\propto(\ket0+e^{2i\lambda x}\ket1)/\sqrt2$ and $\xi$ is the same displacement noise as in Algorithm~\ref{alg:qcs_phase_oracle}. Each coordinate $A_\mu$ is now written into the relative phase of memory qubit $\mu$ with gain $2\lambda$, and no truncation of the memory is needed.

Each memory qubit now holds the same state that a sensor holds at the end of a high-gain sequence of Section~\ref{sec:high_gain_readout} compiled for coordinate $\mu$ with gain $2\lambda$, up to the noise-induced phase error $2\lambda\xi_\mu$. In high-precision randomized spectroscopy, such a sequence costs interrogation time of order $\lambda$ for each coordinate, or order $N\lambda$ for all of them. Since $2\lambda=\alpha T/(s_0C)$ with $C=\widetilde\Theta(\sqrt N)$, the phase oracle instead prepares all $2N$ phases with gain $2\lambda$ in interrogation time $T=2s_0C\lambda/\alpha=\widetilde\Theta(\sqrt N \lambda/\alpha)$. This is precisely the $\sqrt{N}$ speedup of QCS, and the reusable readout below is the readout of Algorithm~\ref{algo:randomized_spectroscopy_fine}, applied to all memory qubits in parallel.

Sensor depolarization enters exactly as before. The noise kicks $e^{\pm iO_k}$ are now products of single-qubit $Z$ rotations, so each kick shifts the relative phases of all memory qubits by known random amounts rather than displacing bosonic momenta, and $\operatorname{Cov}(\xi)$ obeys the same bound $\widetilde O(\gamma/(\alpha^2T))$. The only difference is that instead of causing a random displacement error in the displaced quadrature, a random phase shift reduces the contrast in the readout of each qubit, similar to how a mean-zero random field reduces contrast in Ramsey interferometry. We show in Section~\ref{sec:deferred_qubit_qcs} that this produces the same scaling under noise as the QCS algorithm that uses $D$-level registers. Qubit QCS may be easier to implement on near term platforms since it does not require readout of $D$-level systems and involves simpler, deterministic sensing intervals, and it produces the same asymptotic beyond-Heisenberg-limited advantage under noise at polylogarithmic sensing time overhead relative to our displacement-based QCS implementations.

\subsubsection{Reusable readout}

We read out the state~\eqref{eq:qubit_memory_oracle_state} using the three steps of Section~\ref{sec:high_gain_readout}, applied to every memory qubit. Phase unwrapping determines the branch of each phase, residual sampling isolates the small remaining error, and sine inversion removes the bias of the final estimate. Each step uses fresh copies of the oracle state, and a copy with gain $h$ costs interrogation time $s_0Ch/\alpha$.

\begin{algorithm}[h]
\DontPrintSemicolon
\caption{Reusable readout of qubit phases}
\label{alg:qcs_qubit_reusable_readout}

\KwIn{Known frequencies $\{\omega_j\}_{j=1}^N$, an integer $J$ setting the final gain $H=2^J/4$, the number $L$ of output vectors, the number $R$ of averaged estimates, inversion accuracy $\epsilon_{\rm inv}\in(0,1/2]$, and failure probability $\delta$.}

\KwOut{Vectors $U_1,\ldots,U_L\in\mathbb R^{2N}$ sharing a reference $a_0$ that satisfies $\|H(A-a_0)\|_\infty\leq1/8$, except with probability at most $\delta/4$.}

Let $h_j=2^j/4$ and $r_{\rm p}$ be as in Algorithm~\ref{algo:phase_unwrapping}. For each $j=0,\ldots,J$, prepare $2r_{\rm p}$ independent registers in the state~\eqref{eq:qubit_memory_oracle_state} at gain $h_j$, using Algorithm~\ref{alg:qcs_phase_oracle} with interrogation time $s_0Ch_j/\alpha$. Measure every memory qubit in the $X$ basis on $r_{\rm p}$ of the registers and in the $Y$ basis on the others.\;

For each coordinate $\mu$, compute $\widetilde\phi_{\mu,j}=\mathrm{Arg}(\overline X_{\mu,j}+i\overline Y_{\mu,j})$ from the empirical means of these outcomes, and unwrap these phases as in Algorithm~\ref{algo:phase_unwrapping} to obtain the reference $a_0$.\;

Set $m=\lceil C\log^2(4N/\epsilon_{\rm inv})\rceil$ for a sufficiently large numerical constant $C$. For each $i=1,\ldots,L$, prepare $Rm$ further independent registers at gain $H$. On each, apply $e^{iH(a_0)_\mu Z_\mu/2}$ to every memory qubit, measure every qubit in the $Y$ basis, and record $W_\mu=B_\mu/\chi_\mu$ from each outcome $B_\mu\in\{\pm1\}$, where $\chi_\mu$ is the known visibility of qubit $\mu$.\;

Divide the $Rm$ registers for each $i$ into $R$ groups of $m$. For each group $g$ and each coordinate $\mu$, pass the $m$ recorded values of $W_\mu$ to Algorithm~\ref{algo:sine_inversion} with accuracy $\epsilon_{\rm inv}$, and collect the outputs into a vector $S^{(i,g)}\in\mathbb R^{2N}$.\;

Return $U_i=a_0+\frac{1}{RH}\sum_{g=1}^RS^{(i,g)}$ for $i=1,\ldots,L$.\;
\end{algorithm}

Phase unwrapping proceeds exactly as in randomized spectroscopy, and we show in Section~\ref{sec:deferred_qubit_qcs} that this readout preserves the reusable QCS guarantee.

\begin{prop}
\label{prop:qcs_qubit_memory_noisy}
Replace the $L$ phase-oracle acquisitions and linear readouts in each batch of Algorithm~\ref{alg:qcs_reusable} by one call to Algorithm~\ref{alg:qcs_qubit_reusable_readout} on $2N$ memory qubits, with $H$, $R$, and $\epsilon_{\rm inv}$ chosen as above. The resulting protocol satisfies the guarantee of Algorithm~\ref{alg:qcs_reusable} using the same batch time $T_{\rm batch}$, up to logarithmic factors, and holds only one register of $2N$ memory qubits at a time. With the matched oracle of Algorithm~\ref{alg:qcs_matched_oracle} on $r$ memory qubits, the same holds with $2N$ replaced by $r$; in particular, Proposition~\ref{prop:qcs_target_linear} holds with qubit memories.
\end{prop}

\subsubsection{Matched QCS with one memory qubit}
\label{sec:qcs_one_qubit_matched}

Finally, we combine qubit-phase QCS and matched QCS to show how a quantum computational sensing speedup beyond the Heisenberg limit can be realized with a single sensor and a single memory qubit. This variant may be implemented on existing nitrogen-vacancy, superconducting circuit, trapped ion, and atomic clock sensing platforms.

The matched oracle requires one memory register per dimension of the span of the desired observables, so a single observable $K^TA$ requires only one memory qubit. The construction then simplifies considerably. Since $Z_M$ has eigenvalues $\pm1$, every memory operator is a known multiple of $Z_M$, so the harmonic sampling and truncated arcsine series of Algorithm~\ref{alg:qcs_phase_oracle} are no longer needed. The coupling between memory and sensor reduces to a single two-qubit gate, which we take to be a CNOT. With the sensor in $\ket0$, the CNOT copies the $Z$ value of the memory onto the sensor, so the sensing Hamiltonian $\frac12B(t)Z_s$ acts as $\frac12B(t)Z_M$. Sensing for a duration $d$ and then undoing the CNOT returns the sensor to $\ket0$ and leaves the memory with the exact phase $e^{-i\theta Z_M/2}$, where $\theta$ is the signal phase accumulated during that duration. We refer to this sequence as a controlled sensing interval. To accumulate the opposite phase $e^{+i\theta Z_M/2}$, we additionally apply $X_s$ to the sensor after the CNOT and again before undoing it. The memory retains its phase between controlled sensing intervals.

The controlled sensing intervals only need to be weighted by the matched filter. On a window $I$ of length $W_0$, define $G_I=\frac{1}{W_0}\int_If(t)f(t)^T\,dt$ and the scalar filter $q_I(t)=K^TG_I^{-1}f(t)/\kappa$, which satisfies $\frac{1}{W_0}\int_Iq_I(t)f(t)\,dt=K/\kappa$. As in matched spectroscopy, $q_I$ is small on average, with $\frac{1}{W_0}\int_I|q_I(t)|\,dt\leq2$, but it has peaks as large as $4\sqrt{2m}$ when $K$ is supported on $m$ frequencies. A conventional sensor must divide the filter by its peak to obtain a bounded modulation. Here we instead divide each window into short cells and, in each cell, perform one controlled sensing interval whose duration is proportional to the magnitude of the integral of $q_I$ over the cell and whose sign is the sign of that integral. Interrogation time then scales with the average size of the filter rather than its peak, while the memory accumulates the phases of all intervals.

\begin{algorithm}[h]
\DontPrintSemicolon
\caption{Matched QCS with one memory qubit}
\label{alg:qcs_one_qubit_matched}

\KwIn{Known frequencies $\{\omega_j\}_{j=1}^N$, a linear observable $K$ with $\|K\|_2=\kappa$ supported on $m$ frequencies, accuracy $\epsilon$, and failure probability $\delta$.}

\KwOut{An estimate $\widehat F_K$ satisfying $|\widehat F_K-K^TA|\leq\epsilon$ with probability at least $1-\delta$.}

Set $J=\lceil\log_2(\kappa/(2\epsilon))\rceil$, gains $h_j=2^j/4$ for $j=0,\ldots,J$, and $r_{\rm p}=\lceil512\log(16(J+1)/\delta)\rceil$. Set $W_0=\Theta(\log(2N)/\Delta_\omega)$, run spans $T_j=W_0\lceil4\sqrt{2m}\,h_j/W_0\rceil$, and cell width $\Delta t=1/(64\,\omega_N\sqrt N\,T_J)$.\;

Fix the start time of every run in advance. For each run at gain $h_j$, divide its span into windows $I$ of length $W_0$, compute $q_I$ on each window, and divide the span into cells $\mathcal C$ of width at most $\Delta t$. For each cell, record the sensing duration and sign
\begin{equation}
    d_{\mathcal C}=\frac{h_j}{T_j}\left|\int_{\mathcal C}q_I(t)\,dt\right|,\qquad \sigma_{\mathcal C}=\operatorname{sgn}\left(\int_{\mathcal C}q_I(t)\,dt\right).
\end{equation}

\For{$j=0,\ldots,J$}{
    \For{each of $2r_{\rm p}$ runs at gain $h_j$}{
        Prepare the memory in $\ket+$.\;
        \For{each cell $\mathcal C$ of the run with $d_{\mathcal C}>0$, in chronological order}{
            At the start of the cell, prepare the sensor in $\ket0$ and apply $\operatorname{CNOT}_{M\rightarrow s}$.\;
            If $\sigma_{\mathcal C}=-1$, apply $X_s$.\;
            Let the sensor sense the signal for duration $d_{\mathcal C}$ without control.\;
            If $\sigma_{\mathcal C}=-1$, apply $X_s$. Then apply $\operatorname{CNOT}_{M\rightarrow s}$.\;
        }
        Measure the memory in the $X$ basis for the first $r_{\rm p}$ runs and in the $Y$ basis for the others.\;
    }
    Let $\overline X_j$ and $\overline Y_j$ be the empirical means of the outcomes, and set $\widetilde\phi_j=\mathrm{Arg}(\overline X_j+i\overline Y_j)$.\;
}

Set $\widehat\phi_0=\widetilde\phi_0$. For $j=1,\ldots,J$, let $\widehat\phi_j$ be the $2\pi$-lift of $\widetilde\phi_j$ closest to $2\widehat\phi_{j-1}$. Return $\widehat F_K=\kappa\widehat\phi_J/h_J$.\;
\end{algorithm}

Each run is a matched phase oracle for the scalar $x=K^TA/\kappa$. The controlled sensing interval in cell $\mathcal C$ adds $\sigma_{\mathcal C}$ times the signal phase accumulated over a duration $d_{\mathcal C}$. Summed over cells, this reproduces the phase that the modulation $h_jq_I(t)/T_j$ would accumulate, up to a discretization error of at most $\omega_N\sqrt N\,T_j\Delta t\leq1/64$, as in Algorithm~\ref{algo:discretize_response}. Since $\frac{1}{W_0}\int_Iq_If\,dt=K/\kappa$ on every window, a run leaves the memory in $(\ket0+e^{ih_jx}\ket1)/\sqrt2$ up to this error, so the memory holds the phase that a sensor would hold after a sequence with gain $h_j$. The measurements and the final unwrapping are those of Algorithm~\ref{algo:phase_unwrapping} for a single coordinate. For signals at a large carrier frequency, the carrier can be removed by a known reference drive or $\pi$-pulse train during each controlled sensing interval, as in Figure~\ref{fig:exp}b, with the filter computed from the correspondingly modulated response.

\begin{prop}
\label{prop:jc_single_qubit_matched_oracle}
With noiseless sensing and interface gates, Algorithm~\ref{alg:qcs_one_qubit_matched} returns an estimate satisfying $|\widehat F_K-K^TA|\leq\epsilon$ with probability at least $1-\delta$, using one sensor qubit, one memory qubit, and
\begin{equation}
    T_{\rm int}=\widetilde O\left(\frac{\kappa}{\epsilon}\right),\qquad T_{\rm wall}=\widetilde O\left(\sqrt m\,\frac{\kappa}{\epsilon}+\frac{1}{\Delta_\omega}\right).
    \label{eq:qcs_one_qubit_cost}
\end{equation}
\end{prop}

\noindent The proof is given in Section~\ref{sec:deferred_one_qubit_matched}. The interrogation time is independent of both $N$ and $m$, while the wall-clock time saturates the bound of Theorem~\ref{thm:matched_wall_clock}. At this wall-clock time, any conventional protocol requires interrogation time $\widetilde\Omega(\sqrt m\,\kappa/\epsilon)$ for a worst-case observable, so the single memory qubit provides a speedup by a factor of $\sqrt m$.

During each controlled sensing interval the memory is entangled with the sensor and inherits its dephasing, so Algorithm~\ref{alg:qcs_one_qubit_matched} achieves the optimal scaling for $\gamma\lesssim\epsilon/\kappa$. Under stronger noise, we instead weight each controlled sensing interval by the strength of the coupling rather than by its duration. Each interval is placed at a time sampled as in Algorithm~\ref{alg:qcs_matched_oracle} with $r=1$, and uses the controlled rotation of Algorithm~\ref{alg:qcs_transduction} with memory operator $O_k=\arcsin(\widetilde q_k(t_k))Z_M$. Since $\sin O_k=\widetilde q_k(t_k)Z_M$ exactly, the memory receives the signal in proportion to $\widetilde q_k$ but dephases only in proportion to $\widetilde q_k^2$. Capping each run at duration of order $1/\gamma$ then gives interrogation time $\widetilde O(\max\{\kappa/(\alpha\epsilon),\gamma\kappa^2/(\alpha^2\epsilon^2)\})$, still independent of $N$ and $m$ (Section~\ref{sec:deferred_one_qubit_matched}).

We simulate this single-ancilla-qubit variant of matched QCS in Fig.~\ref{fig:exp}, where we find significant speedups over DD for estimating random matched filters with an NV electronic spin sensor and $^{13}$C nuclear spin memory, when both sensor and memory have finite coherence time. The reset requirement for the protocol can also be relaxed based on the available reset speed and memory coherence time so that sensor reset can happen after many controlled sensing intervals. This illustrates that the QCS sensing advantage is not limited to future platforms with fault-tolerant memories, and that meaningful AC sensing gains beyond the Heisenberg limit can be observed with existing devices.

\subsection{Optimality and noise-robust quantum computational sensing advantage}

Having presented every variation of QCS, we now compare their performance with the fundamental limits on conventional and quantum computational sensing. First consider reusable QCS. For linear observables, our construction achieves
\begin{equation}\label{eq:qcs_noise_optimality}
    T_{\rm QCS}
    =\widetilde O\left(
    \max\left\{
        \frac{\sqrt N\,\kappa}{\epsilon},
        \frac{\gamma\kappa^2}{\epsilon^2}
    \right\}\right).
\end{equation}
In the noiseless limit, this is optimal up to logarithmic factors, even among protocols with arbitrary quantum memory and unrestricted computational power. Indeed, Theorem~\ref{thm:all_memory_lower_bound} establishes a matching lower bound of $\Omega(\sqrt N\,\kappa/\epsilon)$. By contrast, our conventional spectroscopy lower bounds in Sec.~\ref{sec:lower_bounds} require $\widetilde\Omega(N\kappa/\epsilon)$ in the sufficiently fine precision regime, even allowing arbitrary classical feedback and noncommuting single-qubit control. Thus, QCS achieves a $\widetilde\Omega(\sqrt N)$ advantage over optimal conventional spectroscopy. When $\kappa/\epsilon\lesssim\sqrt N$, QCS improves the $1/\epsilon^2$ scaling of Gaussian randomized spectroscopy to $1/\epsilon$. The same fine-precision scaling extends to nonlinear observables of any fixed order $L$. Thus, QCS achieves a quartic speedup over the standard quantum limit.

This advantage persists under constant-strength depolarizing noise acting on the sensing qubit. The conventional lower bound of Theorem~\ref{thm:depolarizing_lower_bound}, established in Sec.~\ref{sec:noisy_SQL_lower_bounds}, implies that learning all spectral quadratures requires, for $\gamma\gtrsim\epsilon/\kappa$,
\begin{equation}
    T_{\rm conventional}
    =\Omega\left(
    \max\left\{
        \gamma,\frac{N\gamma\Delta_\omega}{\gamma+\Delta_\omega}
    \right\}
    \frac{\kappa^2}{\epsilon^2}
    \right).
\end{equation}
For fixed nonzero $\gamma$ and fixed spectral density, this scales as $\Omega(N\kappa^2/\epsilon^2)$. As in DC metrology, no conventional AC sensing strategy without ancillary quantum memory, even with arbitrary quantum error correction of the sensor, can surpass standard quantum limited scaling. This leads to the central advance of QCS: it is the first quantum metrology protocol which retains metrological advantage in the presence of nonvanishing noise, and indeed its advantage over all conventional protocols grows with the noise rate. 

Matched QCS provides a further improvement when the observables are specified before interrogation. For linear observables spanning an $r$-dimensional subspace, the corresponding projection of the spectrum can be encoded into quantum memory, achieving
\begin{equation}
T_{\rm matched\ QCS}
=\widetilde O\left(
\max\left\{
\frac{\sqrt r\,\kappa}{\epsilon},
\frac{\gamma\kappa^2}{\epsilon^2}
\right\}\right).
\end{equation}
In particular, a single specified observable can be estimated in time $\widetilde O(\max\{\kappa/\epsilon,\gamma\kappa^2/\epsilon^2\})$, independent of $N$, using only one memory qubit. Near $\gamma\sim\Delta_\omega=\Theta(1)$, conventional matched spectroscopy instead requires $\widetilde\Theta(\sqrt N\,\gamma\kappa^2/\epsilon^2)$ in the noise-dominated regime. Importantly, reusable QCS already achieves the same $\widetilde\Omega(\sqrt N)$ advantage when its noise-dependent term dominates, despite acquiring information about the entire spectrum. Matched QCS further improves the interrogation time when coherent accumulation remains the limiting cost.

For nonlinear observables, matched QCS likewise restricts the memory to the spectral subspace supporting the specified tensors. The corresponding nonlinear measurements depend on the observables themselves, however, so the resulting classical data cannot generally be reused for subsequent queries.

These noise-robust quantum advantages are possible because the metrological gains of QCS come not from long sensor evolutions or entangled sensor arrays, but from quantum computation. The signal, which couples to the sensor via a single Pauli generator, contains a large amount of multiplexed information. Ancillary quantum computation allows QCS to process this information in superposition. This differentiates the QCS speedup further from existing quantum metrology protocols: its advantage grows with the bandwidth of the signal, not only with the number of sensors. Consequently, asymptotically growing QCS speedups can be observed with fixed, modest sensing resources (see Fig.~\ref{fig:exp}). Experimental demonstrations with even a single ancilla qubit can thus capture the $\sqrt{N}$ sensing advantage, whereas using the ancilla as an additional sensor yields at most a constant-factor improvement: the speedup is genuinely quantum computational.

\newpage
\section{Lower bounds} \label{sec:lower_bounds}

\subsection{Heisenberg limits for reusable and matched AC sensing}
\subsubsection{Heisenberg limit for reusable AC sensing} \label{sec:nonadaptive_lower_bounds}
Here we establish the optimality of RQS (Section~\ref{sec:randomized_spec}) among all conventional spectroscopy protocols that are compatible with post-hoc learning. In the next section we will show that RQS remains optimal even when compared to protocols given prior knowledge of the observables, as long as the number of observables $M$ is comparable to $N$ (such as when one is trying to learn the entire spectrum) or when noise is substantial.

As with the upper bounds from Section~\ref{sec:randomized_spec}, the lower bound proven in this section is comprised of two scalings which dominate in standard-precision and high-precision regimes respectively; these regimes align exactly with the crossover between our two upper bounds. 
\begin{graydefinitionbox}
\begin{theorem}[Heisenberg Limit on reusable AC sensing]
\label{thm:nonadaptive_posthoc_lower_bound}
Fix any $L\geq 1$ and $M \geq 4$, let $s=\min\{N,1+\log M\}$, and let $\eta$ be a universal constant. Then, any nonadaptive spectroscopy protocol with access to arbitrary single-qubit control and classical postprocessing that achieves precision $0<\epsilon\leq\kappa \eta$ with failure probability $0<\delta\leq 1/20$ must interrogate the unknown signal for time at least
\begin{equation}
T\geq
\begin{cases}
\Omega\left(
\min\left\{
\sqrt N\,s\left(\frac{\kappa}{\epsilon}\right)^2,
N\sqrt s\,\frac{\kappa}{\epsilon}
\right\}\right), & L=1,\\[2mm]
\Omega_L\left(
\max\left\{
N^{3/2-1/L}s^{1/L}
\left(\frac{\kappa}{\epsilon}\right)^{2/L},
N\sqrt s\,\frac{\kappa}{\epsilon}
\right\}\right), & L\geq2.
\end{cases}
\label{eq:nonadaptive_posthoc_combined_lower_compact}
\end{equation}
\end{theorem}
\end{graydefinitionbox}

\begin{proof}
Choose $\eta$ smaller than the constants in Theorems~\ref{thm:posthoc_linear_coarse_lower_bound} and \ref{thm:posthoc_codeword_lower_bound}, and also smaller than $1/\sqrt2$. If $\epsilon/\kappa\geq N^{-1/2}\sqrt{s/2}$, then necessarily $s=1+\log M$ and Theorem~\ref{thm:posthoc_linear_coarse_lower_bound} applies. If $\epsilon/\kappa\leq N^{-1/2}\sqrt{s/2}$, then $2N(\epsilon/\kappa)^2\leq s$ and Theorem~\ref{thm:posthoc_codeword_lower_bound} applies. At equality both bounds scale as $\Theta_L(N^{3/2})$, and comparison on either side of the crossover gives \eqref{eq:nonadaptive_posthoc_combined_lower_compact}. 
\end{proof}

We now prove the two lower bounds used in this result, demonstrating that our algorithms are tight for reusable linear and nonlinear spectroscopy.

\subsubsection{Standard-precision lower bound}
We first define a discrimination problem that can be solved using any successful post-hoc spectroscopy algorithm; our approach will be to prove a lower bound on the interrogation time required to solve this discrimination problem, which then lifts to a lower bound on post-hoc spectroscopy.

\begin{definition}[Hidden bias finding problem]
\label{def:posthoc_spectral_discrimination}
\normalfont
Fix $N,M\ge4$, an integer $L\ge1$, $\kappa>0$, and $0<\epsilon\le c_{0,L}\kappa$, where $c_{0,L}>0$ is a sufficiently small constant depending only on $L$. Draw independent $X_\mu\sim\mathrm{Unif}[-1,1]$, for $\mu=1,\ldots,2N$, and set
\begin{equation}
A=\frac{X}{\sqrt{2N}},
\qquad
S_\mu=\operatorname{sgn}(X_\mu).
\label{eq:posthoc_discrimination_prior}
\end{equation}
The learner has query access to time-evolution under $H_A(t)=B_A(t)Z/2+H_c(t)$, with any single-qubit initial state, any choice of $H_c$, and any final single-qubit measurement.

The discrimination task is as follows. An unknown index $J$ is sampled uniformly from $\{1,\ldots,M-1\}$. For $L=1$, set $\rho=64\epsilon/\kappa$, draw $2N$ independent signs $E_\mu$ with $\Pr(E_\mu=1)=(1+\rho)/2$, and set $R_J=S\odot E$, where $\odot$ denotes coordinatewise multiplication. All other $R_j$, for $j<M$ and $j\ne J$, are drawn independently and uniformly from $\{\pm1\}^{2N}$. Define $K_j=\kappa R_j/\sqrt{2N}$ for $j<M$. For $L\ge2$, choose a sufficiently large constant $C_L$ depending only on $L$, set $\rho=8C_L4^L\epsilon/\kappa$, and draw independent rank-$L$ tensors $G_1,\ldots,G_{M-1}$ whose entries are independent standard Gaussian variables. Define
\begin{equation}
R_j=G_j+\indicator_{\{j=J\}}\frac{\rho}{(2N)^{(L-1)/2}}S^{\otimes L},
\qquad
T_j=\frac{\operatorname{Sym}(R_j)}{\sqrt{2N}},
\qquad
K_j=\frac{\kappa T_j}{\max\{C_L,\|T_j\|_\infty\}},
\quad j<M,
\label{eq:posthoc_discrimination_queries}
\end{equation}
where $\operatorname{Sym}$ averages over permutations of the tensor indices. In both cases, set $K_M=\kappa e_1^{\otimes L}$, where $e_1$ is the first coordinate vector. The learner may perform nonadaptive data acquisition with arbitrary control, and can store any classical data. After acquisition, the list $K_1,\ldots,K_M$ is revealed. The $(L, N, M)$-hidden bias finding problem is to identify $J$ with success probability at least $9/10$. 
\end{definition}

In this problem, the signal is constructed from random signs, and only a secret hidden index $J$ provides an $R_J$ whose conditional mean is correlated with the signal's randomness. The goal is to determine which index was marked.

Observe that $\|A\|_2\le1$ and every supplied observable has norm at most $\kappa$, with $\max_j\|K_j\|_\infty=\kappa$. Moreover, we show in Lemma~\ref{lem:posthoc_observable_separation} that the value of the true $K_J$ differs from all other random observables by more than $2\epsilon$ with high probability over realizations of the problem. Therefore, any algorithm that solves the post-hoc spectroscopy task to additive error $\epsilon$ can solve this discrimination problem without additional interrogation, by returning the largest estimated value among the first $M-1$ observables. A lower bound on the interrogation time required to solve this task therefore implies a lower bound on the post-hoc spectroscopy problem.

\begin{lemma}
\label{lem:posthoc_observable_separation}
Under \eqref{eq:posthoc_coarse_regime}, the instance in Definition~\ref{def:posthoc_spectral_discrimination} satisfies, with probability at least $99/100$,
\begin{equation}
K_J[A^{\otimes L}]\ge7\epsilon,
\qquad
\max_{\substack{j<M\\j\ne J}}|K_j[A^{\otimes L}]|\le4\epsilon.
\label{eq:posthoc_hidden_separation}
\end{equation}
Hence a learner satisfying
\begin{equation}
\Pr\left(\max_{j\le M}|\widehat F_j-K_j[A^{\otimes L}]|\le\epsilon\right)
\ge1-\delta
\label{eq:posthoc_linear_learning_guarantee}
\end{equation}
with $\delta\le1/20$ solves the discrimination task with error at most $\delta+1/100<1/10$.
\end{lemma}

\begin{proof}
For $L=1$ and the true hidden index $J$,
\begin{equation}
K_J^TA=\frac{\kappa}{2N}\sum_{\mu=1}^{2N}E_\mu|X_\mu|,
\qquad
\mathbb E[K_J^TA]=\frac{\kappa\rho}{2}=32\epsilon.
\label{eq:posthoc_hidden_mean}
\end{equation}
The summands $E_\mu|X_\mu|$ are independent and lie in $[-1,1]$. By Hoeffding's inequality,
\begin{equation}
\Pr(K_J^TA<16\epsilon)
\le\exp\left(-256N\frac{\epsilon^2}{\kappa^2}\right).
\label{eq:posthoc_hidden_tail}
\end{equation}
Conditional on $A$ and $J$, every other $R_j$ is an independent uniform sign vector, and the squared coefficients of its contraction satisfy $\sum_\mu(\kappa X_\mu/(2N))^2\le\kappa^2/(2N)$. Hoeffding then gives
\begin{equation}
\Pr\left(|K_j^TA|>4\epsilon\mid A,J\right)
\le2\exp\left(-16N\frac{\epsilon^2}{\kappa^2}\right),
\qquad j<M,\quad j\ne J.
\end{equation}
A union bound shows that the desired separation fails with probability at most
\begin{equation}
\exp\left(-256N\frac{\epsilon^2}{\kappa^2}\right)
+2(M-1)\exp\left(-16N\frac{\epsilon^2}{\kappa^2}\right)
\le e^{-128}+2e^{-7}<\frac1{100},
\end{equation}
where we used \eqref{eq:posthoc_coarse_regime}. Now consider the case $L\ge2$. We first control the denominator of $K_J$ using a covering net argument. Write $u=S/\sqrt{2N}$ and $W_j=\operatorname{Sym}(G_j)/\sqrt{2N}$, so that $T_J=W_J+\rho u^{\otimes L}$ and $T_j=W_j$ for $j\ne J$. Since
\begin{equation}
u^TA=\frac1{2N}\sum_{\mu=1}^{2N}|X_\mu|,
\qquad
\mathbb E[u^TA]=\frac12,
\end{equation}
Hoeffding's inequality gives $\Pr(u^TA<1/4)\le e^{-N/4}$. Relative to Euclidean norm, there exists a $\delta$-covering net of $S^{d-1}$ with cardinality $\leq (1+2/\delta)^d$. Over the $2N$-dimensional unit sphere we then have a $1/(2L)$ covering net containing at most $(1+4L)^{2N}$ points. For arbitrary unit vectors $x_1,...,x_L$ on the sphere, we can choose $y_1,...,y_L$ such that for each $j = 1,...,L$, $\|x_i - y_i\|_2 \leq \delta$. Then, for a rank-$L$ tensor $T$, we have
\begin{equation}
T[x_1,\ldots,x_L]-T[y_1,\ldots,y_L] = \sum_{i=1}^L T[y_1,\ldots,y_{i-1},x_i-y_i,x_{i+1},\ldots,x_L].
\end{equation}
Each term is bounded by $\|T\|_\infty \|x_i - y_i\|_2 \leq \delta \|T\|_\infty$, so the difference is bounded by $L\delta\|T\|_\infty$. Setting $\delta = 1/2L$ and applying reverse triangle inequality, $|T[x_1,\ldots,x_L]|\leq |T[y_1,\ldots,y_L]|+\|T\|_\infty/2$. Taking the sup over all choices of $x_i$, the left hand side becomes $\|T\|_\infty$, and we obtain $\|T\|_\infty \leq 2\max_{y_1...y_L\in \mathcal{N}_{1/2L}} |T[y_1,...y_L]|$ where $\mathcal{N}_{1/2L}$ denotes the covering net.

We use this identity to bound the norm of the hidden multilinear observable. Note that any contraction of $W_J$ with $L$ unit vectors is Gaussian with variance at most $1/(2N)$. By the above, for a $1/(2L)$-covering net $\mathcal{N}$, $\|W_J\|_\infty \leq 2\max_{x_1,...,x_L\in\mathcal{N}}|W_J[x_1,...,x_L]|$. Therefore, if $\|W_J\|_\infty > C_L-1$, there must exist some set of $L$ points in $\mathcal{N}$ for which the contraction against $W_J$ is larger than $(C_L-1)/2$ in magnitude. For a random set of $L$ points, the standard Gaussian tail bound gives
\begin{equation}
\textnormal{Pr}\left(|W_J[x_1,\ldots,x_L]|>\frac{C_L-1}{2}\right)
\leq 2\exp\left(-\frac{N(C_L-1)^2}{4}\right)
\end{equation}
A union bound over all $|\mathcal{N}|^L\leq (1+4L)^{2NL}$ points in the covering net gives
\begin{equation}
\Pr(\|W_J\|_\infty>C_L-1)
\le2\exp\left(2NL\log(1+4L)-\frac{2N(C_L-1)^2}{8}\right)
\le2e^{-20N}
\label{eq:posthoc_gaussian_norm}
\end{equation}
for a sufficiently large choice of $C_L$. Taking $c_{0,L}$ small enough that $\rho\le1$, the complementary event ensures $\|T_J\|_\infty\le C_L$, so $K_J=\kappa T_J/C_L$.

Finally, we control the contraction of the Gaussian random observables against $A$. Conditional on $A$, each $W_j[A^{\{L\}}]$ is a centered Gaussian with variance $\|A\|_2^{2L}/(2N)\le1/(2N)$. Thus
\begin{equation}
\Pr\left(\max_{j<M}|W_j[A^{\{L\}}]|>C_L\frac{\epsilon}{\kappa}\right)
\le2(M-1)\exp\left(-C_L^2N\frac{\epsilon^2}{\kappa^2}\right).
\label{eq:posthoc_gaussian_contractions}
\end{equation}
On the complementary events,
\begin{equation}
K_J[A^{\otimes L}]
\ge\frac{\kappa}{C_L}\left(\rho4^{-L}-C_L\frac{\epsilon}{\kappa}\right)
=7\epsilon,
\qquad
|K_j[A^{\otimes L}]|\le\epsilon\quad(j<M,\ j\ne J).
\end{equation}
The sum of the three failure probabilities is at most $1/100$ under \eqref{eq:posthoc_coarse_regime}, after taking $c_{0,L}\le1/128$ and $C_L$ sufficiently large. In either case, $\epsilon$-accurate estimates satisfy $\widehat F_J\ge6\epsilon$ and $\widehat F_j\le5\epsilon$ for $j<M$ and $j\ne J$.
\end{proof}

We now state the main lower bound on the hidden bias finding task.
\begin{theorem}
\label{thm:posthoc_linear_coarse_lower_bound}
For every fixed integer $L\ge1$, there is a constant $c_{0,L}>0$ such that the following holds. Let $N,M\ge4$, $\kappa>0$, $0<\delta\le1/20$, and
\begin{equation}
0<\epsilon\le c_{0,L}\kappa,
\qquad
1+\log M\le2N\left(\frac{\epsilon}{\kappa}\right)^2.
\label{eq:posthoc_coarse_regime}
\end{equation}
For fixed frequency separation scale $\Delta_\omega>0$, any nonadaptive single-qubit protocol solving Definition~\ref{def:posthoc_spectral_discrimination} must satisfy
\begin{equation}
T_{\rm int}\ge c_L N^{3/2-1/L}(1+\log M)^{1/L}
\left(\frac{\kappa}{\epsilon}\right)^{2/L},
\label{eq:posthoc_linear_coarse_lower_bound}
\end{equation}
where $c_L>0$ depends only on $L$ and the fixed frequency parameters. The same bound holds for any nonadaptive post-hoc spectroscopy algorithm satisfying \eqref{eq:posthoc_linear_learning_guarantee} for every admissible signal and every fixed list of rank-$L$ observables.
\end{theorem}

To begin, we first bound the mutual information $I(D;R)$ between any random variable $D$ and the vector $R$ containing the hidden bias when $L=1$.

\begin{lemma}
\label{lem:posthoc_noisy_sign_contraction}
Let $S$ be uniform on $\{\pm1\}^{2N}$ and let $D$ be any random variable jointly distributed with $S$. Independently of $(S,D)$, draw independent signs $E_\mu$ with mean $\rho\in[0,1]$, and set $R=S\odot E$. Then
\begin{equation}
I(D;R)\le\rho^2 I(D;S).
\label{eq:posthoc_noisy_sign_contraction}
\end{equation}
\end{lemma}

\begin{proof}
First consider one uniform sign $S$. Write $m(d)=\mathbb E[S\mid D=d]$ and define
\begin{equation}
\phi(t)=\frac12\big[(1+t)\log(1+t)+(1-t)\log(1-t)\big]
=\sum_{k=1}^{\infty}\frac{t^{2k}}{(2k)(2k-1)}.
\label{eq:posthoc_binary_information_function}
\end{equation}
The conditional probabilities of $S$ are $(1\pm m(d))/2$, while those of $R=SE$ are $(1\pm\rho m(d))/2$. Both unconditional signs are uniform, so
\begin{equation}
I(D;S)=\mathbb E[\phi(m(D))],
\qquad
I(D;R)=\mathbb E[\phi(\rho m(D))].
\end{equation}
Every coefficient in \eqref{eq:posthoc_binary_information_function} is nonnegative. Since $\rho^{2k}\le\rho^2$ for $k\ge1$, we have $\phi(\rho t)\le\rho^2\phi(t)$, proving the inequality for the scalar case. To extend to vectors $S,R$, let $S_{<\mu}=(S_1,\ldots,S_{\mu-1})$, and define $R_{<\mu}$ analogously. Conditional on $R_{<\mu}$, the sign $S_\mu$ is still uniform, and $E_\mu$ is independent of $(D,S,R_{<\mu})$. Applying the scalar inequality conditionally gives
\begin{equation}
I(D;R_\mu\mid R_{<\mu})
\le\rho^2 I(D;S_\mu\mid R_{<\mu}).
\end{equation}
Moreover,
\begin{align}
I(D;S_\mu\mid R_{<\mu})
&=I(D,R_{<\mu};S_\mu)\nonumber\\
&\le I(D,S_{<\mu};S_\mu)\nonumber\\
&=I(D;S_\mu\mid S_{<\mu}).
\label{eq:posthoc_noisy_sign_tensorization}
\end{align}
The equalities use independence of $S_\mu$ from the preceding signs. The inequality is data processing: $(D,R_{<\mu})$ is obtained from $(D,S_{<\mu})$ by adding $E_{<\mu}$ independent of $(D,S)$. Summing over $\mu$ and applying the chain rule of mutual information proves \eqref{eq:posthoc_noisy_sign_contraction}.
\end{proof}

For $L\ge2$, we instead bound the relative entropy of the Gaussian tensor containing the hidden bias. We first do this when the distribution of the signal's signs has a bounded density.

\begin{lemma}
\label{lem:posthoc_gaussian_bias_information}
Let $L\ge2$, let $\pi$ be uniform probability measure on $\{\pm1\}^{2N}$, and let $g$ satisfy $\mathbb E_\pi g=1$ and $0\le g\le e^B$. Draw $S$ from the distribution with density $g$ relative to $\pi$, and independently draw a rank-$L$ tensor $G$ with independent standard Gaussian entries. Let $P_g$ be the law of
\begin{equation}
R=G+\frac{\rho}{(2N)^{(L-1)/2}}S^{\otimes L},
\end{equation}
and let $\Gamma$ be the law of $G$. For $0<\rho\le(100\cdot2^{L-1})^{-1/2}$,
\begin{equation}
D_{\rm KL}(P_g\|\Gamma)
\le\frac{2^{2L-1}\rho^2B^L}{(2N)^{L-1}}+\frac1{20}.
\label{eq:posthoc_gaussian_bias_information}
\end{equation}
\end{lemma}

\begin{proof}
The density of $P_g$ relative to $\Gamma$ is
\begin{equation}
\ell_g(y)=\mathbb E_{S\sim g\pi}\exp\left(
\frac{\rho\langle y,S^{\otimes L}\rangle}{(2N)^{(L-1)/2}}-N\rho^2
\right),
\label{eq:posthoc_gaussian_likelihood}
\end{equation}
where the normalization uses $\|S^{\otimes L}\|_F^2=(2N)^L$. For independent $S,S'\sim g\pi$, Gaussian integration gives
\begin{equation}
\int\ell_g(y)^2\,d\Gamma(y)
=\mathbb E\exp\left(\frac{\rho^2\langle S^{\otimes L},(S')^{\otimes L}\rangle}{(2N)^{L-1}}\right)
=\mathbb E\exp\left(\frac{\rho^2(S\cdot S')^L}{(2N)^{L-1}}\right).
\label{eq:posthoc_gaussian_second_moment}
\end{equation}
We therefore need to control the inner product $V=S\cdot S'$. For $U\subseteq\{1,\ldots,2N\}$, write $\widehat g(U)=\mathbb E_\pi[g(S)\prod_{\mu\in U}S_\mu]$. The Bonami--Beckner hypercontractive inequality~\cite{ben-aroya_hypercontractive_2008} gives, for $0\le a\le1$,
\begin{equation}
\sum_U a^{|U|}\widehat g(U)^2
\le\left(\mathbb E_\pi g^{1+a}\right)^{2/(1+a)}
\le e^{2Ba},
\label{eq:posthoc_boolean_hypercontractivity}
\end{equation}
where the second inequality uses $g\le e^B$ and $\mathbb E_\pi g=1$. Expanding $e^{\pm tS_\mu S'_\mu}=\cosh t(1\pm S_\mu S'_\mu\tanh t)$ gives, for $t\ge0$,
\begin{align}
\mathbb E e^{\pm tV}
&=(\cosh t)^{2N}\sum_U(\pm\tanh t)^{|U|}\widehat g(U)^2\nonumber\\
&\le(\cosh t)^{2N}\sum_U(\tanh t)^{|U|}\widehat g(U)^2
\le e^{Nt^2+2Bt}.
\end{align}
The last inequality uses \eqref{eq:posthoc_boolean_hypercontractivity}, $\log\cosh t\le t^2/2$, and $\tanh t\le t$. Exponential Markov's inequality, optimized over $t$, therefore yields
\begin{equation}
\Pr(|V|\ge2B+v)\le2e^{-v^2/(4N)},
\qquad v\ge0.
\label{eq:posthoc_sign_inner_product_tail}
\end{equation}
Set $U_0=(|V|-2B)_+$ and $a=2^{L-1}\rho^2\le1/100$. Since $U_0\le2N$ and $L\ge2$,
\begin{equation}
\frac{\rho^2|V|^L}{(2N)^{L-1}}
\le\frac{2^{2L-1}\rho^2B^L}{(2N)^{L-1}}
+\frac{aU_0^L}{(2N)^{L-1}}
\le\frac{2^{2L-1}\rho^2B^L}{(2N)^{L-1}}
+\frac{aU_0^2}{2N}.
\end{equation}
Integrating the tail in \eqref{eq:posthoc_sign_inner_product_tail} gives
\begin{equation}
\mathbb E\left[e^{aU_0^2/(2N)}-1\right]
\le\int_0^\infty\frac{4av}{2N}e^{-(1/2-a)v^2/(2N)}\,dv
=\frac{4a}{1-2a}.
\end{equation}
Consequently,
\begin{equation}
\log\mathbb E\exp\left(\frac{\rho^2V^L}{(2N)^{L-1}}\right)
\le\frac{2^{2L-1}\rho^2B^L}{(2N)^{L-1}}
+\log\left(1+\frac{4a}{1-2a}\right)
\le\frac{2^{2L-1}\rho^2B^L}{(2N)^{L-1}}+\frac1{20}.
\end{equation}
Finally, Jensen's inequality gives $D_{\rm KL}(P_g\|\Gamma)=\mathbb E_{P_g}\log\ell_g\le\log\mathbb E_{P_g}\ell_g=\log\int\ell_g^2\,d\Gamma$. Combining this with \eqref{eq:posthoc_gaussian_second_moment} proves the result.
\end{proof}

We now apply these lemmas to the hidden bias finding problem.

\begin{lemma}
\label{lem:posthoc_required_information}
Under \eqref{eq:posthoc_coarse_regime}, let $D$ be the learner's complete classical dataset before the list in Definition~\ref{def:posthoc_spectral_discrimination} is revealed. If there exists a decoder solving the hidden bias finding problem with probability at least $9/10$, it must hold that
\begin{equation}
I(S;D)\ge c_L(2N)^{1-1/L}(1+\log M)^{1/L}
\left(\frac{\kappa}{\epsilon}\right)^{2/L},
\label{eq:posthoc_required_information}
\end{equation}
where $c_L>0$ depends only on $L$.
\end{lemma}

\begin{proof}
First we take $L=1$. Let $\nu$ be the uniform distribution on $\{\pm1\}^{2N}$. The vector $S\odot E$ has marginal distribution $\nu$, because $S$ is uniform. More strongly, conditional on any fixed $J$, the list $R_1,\ldots,R_{M-1}$ has law $\nu^{\otimes(M-1)}$. Thus the list of observables, on its own, gives no information about $J$.

Now let $P_j$ be the joint law of $Y=(D,R_1,\ldots,R_{M-1})$ conditional on $J=j$, and compare it with $Q=P_D\otimes\nu^{\otimes(M-1)}$, where $P_D$ denotes the marginal of $D$. Under $P_j$, only $R_j$ is correlated with $D$; all other vectors are independent of $(D,R_j)$. Hence
\begin{equation}
D_{\rm KL}(P_j\|Q)
=I(D;S\odot E)
\le\rho^2 I(D;S),
\label{eq:posthoc_reference_divergence}
\end{equation}
where the last step is Lemma~\ref{lem:posthoc_noisy_sign_contraction}. Set $\overline P=(M-1)^{-1}\sum_{j<M}P_j$; $\overline P$ is the unconditional law of $Y$. Using $\log(P_j(y)/Q(y))=\log(P_j(y)/\overline P(y))+\log(\overline P(y)/Q(y))$, we have
\begin{align}
\frac1{M-1}\sum_{j<M}D_{\rm KL}(P_j\|Q)
&=\frac1{M-1}\sum_{j<M}\int P_j(y)\log\left(\frac{P_j(y)}{\overline P(y)}\right)dy\nonumber\\
&\quad+\frac1{M-1}\sum_{j<M}\int P_j(y)\log\left(\frac{\overline P(y)}{Q(y)}\right)dy\nonumber\\
&=I(J;Y)+D_{\rm KL}(\overline P\|Q).
\label{eq:posthoc_reference_identity}
\end{align}
The last equality follows by recognizing the first term as the mutual information and substituting $(M-1)^{-1}\sum_{j<M}P_j=\overline P$ in the second term. Nonnegativity of the second term and Equation~\eqref{eq:posthoc_reference_divergence} then imply
\begin{equation}
I(J;D,R_1,\ldots,R_{M-1})\le\rho^2 I(D;S).
\label{eq:posthoc_index_information_upper}
\end{equation}
Fano's inequality~\cite{cover_elements_2005} states that decoding a uniform label on $M-1$ possibilities with error probability $p_{\rm err}$ requires
\begin{equation}
I(J;D,R_1,\ldots,R_{M-1})
\ge(1-p_{\rm err})\log(M-1)-\log2.
\label{eq:posthoc_fano}
\end{equation}
For $p_{\rm err}\le1/10$ and $M\ge4$, the right-hand side is at least $\frac14\log(M-1)$. Combining with \eqref{eq:posthoc_index_information_upper}, using $\log(M-1)\asymp1+\log M$ and $\rho=64\epsilon/\kappa$, proves the claim for $L=1$.

Now suppose $L\ge2$. The observables are known functions of $R_1,\ldots,R_{M-1}$, so the same decoder succeeds if it is given these tensors instead. Write $I=I(S;D)$, let $\pi$ be uniform on $\{\pm1\}^{2N}$, and define the conditional density
\begin{equation}
g_d(s)=\frac{\Pr(S=s\mid D=d)}{\pi(s)},
\qquad
I=\mathbb E_D\mathbb E_\pi[g_D\log g_D].
\end{equation}
To apply Lemma~\ref{lem:posthoc_gaussian_bias_information}, we first restrict to an event on which this density is bounded. Set $b=100(I+1)$. Since $x\log x\ge-1/e$,
\begin{equation}
\Pr(g_D(S)>e^b)
=\mathbb E_D\mathbb E_\pi[g_D\indicator_{\{g_D>e^b\}}]
\le\frac{I+1/e}{b}\le\frac1{100}.
\label{eq:posthoc_posterior_tail}
\end{equation}
Let $m_d=\mathbb E_\pi[g_d\indicator_{\{g_d\le e^b\}}]$ and define the event $\mathcal E=\{m_D\ge1/2,\ g_D(S)\le e^b\}$. Equation~\eqref{eq:posthoc_posterior_tail} gives $\mathbb E_D(1-m_D)\le1/100$, so $\Pr(m_D<1/2)\le2/100$ and $\Pr(\mathcal E)\ge97/100$. Condition the joint distribution on $\mathcal E$, and denote mutual information under this distribution by $I_{\mathcal E}$. The decoder's conditional error is at most $(1/10)/(97/100)<1/8$.

Under this conditioning, $J$ remains uniform and independent of $D$, and the Gaussian tensors remain independent of $(S,D,J)$. For each value $D=d$, the conditional density of $S$ is
\begin{equation}
\widetilde g_d(s)=\frac{g_d(s)\indicator_{\{g_d(s)\le e^b\}}}{m_d},
\qquad
\mathbb E_\pi\widetilde g_d=1,
\qquad
0\le\widetilde g_d\le2e^b.
\end{equation}
Thus Lemma~\ref{lem:posthoc_gaussian_bias_information} applies with $B=b+\log2\le101(I+1)$, after taking $c_{0,L}$ small enough that $\rho\le(100\cdot2^{L-1})^{-1/2}$.

For fixed $D=d$, let $P_{j,d}$ be the law of $(R_1,\ldots,R_{M-1})$ conditional on $J=j$ and $\mathcal E$, and compare it with $Q=\Gamma^{\otimes(M-1)}$. Under $P_{j,d}$, only $R_j$ differs from an independent standard Gaussian tensor. Hence
\begin{equation}
D_{\rm KL}(P_{j,d}\|Q)
=D_{\rm KL}(P_{\widetilde g_d}\|\Gamma)
\le C'_L\frac{\rho^2(I+1)^L}{(2N)^{L-1}}+\frac1{20}.
\end{equation}
Applying the same relative entropy identity \eqref{eq:posthoc_reference_identity} conditional on $D=d$, and then averaging over $D$ under $\mathcal E$, gives
\begin{equation}
I_{\mathcal E}(J;D,R_1,\ldots,R_{M-1})
\le C'_L\frac{\rho^2(I+1)^L}{(2N)^{L-1}}+\frac1{20}.
\label{eq:posthoc_nonlinear_index_information}
\end{equation}
This includes the information in the entire revealed list. Fano's inequality and the conditional error bound give
\begin{equation}
I_{\mathcal E}(J;D,R_1,\ldots,R_{M-1})
\ge\frac78\log(M-1)-\log2
\ge\frac15\log(M-1).
\end{equation}
Combining the last two bounds, and absorbing $1/20$ using $M-1\ge3$, yields
\begin{equation}
I(S;D)+1
\ge c_L(2N)^{1-1/L}\rho^{-2/L}(1+\log M)^{1/L}
=c'_L(2N)^{1-1/L}(1+\log M)^{1/L}
\left(\frac{\kappa}{\epsilon}\right)^{2/L}.
\label{eq:posthoc_nonlinear_required_information}
\end{equation}
By \eqref{eq:posthoc_coarse_regime}, the expression after $c'_L$ is at least $(1+\log M)(\kappa/\epsilon)^2$. Taking $c_{0,L}$ sufficiently small proves \eqref{eq:posthoc_required_information} for the original, unconditioned dataset.
\end{proof}

Finally, we bound the mutual information acquired per unit interrogation time.

\begin{lemma}
\label{lem:posthoc_information_time}
Let $X$ be uniform on $[-1,1]^{2N}$ and $A=X/\sqrt{2N}$. For positive distinct frequencies, define
\begin{equation}
D_\omega=\max_{j\le N}\left[
\frac1{\omega_j}
+2\sum_{k\ne j}\left(
\frac1{|\omega_j-\omega_k|}+\frac1{\omega_j+\omega_k}
\right)\right].
\label{eq:posthoc_frequency_geometry}
\end{equation}
Every nonadaptive single-qubit acquisition with record $D$ and deterministic interrogation-time cap $T_{\rm int}$ satisfies
\begin{equation}
I(X;D)\le C T_{\rm int}\left(
\frac1{\sqrt{2N}}+\frac{D_\omega}{2N}
\right).
\label{eq:posthoc_information_time_general}
\end{equation}
For fixed frequency separation scale $\Delta_\omega>0$,
\begin{equation}
I(X;D)\le C\frac{T_{\rm int}}{\sqrt N},
\label{eq:posthoc_information_time_fixed_density}
\end{equation}
where the constant depends only on the fixed frequency parameters.
\end{lemma}

\begin{proof}
First consider one shot on an interval $I=[t_0,t_0+\tau]$. Recall the notation
\begin{equation}
f(t)=(\cos\omega_1t,\ldots,\cos\omega_Nt,
\sin\omega_1t,\ldots,\sin\omega_Nt)^T,
\qquad
G_I=\int_I f(t)f(t)^T\,dt.
\end{equation}
By the same product-to-sum estimate used in the preliminaries, the absolute row sums of $G_I-\tau I_{2N}/2$ are bounded by $D_\omega$. Since this matrix is symmetric,
\begin{equation}
\lambda_{\max}(G_I)\le\frac\tau2+D_\omega.
\label{eq:posthoc_gram_bound}
\end{equation}
This bound holds for any starting time $t_0$. Now let $V_A(t,t_0)$ be the propagator of any particular controlled protocol. Its derivative with respect to $A_\mu$ is generated by
\begin{equation}
\mathcal G_\mu
:=iV_A(t_0+\tau,t_0)^\dagger\partial_{A_\mu}V_A(t_0+\tau,t_0)
=\frac12\int_I f_\mu(t)V_A(t,t_0)^\dagger ZV_A(t,t_0)\,dt.
\end{equation}
Write $V_A(t,t_0)^\dagger ZV_A(t,t_0)=n_A(t)\cdot\boldsymbol{\sigma}$, where $\boldsymbol{\sigma}$ is the Pauli vector and $\|n_A(t)\|_2=1$. Then $\mathcal G_\mu=g_\mu\cdot\boldsymbol{\sigma}/2$, with $g_\mu=\int_I f_\mu(t)n_A(t)\,dt$. Any single-qubit spectroscopy protocol has a propagator of this form, even when controls along all Pauli axes are used. We next use the form of this generator to control the classical Fisher information of measurement outcomes.

For a POVM with outcome probabilities $p_y(A)$, define its classical Fisher information matrix by
\begin{equation}
J_A(A)=\sum_{y:p_y(A)>0}
\frac{\nabla_Ap_y(A)\nabla_Ap_y(A)^T}{p_y(A)}.
\end{equation}
For a pure initial state $|\psi_0\rangle$, the measurement-Fisher-information bound~\cite{braunstein_statistical_1994} gives
\begin{equation}
[J_A(A)]_{\mu\mu}
\le4\operatorname{Var}_{\psi_0}(\mathcal G_\mu)
\le\|g_\mu\|_2^2,
\label{eq:posthoc_measurement_fisher}
\end{equation}
with the same bound for mixed preparations by convexity. The map $v\mapsto\int_I f(t)v(t)\,dt$ has squared operator norm $\lambda_{\max}(G_I)$. Applying it separately to the three components of $n_A(t)$ yields
\begin{equation}
\operatorname{tr}J_A(A)
\le\sum_{\mu=1}^{2N}\|g_\mu\|_2^2
\le\lambda_{\max}(G_I)\int_I\|n_A(t)\|_2^2\,dt
=\tau\lambda_{\max}(G_I).
\label{eq:posthoc_fisher_trace}
\end{equation}
Moreover, for the parameter $X$, the chain rule gives $J_X(x)=J_A(x/\sqrt{2N})/(2N)$. We next convert this Fisher information bound into mutual information. Let $\mu$ be uniform probability measure on $[-1,1]^{2N}$ and define $\operatorname{Ent}_\mu(q)=\mathbb E_\mu[q\log q]-\mathbb E_\mu[q]\log\mathbb E_\mu[q]$. We will use
\begin{equation}
\operatorname{Ent}_\mu(u^2)
\le\frac4\pi\mathbb E_\mu\|\nabla u\|_2^2.
\label{eq:posthoc_cube_log_sobolev}
\end{equation}
This follows from Gross's Gaussian logarithmic Sobolev inequality~\cite{gross_logarithmic_1975}. For a POVM with outcome distribution $p_y(x)=\Pr(Y=y\mid X=x)$,
\begin{align}
I(X;Y)
&=\sum_y\operatorname{Ent}_\mu(p_y)\nonumber\\
&\le\frac4\pi\mathbb E_X\sum_y\|\nabla_X\sqrt{p_y(X)}\|_2^2\nonumber\\
&=\frac1\pi\mathbb E_X\operatorname{tr}J_X(X)
\le\frac{\tau\lambda_{\max}(G_I)}{2\pi N}.
\label{eq:posthoc_information_fisher}
\end{align}
The first inequality uses Equation~\eqref{eq:posthoc_cube_log_sobolev}, the last equality uses the definition of Fisher information, and the final inequality uses Equation~\eqref{eq:posthoc_fisher_trace} together with the chain rule. We can extend this result to general continuous POVMs by applying the finite-outcome argument to every finite partition of its outcome space and taking the supremum.

Independently, Holevo's bound~\cite{holevo_bounds_1973} implies
\begin{equation}
I(X;Y)\le S_{\rm vN}(\mathbb E_X\rho_X)
-\mathbb E_X S_{\rm vN}(\rho_X)\le\log2,
\label{eq:posthoc_holevo}
\end{equation}
where $\rho_X$ is the terminal qubit state and $S_{\rm vN}(\rho)=-\operatorname{tr}(\rho\log\rho)$. Combining \eqref{eq:posthoc_gram_bound}, \eqref{eq:posthoc_information_fisher}, and \eqref{eq:posthoc_holevo}, we obtain
\begin{align}
I(X;Y)
&\le\min\left\{\log2,
\frac{\tau^2}{4\pi N}+\frac{D_\omega\tau}{2\pi N}\right\}\nonumber\\
&\le\min\left\{\log2,\frac{\tau^2}{4\pi N}\right\}
+\frac{D_\omega\tau}{2\pi N}\nonumber\\
&\le C\tau\left(\frac1{\sqrt{2N}}+\frac{D_\omega}{2N}\right).
\label{eq:posthoc_single_shot_information_rate}
\end{align}
The last step uses $\min\{a,b\}\le\sqrt{ab}$ for nonnegative $a,b$. Thus very short shots have little information because their Fisher information is small, whereas long shots are limited by the information capacity of one qubit. By nonadaptivity, after conditioning on the prechosen settings, the fresh-shot outcomes are independent given $X$, so the dataset's mutual information is at most the sum of the single-shot mutual informations. Applying \eqref{eq:posthoc_single_shot_information_rate} at each shot and summing the durations proves \eqref{eq:posthoc_information_time_general}. For fixed $\Delta_\omega$, $D_\omega\le C\Delta_\omega^{-1}\log(2N)$. Thus Equation~\eqref{eq:posthoc_information_time_fixed_density} follows.
\end{proof}

We can now complete the proof of Theorem~\ref{thm:posthoc_linear_coarse_lower_bound}.

\begin{proof}
For any successful discrimination protocol, Lemma~\ref{lem:posthoc_required_information} gives
\begin{equation}
I(S;D)\ge c_L(2N)^{1-1/L}(1+\log M)^{1/L}
\left(\frac{\kappa}{\epsilon}\right)^{2/L}.
\end{equation}
Since $S$ is a deterministic function of $X$, data processing and Lemma~\ref{lem:posthoc_information_time} give
\begin{equation}
I(S;D)\le I(X;D)\le C\frac{T_{\rm int}}{\sqrt N}.
\end{equation}
Combining these inequalities proves \eqref{eq:posthoc_linear_coarse_lower_bound}. Lemma~\ref{lem:posthoc_observable_separation} then lifts this to a lower bound against all nonadaptive strategies for post-hoc spectroscopy.
\end{proof}

\subsubsection{High-precision lower bound}
\label{sec:posthoc_codeword_lower_bound}

The previous lower bound established optimality of the Gaussian randomized spectroscopy algorithm (Algorithm \ref{algo:randomized_spectroscopy_coarse}) in the $\epsilon/\kappa \gtrsim 1/\sqrt{N}$ regime. However, recall that in the fine-precision regime $\epsilon/\kappa \lesssim 1/\sqrt{N}$, we gave a phase-unwrapping algorithm (Algorithm \ref{algo:randomized_spectroscopy_fine}) making longer coherent queries to achieve better performance. Here, we introduce a separate lower bound for this regime, which establishes the optimality of Algorithm \ref{algo:randomized_spectroscopy_fine}, and proving that our combined RQS algorithm is optimal.

Once again, we define a discrimination problem that can be solved by any algorithm for post-hoc AQS, this time in the fine-precision regime $\epsilon/\kappa \lesssim 1/\sqrt{N}$.

\begin{definition}[Codeword detection problem]
\label{def:posthoc_codeword_detection}
\normalfont
Fix $N,M,L\ge1$, $\kappa>0$, and $0<\epsilon\le\kappa/(4C_L)$, where $C_L=2^{L+11}$. Set $k=\min\{2N-1,1+\lfloor\log_2 M\rfloor\}$, $h=\frac{C_L\epsilon}{\kappa\sqrt{k}}$, and
\begin{align}
 q&=k\left\lfloor\frac1k\min\left\{2N-1,\frac1{4h^2}\right\}\right\rfloor.
\label{eq:posthoc_codeword_parameters}
\end{align}
Let $e_0,\ldots,e_{2N-1}$ be the standard basis of $\mathbb R^{2N}$. Draw independent $X_i\sim\mathrm{Unif}[-1,1]$, for $i=1,\ldots,q$, and set
\begin{equation}
 A=\frac12e_0+h\sum_{i=1}^qX_ie_i,
 \qquad
 S_i=\operatorname{sgn}(X_i).
\label{eq:posthoc_codeword_signal}
\end{equation}
The learner has query access to time-evolution under $H_A(t)=B_A(t)Z/2+H_c(t)$,  with arbitrary preparation, control, and measurement of the qubit. The learner may retain any classical dataset $D$.

The discrimination task is as follows. After acquisition, an index $J$ is sampled uniformly from $\{1,\ldots,q\}$, independently of the signal and acquisition, and revealed to the learner. The codeword detection problem is to return $S_J$ with success probability at least $7/8$, averaged over the signal, the requested index, and the learner's randomness.
\end{definition}

In this problem, the unknown signs $S$ form a binary codeword, and the learner must predict a randomly requested entry of $S$ using data collected before being told which entry will be requested. The choice of $q$ ensures $q\ge k$ and
\begin{equation}
 \|A\|_2^2\le\frac14+qh^2\le\frac12.
\end{equation}
Moreover, we show below that a suitable list of at most $M$ rank-$L$ observables determines most signs in any requested block of $k$ coordinates. Therefore, any algorithm solving post-hoc spectroscopy can solve this discrimination problem without additional interrogation.

\begin{lemma}
\label{lem:posthoc_codeword_reduction}
Suppose a list-independent acquisition produces estimates satisfying
\begin{equation}
 \Pr\left(\max_{j\le M}|\widehat F_j-F_{K_j}(A)|\le\epsilon\right)\ge1-\delta
\label{eq:posthoc_codeword_learning_guarantee}
\end{equation}
for every fixed $\|A\|_2\le1$ and every fixed list with $\|K_j\|_\infty\le\kappa$, where $0<\delta\le1/20$. Then it solves Definition~\ref{def:posthoc_codeword_detection} without additional interrogation.
\end{lemma}

\begin{proof}
Partition $\{1,\ldots,q\}$ into blocks of size $k$. For a block $B=\{i_1,\ldots,i_k\}$ and a sign vector $w\in\{\pm1\}^k$ with $w_1=1$, define
\begin{equation}
 v_{B,w}=\frac1{\sqrt{k}}\sum_{a=1}^k w_ae_{i_a},
 \qquad
 K_{B,w}=\frac{\kappa}{\nu_L}\operatorname{Sym}\left(e_0^{\otimes(L-1)}\otimes v_{B,w}\right),
\label{eq:posthoc_codeword_queries}
\end{equation}
where
\begin{equation}
 \nu_L=\left\|\operatorname{Sym}\left(e_0^{\otimes(L-1)}\otimes e_1\right)\right\|_\infty,
 \qquad
 \alpha_L=\frac{2^{-(L-1)}}{\nu_L}.
\label{eq:posthoc_codeword_normalization}
\end{equation}
Here $\operatorname{Sym}$ averages over permutations of the tensor slots; the empty tensor power is omitted when $L=1$, giving $\nu_1=\alpha_1=1$. Let $T = {\rm Sym}\left(e_0^{\otimes(L-1)}\otimes e_1\right)$, and note $\|T\|_\infty$. Since $|T[e_1, e_0, e_0,...,e_0]| = 1/L$,  $1/L\le\nu_L = \|T\|_\infty \le1$. For each block $B=\{i_1,\ldots,i_k\}$ and sign vector $w\in\{\pm1\}^k$ with $w_1=1$, recall
\begin{equation}
v_{B,w}=\frac1{\sqrt{k}}\sum_{a=1}^k w_ae_{i_a},
\qquad
K_{B,w}=\frac{\kappa}{\nu_L}
\operatorname{Sym}\left(e_0^{\otimes(L-1)}\otimes v_{B,w}\right).
\end{equation}
The vector $v_{B,w}$ has unit norm and is orthogonal to $e_0$. Hence there is an orthogonal transformation $Q$ satisfying $Qe_0=e_0$ and $Qe_1=v_{B,w}$. Since applying the same orthogonal transformation to every slot preserves the full multilinear norm,
\begin{equation}
\left\|
\operatorname{Sym}\left(e_0^{\otimes(L-1)}\otimes v_{B,w}\right)
\right\|_\infty
=
\left\|
\operatorname{Sym}\left(e_0^{\otimes(L-1)}\otimes e_1\right)
\right\|_\infty
=
\nu_L.
\end{equation}
Thus every $K_{B,w}$ has norm exactly $\kappa$.  Now write $X_B=(X_{i_1},\ldots,X_{i_k})$. Because
\begin{equation}
A=\frac12e_0+h\sum_{i=1}^qX_ie_i,
\end{equation}
we have
\begin{equation}
e_0^TA=\frac12,
\qquad
v_{B,w}^TA=\frac{h}{\sqrt{k}}w^TX_B.
\end{equation}
Since all $L$ arguments of the tensor are equal to $A$, symmetrization does not change its value, and therefore
\begin{equation}
F_{K_{B,w}}(A)
=
\frac{\kappa}{\nu_L}(e_0^TA)^{L-1}v_{B,w}^TA
=
\frac{\alpha_L\kappa h}{\sqrt{k}}w^TX_B.
\label{eq:posthoc_codeword_values}
\end{equation}
Thus the answers to this list are exactly proportional to $w^TX_B$. After the index $J$ is revealed, let $B$ be the block containing $J$ and supply the post-hoc estimator with the list of observables. Then, given the estimates $\widehat F_{B,w}$, choose any $\widehat x\in[-1,1]^k$ satisfying
\begin{equation}
\left|
\frac{\alpha_L\kappa h}{\sqrt{k}}w^T\widehat x-\widehat F_{B,w}
\right|
\le\epsilon
\qquad\text{for every listed }w.
\label{eq:posthoc_codeword_fit}
\end{equation}
By triangle inequality, for every listed $w$,
\begin{equation}
\left|
\frac{\alpha_L\kappa h}{\sqrt{k}}
w^T(\widehat x-X_B)
\right|
\le2\epsilon.
\end{equation}
The event that no such vector exists only occurs with probability at most $\delta$. The error in the estimate $\widehat x$ is
\begin{align}
 \|\widehat x-X_B\|_1
 &=\max_{w\in\{\pm1\}^k}|w^T(\widehat x-X_B)|\nonumber\\
 &\le\frac{2\epsilon\sqrt{k}}{\alpha_L\kappa h}
 =\frac{2k}{\alpha_LC_L}
 \le\frac{k}{2048}.
\label{eq:posthoc_codeword_l1_error}
\end{align}
The final inequality uses $\nu_L\le1$ and $C_L=2^{L+11}$. Estimate a coordinate's sign as $+1$ when its fitted value is nonnegative, and $-1$ otherwise. A wrong sign on a coordinate with $|X_i|>1/16$ contributes at least $1/16$ to the error in \eqref{eq:posthoc_codeword_l1_error}. On $\mathcal E_B^c$ there are at most $k$ sign errors. Since $|X_i|$ is uniform on $[0,1]$, averaging gives
\begin{align}
 \frac1k\sum_{i\in B}\Pr(\widehat S_i\ne S_i)
 &\le\Pr(\mathcal E_B^c)+\frac1k\sum_{i\in B}\Pr(|X_i|\le1/16)+\frac{16}{2048}\nonumber\\
 &\le\delta+\frac1{16}+\frac1{128}<\frac18.
\label{eq:posthoc_codeword_sign_error}
\end{align}
This bound holds for every block list. Averaging over the uniformly requested index $J$ proves the required success probability.
\end{proof}

\begin{theorem}
\label{thm:posthoc_codeword_lower_bound}
Fix $L\ge1$, $N,M\ge1$, $\kappa>0$, and
\begin{equation}
 s=\min\{N,1+\log M\},
 \qquad
 0<\epsilon\le\frac{\kappa}{4C_L},
 \qquad
 2N\left(\frac{\epsilon}{\kappa}\right)^2\le s.
\label{eq:posthoc_codeword_regime}
\end{equation}
Any nonadaptive single-qubit protocol solving Definition~\ref{def:posthoc_codeword_detection} must use interrogation time
\begin{equation}
 T_{\rm int}\ge c_LN\sqrt{s}\frac{\kappa}{\epsilon},
\label{eq:posthoc_codeword_lower_bound}
\end{equation}
where $c_L>0$ depends only on $L$ and the fixed frequency-class parameters. The same lower bound holds for post-hoc spectroscopy with failure probability at most $1/20$.
\end{theorem}

To begin, we bound how much information the dataset must retain about the hidden codeword.

\begin{lemma}
\label{lem:posthoc_codeword_information}
Let $D$ be the learner's complete classical dataset before the index in Definition~\ref{def:posthoc_codeword_detection} is revealed. If there exists a decoder solving the codeword detection problem (with probability at least $7/8$), then
\begin{equation}
 I(S;D)\ge q\left[\log2-h_2(1/8)\right],
\label{eq:posthoc_codeword_information}
\end{equation}
where $h_2(p)=-p\log p-(1-p)\log(1-p)$ is binary entropy.
\end{lemma}

\begin{proof}
Let $\widehat S_i(D)$ be the decoder's answer when the requested index is $i$ and let $E_i = \textnormal{Pr}(\widehat S_i(D) \neq S_i)$ denote the probability of error when predicting the $i^{\rm th}$ sign. Since $J$ is uniform and independent of $(S,D)$, assuming the decoder succeeds with probability at least $7/8$ gives
\begin{equation}
 \frac1q\sum_{i=1}^q E_i\le\frac18.
\label{eq:posthoc_codeword_average_error}
\end{equation}
Binary Fano's inequality gives $H(S_i\mid D)\le h_2(E_i)$~\cite{cover_elements_2005}. The signs $S_i$ are independent and uniform. Conditional entropy subadditivity and concavity of $h_2$ therefore imply
\begin{align}
 I(S;D)
 &=q\log2-H(S\mid D)
 \ge q\log2-\sum_{i=1}^qH(S_i\mid D)\nonumber\\
 &\ge q\log2-\sum_{i=1}^qh_2(E_i)\nonumber\\
 &\ge q\left[\log2-h_2\left(\frac1q\sum_{i=1}^qE_i\right)\right]
 \ge q\left[\log2-h_2(1/8)\right].
\end{align}
\end{proof}

Finally, we extend the information-per-time bound from the previous section to the signal in \eqref{eq:posthoc_codeword_signal}.

\begin{lemma}
\label{lem:posthoc_codeword_information_time}
For the signal in Definition~\ref{def:posthoc_codeword_detection}, every nonadaptive single-qubit acquisition with dataset $D$ and deterministic interrogation-time cap $T_{\rm int}$ satisfies
\begin{equation}
 I(X;D)\le CT_{\rm int}(h+h^2D_\omega),
\label{eq:posthoc_codeword_information_time}
\end{equation}
where $D_\omega$ is defined in \eqref{eq:posthoc_frequency_geometry}.
\end{lemma}

\begin{proof}
Let $U$ have columns $e_1,\ldots,e_q$, so $A=e_0/2+hUX$ and $U^TU=I_q$. For one shot of duration $\tau$, the Fisher information matrices satisfy
\begin{equation}
 J_X=h^2U^TJ_AU,
 \qquad
 \operatorname{tr}J_X\le h^2\operatorname{tr}J_A
 \le h^2\tau\lambda_{\max}(G_I),
\end{equation}
where the last inequality is \eqref{eq:posthoc_fisher_trace}. The hypercube entropy inequality \eqref{eq:posthoc_cube_log_sobolev} is dimension-independent, so the same calculation as in \eqref{eq:posthoc_information_fisher} gives
\begin{equation}
 I(X;Y)\le\frac1\pi\mathbb E\operatorname{tr}J_X
 \le\frac{h^2\tau}{\pi}\left(\frac\tau2+D_\omega\right).
\end{equation}
Combining this with Holevo's bound \eqref{eq:posthoc_holevo},
\begin{align}
 I(X;Y)
 &\le\min\left\{\log2,\frac{h^2\tau^2}{2\pi}+\frac{h^2D_\omega\tau}{\pi}\right\}\nonumber\\
 &\le\min\left\{\log2,\frac{h^2\tau^2}{2\pi}\right\}+\frac{h^2D_\omega\tau}{\pi}
 \le C\tau(h+h^2D_\omega),
\label{eq:posthoc_codeword_single_shot_rate}
\end{align}
using $\min\{a,b\}\le\sqrt{ab}$. 
\end{proof}
We now complete the proof of the nonadaptive lower bound.
\begin{proof}[Proof of Theorem~\ref{thm:posthoc_codeword_lower_bound}]
For any successful codeword detection protocol, Lemmas~\ref{lem:posthoc_codeword_information} and \ref{lem:posthoc_codeword_information_time}, together with data processing, give
\begin{equation}
 cq\le I(S;D)\le I(X;D)\le CT_{\rm int}h(1+hD_\omega).
\label{eq:posthoc_codeword_combined_information}
\end{equation}
The definition of $k$ gives $s/2\le k\le2s$. The quantity $\left\lfloor\frac1k\min\left\{2N-1,\frac1{4h^2}\right\}\right\rfloor$ is at least one, so $\lfloor x\rfloor\ge x/2$ gives
\begin{align}
 q&\ge\frac12\min\left\{2N-1,\frac1{4h^2}\right\}\nonumber\\
 &=\frac12\min\left\{2N-1,\frac{k\kappa^2}{4C_L^2\epsilon^2}\right\}
 \ge c_LN.
\label{eq:posthoc_codeword_dimension}
\end{align}
The last inequality uses $k\ge s/2$ and \eqref{eq:posthoc_codeword_regime}. The same conditions imply $h\le C_L/\sqrt N$. Under the fixed-density frequency assumptions, $D_\omega\le C\log(2N)$, and hence $hD_\omega$ is bounded by a constant depending only on $L$ and the frequency-class parameters. Substituting into \eqref{eq:posthoc_codeword_combined_information} yields
\begin{equation}
 T_{\rm int}\ge c_L\frac qh
 \ge c_LN\sqrt{k}\frac{\kappa}{\epsilon}
 \ge c_LN\sqrt{s}\frac{\kappa}{\epsilon}.
\end{equation}
Lemma~\ref{lem:posthoc_codeword_reduction} then lifts this to a lower bound against all nonadaptive strategies for post-hoc spectroscopy.
\end{proof}

\subsubsection{Heisenberg limit for matched AC sensing}
\label{sec:adaptive_conventional_lower_bound}

Next we turn to the matched setting, where the observables $K_1, \ldots, K_M$ are supplied before sensing, and the protocol may tailor its single-qubit control to them and adapt later shots to earlier outcomes. We prove a lower bound for every $1 \leq M \leq N$ that allows noncommuting control and classical feedback, first for linear observables and then for fixed tensor order $\geq 2$. When $M$ is proportional to $N$ and $\kappa / \epsilon$ is large enough, the bound matches reusable randomized spectroscopy up to logarithmic factors.

Each sensing shot begins with a freshly prepared qubit, evolves over a single contiguous interaction interval with the unknown field, and ends with an arbitrary single-qubit POVM. The preparation, control, starting time, interrogation duration, and terminal measurement may all depend on the preceding classical record. Only classical information is retained between shots. Here we count only the time during which the sensor interacts with the unknown field, not idle time between shots, so the total interrogation time is $T_{\rm int} = \sum_r \tau_r$, where $\tau_r$ is the duration of the $r$-th shot; the wall-clock bound which accounts for waiting is given later. 

We take the equally spaced frequencies
\begin{equation}
\omega_j = j \Delta_\omega, \quad 1 \leq j \leq N, \quad \Delta_\omega > 0,
\label{eq:adaptive_frequency_grid}
\end{equation}
with $\Delta_\omega$ fixed as $N$ grows, so the largest frequency is $N \Delta_\omega$. Constants in the asymptotic bounds may depend on $\Delta_\omega$. 

\begin{graydefinitionbox}
\begin{theorem}[Heisenberg Limit on adaptive AC sensing]
\label{thm:adaptive_conventional_linear_lb}
Let $\Delta_\omega > 0$, $1 \leq M \leq N$, $\kappa > 0$, $0 < \epsilon \leq \kappa / (8 \sqrt{M})$, and $0 < \delta \leq 1 / 20$. There exists a list of $M$ linear observables $K_1,\ldots,K_M$ with $\|K_j\|_2 = \kappa$ for the frequency grid \eqref{eq:adaptive_frequency_grid} such that any adaptive conventional protocol that, for every $\|A\|_2 \leq 1$, returns estimates satisfying
\begin{equation}
\Pr\left[\max_{j \leq M} |\widehat F_j - K_j^T A| \leq \epsilon\right] \geq 1 - \delta
\label{eq:adaptive_linear_success}
\end{equation}
must satisfy
\begin{equation}
\sup_{\|A\|_2 \leq 1} \mathbb E_A T_{\rm int} \geq c_{\Delta_\omega} \min\left\{ N^{1/3} M^{2/3} \left( \frac{\kappa}{\epsilon} \right)^{4/3}, \sqrt{N M} \, \frac{\kappa}{\epsilon} \right\}.
\label{eq:adaptive_linear_lower_bound}
\end{equation}
Here the probability and expectation are over the measurement outcomes and private randomness of the protocol for fixed $A$, and $c_{\Delta_\omega} > 0$ depends only on $\Delta_\omega$. The list may depend on $N$, $M$, $\kappa$, and $\epsilon$, and is known before sensing.
\end{theorem}
\end{graydefinitionbox}

The two terms in \eqref{eq:adaptive_linear_lower_bound} are equal when $\kappa / \epsilon = \sqrt{N / M}$. Above this crossover the bound is $\sqrt{N M} \, \kappa / \epsilon$, which equals $N \kappa / \epsilon$ when $M = N$. Since the theorem requires $\kappa / \epsilon \geq 8 \sqrt{M}$, the regime below the crossover exists only when $M \leq \sqrt{N} / 8$. For $M$ proportional to $N$, the bound is always $\sqrt{N M} \, \kappa / \epsilon$. The theorem is a worst-case statement over the supplied observables, and lists with redundant or repeated observables can be easier.

The proof uses $M$ orthogonal spectral directions whose waveforms share a common nonnegative envelope, so that estimating the observables amounts to estimating $M$ coefficients. The envelope limits how fast a qubit under arbitrary control can gain posterior precision about these coefficients. This rate bound holds conditional on any earlier measurement record, so it applies to adaptive protocols.

\subsubsection{Hard signals with a common envelope}

Choose positive integers $m$ and $s$ with $s M (2m - 1) \leq N$. Each waveform occupies a block of $2m - 1$ frequencies, and $s$ sets the spacing between the frequencies within a block. Define the \textit{Fej\'{e}r kernel}
\begin{equation}
\mathcal F_m(\theta) = \frac{1}{m} \left| \sum_{r = 0}^{m - 1} e^{i r \theta} \right|^2 = 1 + 2 \sum_{r = 1}^{m - 1} \left( 1 - \frac{r}{m} \right) \cos(r \theta).
\label{eq:adaptive_fejer_kernel}
\end{equation}
It is nonnegative and contains only Fourier modes $|r| < m$. Let $c_m^2 = 1 + 2 \sum_{r = 1}^{m - 1} \left( 1 - \frac{r}{m} \right)^2 = (2m^2 + 1) / (3m)$ and $k_j = m + (j - 1) (2m - 1)$, and define
\begin{equation}
g_j(t) = \frac{\mathcal F_m(s \Delta_\omega t)}{c_m} \cos(k_j s \Delta_\omega t).
\label{eq:adaptive_packet_waveforms}
\end{equation}
The $g_j$ share the envelope $\mathcal F_m(s \Delta_\omega t) / c_m$ and differ only in their carrier frequencies $k_j s \Delta_\omega$. Expanding the product gives
\begin{equation}
g_j(t) = \frac{1}{c_m} \left[ \cos(k_j s \Delta_\omega t) + \sum_{r = 1}^{m - 1} \left( 1 - \frac{r}{m} \right) \left\{ \cos((k_j - r) s \Delta_\omega t) + \cos((k_j + r) s \Delta_\omega t) \right\} \right].
\end{equation}
Each $g_j$ therefore contains only a block of $2m - 1$ grid frequencies. The choice of $k_j$ makes these blocks disjoint, and $s M (2m - 1) \leq N$ places all of them inside the available frequency range. Let $u_j \in \mathbb R^{2N}$ be the corresponding cosine coefficient vector, with all sine coefficients zero. The vectors $u_j$ have disjoint supports, and the choice of $c_m$ gives $u_j^T u_l = \delta_{jl}$. We restrict the signal and observables to
\begin{equation}
A_x = \sum_{j = 1}^M x_j u_j, \quad K_j = \kappa u_j, \quad \|A_x\|_2 = \|x\|_2, \quad K_j^T A_x = \kappa x_j.
\label{eq:adaptive_hard_family}
\end{equation}
Estimating the supplied observables is therefore the same as estimating the coefficients $x_1,\ldots,x_M$. The corresponding field is $B_x(t) = \sum_{j = 1}^M x_j g_j(t)$.

The envelope contains only Fourier modes $|r| < m$, while $|k_j - k_l| \geq 2m - 1$ for $j \neq l$ and $k_j + k_l \geq 2m$ for all $j, l$. Hence
\begin{equation}
\int_0^{2 \pi} \mathcal F_m(\theta) \cos(k_j \theta) \cos(k_l \theta) \, d\theta = \pi \delta_{jl}, \quad \int_0^{2 \pi} \mathcal F_m(\theta) \, d\theta = 2 \pi.
\label{eq:adaptive_weighted_orthogonality}
\end{equation}
Each $g_j$ has period $2 \pi / (s \Delta_\omega)$. Write $g(t) = (g_1(t),\ldots,g_M(t))^T$. If $I$ is an interval of length at most one period and $q$ is a probability distribution supported on $\|x\|_2 \leq 1$, then
\begin{equation}
\int_I \sqrt{\mathbb E_q B_x(t)^2} \, dt \leq \frac{\sqrt{2} \, \pi}{s \Delta_\omega c_m}, \quad \left\| \int_I \beta(t) g(t) \, dt \right\|_2 \leq \frac{\sqrt{2} \, \pi}{s \Delta_\omega c_m} \quad \text{whenever } |\beta(t)| \leq 1.
\label{eq:adaptive_joint_waveform_bounds}
\end{equation}
For the first inequality, enlarge $I$ to a full period, write $C_x(\theta) = \sum_{j = 1}^M x_j \cos(k_j \theta)$, and change variables to $\theta = s \Delta_\omega t$. Cauchy--Schwarz and \eqref{eq:adaptive_weighted_orthogonality} then give
\begin{equation}
\int_I \sqrt{\mathbb E_q B_x(t)^2} \, dt \leq \frac{1}{s \Delta_\omega c_m} \left( \int_0^{2 \pi} \mathcal F_m \, d\theta \right)^{1/2} \left( \int_0^{2 \pi} \mathcal F_m \mathbb E_q C_x^2 \, d\theta \right)^{1/2} \leq \frac{\sqrt{2} \, \pi}{s \Delta_\omega c_m}.
\end{equation}
For the second inequality, every unit vector $x$ satisfies $x^T \int_I \beta(t) g(t) \, dt \leq \int_I |B_x(t)| \, dt$, and the first inequality with $q$ a point mass at $x$ bounds the right-hand side. Both bounds hold for intervals beginning at any time, because the integrands are periodic. The disjoint Fourier supports of the $g_j$ give 
\begin{equation}
\int_{t_0}^{t_0 + 2 \pi / (s \Delta_\omega)} g_j(t) g_l(t) \, dt = \frac{\pi}{s \Delta_\omega} \delta_{jl}.
\label{eq:adaptive_unweighted_orthogonality}
\end{equation}
Let $\mathcal H$ be a real Hilbert space, $\xi(t)$ an $\mathcal H$-valued function, and $I$ an interval of length at most $2 \pi / (s \Delta_\omega)$. Extending $\xi$ by zero to a full period and applying Bessel's inequality to each coordinate in an orthonormal basis of $\mathcal H$ gives
\begin{equation}
\sum_{j = 1}^M \left\| \int_I g_j(t) \xi(t) \, dt \right\|_{\mathcal H}^2 \leq \frac{\pi}{s \Delta_\omega} \int_I \|\xi(t)\|_{\mathcal H}^2 \, dt.
\label{eq:adaptive_vector_bessel}
\end{equation}

\subsubsection{Precision gained in one sensing shot}

We measure what a shot reveals about $x$ through the Fisher matrix of the posterior. For a posterior density $q$ on $x$, let
\begin{equation}
J_q = 4 \int \nabla \sqrt{q(x)} \, \nabla \sqrt{q(x)}^T \, dx, \quad \mathcal P_\lambda(q) = \operatorname{tr} \sqrt{J_q + \lambda I_M}, \quad \lambda > 0.
\label{eq:adaptive_posterior_precision}
\end{equation}
Taking the square root makes each Fisher-information eigenvalue contribute at the scale of inverse estimation error rather than inverse squared error. The shift by $\lambda$ bounds how fast $\mathcal P_\lambda$ can grow in directions where little information has been acquired.

Fix the classical record from all previous sensing shots, together with any private randomness used to choose the next shot. Given this information, the preparation, control, and final measurement are fixed. Suppose the qubit is prepared in a pure state $|\psi_0\rangle$ that does not depend on $x$. Mixed preparations are treated at the end of the proof of Theorem~\ref{thm:adaptive_conventional_linear_lb}. Combine the posterior and the conditional qubit state into the amplitude $\Psi(x) = \sqrt{q(x)} |\psi_x\rangle$, regarded as a vector in the real Hilbert space $\mathcal H$ below, and let $\Gamma_\Psi : \mathbb R^M \to \mathcal H$ be its derivative map,
\begin{equation}
\mathcal H = L^2(dx; \mathbb C^2), \quad \langle \Phi, \Psi \rangle_{\mathcal H} = \operatorname{Re} \int \Phi(x)^\dagger \Psi(x) \, dx, \quad \Gamma_\Psi e_j = 2 \partial_{x_j} \Psi.
\label{eq:adaptive_amplitude_space}
\end{equation}
At the beginning of the shot, $|\psi_0\rangle$ carries no dependence on $x$, so $\Gamma_{\Psi_0}^* \Gamma_{\Psi_0} = J_q$.

Consider an arbitrary POVM $\{E_y\}$ at the end of the shot. If outcome $y$ occurs with probability $p_y$ and produces the posterior $q_y$, then $\|E_y^{1/2} \Psi(x)\|^2 = p_y q_y(x)$. Differentiating this relation and summing over the outcomes gives
\begin{equation}
\sum_y p_y J_{q_y} \preceq \Gamma_\Psi^* \Gamma_\Psi, \quad \mathbb E_y \mathcal P_\lambda(q_y) \leq \operatorname{tr} \sqrt{\Gamma_\Psi^* \Gamma_\Psi + \lambda I_M}.
\label{eq:adaptive_measurement_contraction}
\end{equation}
The second inequality follows from the first because $X \mapsto \operatorname{tr} \sqrt{X}$ is monotone and concave on positive matrices, as the variational formula $\operatorname{tr} \sqrt{X} = \tfrac{1}{2} \inf_{Y > 0} \operatorname{tr}(Y + Y^{-1} X)$ shows.

We now separate the dependence on $x$ that the shot generates. Let $U_x(t)$ be the propagator, including the control, from the start of the shot at absolute time $t_0$ to time $t_0 + t$, and let $\tau$ be the duration of the shot. Differentiating the evolution with respect to $x_j$ gives
\begin{equation}
2 \partial_{x_j} U_x(\tau) = - i U_x(\tau) G_j(x), \quad G_j(x) = \int_0^\tau g_j(t_0 + t) U_x(t)^\dagger Z U_x(t) \, dt.
\label{eq:adaptive_duhamel_source}
\end{equation}
Define the source operator $S : \mathbb R^M \to \mathcal H$ by $S e_j = \sqrt{q} \, G_j(x) |\psi_0\rangle$. It collects the dependence on $x$ generated during the shot. Each $G_j$ is Hermitian and the derivatives of $\sqrt{q}$ are real, so the cross terms between the initial posterior dependence and the newly generated dependence vanish in the real inner product. Therefore
\begin{equation}
\Gamma_{\Psi_{\rm out}} = U_x(\tau) (\Gamma_{\Psi_0} - i S), \quad \Gamma_{\Psi_{\rm out}}^* \Gamma_{\Psi_{\rm out}} = J_q + S^* S.
\label{eq:adaptive_fresh_source_identity}
\end{equation}
The cancellation requires only that the qubit be independent of $x$ at the start of the shot. The state may acquire $x$-dependence during the evolution.

\begin{lemma}[Precision gained in one sensing shot]
\label{lem:adaptive_source_precision}
Let $\|S\|_*$ and $\|S\|_{\rm HS}$ denote the nuclear and Hilbert--Schmidt norms of $S$. Then
\begin{equation}
\mathbb E_y \mathcal P_\lambda(q_y) - \mathcal P_\lambda(q) \leq \min\left\{ \|S\|_*, \frac{\|S\|_{\rm HS}^2}{2 \sqrt{\lambda}} \right\}.
\label{eq:adaptive_source_precision_bound}
\end{equation}
\end{lemma}

\begin{proof}
Equations \eqref{eq:adaptive_measurement_contraction} and \eqref{eq:adaptive_fresh_source_identity} bound the increase in precision by $\operatorname{tr} \sqrt{J_q + \lambda I_M + S^* S} - \operatorname{tr} \sqrt{J_q + \lambda I_M}$. For a positive definite matrix $X$, $d \operatorname{tr} \sqrt{X} = \tfrac{1}{2} \operatorname{tr}(X^{-1/2} \, dX)$. The map $X \mapsto X^{-1/2}$ reverses the positive-matrix order, by the integral representation
\begin{equation}
X^{-1/2} = \frac{1}{\pi} \int_0^\infty t^{-1/2} (X + t I_M)^{-1} \, dt.
\end{equation}
Since $J_q \succeq 0$, integrating along $J_q + \lambda I_M + t S^* S$ for $0 \leq t \leq 1$ gives
\begin{equation}
\operatorname{tr} \sqrt{J_q + \lambda I_M + S^* S} - \operatorname{tr} \sqrt{J_q + \lambda I_M} \leq \frac{1}{2} \int_0^1 \operatorname{tr}\left[ (\lambda I_M + t S^* S)^{-1/2} S^* S \right] dt = \sum_j \left( \sqrt{\lambda + \sigma_j(S)^2} - \sqrt{\lambda} \right).
\label{eq:adaptive_trace_root_increment}
\end{equation}
This argument does not require $J_q$ and $S^* S$ to commute. Each term $\sqrt{\lambda + \sigma_j(S)^2} - \sqrt{\lambda}$ is at most $\sigma_j(S)$ and at most $\sigma_j(S)^2 / (2 \sqrt{\lambda})$. Summing each bound over $j$ gives $\|S\|_*$ and $\|S\|_{\rm HS}^2 / (2 \sqrt{\lambda})$, which proves the lemma.
\end{proof}

\subsubsection{Precision growth under arbitrary control}

A general control rotates the signal Hamiltonian $B_x(t) Z / 2$ through different Pauli directions, while the waveform bounds still how fast posterior precision grows. Consider first an interval of duration $h \leq 1 / (s \Delta_\omega)$ beginning at absolute time $t_a$. Let $\Psi_a(x)$ be the amplitude at the beginning of this interval, with $\|\Psi_a(x)\|^2 = q(x)$. Unlike at the start of a shot, the qubit state inside $\Psi_a$ may already depend on $x$. Write the propagator over $[t_a, t_a + t]$ as $U_c(t) V_x(t)$, where $U_c$ is the known control. The control rotates $Z$ to a unit vector $n(t)$ on the Bloch sphere, so
\begin{equation}
U_c(t)^\dagger Z U_c(t) = n(t) \cdot \boldsymbol\sigma, \quad \|n(t)\|_2 = 1, \quad \dot V_x(t) = - \frac{i}{2} B_x(t_a + t) [n(t) \cdot \boldsymbol\sigma] V_x(t), \quad V_x(0) = I.
\label{eq:adaptive_control_frame}
\end{equation}
The contribution of this interval to the source operator has columns
\begin{equation}
S_a e_j = \int_0^h g_j(t_a + t) V_x(t)^\dagger [n(t) \cdot \boldsymbol\sigma] V_x(t) \Psi_a(x) \, dt.
\label{eq:adaptive_block_source}
\end{equation}
For $\alpha = 1,2,3$, define $\eta_\alpha(t,x) = V_x(t)^\dagger \sigma_\alpha V_x(t) \Psi_a(x)$ and $\rho_\alpha(t) = \int_t^h n_\alpha(t') g(t_a + t') \, dt'$. Integrating by parts gives
\begin{equation}
S_a = \sum_{\alpha = 1}^3 \left[ \eta_\alpha(0) \rho_\alpha(0)^T + \int_0^h \dot \eta_\alpha(t) \rho_\alpha(t)^T \, dt \right].
\label{eq:adaptive_three_pauli_decomposition}
\end{equation}
The vectors $\rho_\alpha(t)$ do not depend on $x$, and \eqref{eq:adaptive_joint_waveform_bounds} with $\beta = n_\alpha$ gives $\|\rho_\alpha(t)\|_2 \leq \sqrt{2} \, \pi / (s \Delta_\omega c_m)$ for every $t$. Each $\eta_\alpha(0)$ has unit norm in $\mathcal H$. Differentiating the signal-dependent evolution gives
\begin{equation}
\dot \eta_\alpha(t) = \frac{i}{2} B_x(t_a + t) V_x(t)^\dagger [n(t) \cdot \boldsymbol\sigma,\sigma_\alpha] V_x(t) \Psi_a(x), \quad \sum_{\alpha = 1}^3 \|\dot \eta_\alpha(t)\|_{\mathcal H}^2 = 2 \mathbb E_q B_x(t_a + t)^2.
\label{eq:adaptive_pauli_derivative}
\end{equation}
For the second identity we used $[n \cdot \boldsymbol\sigma,\sigma_\alpha] = 2 i (n \times e_\alpha) \cdot \boldsymbol\sigma$ and $\sum_\alpha \|n \times e_\alpha\|_2^2 = 2$. The nuclear norm of a rank-one operator is the product of the norms of its two vectors, so \eqref{eq:adaptive_three_pauli_decomposition}, Cauchy--Schwarz over the three Pauli directions, and \eqref{eq:adaptive_joint_waveform_bounds} give
\begin{equation}
\|S_a\|_* \leq \frac{3 \sqrt{2} \, \pi}{s \Delta_\omega c_m} + \frac{\sqrt{12} \, \pi}{s \Delta_\omega c_m} \int_0^h \sqrt{\mathbb E_q B_x(t_a + t)^2} \, dt \leq \frac{3 \sqrt{2} \, \pi}{s \Delta_\omega c_m} + \frac{2 \sqrt{6} \, \pi^2}{s^2 \Delta_\omega^2 c_m^2}.
\label{eq:adaptive_block_nuclear_bound}
\end{equation}
The integration by parts differentiates only $V_x(t)$, never the control direction $n(t)$. The estimate therefore needs no smoothness or rate bound on the control and holds for arbitrary time-dependent single-qubit control, including rotations about rapidly changing, noncommuting axes. It makes no assumption that the coordinates of $x$ are independent under $q$, and it allows the incoming qubit state to depend on $x$.

For a shot with $\tau \geq 1 / (s \Delta_\omega)$, divide the sensing interval into $\lceil s \Delta_\omega \tau \rceil \leq 2 s \Delta_\omega \tau$ pieces of length at most $1 / (s \Delta_\omega)$. The division is only an accounting device, and no measurement or reset occurs between pieces. The qubit state entering a later piece depends on $x$, which is why the preceding estimate allowed an $x$-dependent input. If piece $a$ begins at absolute time $t_a$, let $U_a(x)$ be the propagator from the start of the shot to $t_a$, and let $\mathcal U_a$ denote pointwise multiplication by $U_a(x)$ on $\mathcal H$. The amplitude entering piece $a$ is $\Psi_a = \mathcal U_a \Psi_0$, and the source operator of the complete shot decomposes exactly as $S = \sum_a \mathcal U_a^* S_a$, with $S_a$ as in \eqref{eq:adaptive_block_source}. Each $\mathcal U_a$ is unitary on $\mathcal H$, even though $U_a$ depends on $x$, so $\|\mathcal U_a^* S_a\|_* = \|S_a\|_*$. The triangle inequality and \eqref{eq:adaptive_block_nuclear_bound} therefore give
\begin{equation}
\|S\|_* \leq \left( \frac{6 \sqrt{2} \, \pi}{c_m} + \frac{4 \sqrt{6} \, \pi^2}{s \Delta_\omega c_m^2} \right) \tau.
\label{eq:adaptive_long_shot_bound}
\end{equation}
The identity \eqref{eq:adaptive_fresh_source_identity} is applied once, to the complete shot, whose initial qubit state is independent of $x$ given the earlier record. For short shots the nuclear-norm estimate is too weak, because its bound for a single piece does not vanish as $\tau \to 0$. When $\tau \leq 1 / (s \Delta_\omega)$, define the $\mathcal H$-valued function $\xi(t)$ by $\xi(t,x) = \sqrt{q(x)} \, U_x(t)^\dagger Z U_x(t) |\psi_0\rangle$. It has $\|\xi(t)\|_{\mathcal H} = 1$, and the source columns are $S e_j = \int_0^\tau g_j(t_0 + t) \xi(t) \, dt$. Equation~\eqref{eq:adaptive_vector_bessel} therefore gives
\begin{equation}
\|S\|_{\rm HS}^2 \leq \frac{\pi}{s \Delta_\omega} \tau.
\label{eq:adaptive_short_shot_bound}
\end{equation}
This bound is linear in $\tau$, so splitting the interrogation time into many short shots does not help.

Lemma~\ref{lem:adaptive_source_precision}, with the nuclear-norm estimate for long shots and the Hilbert--Schmidt estimate for short shots, gives for every sensing shot and every incoming posterior supported on $\|x\|_2 \leq 1$
\begin{equation}
\mathbb E_y \mathcal P_\lambda(q_y) - \mathcal P_\lambda(q) \leq \left( \frac{6 \sqrt{2} \, \pi}{c_m} + \frac{4 \sqrt{6} \, \pi^2}{s \Delta_\omega c_m^2} + \frac{\pi}{2 s \Delta_\omega \sqrt{\lambda}} \right) \tau.
\label{eq:adaptive_one_shot_progress}
\end{equation}
Since $c_m^2 \asymp m$ and $s \geq 1$, and since $\Delta_\omega$ is fixed in the theorem, this simplifies to
\begin{equation}
\mathbb E_y \mathcal P_\lambda(q_y) - \mathcal P_\lambda(q) \leq C_{\Delta_\omega} \left( \frac{1}{\sqrt{m}} + \frac{1}{s \Delta_\omega \sqrt{\lambda}} \right) \tau.
\label{eq:adaptive_fixed_spacing_rate}
\end{equation}
If $\Delta_\omega$ varies with $N$, the explicit form \eqref{eq:adaptive_one_shot_progress} should be used instead.

\subsubsection{From estimation accuracy to interrogation time}

To convert the accuracy requirement into a bound on interrogation time, we use a prior concentrated at the scale $\epsilon / \kappa$,
\begin{equation}
q_0(x) = \prod_{j = 1}^M \frac{\kappa}{8 \epsilon} \cos^2\left( \frac{\pi \kappa x_j}{16 \epsilon} \right) \mathbf 1_{\{|x_j| \leq 8 \epsilon / \kappa\}}, \quad \lambda = \frac{\kappa^2}{64 \epsilon^2}.
\label{eq:adaptive_smooth_prior}
\end{equation}
Every $x$ in the support of this prior satisfies $\|x\|_2 \leq 8 \epsilon \sqrt{M} / \kappa \leq 1$, so the corresponding signal obeys the normalization assumed in the theorem. The square root of the prior vanishes at the boundary of its support, so the integrations by parts below produce no boundary terms. A direct calculation gives
\begin{equation}
J_{q_0} = \frac{\pi^2 \kappa^2}{64 \epsilon^2} I_M, \quad \mathcal P_\lambda(q_0) = \frac{\sqrt{\pi^2 + 1}}{8} M \frac{\kappa}{\epsilon}.
\label{eq:adaptive_initial_precision}
\end{equation}

Small mean-squared error requires large posterior precision. Let a posterior $q$ have mean $\mu$ and covariance $\Sigma$. Integration by parts gives $\mathbb E_q[(x - \mu) \nabla \log q(x)^T] = - I_M$. For almost every finite measurement record, the posterior is absolutely continuous with respect to Lebesgue measure, so $\Sigma$ is positive definite. The expectation of $(\nabla \log q + \Sigma^{-1}(x - \mu))(\nabla \log q + \Sigma^{-1}(x - \mu))^T$ is positive semidefinite, which gives $0 \preceq J_q - 2 \Sigma^{-1} + \Sigma^{-1} \Sigma \Sigma^{-1} = J_q - \Sigma^{-1}$. Hence $J_q \succeq \Sigma^{-1}$, and
\begin{equation}
\mathcal P_\lambda(q) \geq \operatorname{tr} \Sigma^{-1/2} \geq \frac{M^{3/2}}{\sqrt{\operatorname{tr} \Sigma}}.
\label{eq:adaptive_covariance_precision}
\end{equation}
For a classical record $D$, define the Bayes risk $\mathcal R_D = \mathbb E \|x - \mathbb E[x \mid D]\|_2^2$. Averaging \eqref{eq:adaptive_covariance_precision} over $D$ and applying Jensen's inequality gives
\begin{equation}
\mathbb E \mathcal P_\lambda(q_D) \geq \frac{M^{3/2}}{\sqrt{\mathcal R_D}}.
\label{eq:adaptive_bayes_precision}
\end{equation}
The posterior mean has the smallest mean-squared error of any estimator, so this bound requires no assumption that the estimator is unbiased.

To turn the success guarantee into a bound on mean-squared error, run the protocol three times independently and take, for each observable, the median of the three reported values. If at least two runs succeed, every median is $\epsilon$-accurate, so the medians fail with probability at most $3 \delta^2 - 2 \delta^3 < 1 / 128$. Shots begin at prescribed absolute times, so a run started later would see a time-shifted field, which corresponds to a different signal $A$. Every field on the grid \eqref{eq:adaptive_frequency_grid} has period $2 \pi / \Delta_\omega$, so each shot of a later run can be delayed by an integer number of periods. The run then sees the same field, and the delay adds only idle time, which does not count toward $T_{\rm int}$. If a run can contain unboundedly many shots, it may never finish, so a later run could never start. In that case the three runs are interleaved shot by shot, with each shot again delayed by an integer number of periods.

Divide the median estimates by $\kappa$ to obtain estimates of the coordinates $x_j$, and clip each coordinate to $[-8 \epsilon / \kappa,8 \epsilon / \kappa]$. On the success event, the total squared error is at most $M \epsilon^2 / \kappa^2$. On the failure event, both the true and clipped coordinates lie in the support of the prior, so the total squared error is at most $256 M \epsilon^2 / \kappa^2$. Since the posterior mean minimizes mean-squared error, the complete record $D$ from the three repetitions satisfies
\begin{equation}
\mathcal R_D \leq M \frac{\epsilon^2}{\kappa^2} + \frac{256 M}{128} \frac{\epsilon^2}{\kappa^2} = 3 M \frac{\epsilon^2}{\kappa^2}.
\label{eq:adaptive_amplified_risk}
\end{equation}
Combining this estimate with \eqref{eq:adaptive_initial_precision} gives
\begin{equation}
\frac{M^{3/2}}{\sqrt{\mathcal R_D}} - \mathcal P_\lambda(q_0) \geq \left( \frac{1}{\sqrt{3}} - \frac{\sqrt{\pi^2 + 1}}{8} \right) M \frac{\kappa}{\epsilon}.
\label{eq:adaptive_required_progress}
\end{equation}
The coefficient in parentheses is positive. By \eqref{eq:adaptive_bayes_precision}, accurate estimation therefore requires the expected posterior precision to grow by order $M \kappa / \epsilon$.

The one-shot bound limits how fast an adaptive protocol can supply this growth. Before each sensing shot, condition on the complete record so far and on the private randomness used to choose that shot. Its preparation, control, duration, and measurement are then fixed, and the earlier outcomes are held fixed when differentiating the conditional likelihood. Conditioning cannot move mass outside the support of the prior, so every posterior is supported on $\|x\|_2 \leq 1$ and \eqref{eq:adaptive_fixed_spacing_rate} applies to every shot. If $D_k$ denotes the record after the first $k$ shots, summing the one-shot bound and using $\lambda = \kappa^2 / (64 \epsilon^2)$ gives
\begin{equation}
\mathbb E \mathcal P_\lambda(q_{D_k}) - \mathcal P_\lambda(q_0) \leq C_{\Delta_\omega} \left( \frac{1}{\sqrt{m}} + \frac{\epsilon}{s \Delta_\omega \kappa} \right) \mathbb E \sum_{r \leq k} \tau_r.
\label{eq:adaptive_prefix_progress}
\end{equation}

A protocol may stop after a random number of shots. After a run stops, we set the duration of every later shot to zero. For each finite $k$, \eqref{eq:adaptive_bayes_precision} and \eqref{eq:adaptive_prefix_progress} give
\begin{equation}
\frac{M^{3/2}}{\sqrt{\mathcal R_{D_k}}} - \mathcal P_\lambda(q_0) \leq C_{\Delta_\omega} \left( \frac{1}{\sqrt{m}} + \frac{\epsilon}{s \Delta_\omega \kappa} \right) \mathbb E \sum_{r \leq k} \tau_r.
\label{eq:adaptive_finite_prefix_risk}
\end{equation}
Because the prior has bounded support, the posterior means form a bounded $L^2$ martingale, so $\mathcal R_{D_k}$ decreases to $\mathcal R_D$ for the complete record $D$. The accumulated interrogation time increases to $T_{\rm int}$, so by monotone convergence its expectation tends to $\mathbb E T_{\rm int}$. Letting $k \to \infty$ in \eqref{eq:adaptive_finite_prefix_risk} therefore replaces $\mathcal R_{D_k}$ by $\mathcal R_D$ and the sum by $T_{\rm int}$. Only the Bayes risks pass to the limit, so the argument needs no convergence of the posterior Fisher matrices.

The expected interrogation time of the three repetitions, averaged over the prior, is at most three times the original worst-case expected interrogation time. Comparing \eqref{eq:adaptive_required_progress} with the limit of \eqref{eq:adaptive_finite_prefix_risk} therefore gives, after absorbing numerical constants,
\begin{equation}
\sup_{\|A\|_2 \leq 1} \mathbb E_A T_{\rm int} \geq c_{\Delta_\omega} M \min\left\{ \sqrt{m} \, \frac{\kappa}{\epsilon}, s \Delta_\omega \left( \frac{\kappa}{\epsilon} \right)^2 \right\}.
\label{eq:adaptive_packet_lower_bound}
\end{equation}

To optimize \eqref{eq:adaptive_packet_lower_bound}, take
\begin{equation}
m = \left\lfloor \frac{1}{2} \left( 1 + \min\left\{ \frac{N}{M}, \left( \frac{N \kappa}{M \epsilon} \right)^{2/3} \right\} \right) \right\rfloor, \quad s = \left\lfloor \frac{N}{M (2m - 1)} \right\rfloor.
\label{eq:adaptive_packet_choice}
\end{equation}
These choices satisfy $m,s \geq 1$ and $s M (2m - 1) \leq N$, with $s \asymp N / (M m)$. If $\kappa / \epsilon \leq \sqrt{N / M}$, then $m \asymp (N \kappa / (M \epsilon))^{2/3}$, and \eqref{eq:adaptive_packet_lower_bound} gives
\begin{equation}
\sup_{\|A\|_2 \leq 1} \mathbb E_A T_{\rm int} \geq c_{\Delta_\omega} N^{1/3} M^{2/3} \left( \frac{\kappa}{\epsilon} \right)^{4/3}.
\end{equation}
If $\kappa / \epsilon \geq \sqrt{N / M}$, then $m \asymp N / M$, giving
\begin{equation}
\sup_{\|A\|_2 \leq 1} \mathbb E_A T_{\rm int} \geq c_{\Delta_\omega} \sqrt{N M} \, \frac{\kappa}{\epsilon}.
\end{equation}
Together these two regimes prove Theorem~\ref{thm:adaptive_conventional_linear_lb}.

The argument above assumed pure preparations and finitely many measurement outcomes. Mixed preparations, general POVMs, and measurable controls require no change to the bound. A mixed input state can be decomposed into pure states, with the decomposition label, which does not depend on $x$, included in the classical record. Revealing this label to the protocol can only make estimation easier. For a general POVM, use $\operatorname{tr} E(dy)$ as a finite dominating measure and replace the sums in \eqref{eq:adaptive_measurement_contraction} by integrals. At points where the likelihood vanishes, the weak derivative of its norm is bounded by the weak derivative of the underlying vector amplitude, so the same measurement inequality remains valid. The square root of the prior lies in $H_0^1$ on its support, and propagators and measurement operators preserve the weak differentiability and vanishing boundary values needed for the integrations by parts. The control frame \eqref{eq:adaptive_control_frame} is defined for measurable control axes and has well-defined limits for instantaneous pulses.

\subsubsection{Fixed tensor order and large observable lists}

For fixed tensor order, we fill $L - 1$ tensor slots with a known spectral component and the remaining slot with a hard direction $u_j$. After symmetrization, contracting the tensor with $L$ copies of the signal gives a fixed multiple of $x_j$. The tensors are normalized in the multilinear norm $\|\cdot\|_\infty$ defined in the preliminaries.

\begin{corollary}[Fixed tensor order]
\label{cor:adaptive_fixed_order_lb}
For each fixed integer $L \geq 2$ and fixed $\Delta_\omega > 0$, there exist constants $c_{L,\Delta_\omega}, C_L > 0$ such that, for $1 \leq M \leq N$ and $0 < \epsilon \leq \kappa / (C_L \sqrt{M})$, there is a list of $M$ symmetric order-$L$ tensors with $\|K_j\|_\infty = \kappa$ for which any adaptive conventional protocol achieving simultaneous $\epsilon$-accuracy with failure probability at most $\delta$ for every allowed signal, where $0 < \delta \leq 1 / 20$, must satisfy
\begin{equation}
\sup_{\|A\|_2 \leq 1} \mathbb E_A T_{\rm int} \geq c_{L,\Delta_\omega} \min\left\{ N^{1/3} M^{2/3} \left( \frac{\kappa}{\epsilon} \right)^{4/3}, \sqrt{N M} \, \frac{\kappa}{\epsilon} \right\}.
\label{eq:adaptive_fixed_order_lower_bound}
\end{equation}
\end{corollary}

\begin{proof}
The vectors $u_j$ have only cosine components. Let $w$ select the sine quadrature at $\omega_1$, so $\|w\|_2 = 1$ and $w^T u_j = 0$. Define
\begin{equation}
T_j = \frac{1}{L} \sum_{r = 1}^L w^{\otimes(r - 1)} \otimes u_j \otimes w^{\otimes(L - r)}, \quad \alpha_L = \|T_j\|_\infty, \quad K_j = \frac{\kappa}{\alpha_L} T_j.
\label{eq:adaptive_anchored_tensors}
\end{equation}
The tensor $T_j$ is supported on $\mathrm{span}\{w, u_j\}$, and an orthogonal transformation fixing $w$ maps $u_j$ to any other unit vector orthogonal to $w$ while preserving the multilinear norm. Hence $\alpha_L$ depends only on $L$. The triangle inequality gives $\alpha_L \leq 1$, and evaluating $T_j$ on $(u_j,w,\ldots,w)$ gives $\alpha_L \geq 1 / L > 0$, so $K_j$ is well defined and $\|K_j\|_\infty = \kappa$.

Restrict the signal to
\begin{equation}
A_x = \frac{1}{2} w + \sum_{j = 1}^M x_j u_j, \quad \|x\|_2 \leq \frac{1}{2}.
\label{eq:adaptive_anchored_signal}
\end{equation}
Since $w$ is orthogonal to every $u_j$, these signals satisfy $\|A_x\|_2^2 = 1 / 4 + \|x\|_2^2 \leq 1 / 2$. Contracting $K_j$ with $L$ copies of $A_x$ gives
\begin{equation}
F_{K_j}(A_x) = \frac{\kappa}{2^{L - 1} \alpha_L} x_j.
\label{eq:adaptive_anchored_outputs}
\end{equation}
Estimating $F_{K_j}(A_x)$ to accuracy $\epsilon$ therefore determines $x_j$ to accuracy $2^{L - 1} \alpha_L \epsilon / \kappa$.

The linear proof applies after replacing the coordinate accuracy $\epsilon / \kappa$ by $2^{L - 1} \alpha_L \epsilon / \kappa$. Since $\alpha_L \leq 1$, the rescaled prior is supported on $\|x\|_2 \leq 1 / 2$ whenever $\kappa / \epsilon \geq 2^{L + 3} \sqrt{M}$.

We reveal the fixed component $w / 2$ to the protocol, which can only make estimation easier. Its field is a known sinusoid at frequency $\omega_1$ and can be absorbed into the control Hamiltonian before applying \eqref{eq:adaptive_control_frame}. The unknown part of the field is $\sum_{j = 1}^M x_j g_j(t)$, as in the linear case, so the precision-growth and Bayes-risk bounds apply unchanged. The required coordinate accuracy differs from $\epsilon / \kappa$ by a factor that depends only on $L$, so the linear lower bound gives \eqref{eq:adaptive_fixed_order_lower_bound}.
\end{proof}

The construction extends to longer lists by repeating observables from a hard sublist. Success on the full list requires success on every observable in the sublist, and every tensor in the list still has norm $\kappa$. Taking a hard sublist of size proportional to $N$ gives, for every fixed $L \geq 1$,
\begin{equation}
\sup_{\|A\|_2 \leq 1} \mathbb E_A T_{\rm int} \geq c_{L,c_0,\Delta_\omega} N \frac{\kappa}{\epsilon}, \quad M \geq c_0 N, \quad \frac{\kappa}{\epsilon} \geq C_{L,c_0} \sqrt{N},
\label{eq:adaptive_rich_list_lower_bound}
\end{equation}
where $c_0 > 0$ is fixed.

For fixed confidence and fixed tensor order, reusable randomized spectroscopy uses interrogation time $\widetilde O_L(N \kappa / \epsilon)$ in this precision regime when $M$ is polynomial in $N$. Together with \eqref{eq:adaptive_rich_list_lower_bound}, this shows that the worst-case interrogation time of conventional protocols, over lists of $M \geq c_0 N$ observables in this regime, is $\widetilde\Theta_L(N \kappa / \epsilon)$. Reusable randomized spectroscopy achieves this scaling with preplanned commuting control and produces a record that can be reused to estimate observables chosen after the experiment. In the regime above, knowing the list before sensing, adapting later shots to earlier outcomes, and using noncommuting control cannot improve the scaling by a polynomial factor.

This comparison uses both $M \geq c_0 N$ and $\kappa / \epsilon \geq C_{L,c_0} \sqrt{N}$. The lower bound above does not cover the regime $\kappa / \epsilon \ll \sqrt{M}$, and superpolynomial list sizes require separate accounting of the logarithmic dependence on $M$. For a small known list, \eqref{eq:adaptive_linear_lower_bound} is smaller than $N \kappa / \epsilon$, so it leaves room for protocols tailored to those observables to use less interrogation time.

\subsection{Wall-clock Heisenberg and standard quantum limits}
Here we establish fundamental limits on the wall-clock time required for conventional and quantum computational AC sensing protocols. Importantly, by showing that conventional matched protocols with beyond-Heisenberg interrogation time must incur standard quantum limited wall-clock time, we pinpoint the source of quantum computational advantage in matched filtering tasks, which appears in the tradeoff between interrogation and wall-clock time rather than directly in the interrogation time as was the case for sensing many observables. 

\begin{theorem}[Wall-clock Heisenberg limit]
\label{thm:matched_wall_clock}
Fix $\Delta_\omega > 0$ and $\omega_j = j\Delta_\omega$ for $j = 1,...,N$. There exists a signal with spectrum vector $A$ and a linear observable $K$ with $\|K\|_2 = \kappa$ supported on at most $m\leq N$ frequencies, such that any protocol with one sensor qubit requires
\begin{equation} \label{eq:matched_wall_universal}
T_{\rm wall}=\Omega\left(\sqrt m\,\kappa/\epsilon\right)
\end{equation}
total wall-clock time. This applies with arbitrary coherent quantum memory and adaptivity. Moreover, any protocol without persistent quantum memory across measurements requires
\begin{equation}
T_{\rm int}^2+T_{\rm wall}+1=\Omega\left(m\kappa^2/\epsilon^2\right).
\label{eq:matched_joint_resource}
\end{equation}
In particular, if $T_{\rm int}=o(\sqrt m\,\kappa/\epsilon)$, then $T_{\rm wall}=\Omega(m\kappa^2/\epsilon^2)$.
\end{theorem}

\begin{proof}
We construct a broadband observable whose time-domain response is concentrated into narrow peaks. Let $h=\lfloor(m+1)/2\rfloor$, and consider the Fej\'er kernel
\begin{equation}
F_h(\theta)=\frac1h\left|\sum_{r=0}^{h-1}e^{ir\theta}\right|^2
=1+2\sum_{r=1}^{h-1}\left(1-\frac rh\right)\cos(r\theta).
\end{equation}
Define
\begin{equation}
g(t)=\frac{F_h(\Delta_\omega t)\cos(h\Delta_\omega t)}{c_h},
\qquad
c_h^2=1+2\sum_{r=1}^{h-1}\left(1-\frac rh\right)^2
=\frac{2h^2+1}{3h}.
\end{equation}
Expanding the product of cosines shows that $g(t)=u^Tf(t)$ for a vector $u$ supported on the $2h-1\leq m$ frequencies $\Delta_\omega,\ldots,(2h-1)\Delta_\omega$. The normalization gives $\|u\|_2=1$, while $c_h=\Theta(\sqrt m)$. Because the kernel is nonnegative and normalized, we have
\begin{equation}
\int_0^P |g(t)|\,dt
\leq\frac1{c_h\Delta_\omega}\int_0^{2\pi}F_h(\theta)\,d\theta
=\frac{2\pi}{\Delta_\omega c_h}.
\end{equation}
Since $g$ is periodic, every interval $I$ of total length $\ell$ therefore satisfies
\begin{equation}
\int_I |g(t)| \ dt \leq \frac{\ell +2\pi/\Delta_\omega}{c_h}
\label{eq:matched_packet_envelope}
\end{equation}
This bound holds regardless of the absolute starting time. Now we construct the hard instance as a discrimination problem. Let $z = \kappa/\epsilon$. We choose the observable $K=\kappa u$ and consider discriminating $A_\pm=\pm 2u/z$ through the values $K^TA_\pm=\pm2\epsilon$. Any protocol estimating $K^TA$ to accuracy $\epsilon$ with constant success probability must distinguish these signals with constant advantage.

We first establish the wall-clock bound for arbitrary quantum protocols. Under the same controls, the difference between the two sensing Hamiltonians is
\begin{equation}
H_{A_+}(t)-H_{A_-}(t)=\frac{2g(t)}{z}Z.
\end{equation}
Let $\mathcal{E}_+, \mathcal{E}_-$ denote the quantum channels corresponding to a single sensing round, immediately before measurement. Duhamel's formula bounds the diamond norm between these channels by
\begin{equation}
\|\mathcal{E}_+ - \mathcal{E}_-\|_\diamond \leq \int_{\rm active} \|H_{A_+}(t) - H_{A_-}(t)\|_\infty \ dt\leq\frac2z\int_{\rm active}|g(t)|\,dt
\leq\frac{2(T_{\rm wall}+P)}{zc_h}.
\end{equation}
This allows arbitrary coherent memory, intermediate operations, and adaptive control, since none can increase the norm of the signal-dependent Hamiltonian difference. Constant distinguishing advantage consequently requires $T_{\rm wall}+P=\Omega(c_hz)$, proving Eq.~\eqref{eq:matched_wall_universal} at fixed $\Delta_\omega$.

For the stronger conventional bound, consider an arbitrary adaptive protocol with sensing intervals $I_1,I_2,\ldots$ and measurements at the end of each interval. Condition on the classical record preceding shot $r$, so that its preparation, control, duration, and final POVM are fixed. The lack of quantum memory is already implied by imposing that only a classical transcript is retained between shots. Define $a_r=\int_{I_r}|g(t)|\,dt$. Let $p_{r,+}$ and $p_{r,-}$ be the conditional outcome distributions under $A_+$ and $A_-$, and write
\begin{equation}
h_r^2=\sum_y\left(\sqrt{p_{r,+}(y)}-\sqrt{p_{r,-}(y)}\right)^2
\end{equation}
for their squared Hellinger distance. First consider pure sensor states. Again applying Duhamel's formula, for the final sensor states $\ket{\psi_+}$ and $\ket{\psi_-}$ in round $r$, $\|\ket{\psi_+} - \ket{\psi_-}\|_2\leq 2a_r/z$. Moreover, for any common POVM $\{E_y\}$, $p_{r, \pm}(y) = \Tr(\ketbra{\psi_{\pm}}{\psi_{\pm}}E_y)$. By reverse triangle inequality,
\begin{equation}
    \left|\sqrt{p_{r,+}(y)}-\sqrt{p_{r,-}(y)}\right| \leq \|E_y^{1/2}(\ket{\psi_+} -\ket{\psi_-})\|_2\  .
\end{equation}
Applying this to each term of $h_r^2$ gives $h_r^2 \leq \|\ket{\psi_+} - \ket{\psi_-}\|_2 \leq 4a_r^2/z^2$, where we use POVM completeness and plug in the Duhamel bound. The inequality holds for mixed states by purification, and it holds for arbitrary noncommuting control and every preceding classical record.

Let $P_\pm$ denote the distributions of the complete classical transcript and $P_\pm^{(r)}$ their distributions through shot $r$. The definition of Hellinger distance gives the telescoping identity
\begin{equation}
\begin{aligned}
H^2(P_+,P_-)
&=\sum_r\sum_{\eta}
\sqrt{P_+^{(r-1)}(\eta)P_-^{(r-1)}(\eta)}\,h_r^2(\eta)\\
&\leq\frac12\sum_{\sigma=\pm}\mathbb E_{P_\sigma}\sum_r h_r^2.
\end{aligned}
\label{eq:matched_sequential_hellinger}
\end{equation}
Here $\eta$ denotes the preceding classical record, and the inequality follows from AM-GM. Since successful discrimination requires $\operatorname{TV}(P_+,P_-)=\Omega(1)$ and $\operatorname{TV}(P_+,P_-)\leq H(P_+,P_-)$, we have $\sum_r a_r^2=\Omega(z^2)$.

It remains to relate this quantity to interrogation and wall-clock time. Separate the shots into those of duration at most one period $P$, and those longer than $P$. For every short shot, Eq.~\eqref{eq:matched_packet_envelope} gives $a_r\leq2P/c_h$. Since the short-shot intervals are disjoint and contained in a total time period of length $T_{\rm wall}$,
\begin{equation}
\sum_{\rm short}a_r^2
\leq\frac{2P}{c_h}\sum_{\rm short}a_r
\leq\frac{2P(T_{\rm wall}+P)}{c_h^2}.
\end{equation}
For a long shot of duration $\tau_r>P$, the same bound gives $a_r\leq2\tau_r/c_h$. Consequently,
\begin{equation}
\sum_{\rm long}a_r^2
\leq\frac4{c_h^2}\sum_r\tau_r^2
\leq\frac{4T_{\rm int}^2}{c_h^2}.
\end{equation}
Combining the short- and long-shot contributions,
\begin{equation}
\sum_r a_r^2
\leq\frac{4T_{\rm int}^2+2P(T_{\rm wall}+P)}{c_h^2}.
\end{equation}
Together with the bound on per-shot Hellinger distance and $c_h^2=\Theta(m)$, this yields
\begin{equation}
T_{\rm int}^2+P T_{\rm wall}+P^2=\Omega(mz^2).
\end{equation}
At fixed $\Delta_\omega$, the period $P$ is constant, which proves Eq.~\eqref{eq:matched_joint_resource}. Finally, when $T_{\rm int}=o(\sqrt m\,z)$, its squared contribution is $o(mz^2)$, so necessarily $T_{\rm wall}=\Omega(mz^2)$.
\end{proof}

\subsection{Standard quantum limits on noisy conventional AC sensing}\label{sec:noisy_SQL_lower_bounds}

\subsubsection{SQL for conventional sensing of many observables}
\label{sec:noisy_conventional_lower_bound}

We now prove the noise-aware lower bound stated in Section~\ref{sec:randomized_spec}, which shows that the subroutines of randomized quantum spectroscopy are optimal among all conventional protocols once $\gamma\geq\epsilon/\kappa$. We use the model of adaptive conventional protocols from Section~\ref{sec:adaptive_conventional_lower_bound}, in which the preparation, start time, duration, single-qubit control, and terminal POVM of every shot may depend on all earlier outcomes. The only change is that during each shot the sensing qubit now evolves under
\begin{equation}
    \frac{d\rho}{dt}=-i\left[H_A(t),\rho\right]+\frac{\gamma}{4}\sum_{J\in\{X,Y,Z\}}\left(J\rho J-\rho\right),
    \label{eq:depolarizing_master_equation}
\end{equation}
where $H_A(t)$ is the sensing Hamiltonian \eqref{eq:general_sensing_hamiltonian} with arbitrary control $H_c(t)$. As in Section~\ref{sec:adaptive_conventional_lower_bound}, we take the evenly spaced frequencies \eqref{eq:adaptive_frequency_grid}.

\begin{graydefinitionbox}
\begin{theorem}[Generalized Standard Quantum Limit for noisy AC sensing]
\label{thm:depolarizing_lower_bound}
Let $\Delta_\omega>0$, $\kappa>0$, $0<\epsilon\leq\kappa/3$, and $0<\delta\leq1/20$, and suppose $\gamma\geq\epsilon/\kappa$. Let $K_\mu=\kappa e_\mu$ for $\mu=1,\ldots,2N$. Any adaptive conventional protocol under \eqref{eq:depolarizing_master_equation} which, for every $\|A\|_2\leq1$, returns estimates satisfying $\Pr[\max_{\mu\leq2N}|\widehat F_\mu-K_\mu^TA|\leq\epsilon]\geq1-\delta$ must satisfy
\begin{equation}
    \sup_{\|A\|_2\leq1}\mathbb E_AT_{\rm int}\geq c\max\left\{\gamma,\frac{N\gamma\Delta_\omega}{\gamma+\Delta_\omega}\right\}\frac{\kappa^2}{\epsilon^2},
    \label{eq:depolarizing_lower_bound}
\end{equation}
where $c>0$ is a numerical constant.
\end{theorem}
\end{graydefinitionbox}

When $\gamma\lesssim\Delta_\omega$ the bound reads $N\gamma\kappa^2/\epsilon^2$, while for $\gamma\gtrsim\Delta_\omega$ it reads $\max\{\gamma,N\Delta_\omega\}\kappa^2/\epsilon^2$. Both are matched up to logarithmic factors by the noise-aware upper bound for randomized quantum spectroscopy. The theorem concerns the $2N$ coordinate observables, so it applies to any list containing them. The term $\gamma\kappa^2/\epsilon^2$ uses only one of these observables and therefore holds for every $M\geq1$. This result is a generalized illustration of the concept that the conventional Heisenberg limit is not a stable asymptotic bound under realistic noise, manifested for the AC sensing problem with many constituent frequencies. Later, we will establish the Generalized Heisenberg Limit (GHL) on quantum computational sensors, which quadratically outperforms the conventional Heisenberg limits established previously. Unlike the conventional HL, the GHL will be shown to be a \textit{stable} asymptotic limit set by quantum mechanics even when noise is present, suggesting that the canonical picture of quantum metrology fits within a broader perspective that truly admits stable quantum advantage under realistic noise.

\begin{proof}
The proof compares the zero signal with $2N$ signals that each perturb a single quadrature by just enough for an $\epsilon$-accurate estimate to detect. Depolarization either leaves a shot of duration $\tau$ untouched or erases it, with survival probability $e^{-\gamma\tau}$, while even a noiseless shot can only spread a limited total sensitivity across the $2N$ spectral directions. Together these bound the distinguishability that any shot can generate per unit interrogation time, and summing over shots gives the result.

We begin by reducing learning to hypothesis testing. Set $\alpha=3\epsilon/\kappa$ and define the $2N$ candidate signals $A_0=0$ and $A_\mu=\alpha e_\mu$ for $\mu=1,\ldots,2N$. Since $\epsilon\leq\kappa/3$, each $A_\mu$ is an admissible signal. The $\mu$th observable takes the values $K_\mu^TA_0=0$ and $K_\mu^TA_\mu=3\epsilon$, so declaring the signal to be $A_\mu$ whenever $\widehat F_\mu>3\epsilon/2$ errs with probability at most $\delta$ under either signal. Any algorithm which can solve the learning task can therefore correctly solve all $2N-1$ hypothesis testing problems. Let $P_\mu$ denote the law of the complete classical record of the protocol when the signal is $A_\mu$. It follows that $\operatorname{TV}(P_0,P_\mu)\geq1-2\delta$ for every $\mu$.

Next we decompose this distinguishability over shots. We may assume every experiment consists of $R$ total shots by padding with shots of zero duration if the true number of shots is smaller. Protocols with an unbounded number of shots are also included in our bound, and are handled by applying the argument to the first $R$ shots and letting $R\to\infty$, since none of the bounds below depend on $R$. Now let $H_{r-1}$ denote the collected classical transcript preceding the $r$th shot, and let $h_{r,\mu}^2$ be the unnormalized squared Hellinger distance between the conditional outcome distributions of that shot under $A_\mu$ and $A_0$, given $H_{r-1}$. Lemma~\ref{lem:sequential_hellinger} with $Q=P_0$ gives $(1-2\delta)^2\leq2\,\mathbb E_{P_0}\sum_rh_{r,\mu}^2$ for each $\mu$. Every comparison is made against the same reference distribution $P_0$, so summing over $\mu$ yields
\begin{equation}
    2N(1-2\delta)^2\leq2\,\mathbb E_{P_0}\left[\sum_{r=1}^R\sum_{\mu=1}^{2N}h_{r,\mu}^2\right],
    \label{eq:depolarizing_accumulated}
\end{equation}
where the inner sum over $\mu$ is evaluated on a single adaptive history. Next, we bound this inner sum for one shot.

Condition on $H_{r-1}$, so that the start time $t_0$, duration $\tau$, initial state $\rho$, control, and POVM $\{M_y\}$ of the shot are fixed. Let $U_\mu(u)$ be the noiseless propagator from $t_0$ to $t_0+u$ under the signal $A_\mu$ and the chosen control, so that $U_0$ is the unitary generated by the control operations alone. Since $\sum_JJ\rho J=2\operatorname{tr}(\rho)\mathds{1}-\rho$, the dissipator in \eqref{eq:depolarizing_master_equation} equals $\gamma(\operatorname{tr}(\rho)\mathds{1}/2-\rho)$, and because $\mathds{1}/2$ is invariant under any unitary evolution this gives
\begin{equation}
     \rho(\tau)=e^{-\gamma\tau}U_A(\tau)\rho(0)U_A^\dagger(\tau)+(1-e^{-\gamma\tau})\frac{\mathds{1}}{2}.
\end{equation}
for arbitrary control. The outcome distribution under $A_\mu$ is therefore
\begin{equation}
    p_y^{(\mu)}=e^{-\gamma\tau}\operatorname{tr}\left[M_yU_\mu(\tau)\rho\,U_\mu(\tau)^\dagger\right]+\left(1-e^{-\gamma\tau}\right)\frac{\operatorname{tr}M_y}{2},
\end{equation}
where the second term does not depend on the signal. Now decompose $\rho=\sum_kp_k\ket{\phi_k}\!\bra{\phi_k}$ into pure states. Squared Hellinger distance is an $f$-divergence and hence jointly convex, so comparing two mixtures with the same signal-independent weights bounds $h^2_{r,\mu}$ by $e^{-\gamma\tau}$ times an average of the noiseless distances for the pure inputs $\ket{\phi_k}$. For pure states, the reverse triangle inequality argument in the proof of Lemma~\ref{lem:commuting_outcome_hellinger} shows that the squared Hellinger distance between outcome distributions is at most the squared Euclidean distance between the final states. Hence
\begin{equation}
    \sum_{\mu=1}^{2N}h_{r,\mu}^2\leq e^{-\gamma\tau}\max_{\ket{\phi}}\sum_{\mu=1}^{2N}\left\|\left(U_\mu(\tau)-U_0(\tau)\right)\ket{\phi}\right\|_2^2.
    \label{eq:depolarizing_noise_reduction}
\end{equation}

We now bound the noiseless sum for a fixed pure state $\ket{\phi}$. In the frame of the control, as in \eqref{eq:adaptive_control_frame}, we have $U_0(u)^\dagger ZU_0(u)=n(u)\cdot\boldsymbol\sigma$ for a unit vector $n(u)$, and $V_\mu(u)=U_0(u)^\dagger U_\mu(u)$ obeys 
\begin{equation}
    \frac{d}{du}V_\mu(u)=-\frac{i\alpha}{2}f_\mu(t_0+u)[n(u)\cdot\boldsymbol\sigma]V_\mu(u)    
\end{equation}
with $V_\mu(0)=\mathds{1}$. Unitarity of $U_0$ gives $\|(U_\mu(\tau)-U_0(\tau))\ket{\phi}\|_2=\|(V_\mu(\tau)-\mathds{1})\ket{\phi}\|_2$. Define the accumulated response of the control frame to the $\mu$th quadrature, $g_\mu(u)=\int_0^uf_\mu(t_0+s)n(s)\,ds\in\mathbb R^3$, so that $\frac{d}{du}[g_\mu(u)\cdot\boldsymbol\sigma]=f_\mu(t_0+u)\,n(u)\cdot\boldsymbol\sigma$. Integrating the equation of motion by parts gives
\begin{equation}
    \left(V_\mu(\tau)-\mathds{1}\right)\ket{\phi}=-\frac{i\alpha}{2}\left[g_\mu(\tau)\cdot\boldsymbol\sigma\right]V_\mu(\tau)\ket{\phi}+\frac{\alpha^2}{4}\int_0^\tau f_\mu(t_0+u)\left[g_\mu(u)\cdot\boldsymbol\sigma\right]\left[n(u)\cdot\boldsymbol\sigma\right]V_\mu(u)\ket{\phi}\,du.
\end{equation}
Using $\|g\cdot\boldsymbol\sigma\|_{\rm op}=\|g\|_2$, $|f_\mu|\leq1$, and Cauchy--Schwarz on the integral, we obtain
\begin{equation}
    \left\|\left(V_\mu(\tau)-\mathds{1}\right)\ket{\phi}\right\|_2^2\leq\frac{\alpha^2}{2}\|g_\mu(\tau)\|_2^2+\frac{\alpha^4\tau}{8}\int_0^\tau\|g_\mu(u)\|_2^2\,du.
    \label{eq:depolarizing_single_direction}
\end{equation}
To sum over $\mu$, note that for each Bloch component $a$ the vector with entries $g_{\mu,a}(u)$ is $\int_0^uf(t_0+s)n_a(s)\,ds$. Exactly as in \eqref{eq:posthoc_fisher_trace}, Cauchy--Schwarz bounds its squared norm by $\lambda_{\max}(\int_{t_0}^{t_0+u}f(t)f(t)^Tdt)\int_0^un_a(s)^2ds$, and summing over $a$ with $\|n(s)\|_2=1$ gives $\sum_\mu\|g_\mu(u)\|_2^2\leq u\,\lambda_{\max}(\int_{t_0}^{t_0+u}ff^Tdt)$. The proof of Lemma~\ref{lem:harmonic_bin_response}, applied to a bin containing all $N$ frequencies, bounds this eigenvalue by $(u+2\pi/\Delta_\omega)/2$ for every start time. Substituting into \eqref{eq:depolarizing_single_direction} and using $u\leq\tau$,
\begin{equation}
    \sum_{\mu=1}^{2N}\left\|\left(U_\mu(\tau)-U_0(\tau)\right)\ket{\phi}\right\|_2^2\leq\frac{\alpha^2}{4}\tau\left(\tau+\frac{2\pi}{\Delta_\omega}\right)\left(1+\frac{\alpha^2\tau^2}{4}\right).
    \label{eq:depolarizing_noiseless_shot}
\end{equation}
The leading term has a simple interpretation. A shot shorter than one period $2\pi/\Delta_\omega$ cannot resolve individual frequencies, and its total sensitivity to all $2N$ directions grows only like $\tau/\Delta_\omega$. A longer shot does resolve them, but a single qubit then shares one Heisenberg sensitivity $\tau^2$ among all directions rather than gaining it in each.

We can now combine the two effects. Since $\gamma\geq\epsilon/\kappa$, we have $\alpha\leq3\gamma$, so both $e^{-\gamma\tau}(1+\alpha^2\tau^2/4)$ and $\gamma\tau\,e^{-\gamma\tau}(1+\alpha^2\tau^2/4)$ are bounded by a numerical constant $C$, uniformly in $\gamma\tau$. Inserting \eqref{eq:depolarizing_noiseless_shot} into \eqref{eq:depolarizing_noise_reduction} and writing $\tau_r$ for the duration of the $r$th shot therefore gives
\begin{equation}
    \sum_{\mu=1}^{2N}h_{r,\mu}^2\leq C\alpha^2\tau_r\left(\frac1\gamma+\frac{2\pi}{\Delta_\omega}\right).
\end{equation}
Noise thus makes a shot longer than $1/\gamma$ no more useful per unit time than a shot of duration $1/\gamma$. Summing over shots in \eqref{eq:depolarizing_accumulated} gives $N(1-2\delta)^2\leq C\alpha^2(1/\gamma+2\pi/\Delta_\omega)\,\mathbb E_{P_0}T_{\rm int}$. Since $P_0$ is the law of the record under the admissible signal $A=0$, its expected interrogation time is at most the worst case, and substituting $\alpha=3\epsilon/\kappa$ yields
\begin{equation}
    \sup_{\|A\|_2\leq1}\mathbb E_AT_{\rm int}\geq\frac{(1-2\delta)^2}{9C}\frac{N\gamma\Delta_\omega}{\Delta_\omega+2\pi\gamma}\frac{\kappa^2}{\epsilon^2}.
\end{equation}

It remains to obtain the term $\gamma\kappa^2/\epsilon^2$, which dominates when $\gamma\gg N\Delta_\omega$. For this we compare $A_0$ with $A_1$ alone. Integrating the equation of motion directly gives $\|(V_1(\tau)-\mathds{1})\ket{\phi}\|_2\leq\alpha\tau/2$, so the same noise reduction gives $h_{r,1}^2\leq e^{-\gamma\tau_r}\alpha^2\tau_r^2/4\leq\alpha^2\tau_r/(4e\gamma)$, where we used $xe^{-x}\leq1/e$. Lemma~\ref{lem:sequential_hellinger} then gives $(1-2\delta)^2\leq\alpha^2\,\mathbb E_{P_0}T_{\rm int}/(2e\gamma)$, and hence
\begin{equation}
    \sup_{\|A\|_2\leq1}\mathbb E_AT_{\rm int}\geq\frac{2e(1-2\delta)^2}{9}\gamma\frac{\kappa^2}{\epsilon^2}.
\end{equation}
This is the standard quantum limit for a single parameter under depolarization. Taking the larger of the two bounds and using $\delta\leq1/20$ proves \eqref{eq:depolarizing_lower_bound}.
\end{proof}

Since the argument conditions on the complete record before every shot, the bound holds for arbitrary classical feedback, shot durations, start times, and noncommuting single-qubit control, as well as for protocols that know the observables in advance. For general frequencies with minimum separation $\Delta_\omega$, the same argument applies with $2\pi/\Delta_\omega$ replaced by $D_\omega$ from \eqref{eq:posthoc_frequency_geometry}, at the cost of a factor $\log(2N)$. The anchoring construction of Corollary~\ref{cor:adaptive_fixed_order_lb} also extends the bound to observables of any fixed order $L$. There the known anchoring component is absorbed into the control, which changes only $U_0$ and leaves every step above unchanged.

\subsubsection{SQL for conventional matched sensing of few observables}\label{sec:noisy_matched_lower_bound}

We now extend the adaptive lower bound of Section~\ref{sec:adaptive_conventional_lower_bound} to a single specified linear observable under Markovian depolarizing noise,
\begin{equation}
    \dot \rho = -i[H_A(t),\rho]
    +\frac{\gamma}{4}\sum_{P\in\{X,Y,Z\}}(P\rho P-\rho).
\end{equation}

\begin{graydefinitionbox}
\begin{theorem}[Noisy matched AC sensing]
\label{thm:noisy_matched_lower_bound}
Fix $\Delta_\omega>0$, $\kappa>0$,
$\kappa/\epsilon\geq C$, and $0<\delta\leq1/20$.
For the frequency grid \eqref{eq:adaptive_frequency_grid},
there exists a specified linear observable $K$ with
$\|K\|_2=\kappa$ such that every adaptive conventional protocol
estimating $K^TA$ to accuracy $\epsilon$ and failure probability
at most $\delta$ for every $\|A\|_2\leq1$ satisfies
\begin{equation}
\begin{aligned}
    \sup_{\|A\|_2\leq1}\mathbb E_A T_{\rm int}
    \geq c_{\Delta_\omega}\max\Bigg\{&
    \min\left\{
        N^{1/3}z^{4/3},\sqrt N\,z
    \right\},
    z^2\max\left\{
        \gamma,\min\left\{
            N\gamma,\sqrt{N\gamma\Delta_\omega}
        \right\}
    \right\}
    \Bigg\},
\end{aligned}
\qquad z=\frac{\kappa}{\epsilon}.
\end{equation}
This matches the scaling in Eq.~\eqref{eq:noisy_matched_total} up to logarithmic factors.
\end{theorem}
\end{graydefinitionbox}

\begin{proof}
The noiseless contribution follows from
Theorem~\ref{thm:adaptive_conventional_linear_lb} with $M=1$, since parameter-independent noise cannot improve estimation. For the noise-dependent contribution, take the localized Fej\'er waveform $g(t)=u^Tf(t)$ of Eq.~\eqref{eq:adaptive_packet_waveforms} with $M=1$, where $\|u\|_2=1$ and $s(2m-1)\leq N$. Set $K=\kappa u$ and compare $A_0=0$ with $A_1=3\epsilon u/\kappa$.

The Fej\'er envelope satisfies $\|g\|_\infty\lesssim\sqrt m$ and has absolute integral $O((s\Delta_\omega\sqrt m)^{-1})$ over one period. Consequently, on any interval $I$ of length $\ell$,
\begin{equation}
    \left(\int_I|g(t)|dt\right)^2
    \leq C\left(
        \frac{\ell^2}{m}
        +\frac{\ell}{s\Delta_\omega}
    \right).
    \label{eq:noisy_matched_segment_bound}
\end{equation}

The depolarizing generator decomposes into $Z$-dephasing at rate $\gamma/2$ and independent $X$ and $Y$ Pauli errors, each occurring at rate $\gamma/4$. Revealing and allowing correction of the latter errors can only make estimation easier, so it suffices to consider the remaining $Z$-dephasing. This has the representation $(\gamma/2)(\mathcal P_Z-\mathrm{id})$, where $\mathcal P_Z(\rho)=(\rho+Z\rho Z)/2$. We may therefore reveal projective $Z$ measurements occurring at Poisson rate $\gamma/2$. Conditioned on their outcomes, each subsequent interval begins in a known $Z$ eigenstate. Duhamel's inequality and Eq.~\eqref{eq:noisy_matched_segment_bound} bound the conditional squared Hellinger distance generated by a segment of length $\ell$ by
\begin{equation}
    h_{\rm seg}^2
    \leq
    C\left(\frac{\epsilon}{\kappa}\right)^2
    \left(
        \frac{\ell^2}{m}
        +\frac{\ell}{s\Delta_\omega}
    \right).
\end{equation}
For the Poisson-separated segment lengths $\ell_j$ in a shot of duration $\tau$,
\begin{equation}
    \mathbb E\sum_j\ell_j^2
    \leq\frac{4\tau}{\gamma},
    \qquad
    \sum_j\ell_j=\tau.
\end{equation}
Summing over shots and applying Lemma~\ref{lem:sequential_hellinger} therefore gives
\begin{equation}
    T_{\rm int}
    \gtrsim
    \left(\frac{\kappa}{\epsilon}\right)^2
    \left(
        \frac{1}{m\gamma}
        +\frac{1}{s\Delta_\omega}
    \right)^{-1}
    \gtrsim
    \left(\frac{\kappa}{\epsilon}\right)^2
    \min\{m\gamma,s\Delta_\omega\}.
\end{equation}
Optimizing over $m$ and $s$ yields the $N\gamma$ and $\sqrt{N\gamma\Delta_\omega}$ branches of Eq.~\eqref{eq:noisy_matched_total}.

For the final branch, the remaining $Z$-dephasing can be represented by the classical stochastic field
$dY_t=B_A(t)dt+\sqrt{\gamma}\,dW_t$, observed during the charged sensing intervals. For two signals differing by $O(\epsilon/\kappa)$ in one spectral coordinate, the relative entropy of this more informative observation satisfies
\begin{equation}
    D(P_A\|P_{A'})
    \leq
    \frac{1}{2\gamma}
    \mathbb E_A\int_{\rm active}
    (B_A(t)-B_{A'}(t))^2dt
    \lesssim
    \frac{\epsilon^2}{\gamma\kappa^2}
    \mathbb E_A T_{\rm int}.
\end{equation}
Binary distinguishability then requires $T_{\rm int}\gtrsim\gamma(\kappa/\epsilon)^2$. Taking the worst supplied target across these bounds proves the theorem.
\end{proof}

\subsection{Fundamental limit on commuting control for nonlinear AC sensing}\label{sec:pooled_power_lower_bound}

The pooled spectroscopy algorithm, introduced in Section~\ref{sec:pooled_spectroscopy}, measures the cumulative values of nonlinear spectroscopic observables within chosen frequency bins. It avoids directly learning every spectral line and post-processing to obtain nonlinear transformations by applying noncommuting sequences of pulses that coherently transform the signal before readout. In this section, we rigorously prove that such noncommuting control Hamiltonians enable a speedup for spectroscopy that cannot be matched by \textit{any} commuting protocol like dynamical decoupling or even randomized quantum spectroscopy. We accomplish this by establishing a lower bound on the total interrogation time used by any protocol whose control Hamiltonian only acts along a single Pauli direction, and correspondingly, whose interaction-picture Hamiltonian takes the form in Equation \eqref{eq:Commuting_interaction_pic_ham}, for the task of learning the power of a signal within a chosen frequency bin. A consequence of this bound, discussed in Section~\ref{sec:nc_advantage_pooled}, is that pooled spectroscopy is provably more efficient than any existing conventional spectroscopy protocol for learning the binned power of a time-varying signal.

First, we record a number of technical lemmas used in our lower bound. The first allows us to control the divergence between classical measurement transcripts obtained from running slightly different commuting spectroscopy experiments. For probability densities $p$ and $q$ with respect to a common measure, we use the unnormalized squared Hellinger distance $h^2(p,q)=\int(\sqrt p-\sqrt q)^2$.

\begin{lemma}
\label{lem:commuting_outcome_hellinger}
Consider a signal interrogation with a commuting control Hamiltonian and response $x=v^TA$, followed by an arbitrary single-qubit POVM. After absorbing all known initial and final rotations into the state preparation and measurement, the probability density for any outcome $y$ can be written as
$p_y(x)=a_y+b_y\cos x+c_y\sin x$,
where
$a_y\geq\sqrt{b_y^2+c_y^2}$.
In particular, $|\partial_xp_y(x)|\leq a_y$. Moreover, for every $x,\theta\in\mathbb R$,
$h^2(p(x),p(x+\theta))\leq\min\{2,\theta^2/4\}$.
\end{lemma}

\begin{proof}
A commuting interrogation produces a signal-dependent unitary $R_Z(x)=e^{-ixZ/2}$ after a known change of basis due to \eqref{eq:commuting_propagator}. Let $\rho$ be the initial sensor state after this change of basis and let $M_y$ denote the POVM. The outcome density is
$p_y(x)=\operatorname{tr}[M_yR_Z(x)\rho R_Z(-x)]$.

Conjugation by $R_Z(x)$ leaves the diagonal entries of $\rho$ unchanged and multiplies its two off-diagonal entries by $e^{-ix}$ and $e^{ix}$. The outcome probability is therefore a linear combination of $1,e^{ix},e^{-ix}$, and since it is real it has the form
$p_y(x)=a_y+b_y\cos x+c_y\sin x$.
Since this expression is nonnegative for every $x$, its minimum over $x$ must be nonnegative, which gives $a_y\geq\sqrt{b_y^2+c_y^2}$. It follows immediately that
$|\partial_xp_y(x)|\leq\sqrt{b_y^2+c_y^2}\leq a_y$.

Now we bound how much the measurement distribution can change when the accumulated phase changes by $\theta$. Purify $\rho$ with a reference system which is not acted on by the signal, and denote the resulting pure states after phases $x$ and $x+\theta$ by $\ket{\psi_x}$ and $\ket{\psi_{x+\theta}}$. For every POVM element, the reverse triangle inequality gives
\begin{equation}
    \left|
    \sqrt{\bra{\psi_x}M_y\ket{\psi_x}}
    -
    \sqrt{\bra{\psi_{x+\theta}}M_y\ket{\psi_{x+\theta}}}
    \right|
    \leq
    \sqrt{
    \bra{\psi_x-\psi_{x+\theta}}
    M_y
    \ket{\psi_x-\psi_{x+\theta}}
    }.
\end{equation}
Squaring and integrating over the outcomes, using $\int M_y dy=I$, gives
$h^2(p(x),p(x+\theta))\leq\|\ket{\psi_x}-\ket{\psi_{x+\theta}}\|_2^2$.
The two states differ only by the additional phase rotation $e^{-i\theta Z/2}$ on the sensor, so
\begin{equation}
    \|\ket{\psi_x}-\ket{\psi_{x+\theta}}\|_2
    \leq
    \|e^{-i\theta Z/2}-I\|_{\rm op}
    \leq
    \frac{|\theta|}{2}.
\end{equation}
This proves $h^2(p(x),p(x+\theta))\leq\theta^2/4$. The general bound $h^2(p,q)\leq2$ holds for any two probability distributions, completing the proof.
\end{proof}
Next, we constrain the kinds of responses available to a commuting power spectroscopy protocol which uses interrogation time $\tau$.
\begin{lemma}
\label{lem:harmonic_bin_response}
Suppose the spectrum of $B(t)$ is supported on the evenly spaced grid $\omega_j=j\Delta_\omega$, and consider a commuting protocol with response $v^TA$ realized by applying a control with interaction-picture modulation $|y(t)|\leq 1$ for time $\tau$. Let $b$ denote a subset of $m$ frequencies, and let $\Pi_b$ denote the projector of vectors in $\mathbb R^{2N}$ onto this $2m$-dimensional subspace; let $v_b = \Pi_b v$. Then
\begin{equation}
    \|v_b\|_2^2
    \leq
    \min\left\{
        m\tau^2,
        \frac{\tau}{2}\left(\tau+\frac{2\pi}{\Delta_\omega}\right)
    \right\}.
    \label{eq:harmonic_bin_response}
\end{equation}
In particular, if $\tau\leq2\pi/\Delta_\omega$, then $\|v_b\|_2^2\leq\pi\tau/\Delta_\omega$, while if $\tau>2\pi/\Delta_\omega$, then $\|v_b\|_2\leq\tau$. These bounds hold for any starting time of the interrogation.
\end{lemma}

\begin{proof}
Let $f_b(t)\in\mathbb R^{2m}$ contain the cosine and sine quadratures corresponding to the frequencies in the chosen bin. The response is
$v_b=\int_I y(t)f_b(t)dt$,
where $I$ is the interrogation interval of length $\tau$. We first use the exact orthogonality of the chosen frequency modes over one period. For every starting time $a$,
\begin{equation}
    \int_a^{a+2\pi/\Delta_\omega}
    f_b(t)f_b(t)^Tdt
    =
    \frac{\pi}{\Delta_\omega}I_{2m}.
    \label{eq:harmonic_bin_period}
\end{equation}
Then consider an arbitrary interval $I$ of duration $\tau$. Partition it into complete periods of length $2\pi/\Delta_\omega$ and one remaining interval shorter than one period, so that $\tau = q\frac{2\pi}{\Delta_\omega} + r, q\in \{0,1,..,\lfloor \tau \Delta_\omega/(2\pi)\rfloor\}$. Since $f_b(t)f_b(t)^T$ is positive semidefinite, Equation~\eqref{eq:harmonic_bin_period} implies
\begin{equation}
    \lambda_{\max}\left(
        \int_I f_b(t)f_b(t)^Tdt
    \right)
    \leq
    \frac12\left(
        \tau+\frac{2\pi}{\Delta_\omega}
    \right),
    \label{eq:harmonic_bin_gram}
\end{equation}
which follows because the integrand is positive semidefinite so we can bound the integral by $(q+1)\pi I_{2m}/\Delta_\omega$ and use $(q+1)2\pi/\Delta_\omega \leq \tau + 2\pi/\Delta_\omega$. For any unit vector $u\in\mathbb R^{2m}$, Cauchy--Schwarz now gives
\begin{align}
    |u^Tv_b|^2
    &=
    \left|
        \int_I y(t)u^Tf_b(t)dt
    \right|^2\\
    &\leq
    \tau
    \int_I
    (u^Tf_b(t))^2dt\\
    &\leq
    \frac{\tau}{2}
    \left(
        \tau+\frac{2\pi}{\Delta_\omega}
    \right).
\end{align}
Taking the supremum over unit vectors $u$ gives the second term in Equation~\eqref{eq:harmonic_bin_response}. There is also the direct bound $\|f_b(t)\|_2=\sqrt m$ for every $t$. Therefore
$\|v_b\|_2\leq\int_I\|f_b(t)\|_2dt=\sqrt m\tau$,
which gives the first term in Equation~\eqref{eq:harmonic_bin_response}.

If $\tau\leq2\pi/\Delta_\omega$, the Gram matrix of the interval is bounded by the Gram matrix of one full period, so the preceding argument sharpens to
$\|v_b\|_2^2\leq\pi\tau/\Delta_\omega$.
If $\tau>2\pi/\Delta_\omega$, Equation~\eqref{eq:harmonic_bin_response} gives
$\|v_b\|_2^2\leq\tau(\tau+2\pi/\Delta_\omega)/2\leq\tau^2$.
This proves the stated bounds.
\end{proof}
Finally, we provide a bound on the total variation distance between two adaptive processes based on the Hellinger distance between each random variable comprising those processes; this is separate from the spectroscopy task and applies to general distributions over processes $P, Q$. 
\begin{lemma}
\label{lem:sequential_hellinger}
Let $P$ and $Q$ be probability distributions on a sequence of random variables $Y_1,\ldots,Y_R$. For each $r$, let $H_{r-1}=(Y_1,\ldots,Y_{r-1})$, and let $p_r(\cdot\mid H_{r-1})$ and $q_r(\cdot\mid H_{r-1})$ denote the corresponding conditional distributions under $P$ and $Q$. Then
\begin{equation}
    \operatorname{TV}(P,Q)^2
    \leq
    2\sum_{r=1}^R
    \E_Q\left[
        h^2\left(
            p_r(\cdot\mid H_{r-1}),
            q_r(\cdot\mid H_{r-1})
        \right)
    \right].
    \label{eq:sequential_hellinger}
\end{equation}
\end{lemma}

\begin{proof}
Let $L_r = p(Y_1, ..., Y_{r})/q(Y_1, ..., Y_{r})$ at step $r$, and define the one-step likelihood ratio
\begin{equation}
    u_r
    =
    \frac{
        p_r(Y_r\mid H_{r-1})
    }{
        q_r(Y_r\mid H_{r-1})
    }.
\end{equation}
Then $L_r=L_{r-1}u_r$ and $\E_Q[u_r\mid H_{r-1}]=1$. Next, to track how distance between $P$ and $Q$ accumulates in each step, we introduce a divergence metric whose expectation controls the total variation, and whose growth in each step can be bounded by the Hellinger distance of conditional distributions in that step. For this, consider the function $F(t)=(t-1)^2/(t+1)$, which for $t,u>0$ has the property
\begin{equation}
    F(tu)-F(t)-F'(t)t(u-1)
    =
    \frac{
        4t^2(u-1)^2
    }{
        (1+tu)(1+t)^2
    }.
    \label{eq:sequential_hellinger_identity}
\end{equation}
Moreover, since $(\sqrt u+1)^2\leq(1+tu)\left(1+\frac1t\right)$ by Cauchy-Schwarz, and $(u-1)^2=(\sqrt u-1)^2(\sqrt u+1)^2$, we have 
\begin{equation}
    F(tu)-F(t)-F'(t)t(u-1)
    \leq
    4(\sqrt u-1)^2.
    \label{eq:sequential_hellinger_step}
\end{equation}
Now, plug in $t = L_{r-1}, u = u_r$. Then, taking expectation over $Q$ and conditioning on $H_{r-1}$,
\begin{align}
   \E_Q [F(u_rL_{r-1}) - F(L_{r-1}) - F'(L_{r-1})L_r(u_r-1)|H_{r-1}] \leq 4\E_Q [(\sqrt{u}-1)^2|H_{r-1}] \ .
\end{align}
Note that $L_r$ is a fixed number, so it factors out of the expectation. Applying this observation to the third term, $\E_Q[F'(L_{r-1})L_r(u_r-1)] =F'(L_{r-1})L_r\E_Q[u_r-1]$. But $\E_Q[u_r-1|H_{r-1}] = 0$, so the inequality simplifies to
\begin{equation}
    \E_Q [F(u_rL_{r-1}) - F(L_{r-1})|H_{r-1}]\leq 4\E_Q [(\sqrt{u}-1)^2|H_{r-1}] 
\end{equation}
Moreover, directly integrating the right hand side shows that we can replace the expectation with the Hellinger distance:
\begin{equation}
    \E_Q\left[
        (\sqrt{u_r}-1)^2
        \mid H_{r-1}
    \right]
    =
    h^2\left(
        p_r(\cdot\mid H_{r-1}),
        q_r(\cdot\mid H_{r-1})
    \right).
\end{equation}
Therefore
\begin{equation}
    \E_Q[F(L_r)]
    -
    \E_Q[F(L_{r-1})]
    \leq
    4\E_Q\left[
        h^2\left(
            p_r(\cdot\mid H_{r-1}),
            q_r(\cdot\mid H_{r-1})
        \right)
    \right].
\end{equation}
Summing over $r$ and using $L_0=1$, so that $F(L_0)=0$, gives
\begin{equation}
    \E_Q[F(L_R)]
    \leq
    4\sum_{r=1}^R
    \E_Q\left[
        h^2\left(
            p_r(\cdot\mid H_{r-1}),
            q_r(\cdot\mid H_{r-1})
        \right)
    \right].
    \label{eq:sequential_hellinger_sum}
\end{equation}
Let $p$ and $q$ denote the joint densities of the complete sequence under $P$ and $Q$. Since $L_R=p/q$,
\begin{equation}
    \E_Q[F(L_R)]
    =
    \int
    \frac{(p-q)^2}{p+q}.
\end{equation}
Finally, Cauchy--Schwarz gives
\begin{align}
    \operatorname{TV}(P,Q)^2
    &=
    \frac14
    \left(
        \int |p-q|
    \right)^2\\
    &\leq
    \frac12
    \int
    \frac{(p-q)^2}{p+q}.
\end{align}
Combining this inequality with Equation~\eqref{eq:sequential_hellinger_sum} proves Equation~\eqref{eq:sequential_hellinger}.
\end{proof}

Now we are prepared to define a discrimination task that can be solved by any algorithm for binned power spectroscopy, for which proving a lower bound on interrogation time immediately implies a minimax lower bound on the broader task.

\begin{definition}[Power discrimination problem]
\label{def:power_discrimination}
\normalfont
Fix a bin $\mathcal B_b$ containing $m$ frequencies, amplitude scale $\kappa>0$ and learning precision $0<\epsilon<\kappa/2$. Let $\rho$ be a real value satisfying
\begin{equation}
    \frac{8\epsilon}{\kappa}
    \leq
    \rho
    \leq
    \frac14,
    \qquad
    \frac{\kappa\rho}{\epsilon}
    \leq
    \frac{\sqrt{2m}}{100}.
    \label{eq:power_discrimination_rho}
\end{equation}
Set $\rho_0=\rho$ and $\rho_1=\rho+8\epsilon/\kappa$. For $i\in\{0,1\}$, let $G_i$ be the Gaussian distribution
\begin{equation}
    G_i
    =
    \mathcal N\left(
        0,
        \frac{\rho_i}{2m}I_{2m}
    \right)
    \label{eq:power_discrimination_gaussians}
\end{equation}
on the $2m$ real quadratures in bin $\mathcal B_b$, and define
\begin{equation}
    E_i
    =
    \left\{
        x\in\mathbb R^{2m}:
        \left|
            \|x\|_2^2-\rho_i
        \right|
        \leq
        \frac{\epsilon}{\kappa}
    \right\}.
    \label{eq:power_discrimination_event}
\end{equation}
Let $\pi_i$ denote the distribution $G_i$ conditioned on $E_i$.

Choose $J$ uniformly from $\{0,1\}$, draw the real quadratures $\Pi_bA_{\rm real}$ in bin $\mathcal B_b$ according to $\pi_J$, and set all amplitudes outside the bin to zero; the signal quadrature vector $A$ is fixed in all experiments. The learner is given the frequencies, the bin, and all parameters defining $\pi_0$ and $\pi_1$, but is not given $J$. It may perform an arbitrary adaptive commuting single-qubit spectroscopy protocol, including arbitrary preparations, terminal POVMs, interrogation durations, and classical feedback between shots.

The power discrimination problem is to classify $J$ with success probability at least $9/10$.
\end{definition}

This hypothesis testing task is constructed so that the two distributions in Definition~\ref{def:power_discrimination} differ only slightly in their typical microscopic amplitudes, but their total powers differ by more than twice the estimation accuracy. The conditioning in Equation~\eqref{eq:power_discrimination_event} is introduced to ensure that the hypothesis signal powers are deterministically separated, which may not be the case if they were unconditionally drawn from the two Gaussians. Next, we show that any protocol for binned power spectroscopy can solve the power discrimination problem.

\begin{prop}
\label{prop:power_discrimination_reduction}
Any spectroscopy protocol which, for every $\|A\|_2\leq1$, returns an estimate $\widehat P_b$ satisfying
\begin{equation}
    \Pr\left[
        |\widehat P_b-\kappa P_b|
        \leq
        \epsilon
    \right]
    \geq
    \frac9{10}
    \label{eq:power_estimation_guarantee}
\end{equation}
can solve the power discrimination problem with the same interrogation time and success probability. Consequently, any interrogation-time lower bound for Definition~\ref{def:power_discrimination} is also a lower bound for estimating the known bin power $\kappa P_b$.
\end{prop}

\begin{proof}
There are two points to verify. First, the conditioning used to define the two signal distributions must keep every possible realization inside the physical norm ball $\|A\|_2\leq 1$. Second, the two possible ranges of $P_b$ must be sufficiently separated that an $\epsilon$-accurate power estimate determines the hidden label $J$.

We begin with the norm constraint. Since $8\epsilon/\kappa\leq\rho$, we have $\rho_1=\rho+8\epsilon/\kappa\leq2\rho$. For either $i\in\{0,1\}$, every vector in the support of $\pi_i$ satisfies
\begin{align}
    \|A\|_2^2
    &=
    \|\Pi_bA_{\rm real}\|_2^2\leq
    \rho_i+\frac{\epsilon}{\kappa}\\
    &\leq
    \rho+\frac{9\epsilon}{\kappa}\leq
    \frac{17}{8}\rho\leq \frac{17}{32}
    <1.
\end{align}
Thus every signal used in the discrimination problem obeys the same norm constraint as the original spectroscopy problem. We next show that estimating the power distinguishes the two distributions. Under $J=0$, Equation~\eqref{eq:power_discrimination_event} implies
$P_b\leq\rho+\epsilon/\kappa$.
Under $J=1$, it gives
$P_b\geq\rho+7\epsilon/\kappa$.
Thus the two possible power ranges are separated by $6\epsilon/\kappa$. Suppose now that a protocol satisfies Equation~\eqref{eq:power_estimation_guarantee}. After obtaining its estimate, define
\begin{equation}
    \widehat J
    =
    \begin{cases}
        0,
        & \widehat P_b/\kappa\leq\rho+4\epsilon/\kappa,\\
        1,
        & \widehat P_b/\kappa>\rho+4\epsilon/\kappa.
    \end{cases}
    \label{eq:power_discrimination_decoder}
\end{equation}
On the event that the power estimate is $\epsilon$-accurate, if $J=0$ then
\begin{equation}
    \frac{\widehat P_b}{\kappa}
    \leq
    P_b+\frac{\epsilon}{\kappa}
    \leq
    \rho+\frac{2\epsilon}{\kappa},
\end{equation}
so the decoder returns $\widehat J=0$. Similarly, if $J=1$ then
\begin{equation}
    \frac{\widehat P_b}{\kappa}
    \geq
    P_b-\frac{\epsilon}{\kappa}
    \geq
    \rho+\frac{6\epsilon}{\kappa},
\end{equation}
so the decoder returns $\widehat J=1$. Therefore every successful power-estimation event is also a successful discrimination event. Since Equation~\eqref{eq:power_estimation_guarantee} holds for every admissible signal, it continues to hold after averaging over either of the two signal distributions and over the uniformly random label $J$. The resulting discriminator succeeds with probability at least $9/10$ and requires no additional interrogation time, proving the proposition.
\end{proof}
Now we are prepared to state the main theorem of this section.
\begin{graydefinitionbox}
\begin{theorem}
\label{thm:power_discrimination_lower_bound}
Consider the power discrimination problem in Definition~\ref{def:power_discrimination} for a bin containing $m$ frequencies evenly separated by $\Delta_\omega$. There is a universal constant $c>0$ such that every adaptive commuting single-qubit protocol which solves the problem with success probability at least $9/10$ and has deterministic interrogation-time cap $T_{\rm int}$ requires total interrogation time
\begin{equation}
    T_{\rm int}
    =
    \widetilde\Omega\left(
        \min\left\{
            \sqrt m\,\rho^{3/2}
            \left(\frac{\kappa}{\epsilon}\right)^2,
            \frac{m}{\rho}
        \right\}
    \right).
    \label{eq:power_discrimination_lower_bound_compact}
\end{equation}
\end{theorem}
\end{graydefinitionbox}

\begin{proof}
We compare the two signal distributions by first working with the unconditioned Gaussians $G_0$ and $G_1$ from Definition~\ref{def:power_discrimination}. We then distinguish between long and short interrogations, and use our previous lemmas to show that short shots can only reveal small amounts of new information about the unknown signal. For long interrogations we prove a bound on the rate at which information about new amplitudes can be accumulated. We then show that if the total interrogation time is less than Equation \eqref{eq:power_discrimination_lower_bound_compact}, the complete transcript distributions under the two hypotheses remain too close to distinguish with probability $9/10$, irrespective of how the total time is distributed among shots.

We first record that conditioning the Gaussian distributions in Definition~\ref{def:power_discrimination} changes them only slightly. Under $G_i$,
\begin{equation}
    \E_{G_i}\left[\|\Pi_bA_{\rm real}\|_2^2\right]
    =
    \rho_i,
    \qquad
    \Var_{G_i}\left(\|\Pi_bA_{\rm real}\|_2^2\right)
    =
    \frac{\rho_i^2}{m}.
\end{equation}
Since $\rho_i\leq2\rho$, Chebyshev's inequality and the condition $\kappa\rho/\epsilon\leq\sqrt{2m}/100$ give
\begin{align}
    G_i(E_i^c)
    &\leq
    \frac{\rho_i^2}{m}
    \left(\frac{\kappa}{\epsilon}\right)^2\leq\frac{4\rho^2}{m}
    \left(\frac{\kappa}{\epsilon}\right)^2\leq 8\times10^{-4}.
    \label{eq:power_discrimination_conditioning_error}
\end{align}
Conditioning a probability distribution on an event changes it in total variation by at most the probability of the complementary event. Thus it is enough to prove that the transcript distributions generated from $G_0$ and $G_1$ remain separated from total variation distance $0.8$ by a fixed constant margin. Equation~\eqref{eq:power_discrimination_conditioning_error} will then transfer the result to the actual distributions $\pi_0$ and $\pi_1$.

For the remainder of the proof, set
\begin{equation}
    \lambda
    =
    \sqrt{
        2\log\left(
            \left[
                1+
                \frac{\Delta_\omega\sqrt{2m}}{2\pi}
                \left(\frac{\kappa}{\epsilon}\right)^2
            \right]
        \right)
    }.
    \label{eq:power_discrimination_lambda}
\end{equation}
This quantity appears only in the Gaussian tail cutoff below and contributes the logarithmic factor suppressed in the theorem.

Our next step is to define a modified experiment built from the actual physical protocol; we will obtain distinguishability bounds on the outcomes of the modified experiment, then control how much the modified experiment differs from the physical experiment. In the actual physical experiment, after $r$ measurements the learner has observed a transcript $(Y_1,\ldots,Y_r)$, and its choice of the next interrogation may depend arbitrarily on these outcomes. The modified experiment changes this transcript to $(Y_1',Z_1,\ldots,Y_r',Z_r)$, where each $Z_r$ is additional information that we append to the transcript, while each $Y_i'$ corresponds to the outcome of the modified $r^{\rm th}$ experiment. Since the augmented transcript contains the complete physical transcript, giving additional information via $Z_r$ can only make the discrimination problem easier, while we will control the maximum amount of information that can be lost by replacing $Y_j\rightarrow Y_j'$.

First let us describe the data contained in the $Z_j$'s. These variables will keep track of a collection of orthogonal scalar projections of the unknown amplitudes in the target bin. We keep track of a subspace $\mathcal S\subseteq\mathbb R^{2m}$, initially $\mathcal S=\{0\}$. At each round, $Z_r$ will either contain no additional information, in which case $\mathcal S$ is unchanged, or it will contain the exact value $u^T\Pi_bA$ for a unit vector $u$ orthogonal to $\mathcal S$, in which case we enlarge $\mathcal S$ to include $u$. The precise rule determining when this second case occurs will be given below. In particular, short interrogations will never enlarge $\mathcal S$, while a long interrogation will enlarge it only on a certain probabilistic branch of the experiment.

After some number of such additions, let $A_{\mathcal S}=\Pi_{\mathcal S}\Pi_bA$ denote the component of the unknown bin amplitudes along the directions whose exact coefficients have been appended to the auxiliary transcript. Because those coefficients are part of the augmented transcript, $A_{\mathcal S}$ is known given that transcript. We now explain precisely how each $Z_j$ is chosen.

In each round, the learner chooses a commuting interrogation with full response vector $v$, and let $v_b$ be its restriction to the target bin as in Lemma~\ref{lem:harmonic_bin_response}. We decompose the response into the part along the directions already contained in $\mathcal S$ and a new orthogonal component,
\begin{equation}
    v_b^T\Pi_bA =\mu+\theta,\qquad
    \mu=v_b^TA_{\mathcal S},\qquad
    v_\perp=(I-\Pi_{\mathcal S})v_b,\qquad
    \theta=v_\perp^T\Pi_bA.
    \label{eq:power_discrimination_phase_decomposition}
\end{equation}
Thus $\mu$ is completely determined by the augmented transcript, while $\theta$ depends on directions of the signal whose coefficients have not yet been appended. If $v_\perp=0$, the learner's chosen response lies entirely in directions already represented in $\mathcal S$, and no new information needs to be added.

We now specify how the auxiliary transcript is generated. We divide the learner's interrogations into short shots with $\tau\leq2\pi/\Delta_\omega$ and long shots with $\tau>2\pi/\Delta_\omega$. The two cases are treated differently. For short shots, we show that the amount of information recorded about the part of the signal that has not yet been accessed is sufficiently small that we will remove it from the auxiliary experiment (which makes the corresponding $Y_j'$ less informative than $Y_j$), but we will bound the resulting change in the complete transcript using Lemmas~\ref{lem:commuting_outcome_hellinger}--\ref{lem:sequential_hellinger}. For long shots, we use a more sophisticated construction. We now make this precise.

Conditioned on the preceding transcript, the learner's chosen preparation, control, and terminal POVM are fixed. By Lemma~\ref{lem:commuting_outcome_hellinger}, if the accumulated signal phase in this interrogation were $x$, the conditional density of its outcome $y$ would be
$p_y(x)=a_y+b_y\cos x+c_y\sin x$. For the actual signal, $x=\mu+\theta$, so the outcome density of the actual experiment is $p_y(\mu+\theta)$. In the case where the experiment is a short interrogation, we design the modified experiment to draw the outcome $Y_r$ from $p_y(\mu)$ and set $Z_r=\emptyset$ (i.e. recording no additional information beyond what the learner observes). Thus $\mathcal S$ is unchanged. Notice that this is not generally the true physical outcome distribution, which would be $p_y(\mu+\theta)$. We will bound the cumulative error from this replacement after analyzing the long interrogations.

Next, we define how the modified transcript is constructed for a long interrogation. The quantity $\theta=v_\perp^T\Pi_bA$ is the phase arising from the part of the response outside the directions already recorded in $\mathcal S$; it is the piece that the modified experiment has not yet learned. Under either Gaussian hypothesis, we will show below that, conditional on the augmented transcript, this phase is Gaussian with standard deviation at most $\sqrt{\rho_1/(2m)}\|v_\perp\|_2$. We therefore define
\begin{equation}
    w
    =
    \lambda
    \sqrt{\frac{\rho_1}{2m}}
    \|v_\perp\|_2,
    \label{eq:power_discrimination_w}
\end{equation}
so that $|\theta|\leq w$ except with exponentially small probability. Suppose for the moment that $|\theta|\leq w$. For every possible measurement outcome $y$, define
\begin{equation}
    m_y
    =
    \inf_{|\eta|\leq w}p_y(\mu+\eta).
     \label{eq:power_discrimination_common_density}
\end{equation}
The quantity $m_y$ is the portion of the outcome density that is invariant for every possible value of the unknown phase in the interval $[-w,w]$. In particular, it is independent of the actual value of $\theta$, so this probability mass can never contribute to distinguishability. Then, let
\begin{equation}
    q
    =
    1-\int m_y\,dy.
    \label{eq:power_discrimination_q}
\end{equation}
Thus $1-q$ is the total probability mass common to all of the distributions $p_y(\mu+\eta)$ with $|\eta|\leq w$, while $q$ is the remaining phase-dependent probability mass.

These variables let us give an exact decomposition of the physical outcome distribution. For every $|\theta|\leq w$,
\begin{equation}
    p_y(\mu+\theta)
    =
    (1-q)\frac{m_y}{1-q}
    +
    q\frac{p_y(\mu+\theta)-m_y}{q}.
    \label{eq:power_discrimination_mixture}
\end{equation}
The first term on the right-hand side does not depend on $\theta$ at all. Hence, the experiment admits a description as a classical mixture of experiments: on the first branch, the modified experiment can generate the correct physical outcome while only sampling from a distribution with no knowledge of $v_\perp$. The second distribution does depend on $\theta$, so on the second branch we do access information about $v_\perp^T \Pi_b A$.

Thus, the modified experiment obtains its transcript as follows. It flips a Bernoulli coin with success probability $q$. If the Bernoulli variable is zero, it draws $Y_r$ from the density $m_y/(1-q)$, sets $Z_r=\emptyset$, and leaves $\mathcal S$ unchanged. If the Bernoulli variable is one and $v_\perp\neq0$, it appends
\begin{equation}
    Z_r
    =
    X
    =
    \frac{v_\perp^T\Pi_bA}{\|v_\perp\|_2}
    \label{eq:power_discrimination_recorded_coefficient}
\end{equation}
to the augmented transcript and enlarges $\mathcal S$ by the unit direction $v_\perp/\|v_\perp\|_2$. Since the augmented transcript now contains $X$, it knows the phase $\theta=\|v_\perp\|_2X$ along this new direction and can draw the physical outcome from the second density in Equation~\eqref{eq:power_discrimination_mixture}.

We must address one technical detail in the modified experiment construction. The Gaussian variable $\theta$ is unbounded, so there remains a small probability that $|\theta|>w$. Formally, therefore, on the probability-$q$ branch we draw the outcome from
\begin{equation}
    \frac{
        p_y\left(
            \mu+
            \operatorname{clip}
            \left(
                \|v_\perp\|_2X,
                [-w,w]
            \right)
        \right)
        -
        m_y
    }{q}.
    \label{eq:power_discrimination_long_shot_outcome}
\end{equation}
This is a nonnegative normalized density by the definitions of $m_y$ and $q$. When $|\theta|\leq w$, clipping does nothing and Equation~\eqref{eq:power_discrimination_mixture} shows that the visible outcome $Y_r$ has exactly the physical distribution $p_y(\mu+\theta)$. We will later bound the probability that clipping is ever active.

We have now set up the modified experiment; the next goal is to determine how distinguishable the two hypotheses are based on the outcomes of the modified experiment. By Lemma~\ref{lem:commuting_outcome_hellinger}, $|\partial_xp_y(x)|\leq a_y$ and $\int a_y\,dy=1$. Therefore
\begin{equation}
    0
    \leq
    p_y(\mu)-m_y
    \leq
    wa_y.
\end{equation}
Integrating over $y$ gives
\begin{equation}
    0
    \leq
    q
    =
    \int\left[p_y(\mu)-m_y\right]dy
    \leq
    w.
    \label{eq:power_discrimination_q_bound}
\end{equation}
Thus a long interrogation may point partly along a new direction, but the auxiliary transcript needs to record the corresponding coefficient only with probability at most $\lambda\sqrt{\rho_1/(2m)}\|v_\perp\|_2$. This is the property that will limit how quickly long interrogations can distinguish the two Gaussian hypotheses; we use this in the following lemma to bound the total distinguishability.

\begin{lemma}
\label{lem:power_discrimination_long_shots}
Let $Q_i$ denote the distribution of the complete transcript generated by the modified experiment when $\Pi_bA$ is drawn from $G_i$. If $K$ denotes the number of rounds for which $Z_r$ contains a new scalar coefficient rather than $\emptyset$, then
\begin{equation}
    \E_{Q_i}[K]
    \leq
    C\lambda
    \sqrt{\frac{\rho}{m}}
    T_{\rm int}.
    \label{eq:power_discrimination_revelation_count}
\end{equation}
Moreover,
\begin{equation}
    D_{\rm KL}(Q_0\|Q_1)
    \leq
    C
    \frac{
        \lambda T_{\rm int}
    }{
        \sqrt m\,\rho^{3/2}
    }
    \left(\frac{\epsilon}{\kappa}\right)^2.
    \label{eq:power_discrimination_long_kl}
\end{equation}
\end{lemma}

\begin{proof}
We first show that the part of the signal outside $\mathcal S$ remains an isotropic Gaussian after every step of the modified experiment. This is what allows us to analyze every adaptively chosen response as though its new component were a fresh Gaussian direction.

Initially this statement holds by the definition of $G_i$. Suppose it holds before some round. A short-shot outcome is generated from $p_y(\mu)$, which depends only on the already recorded component $A_{\mathcal S}$, so conditioning on that outcome does not change the distribution of the remaining component in $\mathcal S^\perp$. The same is true on the probability-$(1-q)$ branch of a long shot: the Bernoulli probability $q$ and the density $m_y/(1-q)$ are both determined by the modified history and do not depend on the part of $A$ that has not been recorded in the transcript.

Now consider a long-shot outcome which successfully adds data to the modified transcript. The direction $v_\perp/\|v_\perp\|_2$ that gets stored is determined entirely by the existing transcript and chosen response, and is orthogonal to $\mathcal S$. Hence the recorded coefficient
\begin{equation}
    X
    =
    \frac{v_\perp^T\Pi_bA}{\|v_\perp\|_2}
\end{equation}
is a centered Gaussian with variance $\rho_i/(2m)$. Orthogonal components of an isotropic Gaussian are independent, so after conditioning on $X$, the component of $\Pi_bA$ orthogonal to the enlarged subspace remains centered Gaussian with covariance $\rho_iI/(2m)$. The outcome subsequently generated from Equation~\eqref{eq:power_discrimination_long_shot_outcome} depends only on the recorded value $X$ and therefore does not alter this remaining Gaussian distribution. This completes the induction.

We can now bound how often a new coefficient is appended. In any long interrogation, Equation~\eqref{eq:power_discrimination_q_bound} and the definition of $w$ give
\begin{equation}
    q
    \leq
    \lambda
    \sqrt{\frac{\rho_1}{2m}}
    \|v_\perp\|_2.
\end{equation}
Since $\rho_1\leq2\rho$ and $\|v_\perp\|_2\leq\|v_b\|_2$, Lemma~\ref{lem:harmonic_bin_response} gives $\|v_\perp\|_2\leq\tau$ for every long interrogation. Summing the conditional success probabilities over all long shots therefore yields
\begin{align}
    \E_{Q_i}[K]
    &=
    \E_{Q_i}\left[
        \sum_{\rm long}q
    \right]\leq
    C\lambda
    \sqrt{\frac{\rho}{m}}
    \E_{Q_i}\left[
        \sum_{\rm long}\tau
    \right]\leq
    C\lambda
    \sqrt{\frac{\rho}{m}}
    T_{\rm int},
\end{align}
which proves Equation~\eqref{eq:power_discrimination_revelation_count}. We next bound how much information each nontrivial $Z_j$ contains about whether the signal was drawn from $G_0$ or $G_1$. Conditional on the same history, every part of the modified experiment other than the distribution of a newly appended coefficient is identical under the two hypotheses. The learner then runs the same subsequent experiment. The values of $q$, the unsuccessful-branch outcome distribution, and the outcome distribution conditional on a particular recorded value $X$ are also identical. The only difference is therefore that a newly recorded coefficient has variance $\rho_0/(2m)$ under $G_0$ and $\rho_1/(2m)$ under $G_1$.

Set $t=8\epsilon/(\kappa\rho)$, so that $\rho_1=\rho(1+t)$. Equation~\eqref{eq:power_discrimination_rho} gives $0\leq t\leq1$. The relative entropy between the two possible distributions of one recorded $Z_j$ is
\begin{align}
    D_{\rm KL}\left(
        \mathcal N\left(0,\frac{\rho}{2m}\right)
        \middle\|
        \mathcal N\left(0,\frac{\rho_1}{2m}\right)
    \right)
    &=
    \frac12
    \left[
        \log(1+t)
        +
        \frac1{1+t}
        -
        1
    \right]\\
    &\leq
    \frac{t^2}{4}\\
    &\leq
    \frac{16}{\rho^2}
    \left(\frac{\epsilon}{\kappa}\right)^2.
    \label{eq:power_discrimination_coefficient_kl}
\end{align}
The inequality in the second line follows because the derivative of the expression in square brackets, including the prefactor $1/2$, is $t/[2(1+t)^2]\leq t/2$, and the expression vanishes at $t=0$. Combining Equation~\eqref{eq:power_discrimination_coefficient_kl} with the now proven Equation~\eqref{eq:power_discrimination_revelation_count} gives, via chain rule for relative entropy,
\begin{equation}
    D_{\rm KL}(Q_0\|Q_1)
    \leq
    C
    \frac{
        \lambda T_{\rm int}
    }{
        \sqrt m\,\rho^{3/2}
    }
    \left(\frac{\epsilon}{\kappa}\right)^2,
\end{equation}
which proves the lemma.
\end{proof}

We now have a bound on the distinguishability of the two hypotheses under the modified experiment; we next compare the modified transcript to the transcript of the actual physical experiment. There are two differences to control. First, for every short shot the auxiliary experiment replaced the true phase $\mu+\theta$ by $\mu$. Second, for long shots the auxiliary experiment clipped $\theta$ to the interval $[-w,w]$. We treat these errors separately.

Let $P_i$ denote the physical transcript distribution when the signal is drawn from $G_i$. To control the short shots without making any assumption about the posterior distribution generated by the physical experiment, introduce an intermediate experiment. It begins with the same Gaussian signal and uses the same auxiliary construction on every long shot, but on a short shot it generates the exact physical outcome from $p_y(\mu+\theta)$ instead of $p_y(\mu)$. For this comparison only, include the signal vector $A$, which is a random variable as part of the transcript. The two processes then have identical transcript update rules on long shots conditional on the same history, and differ only in their short-shot outcome distributions.

For a short interrogation, Lemma~\ref{lem:commuting_outcome_hellinger} gives
\begin{equation}
    h^2\left(
        p(\mu+\theta),
        p(\mu)
    \right)
    \leq
    \frac{\theta^2}{4}.
\end{equation}
We proved in Lemma~\ref{lem:power_discrimination_long_shots} that conditional on the modified transcript history, the piece of $\Pi_b A$ orthogonal to $\mathcal{S}$ is still a centered and isotropic Gaussian with covariance matrix $\rho_i I/2m$, Thus, 
\begin{equation}
    \E_{Q_i}\left[
        \theta^2
        \mid
        Y_1,Z_1,\ldots,Y_{r-1},Z_{r-1}
    \right]
    =
    \frac{\rho_i}{2m}
    \|v_\perp\|_2^2
    \leq
    \frac{\rho}{m}
    \|v_b\|_2^2.
\end{equation}
Since the shot is short, Lemma~\ref{lem:harmonic_bin_response} gives $\|v_b\|_2^2\leq\pi\tau/\Delta_\omega$. Lemma~\ref{lem:sequential_hellinger} therefore implies
\begin{align}
    \operatorname{TV}^2
    &\leq
    \frac12
    \E_{Q_i}
    \sum_{\rm short}\theta^2\\
    &\leq
    \frac{\pi\rho}{2m\Delta_\omega}
    \E_{Q_i}
    \sum_{\rm short}\tau\leq
    \frac{\pi\rho T_{\rm int}}{2m\Delta_\omega}.
    \label{eq:power_discrimination_short_tv}
\end{align}
Thus replacing all of the short-shot outcome distributions $p_y(\mu +\theta)$ from the intermediate experiment by $p_y(\mu)$ in the original modified experiment changes the complete joint distribution by at most
\begin{equation}
    \sqrt{
        \frac{\pi\rho T_{\rm int}}{2m\Delta_\omega}
    }.
    \label{eq:power_discrimination_short_tv_root}
\end{equation}
Finally, we control the clipping used for long interrogations. Conditional on every modified history, the phase $\theta$ is Gaussian with variance $\rho_i\|v_\perp\|_2^2/(2m)$. Since $\rho_i\leq\rho_1$, Equation~\eqref{eq:power_discrimination_w} and the standard Gaussian tail bound give
\begin{equation}
    \Pr_{Q_i}\left[
        |\theta|>w
        \mid
        Y_1,Z_1,\ldots,Y_{r-1},Z_{r-1}
    \right]
    \leq
    2e^{-\lambda^2/2}.
\end{equation}
Each long shot has duration greater than $2\pi/\Delta_\omega$, so a protocol with interrogation-time cap $T_{\rm int}$ can contain at most $T_{\rm int}\Delta_\omega/(2\pi)$ long shots. A union bound therefore gives
\begin{equation}
    \Pr_{Q_i}\left[
        |\theta|>w
        \text{ during any long shot}
    \right]
    \leq
    \frac{T_{\rm int}\Delta_\omega}{\pi}
    e^{-\lambda^2/2}.
    \label{eq:power_discrimination_long_tail}
\end{equation}

The probability of the same event in the intermediate experiment can exceed this by at most the total variation distance in Equation~\eqref{eq:power_discrimination_short_tv_root}. Until the first time this event occurs, the intermediate experiment has exactly the same visible conditional outcome distributions as the physical experiment: short shots were already generated using their exact physical phase, and on every long shot with $|\theta|\leq w$ the mixture construction in Equation~\eqref{eq:power_discrimination_mixture} reproduces the exact physical outcome distribution. It follows that the total variation distance between the physical transcript $P_i$ and the visible part of the auxiliary transcript is at most
\begin{equation}
    2\sqrt{
        \frac{\pi\rho T_{\rm int}}{2m\Delta_\omega}
    }
    +
    \frac{T_{\rm int}\Delta_\omega}{\pi}
    e^{-\lambda^2/2},
    \label{eq:power_discrimination_physical_reference_tv}
\end{equation}
which upper bounds the probability that clipping affects the outcomes even once. 

Finally, we can compare the two physical transcript distributions obtained from the actual experiment. $Q_0,Q_1$ are the transcript distributions from the modified experiment; let $Q_0', Q_1'$ be the distributions over the same transcripts, but discarding the $Z_j$'s that the learner never actually has access to. Recall that $P_0, P_1$ are the actual experiment transcript distributions. By data processing and Pinsker's inequality applied to Lemma~\ref{lem:power_discrimination_long_shots},
\begin{equation}
    \operatorname{TV}\left(
        Q_0^{\rm vis},
        Q_1^{\rm vis}
    \right)
    \leq
    C
    \sqrt{
        \frac{
            \lambda T_{\rm int}
        }{
            \sqrt m\,\rho^{3/2}
        }
    }
    \frac{\epsilon}{\kappa},
    \label{eq:power_discrimination_reference_tv}
\end{equation}
Applying Equation~\eqref{eq:power_discrimination_physical_reference_tv} for $i=0,1$ and using the triangle inequality gives
\begin{equation}
    \operatorname{TV}(P_0,P_1)
    \leq
    C\sqrt{
        \frac{\rho T_{\rm int}}{m\Delta_\omega}
    }
    +
    C
    \sqrt{
        \frac{
            \lambda T_{\rm int}
        }{
            \sqrt m\,\rho^{3/2}
        }
    }
    \frac{\epsilon}{\kappa}
    +
    C T_{\rm int}\Delta_\omega e^{-\lambda^2/2}.
    \label{eq:power_discrimination_total_tv}
\end{equation}
Suppose now that
\begin{equation}
    T_{\rm int}
    \leq
    c
    \min\left\{
        \frac{
            \sqrt m\,\rho^{3/2}
        }{\lambda}
        \left(\frac{\kappa}{\epsilon}\right)^2,
        \frac{m\Delta_\omega}{\rho}
    \right\}
    \label{eq:power_discrimination_assumed_small_time}
\end{equation}
for a sufficiently small universal constant $c$. The first two terms in Equation~\eqref{eq:power_discrimination_total_tv} are then bounded by constants of order $\sqrt c$. To bound the final
term, write
\begin{equation}
    u=\frac{\Delta_\omega\sqrt{2m}}{2\pi}
    \left(\frac{\kappa}{\epsilon}\right)^2.
\end{equation}
Using the first bound in
Equation~\eqref{eq:power_discrimination_assumed_small_time},
we obtain
\begin{align}
    T_{\rm int}\Delta_\omega e^{-\lambda^2/2}
    &\le c\pi\rho^{3/2}
    \frac{u}{(1+u)\sqrt{\log(1+u)}}\\
    &\le c\pi\rho^{3/2}
    \le \frac{\pi c}{8},
\end{align}
where the second inequality follows from
$\log(1+u)\ge u/(1+u)$ and the last uses
$\rho\le 1/4$.
Thus, choosing $c$ sufficiently small makes
the right-hand side of
Equation~\eqref{eq:power_discrimination_total_tv}
strictly smaller than $0.8$.

Finally, we return from the unconditioned Gaussian distributions $G_i$ to the actual signal distributions $\pi_i$ in Definition~\ref{def:power_discrimination}. Equation~\eqref{eq:power_discrimination_conditioning_error} shows that replacing $G_i$ by $\pi_i$ changes the resulting physical transcript distribution by at most $8\times10^{-4}$ in total variation. Thus, after decreasing $c$ once more if necessary, the transcript distributions generated from $\pi_0$ and $\pi_1$ still have total variation strictly smaller than $0.8$. For two equally likely hypotheses, a test with success probability at least $9/10$ requires total variation distance at least $0.8$. Hence no adaptive commuting protocol with interrogation time satisfying Equation~\eqref{eq:power_discrimination_assumed_small_time} can solve the power discrimination problem. As a result, we have
\begin{equation}
    T_{\rm int}
    =
    \widetilde\Omega\left(
        \min\left\{
            \sqrt m\,\rho^{3/2}
            \left(\frac{\kappa}{\epsilon}\right)^2,
            \frac{m}{\rho}
        \right\}
    \right),
\end{equation}
which proves the theorem.
\end{proof}

\subsection{Generalized Heisenberg limit for quantum computational sensors}
\label{sec:all_memory_lower_bound}

The conventional lower bounds above no longer apply when signal-dependent quantum information can be retained coherently between sensing intervals. We therefore allow arbitrary quantum memory and unrestricted quantum computation, while keeping a single sensing qubit as the only system that couples directly to the unknown field. The following theorem proves that every such protocol for AQS requires interrogation time $\Omega_L(\sqrt N\,\kappa/\epsilon)$.
\begin{graydefinitionbox}
\begin{theorem}[Lower bound with arbitrary quantum memory]
\label{thm:all_memory_lower_bound}
Fix integers $N\geq2$, $L\geq1$, and $\kappa>0$, and consider the evenly-spaced frequency grid $\{\omega_j = j\Delta_\omega\}_{j=1}^{N}$ for $\Delta_\omega > 0$. Consider any sensing protocol in which the unknown field couples to the apparatus through $B_A(t)Z_{\rm s}/2$, where $Z_{\rm s}$ acts on one sensor qubit, with arbitrarily large ancillary quantum memory, control, adaptive measurements, and classical feedback. There are constants $c_L,c_L'>0$, depending only on $L$, such that whenever $0<\epsilon\leq c_L\kappa$, there is a list of $N$ order-$L$ observables with $\|K_j\|_\infty=\kappa$ such that every protocol satisfying
\begin{equation}
\Pr\left[\max_{j\leq N}|\widehat F_j-F_{K_j}(A)|\leq\epsilon\right]\geq1-\delta
\label{eq:all_memory_success}
\end{equation}
for all $\|A\|_2\leq1$ and $0<\delta\leq1/20$ must obey
\begin{equation}
\sup_{\|A\|_2\leq1}\mathbb E_A T_{\rm int}
\geq c_L'\sqrt N\,\frac{\kappa}{\epsilon}.
\label{eq:all_memory_lower_bound}
\end{equation}
\end{theorem}
\end{graydefinitionbox}
This directly lifts to a lower bound on interrogation time for $\textnormal{AQS}[\{\omega_j\}, \epsilon, \delta, \mathcal K]$ for $N$ frequencies, general $\epsilon$, and $|\mathcal{K}| = \Omega(N)$, with constant or sub-constant $\delta$.

\begin{proof}
We prove the lower bound by reducing sensing to a many-vs-one hypothesis-testing problem. Given a reference field $B_0(t)$ and an ensemble of alternative fields $B_1(t),\ldots,B_N(t)$, consider the two hypotheses
\begin{equation}
H_0:\ B(t)=B_0(t),
\qquad
H_1:\ B(t)=B_J(t),\quad J\sim\operatorname{Unif}([N]).
\end{equation}
The index $J$ is drawn once, before the experiment, and the same unknown field $B_J$ is used throughout all sensing intervals. The discrimination task is to distinguish $H_0$ from $H_1$. We first prove a general lower bound for this discrimination problem.

\begin{lemma}[Many-vs-one discrimination]
\label{lem:all_memory_identification}
Suppose the reference and alternative fields satisfy
\begin{equation}
\sup_t\frac1N\sum_{j=1}^N|B_j(t)-B_0(t)|\leq\beta.
\label{eq:all_memory_average_field_bound}
\end{equation}
Any single-sensor protocol that distinguishes $H_0$ from $H_1$ with error probability at most $\delta<1/2$ under each hypothesis, allowing arbitrary coherent quantum memory, adaptive control, measurements, and stopping, must satisfy
\begin{equation}
\mathbb E_0T_{\rm int}\geq\frac{2(1-2\delta)}{\beta},
\label{eq:all_memory_identification_bound}
\end{equation}
where the expectation is taken under $H_0$.
\end{lemma}

\begin{proof}
Let $\mathsf P_0$ denote the distribution of the final classical transcript under $H_0$, and $\mathsf P_j$ its distribution when the field is $B_j$. Under $H_1$, the transcript distribution is the mixture $\overline{\mathsf P}=\frac1N\sum_{j=1}^N\mathsf P_j$. By Le Cam's two-point bound, distinguishing the hypotheses with error at most $\delta$ requires
\begin{equation}
1-2\delta\leq\operatorname{TV}(\mathsf P_0,\overline{\mathsf P})
\leq\frac1N\sum_{j=1}^N\operatorname{TV}(\mathsf P_0,\mathsf P_j).
\end{equation}
We bound each term by a hybrid argument. During any sensing interval $I$, the Hamiltonians under $B_0$ and $B_j$ differ by $(B_j(t)-B_0(t))Z_{\rm s}/2$. Duhamel's formula therefore bounds the trace distance introduced during that interval by
\begin{equation}
\frac12\int_I|B_j(t)-B_0(t)|\,dt,
\end{equation}
regardless of the quantum memory or parameter-independent controls, since $\|Z_{\rm s}\|_{\rm op}=1$.

To handle adaptivity, compare hybrid protocols that use $B_0$ for the first $r-1$ sensing intervals and differ only in whether interval $r$ uses $B_0$ or $B_j$, with all subsequent intervals using $B_j$. The two hybrids have identical states and classical records before interval $r$, distributed as in the reference experiment. The subsequent operations are identical and cannot increase trace distance. Summing over the intervals gives
\begin{equation}
\operatorname{TV}(\mathsf P_0,\mathsf P_j)
\leq\frac12\mathbb E_0\sum_r\int_{I_r}|B_j(t)-B_0(t)|\,dt,
\label{eq:all_memory_hybrid_bound}
\end{equation}
where the intervals $I_r$ are those selected by the protocol under $H_0$. Averaging this inequality over $j$ and applying the assumed uniform bound yields
\begin{equation}
\begin{aligned}
1-2\delta
&\leq\frac1N\sum_{j=1}^N\operatorname{TV}(\mathsf P_0,\mathsf P_j)\\
&\leq\frac12\mathbb E_0\sum_r\int_{I_r}\frac1N\sum_{j=1}^N|B_j(t)-B_0(t)|\,dt\\
&\leq\frac{\beta}{2}\mathbb E_0T_{\rm int}.
\end{aligned}
\end{equation}
This proves the lemma.
\end{proof}

We now construct a many-vs-one discrimination problem for which $\beta=O(\epsilon/(\kappa\sqrt N))$, while successful AQS solves the discrimination task. Consider the Fej\'er kernel
\begin{equation}
\mathcal F_N(\theta)=\frac1N\left|\sum_{k=0}^{N-1}e^{ik\theta}\right|^2
=1+2\sum_{k=1}^{N-1}\left(1-\frac{k}{N}\right)\cos(k\theta).
\end{equation}
It is nonnegative, and its obeys
\begin{equation}
\sum_{j=1}^N\mathcal F_N\left(\theta-\frac{2\pi(j-1)}N\right)=N.
\label{eq:all_memory_fejer_identity}
\end{equation}
To verify this identity, expand the Fej\'er kernel in Fourier modes and sum:
Summing over the $N$ equally spaced translations gives
\begin{align}
\sum_{j=1}^N\mathcal F_N\left(\theta-\frac{2\pi(j-1)}{N}\right)
&=N+2\sum_{k=1}^{N-1}\left(1-\frac{k}{N}\right)
\operatorname{Re}\left[e^{ik\theta}\sum_{j=1}^N e^{-2\pi i k(j-1)/N}\right]\nonumber\\
&=N,
\end{align}
where the last equality follows from the geometric-series identity
\begin{equation}
\sum_{j=1}^N e^{-2\pi i k(j-1)/N}
=\frac{1-e^{-2\pi i k}}{1-e^{-2\pi i k/N}}=0,
\qquad 1\leq k<N.
\end{equation}

Define the waveforms
\begin{equation}
g_j(t)=\frac1{c_N}\left[\mathcal F_N\left(\Delta_\omega t-\frac{2\pi(j-1)}N\right)-1\right],
\qquad
c_N^2=4\sum_{k=1}^{N-1}\left(1-\frac{k}{N}\right)^2
=\frac{2(N-1)(2N-1)}{3N}.
\label{eq:all_memory_fejer_signals}
\end{equation}
The subtraction removes the zero-frequency component, so each $g_j$ uses only the allowed frequencies $\omega_1,\ldots,\omega_{N-1}$. Writing $g_j(t)=B_{u_j}(t)$, the Fourier expansion and the definition of $c_N$ give $\|u_j\|_2=1$.

The central property of this ensemble follows immediately from nonnegativity and Eq.~\eqref{eq:all_memory_fejer_identity}:
\begin{equation}
\begin{aligned}
\frac1N\sum_{j=1}^N|g_j(t)|
&\leq\frac1{Nc_N}\sum_{j=1}^N\left[\mathcal F_N\left(\Delta_\omega t-\frac{2\pi(j-1)}N\right)+1\right]\\
&=\frac2{c_N}\leq\frac{2\sqrt2}{\sqrt N}.
\end{aligned}
\label{eq:all_memory_uniform_field_bound}
\end{equation}
Unlike a bound on a typical field at a typical time, this holds uniformly at every time, including after arbitrary waiting.

For $L=1$, supply the $N$ observables $K_j=\kappa u_j$ and consider the many-vs-one problem
\begin{equation}
H_0:\ A=0,
\qquad
H_1:\ A=a u_J,\quad J\sim\operatorname{Unif}([N]),
\qquad
a=\frac{4\epsilon}{\kappa}.
\end{equation}
All signals are admissible when $\epsilon/\kappa\leq1/4$. Under $H_0$, every requested observable vanishes. Under alternative $J=j$, however,
\begin{equation}
F_{K_j}(a u_j)=\kappa a=4\epsilon.
\end{equation}
Thus any protocol that estimates all $N$ observables to accuracy $\epsilon$ can distinguish $H_0$ from $H_1$: simply output $H_1$ whenever $\max_j|\widehat F_j|>2\epsilon$. The error probability is at most $\delta$ under either hypothesis, since the estimation guarantee holds for every individual signal. The reference field is zero, while alternative $j$ has field $ag_j(t)$. Equation~\eqref{eq:all_memory_uniform_field_bound} therefore gives $\beta\leq2a/c_N$. Applying Lemma~\ref{lem:all_memory_identification}, and using $c_N\geq\sqrt{N/2}$, we obtain
\begin{equation}
\sup_{\|A\|_2\leq1}\mathbb E_AT_{\rm int}
\geq\mathbb E_{A=0}T_{\rm int}
\geq\frac{(1-2\delta)c_N}{a}
\geq\frac{1-2\delta}{4\sqrt2}\sqrt N\,\frac{\kappa}{\epsilon}.
\label{eq:all_memory_linear_lower_bound}
\end{equation}
This proves the linear case.

For fixed $L\geq2$, we embed the same many-vs-one discrimination problem into a nonlinear problem. Let $w$ be the cosine coordinate at frequency $\omega_N$, orthogonal to every $u_j$, and define
\begin{equation}
S_j=\frac1L\sum_{\ell=1}^L w^{\otimes(\ell-1)}\otimes u_j\otimes w^{\otimes(L-\ell)},
\qquad
\nu_L=\|S_j\|_\infty,\qquad K_j=\frac{\kappa}{\nu_L}S_j.
\end{equation}
Because $w$ and $u_j$ are orthonormal, $\nu_L$ is independent of $j$ and satisfies $1/L\leq\nu_L\leq1$, so $\|K_j\|_\infty=\kappa$. Take the reference signal $A^{(0)}=w/2$ and the alternative ensemble
\begin{equation}
A^{(j)}=\frac12w+2^{L-1}a u_j,
\qquad a=\frac{4\epsilon}{\kappa}.
\end{equation}
For $\epsilon/\kappa\leq2^{-L-2}$, all these vectors lie in the unit ball. Every observable vanishes on the reference signal, whereas under alternative $j$,
\begin{equation}
F_{K_j}(A^{(j)})
=\frac{\kappa}{\nu_L}\left(\frac12\right)^{L-1}2^{L-1}a
=\frac{4\epsilon}{\nu_L}\geq4\epsilon.
\end{equation}
Therefore, simultaneous $\epsilon$-accurate estimation again distinguishes the reference hypothesis $H_0$ from the hypothesis $H_1$ in which $J$ is drawn uniformly and the fixed signal $A^{(J)}$ is supplied. The common component $w/2$ cancels from the differences between the reference and alternative fields. These differences are $2^{L-1}ag_j(t)$, so Eq.~\eqref{eq:all_memory_uniform_field_bound} gives $\beta\leq2^La/c_N$. Applying the same discrimination lemma yields
\begin{equation}
\sup_{\|A\|_2\leq1}\mathbb E_AT_{\rm int}
\geq\frac{(1-2\delta)c_N}{2^{L-1}a}
=\Omega_L\left(\sqrt N\,\frac{\kappa}{\epsilon}\right).
\end{equation}
This proves Eq.~\eqref{eq:all_memory_lower_bound} for every fixed $L\geq1$.
\end{proof}

\newpage
\section{Deferred proofs} \label{sec:deferred}
\subsection{Deferred proofs for randomized quantum spectroscopy} \label{sec:deferred_rqs_perf_guarantees}
\subsubsection{Proof of performance guarantee for Gaussian-randomized spectroscopy}

\begin{proof}
Let $\mu=e^{-\|A\|_2^2/2}$, so that $e^{-1/2}\leq\mu\leq1$. From Algorithm \ref{algo:gaussian_response_compiler}, we know that each Gaussian random pulse sequence initialized in $\ket{+}$ and measured in the $Y$ basis has outcome $B$ satisfying $\E[B\mid g]=\sin(g^TA)$. Then the random variable $Z=Bg$ obeys
\begin{align}
    &\E[Z]
    =\E_g[g\sin(g^TA)]
    =-\nabla_A\E_g[\cos(g^TA)]
    =\mu A,\\
    &\E[ZZ^T]
    =\E_g[gg^T]
    =\mathds{1}_{2N}.
\end{align}
We also measure $X$ in the first $r_0$ shots; the mean of these shots is $\E_g[\cos(g^TA)]=\mu$. The algorithm treats these initial shots as a calibration of $\mu$, and uses them to produce unbiased $U_i$ by adjusting the $Y$-basis measurement outcomes accordingly. It follows that each $U_j$ is an unbiased estimate of $A$; the important piece is a bound on the error in estimating downstream observables, where the algorithm outperforms the naive approach of learning $A$ directly.

Recall that we take $L$ disjoint blocks of $m$ $Y-$basis shots to construct the $L$ outcome vectors. We index these $Lm$ shot outcomes by $B_{i, \ell}$ with $i\in \{1,...,L\}, \ell\in\{1,...,m\}$, and the sampled Gaussian vector for each shot by $g_{i,\ell}$. Let $Z_{i, \ell} = B_{i, \ell}g_{i, \ell}$ and $Z_i = m^{-1}\sum_{\ell }B_{i, \ell}g_{i, \ell}$ denote the mean of block $i$, and let $\xi_i=\overline Z_i-\mu A$ denote the error in this mean. For a tensor $K$ with $\|K\|_\infty\leq\kappa$, consider the quantity,
\begin{equation}
    \widetilde F_K=\mu^{-L}K[\overline Z_1,\ldots,\overline Z_L],
    \qquad
    \widehat F_K=K[U_1,\ldots,U_L]
    =\left(\frac{\mu}{\widehat\mu}\right)^L\widetilde F_K.
    \label{eq:gaussian_true_normalization}
\end{equation}
We do not have access to $\widehat F_K$; it is the estimator adjusted by the true unknown value of $\mu$. We introduce this object because while the final vectors $U_j$ are constructed from disjoint $Y$-basis measurement, they are all normalized by the same $\widehat \mu$, so we cannot treat them as unconditionally independent.

Now we bound the fluctuations in the measurement outcomes. For a scalar random variable $V$, write $\|V\|_{L^p}=(\E|V|^p)^{1/p}$. Now, introduce $Z'_{i,\ell}$ as a copy of $Z_{i,\ell}$; note that this is not a measurement that we make but a fictitious random variable with the same distribution as the actual shot $Z_{i, \ell}$. Then,
\begin{equation}
    \E_{Z'}\left[\frac{1}{m}\sum_{\ell=1}^m\left(Z_{i,\ell}-Z'_{i,\ell}\right)\right]
    =
    \xi_i.
\end{equation}
Since our observables are linear with respect to each index and $x\mapsto |x|^p$ is convex, the conditional Jensen's inequality allows us to replace $\xi_i$ by this difference when upper bounding any $L^p$ norm of a tensor contraction involving $\xi_i$. Each difference $Z_{i,\ell}-Z'_{i,\ell}$ also has a mean-zero, symmetric distribution. Multiplying it by a uniformly random sign $\sigma_{i,\ell} \in \{\pm 1\}$ therefore does not each its distribution. 

Using these two tricks alongside the triangle inequality, we have for any rank-$r$ tensor $K$ and any random variables $x_2, ... , x_r$ independent of $Z_i$, 
\begin{equation}
    \left\|K[\xi_i,x_2,\ldots,x_r]\right\|_{L^p}
    \leq
    2\left\|K\left[\frac{1}{m}\sum_{\ell=1}^m\sigma_{i,\ell}Z_{i,\ell},x_2,\ldots,x_r\right]\right\|_{L^p},
    \label{eq:one_slot_symmetrization}
\end{equation}
We will use this identity on each slot of our observables of interest; the next step is to leverage the fact that centered Gaussian random variables are invariant under multiplication by $-1$, which gives us $\sigma_{i,\ell}Z_{i,\ell}\sim\mathcal N(0,\mathds{1}_{2N})$. Consequently,
\begin{equation}
    \frac{1}{m}\sum_{\ell=1}^m\sigma_{i,\ell}Z_{i,\ell}
    \overset{d}{=}
    \frac{1}{\sqrt m}G_i,
\end{equation}
where $G_i\sim\mathcal N(0,\mathds{1}_{2N})$. Applying Equation~\eqref{eq:one_slot_symmetrization} successively to the $r$ independent indices therefore gives
\begin{equation}
    \left\|K[\xi_1,\ldots,\xi_r]\right\|_{L^p}
    \leq
    \left(\frac{2}{\sqrt m}\right)^r
    \left\|K[G_1,\ldots,G_r]\right\|_{L^p},
    \label{eq:gaussian_score_symmetrization}
\end{equation}
where $G_1,\ldots,G_r$ are independent standard Gaussian vectors in $\mathbb R^{2N}$. Conditional on $G_2,\ldots,G_r$, the contraction $K[G_1,\ldots,G_r]$ is therefore a mean-zero one-dimensional Gaussian with standard deviation at most $\|K\|_\infty\prod_{i=2}^r\|G_i\|_2$. Since $p$ is even, its scalar moment contributes the factor $((p-1)!!)^{1/p}$. By AM-GM, $(p-1)!! = 1\cdot 3\cdot...\cdot p-1 \leq (p/2)^{p/2}$, so $((p-1)!!)^{1/p} \leq \sqrt{p/2}$. Moreover,
\begin{equation}
    \E\|G_i\|_2^p=\prod_{\ell=0}^{p/2-1}(2N+2\ell)\leq\left(2N+\frac{p-2}{2}\right)^{p/2}.
\end{equation}
Independence of the remaining Gaussian vectors then yields
\begin{equation}
    \|K[G_1,\ldots,G_r]\|_{L^p}
    \leq\|K\|_\infty\sqrt{\frac p2}\left(2N+\frac{p-2}{2}\right)^{(r-1)/2}.
    \label{eq:gaussian_tensor_moment}
\end{equation}
Equipped with this identity, we return to Equation~\eqref{eq:gaussian_true_normalization}. Expanding $\overline Z_i=\mu A+\xi_i$ gives
\begin{equation}
    \widetilde F_K-K[A^{\otimes L}]
    =
    \sum_{\emptyset\neq S\subseteq\{1,\ldots,L\}}\mu^{-|S|}K[\xi_S,A_{\bar S}],
    \label{eq:gaussian_tensor_error_expansion}
\end{equation}
where the slots indexed by $S$ contain their respective $\xi_s, s\in S$, and the other slots contain copies of $A$. For $|S|=r$, contracting the remaining slots against $A$ gives an $r$-linear tensor of norm at most $\kappa$. Using $\mu^{-1}\leq\sqrt e$ and Equations~\eqref{eq:gaussian_score_symmetrization} and \eqref{eq:gaussian_tensor_moment}, we obtain
\begin{equation}
    \|\widetilde F_K-K[A^{\otimes L}]\|_{L^p}
    \leq
    \kappa\sqrt{\frac p2}\sum_{r=1}^L\binom Lr(2\sqrt e)^r\frac{(2N+(p-2)/2)^{(r-1)/2}}{m^{r/2}}.
    \label{eq:gaussian_full_moment_bound}
\end{equation}
Now consider the case $p\leq2N$ (which corresponds to learning subexponentially many observables). For $r=1,\ldots,L$, define
\begin{equation}
    m_r=\left[p(2N)^{r-1}\left(\frac{\kappa}{\epsilon}\right)^2\right]^{1/r}.
\end{equation}
When $L>1$, these quantities satisfy
\begin{equation}
    m_r=m_1^{(L-r)/(r(L-1))}m_L^{L(r-1)/(r(L-1))}
    \leq\max\{m_1,m_L\}.
\end{equation}
The two exponents are nonnegative and sum to one. For $L=1$, we only have $m_1$; thus the chosen shot count in our algorithm satisfies $m\geq C_Lm_r$ for every $r$. Since $2N+(p-2)/2\leq3N$, this argument gives us
\begin{equation}
    \frac{\sqrt{p/2}(2N+(p-2)/2)^{(r-1)/2}}{m^{r/2}}
    \leq
    \frac{1}{\sqrt2}\left(\frac32\right)^{(r-1)/2}C_L^{-r/2}\frac{\epsilon}{\kappa}.
    \label{eq:gaussian_each_error_order}
\end{equation}
When $p>2N$, the same inequality follows instead from $m\geq C_Lp(\kappa/\epsilon)^2$, $2N+(p-2)/2\leq3p/2$, and $(\epsilon/\kappa)^r\leq\epsilon/\kappa$. Thus, substituting Equation~\eqref{eq:gaussian_each_error_order} into Equation~\eqref{eq:gaussian_full_moment_bound} and summing the binomial expansion gives
\begin{align}
    \|\widetilde F_K-K[A^{\otimes L}]\|_{L^p}
    &\leq
    \frac{\epsilon}{\sqrt3}\left[\left(1+\sqrt{\frac{6e}{C_L}}\right)^L-1\right]\\
    &\leq
    \frac{\epsilon}{\sqrt3}\left(e^{\sqrt{e/24}}-1\right)
    <\frac{63\epsilon}{100e},
\end{align}
where we used $C_L=(12L)^2$. Markov's inequality implies
\begin{align}
    \Pr\left[\left|\widetilde F_K-K[A^{\otimes L}]\right|>\frac{63\epsilon}{100}\right] &=   \Pr\left[\left|\widetilde F_K-K[A^{\otimes L}]\right|^p>\left(\frac{63\epsilon}{100}\right)^p\right] \\
    &\leq \frac{\mathbb E\left[\left|\widetilde F_K-K[A^{\otimes L}]\right|^p\right]}{(63\epsilon/100)^p}   
    \leq e^{-p}.
\end{align}
A union bound over the specified list therefore gives simultaneous error at most $63\epsilon/100$ for the estimators $\widetilde F_{K_j}$, except with probability at most $Me^{-p}\leq\delta/4$.

Finally, we control errors in the estimate of $\mu$. Since the interval $[e^{-1/2},1]$ contains $\mu$, clipping the empirical $X$ mean to this interval cannot increase its distance from $\mu$. Hoeffding's inequality and the chosen value of $r_0$ give
\begin{equation}
    \Pr\left[|\widehat\mu-\mu|>\sqrt{\frac{2}{C_L}}\frac{\epsilon}{\kappa}\right]
    \leq
    2\exp\left(-\frac{r_0}{C_L}\left(\frac{\epsilon}{\kappa}\right)^2\right)
    \leq\frac{\delta}{4}.
\end{equation}
On the complementary event, the logarithm is $\sqrt e$-Lipschitz on $[1/\sqrt{e}, 1]$, so
\begin{equation}
    \left|\log\left(\frac{\mu}{\widehat\mu}\right)^L\right|
    \leq
    L\sqrt{\frac{2e}{C_L}}\frac{\epsilon}{\kappa}
    =
    \sqrt{\frac{e}{72}}\frac{\epsilon}{\kappa}.
\end{equation}
For $0\leq x\leq1$, convexity gives $e^{ax}-1\leq x(e^a-1)$. Since $e^{\sqrt{e/72}}<243/200$ and $\epsilon/\kappa<1$, it follows that
\begin{equation}
    \left|\left(\frac{\mu}{\widehat\mu}\right)^L-1\right|
    \leq\frac{43\epsilon}{200\kappa},
    \qquad
    \left(\frac{\mu}{\widehat\mu}\right)^L\leq\frac{243}{200}.
\end{equation}
Combining the errors from calibration with the error in each observable, and using $|K_j[A^{\otimes L}]|\leq\kappa$, gives
\begin{align}
    |\widehat F_{K_j}-K_j[A^{\otimes L}]|
    &\leq
    \left(\frac{\mu}{\widehat\mu}\right)^L|\widetilde F_{K_j}-K_j[A^{\otimes L}]|
    +
    \left|\left(\frac{\mu}{\widehat\mu}\right)^L-1\right||K_j[A^{\otimes L}]|\\
    &\leq
    \frac{243}{200}\frac{63\epsilon}{100}+\frac{43\epsilon}{200}
    =
    \frac{19609\epsilon}{20000}
    <\epsilon.
\end{align}
The ideal experiment therefore succeeds for the whole specified list except with probability at most $\delta/2$.

Finally, we include the errors from pulse sequence synthesis. The probability that any of the $R$ Gaussian compiler calls fails is at most $R\delta_{\rm syn}=\delta/4$. Conditional on successful compilation, suppose a requested response $g$ is implemented as $g+e$, with $\|e\|_2\leq\epsilon_{\rm syn}$. The ideal and implemented accumulated phases then differ by
\begin{equation}
|e^TA|\leq\|e\|_2\|A\|_2\leq\epsilon_{\rm syn}.
\end{equation}
For an $X$-basis measurement, the probabilities of outcome $+1$ in the ideal and actual experiments are $(1+\cos(g^TA))/2$ and $(1+\cos((g+e)^TA))/2$, respectively. Their total-variation distance is therefore at most $\epsilon_{\rm syn}/2$. The same conclusion holds for a $Y$-basis measurement, with cosine replaced by sine.

Thus the outcome of each actual shot disagrees with the ideal outcome with probability at most $\epsilon_{\rm syn}/2$. A union bound over the $R$ shots shows that the complete implemented measurement record differs from its ideal counterpart with probability at most $R\epsilon_{\rm syn}/2=\delta/4$. Together with the compiler-failure probability and the two statistical failure events, this gives total failure probability at most $\delta$.

Finally, from the guarantee of algorithm \ref{algo:gaussian_response_compiler}, every compiler call has interrogation duration at most
\begin{equation}
C_G\left(\Delta_\omega^{-1}+\sqrt{N\log\left(\frac{CN}{\delta_{\rm syn}}\right)}\right).
\end{equation}
Substituting $\delta_{\rm syn}=\delta/(4R)$ and multiplying by $R$ proves Equation~\eqref{eq:gaussian_spectroscopy_time_cap}. Substituting 
\begin{equation}
    R=O_L\left(\max\left\{p\left(\frac{\kappa}{\epsilon}\right)^2,(2N)^{1-1/L}p^{1/L}\left(\frac{\kappa}{\epsilon}\right)^{2/L}\right\}\right).
\end{equation}
and suppressing constants gives us Equation \ref{eq:gaussian_spectroscopy_sample_count}.
\end{proof}
\subsubsection{Proof of guarantee for high-precision RQS}

\begin{proof}
We first analyze the experiment with exactly implemented response vectors, then incorporate the error due to pulse compilation. The choice of $J$ gives $\sigma^{-1}\leq H<2\sigma^{-1}$; we will show that these choices ensure that phase unwrapping determines the correct branch of each residual phase while estimating $a_0$ to the required accuracy, so that sine inversion produces vectors with small bias and coordinate fluctuations of order $\sigma$.

We start by analyzing the phase unwrapping subroutine. First we determine the error in each iterative step of the unwrapping. Recall from Algorithm \ref{algo:phase_unwrapping} the definition $r_{\rm p}=\left\lceil512\log(32N(J+1)/\delta)\right\rceil$. Hoeffding's inequality gives
\begin{equation}
    \Pr\left[\left|\overline X_{\mu,j}-\cos(h_jA_\mu)\right|>\frac{1}{16}\right]
    \leq2e^{-r_{\rm p}/512},
\end{equation}
and the same bound holds for $\overline Y_{\mu,j}$. There are $4N(J+1)$ empirical means, so the probability that any of them has error greater than $1/16$ is at most
\begin{equation}
    8N(J+1)e^{-r_{\rm p}/512}\leq\frac{\delta}{4}.
\end{equation}
On the complementary event that error is less than $1/16$, $|\overline{X}_{\mu, j} + i\overline{Y}_{\mu, j} - e^{ih_jA_\mu}| \leq \sqrt2/16$.  It follows that, letting $ \widetilde\phi_{\mu,j} = \textnormal{Arg}(\overline{X}_{\mu, j} + i\overline{Y}_{\mu, j})$,
\begin{equation}
    \min_{k\in\mathbb Z}
    \left|
    \widetilde\phi_{\mu,j}+2\pi k-h_jA_\mu
    \right|
    \leq
    \arcsin\left(\frac{\sqrt{2}}{16}\right)
    <\frac18.
\end{equation}
With this bound at each scale, we now show inductively that the algorithm always chooses the correct $2\pi$ branch. The important point is that the estimate at each new scale is obtained from a fresh measurement whose phase error is at most $1/8$; the estimate from the preceding scale is used only to identify which $2\pi$ lift of this new measurement is the correct one.

At the first scale, $|h_0A_\mu|\leq1/4$, so there is no wraparound ambiguity and we may take $\widehat\phi_{\mu,0}=\widetilde\phi_{\mu,0}$. By the bound above,
\begin{equation}
    \left|\widehat\phi_{\mu,0}-h_0A_\mu\right|<\frac18.
\end{equation}
Now suppose that at scale $j-1$ we have already selected the correct branch, so that
\begin{equation}
    \left|\widehat\phi_{\mu,j-1}-h_{j-1}A_\mu\right|<\frac18.
\end{equation}
Since $h_j=2h_{j-1}$, doubling the previous estimate predicts the next unwrapped phase with error at most
\begin{equation}
    \left|2\widehat\phi_{\mu,j-1}-h_jA_\mu\right|
    =
    2\left|\widehat\phi_{\mu,j-1}-h_{j-1}A_\mu\right|
    <\frac14.
\end{equation}
Meanwhile, the new $X$ and $Y$ measurements determine the phase at scale $j$ modulo $2\pi$. By the phase error bound above, there is an integer $k^\star_{\mu,j}$ such that $\widetilde \phi_{\mu, j}$  satisfies
\begin{equation}
    \left|\widetilde\phi_{\mu,j}+2\pi k^\star_{\mu,j}-h_jA_\mu\right|
    <\frac18.
\end{equation}
Therefore the correct lift lies within
\begin{equation}
    \left|\widetilde\phi_{\mu,j}+2\pi k^\star_{\mu,j}-2\widehat\phi_{\mu,j-1}\right|
    <
    \frac18+\frac14
    =
    \frac38
\end{equation}
of the prediction from the previous scale. Every other choice of branch has the form $\widetilde\phi_{\mu,j}+2\pi k$ with $k\neq k^\star_{\mu,j}$, and hence differs from the correct lift by at least $2\pi$. Its distance from $2\widehat\phi_{\mu,j-1}$ is therefore at least $2\pi-3/8>3/8$. Thus the lift nearest to $2\widehat\phi_{\mu,j-1}$ is uniquely the correct one, and Algorithm~\ref{algo:phase_unwrapping} selects $k^\star_{\mu,j}$. Once the correct branch has been selected, the new estimate is $\widehat\phi_{\mu,j}=
\widetilde\phi_{\mu,j}+2\pi k^\star_{\mu,j}$. Its error is determined by the fresh measurement at scale $j$ and not the bound of $3/8$, which is relevant only to the branch selection. The induction consequently preserves the same $1/8$ phase error at every scale. At the final gain $h_J=H$, dividing the unwrapped phase estimate by $H$ gives
\begin{equation}
    \|a_0-A\|_\infty\leq\frac{1}{8H}
\end{equation}
except with probability at most $\delta/4$.

Next we analyze the residual sampling step. The purpose of this subroutine is to use the estimate $a_0$ to remove the known part of the phase $HA_\mu$, leaving only the small residual phase $H(A_\mu-(a_0)_\mu)$. We note that the measurements we make can even be done before phase unwrapping, and do not depend on having learned $a_0$; the actual residual subtraction is done entirely in classical postprocessing, leaving the quantum data-collection stage of the algorithm completely nonadaptive.

Assume that phase unwrapping has succeeded, so that $\|a_0-A\|_\infty\leq1/(8H)$. For a measurement of coordinate $A_\mu$, an $X$-basis outcome $B$ has mean $\cos(HA_\mu)$, while a $Y$-basis outcome has mean $\sin(HA_\mu)$. Since the measurement basis is chosen uniformly at random, variables in Algorithm~\ref{algo:high_gain_residual_sampling} satisfy
\begin{align}
    \E[W_{i,\mu,r}\mid a_0]
    &=
    -\cos(HA_\mu)\sin(H(a_0)_\mu)
    +
    \sin(HA_\mu)\cos(H(a_0)_\mu)\\
    &=
    \sin\left(H(A_\mu-(a_0)_\mu)\right).
\end{align}
Thus the algorithm has removed the large phase $HA_\mu$ and replaced it by a sinusoidal response to the small remaining error in $a_0$. Define $\theta_\mu=H(A_\mu-(a_0)_\mu)$ and $s_\mu=\sin\theta_\mu$. The phase unwrapping guarantee gives $|\theta_\mu|\leq1/8$, and therefore $|s_\mu|\leq1/8$. In particular, $\theta_\mu$ lies in the principal branch of the inverse sine, so $\arcsin(s_\mu)=\theta_\mu=H(A_\mu-(a_0)_\mu)$.

Given this guarantee, we now use Algorithm~\ref{algo:sine_inversion} to invert the sine. In one call, let $q$ be the number of scores in each group and $2P+1$ the number of groups. The construction gives $q\geq\sqrt m-1$ and $2P+3>\sqrt m$. Let $\overline W_r$ denote a group mean before clipping and $Q_r$ its clipped value. Since the individual scores lie in $[-2,2]$, Hoeffding's inequality gives $ \Pr\left[|\overline W_r-s_\mu|>u\mid a_0\right]\leq2e^{-qu^2/8}$. Clipping can change the group mean only when $|\overline W_r-s_\mu|\geq3/8$, and it changes its value by at most two. Consequently, writing $\nu_\mu=\E[Q_r\mid a_0]$, we obtain
\begin{equation}
    |\nu_\mu-s_\mu|
    \leq4e^{-9q/512}.
    \label{eq:high_precision_clipping_bias}
\end{equation}
Since $m\geq C\log^2(2/\epsilon_{\rm inv})$ for a sufficiently large numerical constant $C$, the right-hand side is at most $\epsilon_{\rm inv}/4$, so $|\nu_\mu|\leq1/4$.

Let $c_k=\binom{2k}{k}/[4^k(2k+1)]$ denote the coefficients of the arcsin power series. Independence of the groups gives
\begin{equation}
    \E[S_{i,\mu}\mid a_0]
    =
    \sum_{k=0}^P c_k\E\left[\prod_{r=1}^{2k+1}Q_r\middle|a_0\right]
    =
    \sum_{k=0}^P c_k\nu_\mu^{2k+1}.
\end{equation}
The products for different $k$ may share factors; only the factors within each product must be independent. The derivative of $\arcsin$ is at most two on $[-1/4,1/4]$, and $0<c_k\leq1$. Thus
\begin{align}
    \left|\E[S_{i,\mu}\mid a_0]-\arcsin(s_\mu)\right|
    &\leq
    2|\nu_\mu-s_\mu|+\sum_{k=P+1}^{\infty}|\nu_\mu|^{2k+1}\\
    &\leq
    8e^{-9q/512}+\frac{16}{15}4^{-(2P+3)}
    \leq\epsilon_{\rm inv}.
    \label{eq:high_precision_inversion_bias}
\end{align}
For the last inequality, the first term was bounded above, while $2P+3>\sqrt m\geq160\log(2/\epsilon_{\rm inv})$ makes the second term at most $\epsilon_{\rm inv}/2$. Moreover, every realized $S_{i, \mu}$ satisfies
\begin{equation}
    |S_{i,\mu}|
    \leq\sum_{k=0}^P c_k2^{-(2k+1)}
    \leq\arcsin(1/2)
    =\frac{\pi}{6}.
    \label{eq:high_precision_inversion_bound}
\end{equation}
This proves both guarantees of the sine inversion subroutine. Since $\arcsin(s_\mu)=H(A_\mu-(a_0)_\mu)$, Equation~\eqref{eq:high_precision_inversion_bias} implies
\begin{equation}
    \left|\E[(U_i)_\mu\mid a_0]-A_\mu\right|
    \leq\frac{\epsilon_{\rm inv}}H
    \leq\sigma\epsilon_{\rm inv}.
\end{equation}
Define $a_i=\E[U_i\mid a_0]$ and $\xi_i=U_i-a_i$. The choice $\epsilon_{\rm inv}\leq\sqrt{p/(2N)}$ therefore gives
\begin{equation}
    \|a_i-A\|_2
    \leq\sqrt{2N}\sigma\epsilon_{\rm inv}
    \leq\sigma\sqrt p
    \leq\frac{\epsilon}{16L\kappa}.
    \label{eq:high_precision_vector_bias}
\end{equation}
Write $b=\epsilon/(16L\kappa)$, so that $\|a_i\|_2\leq1+b$. We now bound their fluctuations, making all statements conditional on the fixed $a_0$. Equation~\eqref{eq:high_precision_inversion_bound} confines each coordinate $(U_i)_\mu$ to an interval of length $\pi/(3H)<2\sigma$. Subtracting its conditional mean preserves that interval's length and gives $|(\xi_i)_\mu|\leq2\sigma$. The coordinates are independent conditional on $a_0$, so Hoeffding's exponential-moment bound gives, for every fixed vector $v$,
\begin{equation}
    \E[e^{tv^T\xi_i}\mid a_0]
    =
    \prod_{\mu=1}^{2N}\E[e^{tv_\mu(\xi_i)_\mu}\mid a_0]
    \leq e^{t^2\sigma^2\|v\|_2^2/2},
    \qquad
    \|\xi_i\|_2\leq2\sigma\sqrt{2N}.
    \label{eq:high_precision_vector_fluctuations}
\end{equation}
Exponential Markov's inequality, optimized over $t$ and applied to both signs, implies
\begin{equation}
    \Pr\left[|v^T\xi_i|>u\mid a_0\right]
    \leq2\exp\left(-\frac{u^2}{2\sigma^2\|v\|_2^2}\right).
\end{equation}
Integrating this tail and using that $p$ is even gives
\begin{align}
    \E[|v^T\xi_i|^p\mid a_0]
    &\leq
    2p\int_0^\infty u^{p-1}\exp\left(-\frac{u^2}{2\sigma^2\|v\|_2^2}\right)du\\
    &=2(2\sigma^2\|v\|_2^2)^{p/2}(p/2)!
    \leq(2\sigma\sqrt p\|v\|_2)^p.
    \label{eq:high_precision_scalar_moment}
\end{align}

For a tensor $K$ with $\|K\|_\infty\leq\kappa$, write $\widehat F_K=K[U_1,\ldots,U_L]$. Conditional independence gives $\E[\widehat F_K\mid a_0]=K[a_1,\ldots,a_L]$. Expanding $U_i=a_i+\xi_i$ gives
\begin{equation}
    \widehat F_K-K[a_1,\ldots,a_L]
    =
    \sum_{\emptyset\neq S\subseteq\{1,\ldots,L\}}K[\xi_S,a_{\bar S}],
    \label{eq:high_precision_tensor_expansion}
\end{equation}
where the slots indexed by $S$ contain their respective errors, and the remaining slots contain their respective conditional means. Consider a term with $|S|=r$, and condition on all but one of its vectors. As a function of the remaining vector, this term is a scalar linear functional with coefficient norm at most $\kappa(1+b)^{L-r}\left(2\sigma\sqrt{2N}\right)^{r-1}$. Applying Equation~\eqref{eq:high_precision_scalar_moment} to that vector and then averaging over the others gives
\begin{equation}
    \|K[\xi_S,a_{\bar S}]\|_{L^p\mid a_0}
    \leq
    \kappa(1+b)^{L-r}2^r\sigma^r\sqrt p(2N)^{(r-1)/2},
\end{equation}
where the norm denotes $(\E[|\cdot|^p\mid a_0])^{1/p}$. Summing the terms in Equation~\eqref{eq:high_precision_tensor_expansion}, we obtain
\begin{equation}
    \|\widehat F_K-K[a_1,\ldots,a_L]\|_{L^p\mid a_0}
    \leq
    \kappa\sqrt p
    \sum_{r=1}^L\binom Lr
    (1+b)^{L-r}2^r(2N)^{(r-1)/2}\sigma^r.
    \label{eq:high_precision_tensor_moment}
\end{equation}
The definition of $\sigma$ lets us control each term. For $r=1$ and $r=1$:
\begin{equation}
    \sqrt p\,\sigma\leq\frac{\epsilon}{16L\kappa},
    \qquad
    \sqrt p(2N)^{(L-1)/2}\sigma^L
    \leq\frac{\epsilon}{(16L)^L\kappa}.
\end{equation}
For every other $L>1$, 
\begin{align}
    \sqrt p(2N)^{(r-1)/2}\sigma^r
    &=
    (\sqrt p\,\sigma)^{(L-r)/(L-1)}
    \left(\sqrt p(2N)^{(L-1)/2}\sigma^L\right)^{(r-1)/(L-1)}\\
    &\leq\frac{\epsilon}{(16L)^r\kappa}.
\end{align}
Substituting into Equation~\eqref{eq:high_precision_tensor_moment} and summing the binomial expansion yields
\begin{align}
    \|\widehat F_K-K[a_1,\ldots,a_L]\|_{L^p\mid a_0}
    &\leq
    \epsilon\left[
    \left(1+b+\frac{1}{8L}\right)^L-(1+b)^L
    \right]\\
    &\leq
    \epsilon(1+b)^L
    \left[\left(1+\frac{1}{8L}\right)^L-1\right]\\
    &\leq
    \epsilon e^{1/16}(e^{1/8}-1)
    <\frac{\epsilon}{2e},
\end{align}
where we used $Lb=\epsilon/(16\kappa)<1/16$. Applying Markov's inequality to the $p$th power of the error gives
\begin{align}
    \Pr\left[\left|\widehat F_K-K[a_1,\ldots,a_L]\right|>\frac{\epsilon}{2}\middle|a_0\right]
    &\leq
    \frac{\E[|\widehat F_K-K[a_1,\ldots,a_L]|^p\mid a_0]}{(\epsilon/2)^p}\\
    &\leq e^{-p}.
\end{align}
A union bound over the specified list therefore bounds all these fluctuations by $\epsilon/2$, except with conditional probability at most $Me^{-p}\leq\delta/4$. It remains to control the difference between the conditional mean of each observable and its actual value. By multilinearity, we can use the following telescoping identity,
\begin{equation}
    K[a_1,\ldots,a_L]-K[A^{\otimes L}]
    =
    \sum_{i=1}^L K[A^{\{i-1\}},a_i-A,a_{i+1},\ldots,a_L].
\end{equation}
Using Equation~\eqref{eq:high_precision_vector_bias}, we obtain
\begin{equation}
    |K[a_1,\ldots,a_L]-K[A^{\otimes L}]|
    \leq L\kappa b(1+b)^{L-1}
    \leq\frac{\epsilon e^{1/16}}{16}
    <\frac{\epsilon}{8}.
\end{equation}
On the good events where the phase unwrapping, residual estimation and sine inversion succeeded, every requested observable thus has error at most $\epsilon/2+\epsilon/8<\epsilon$. The bounds hold for every $a_0$ satisfying the phase-unwrapping guarantee, so averaging over $a_0$ gives failure probability at most $\delta/2$ in the ideal experiment.

Finally, we include pulse synthesis errors. Every response $v$ is implemented with response error at most $\epsilon_{\rm syn}$, and hence with phase error at most $\epsilon_{\rm syn}$ because $\|A\|_2\leq1$. For an $X$ measurement, the probability of outcome $+1$ is $(1+\cos(v^TA))/2$; for a $Y$ measurement, it is $(1+\sin(v^TA))/2$. Both probabilities change by at most $\epsilon_{\rm syn}/2$ under the synthesis error. By the same total variation argument as Proposition \ref{prop:gaussian_randomized_spectroscopy}, the probability that any of the $R$ outcomes disagrees with their value when pulse sequences are executed perfectly is at most $R\epsilon_{\rm syn}/2 = \delta/2$.

To conclude we prove the interrogation time guarantee. The phase-unwrapping stage uses $2r_{\rm p}$ shots per coordinate at every step, while the residual stage uses $Lm$ shots per coordinate at gain $H$. Algorithm~\ref{algo:deterministic_response_compiler} implements a coordinate response $he_\mu$ in time at most $C_{\rm comp}(h+\Delta_\omega^{-1})$. Since
\begin{equation}
    \sum_{j=0}^J h_j=\frac{2^{J+1}-1}{4}=2H-\frac14,
\end{equation}
the total interrogation time satisfies
\begin{equation}
    T\leq
    C_{\rm comp}\left[
    4Nr_{\rm p}\left((J+1)\Delta_\omega^{-1}+2H\right)
    +
    2NLm\left(\Delta_\omega^{-1}+H\right)
    \right].
    \label{eq:high_precision_spectroscopy_shot_cost}
\end{equation}
The parameter choices give
\begin{equation}
    J+1=O_L\left(\log\left(\frac{CNp\kappa}{\epsilon}\right)\right),
    \qquad
    r_{\rm p}=O_L\left(\log\left(\frac{CNp\kappa}{\epsilon\delta}\right)\right).
\end{equation}
Also,
\begin{equation}
    \frac{2}{\epsilon_{\rm inv}}
    =
    \max\left\{4,2\sqrt{\frac{2N}{p}}\right\},
    \qquad
    m=O(\log^2(2N)).
\end{equation}
Thus the number of residual measurements per coordinate is independent of $\epsilon$. Substituting these bounds and $H<2/\sigma$ into Equation~\eqref{eq:high_precision_spectroscopy_shot_cost} proves Equation~\eqref{eq:high_precision_spectroscopy_time_cap}. In the high-precision regime $p\geq2N(\epsilon/\kappa)^2$, the first term in the definition of $\sigma$ is the smaller one, so $\sigma^{-1}=16L\kappa\sqrt p/\epsilon$. For fixed $\Delta_\omega$, this gives the leading interrogation-time dependence $N\kappa\sqrt p/\epsilon$. Since $p$ is comparable to $\log(4M/\delta)$, suppressing the remaining log factors gives Equation~\eqref{eq:high_precision_spectroscopy_scaling}.
\end{proof}
\subsubsection{Proof of guarantee for reusable Qdyne spectroscopy}
\begin{proof}
The proof has three steps. We first show that the timestamped broadband measurements form a well-conditioned linear system for the spectral amplitudes. We then bound the bias and variance of the raw estimator $\widetilde A$. Finally, at precisions where the small nonlinear bias of $\widetilde A$ matters, we show that the classical iteration in Algorithm~\ref{algo:large_noise_rqs} removes this bias without changing the statistical error by more than a constant factor.

We begin with the Gram matrix. By definition, $b_r$ is the exact average of $f(t)$ over the $r$th sensing window. Direct integration gives, for each mode,
\begin{equation}
    (b_r)_j=\operatorname{sinc}\left(\frac{\omega_j\tau}{2}\right)\cos(\omega_jt_r),\qquad (b_r)_{N+j}=\operatorname{sinc}\left(\frac{\omega_j\tau}{2}\right)\sin(\omega_jt_r).
\end{equation}
Thus finite interrogation time only multiplies the two quadratures of each frequency by the same known sinc factor. Since $\omega_N=O(N\Delta_\omega)$ and $\tau=c/(N\Delta_\omega+\gamma)$, choosing $c$ sufficiently small keeps every factor $\operatorname{sinc}(\omega_j\tau/2)$ bounded above and below by numerical constants. The short sensing window therefore does not suppress any spectral component.

We next show that the different timestamps separate the frequencies. Temporarily writing the sine and cosine coordinates in the signed complex-frequency basis, consider two distinct frequencies $\nu,\nu'\in\{\pm\omega_1,\ldots,\pm\omega_N\}$. Since $t_r=(r-\frac12)\tau$, their normalized overlap is a geometric sum satisfying
\begin{equation}
    \left|\frac{1}{R}\sum_{r=1}^R e^{i(\nu-\nu')t_r}\right|\leq\frac{C}{R\tau|\nu-\nu'|}.
    \label{eq:qdyne_discrete_overlap}
\end{equation}
Here the choice $\omega_N\tau=O(1)$ ensures that distinct frequencies are not aliased by the sampling interval. The signed frequencies are separated by $\Theta(\Delta_\omega)$, so summing \eqref{eq:qdyne_discrete_overlap} shows that each off-diagonal row sum is at most $C\log N/(R\tau\Delta_\omega)$. The second term in the definition of $R$ makes this a sufficiently small numerical constant. Returning to the real sine--cosine basis and including the bounded sinc factors therefore gives
\begin{equation}
    c\mathds{1}\preceq G\preceq C\mathds{1},
    \label{eq:qdyne_gram_conditioning}
\end{equation}
via the same Gershgorin circle theorem argument used in several earlier bounds. As such, the summed Gram matrix is well conditioned.

We now analyze the raw estimator. The phase accumulated during shot $r$ is exactly $\tau b_r^TA$. Under depolarizing noise, the $Y$-basis measurement therefore has mean
\begin{equation}
    \E[B_r]=e^{-\gamma\tau}\sin(\tau b_r^TA).
\end{equation}
Because $\gamma\tau=O(1)$, correcting the known factor $e^{-\gamma\tau}$ costs only a numerical constant. Taking the expectation of $\widetilde A$ shows that
\begin{equation}
    \E[\widetilde A]=\Phi(A)=\frac{1}{R\tau}G^{-1}\sum_{r=1}^R b_r\sin(\tau b_r^TA).
    \label{eq:qdyne_mean_map}
\end{equation}
We first bound its second moment. For any unit vector $u$,
\begin{equation}
    u^T(\widetilde A-\Phi(A))=\frac{e^{\gamma\tau}}{R\tau}\sum_{r=1}^R(B_r-\E B_r)u^TG^{-1}b_r.
\end{equation}
The summands are independent and bounded. Moreover, using the definition of $G$,
\begin{equation}
    \frac{1}{R}\sum_{r=1}^R(u^TG^{-1}b_r)^2=u^TG^{-1}u=O(1).
\end{equation}
Applying Hoeffding's inequality therefore gives
\begin{equation}
    \Pr\left[|u^T(\widetilde A-\Phi(A))|>s\right]\leq2\exp(-cR\tau^2s^2).
    \label{eq:qdyne_directional_concentration}
\end{equation}

The remaining error in $\widetilde A$ is deterministic and comes only from the nonlinear sine response. If sine were replaced by its argument in \eqref{eq:qdyne_mean_map}, the definition of $G$ would give $\Phi(A)=A$ exactly. Using $|\sin x-x|\leq|x|^3/6$, for any unit vector $u$ we instead obtain
\begin{equation}
    |u^T(\Phi(A)-A)|\leq\frac{C\tau^2}{R}\sum_{r=1}^R|u^TG^{-1}b_r||b_r^TA|^3.
\end{equation}
The first term in each sum is $O(1)$ by \eqref{eq:qdyne_gram_conditioning}. We also have $\|b_r\|_2=O(\sqrt N)$ and $R^{-1}\sum_r(b_r^TA)^2=A^TGA=O(1)$. Cauchy--Schwarz consequently gives
\begin{equation}
    |u^T(\Phi(A)-A)|=O(\tau^2N).
    \label{eq:qdyne_raw_bias}
\end{equation}
For constant $\Delta_\omega$, $\tau^2N=O(1/N)$. Thus when $\epsilon/\kappa\geq1/\sqrt N$, this bias is already below the target accuracy, and Equation~\eqref{eq:qdyne_directional_concentration} proves the proposition directly for the returned value $\widehat A=\widetilde A$.

When higher precision is needed, we must remove the bias. Importantly, $\Phi$ is completely known from the frequencies and timestamps, so this correction is purely classical and requires no additional sensing. Differentiating \eqref{eq:qdyne_mean_map},
\begin{equation}
    D\Phi(x)=G^{-1}\frac{1}{R}\sum_{r=1}^R b_rb_r^T\cos(\tau b_r^Tx).
\end{equation}
The same estimates used above, together with $|1-\cos y|\leq y^2/2$, give for any $x$ with constant magnitude,
\begin{equation}
    \|D\Phi(x)-\mathds{1}\|=O(\tau^2N).
    \label{eq:qdyne_mean_derivative}
\end{equation}
For sufficiently large $N$ this is smaller than a fixed constant less than one. The update $x\mapsto x+\widetilde A-\Phi(x)$ is therefore a contraction. In the high-precision regime $\epsilon/\kappa<1/\sqrt N$, Equation~\eqref{eq:qdyne_directional_concentration} also gives $\|\widetilde A-\Phi(A)\|_2=O(1)$ with arbitrarily high constant probability after increasing the numerical constant in $R$, so the iteration begins and remains inside this neighborhood. It consequently converges to the unique nearby solution $\widehat A$ of $\Phi(\widehat A)=\widetilde A$.

Finally we show that the iteration removes the bias without increasing the variance. Note that $\Phi(\widehat A)-\Phi(A)=\widetilde A-\Phi(A)$, so bias is indeed removed. By the mean-value theorem and \eqref{eq:qdyne_mean_derivative}, the matrix relating $\widehat A-A$ to $\widetilde A-\Phi(A)$ is $\mathds{1}+O(\tau^2N)$. Hence for every unit vector $u$,
\begin{equation}
    |u^T(\widehat A-A)|\leq|u^T(\widetilde A-\Phi(A))|+O\left(\tau^2N\|\widetilde A-\Phi(A)\|_2\right).
    \label{eq:qdyne_corrected_error}
\end{equation}
To bound these two terms, note that for any unit vector $u$,
\begin{equation}
    \textnormal{Var}(u^T(\widetilde A - \Phi(A))) \leq \frac{C}{R^2\tau^2}\sum_{r=1}^R(u^T G^{-1}b_r)^2 = \frac{C}{R\tau^2} u^T G^{-1} u = O(1/R\tau^2) \ .
\end{equation}
The first term thus has size $O((R\tau^2)^{-1/2})$, while the second term sums over $N$ directions and is therefore $O(\tau^2 N \sqrt{N/(R\tau^2)})$. The second term is smaller by a factor $O(\tau^2N^{3/2})=O(N^{-1/2})$. Thus the corrected estimator retains the same leading variance as the raw estimator.

Taking $s=\epsilon/\kappa$ in \eqref{eq:qdyne_directional_concentration}, choosing the first term in $R$ as in Algorithm~\ref{algo:large_noise_rqs}, and union bounding over the $M$ later-supplied directions gives simultaneous error at most $\epsilon$ with probability at least $1-\delta$. Finally, all $R$ sensing windows are consecutive and have duration $\tau$, so
\begin{equation}
    T=R\tau=\widetilde O\left(\frac{1}{\tau}\left(\frac{\kappa}{\epsilon}\right)^2\right)=\widetilde O\left((N\Delta_\omega+\gamma)\left(\frac{\kappa}{\epsilon}\right)^2\right).
\end{equation}
\end{proof}

\subsubsection{Proof of the noise-aware randomized spectroscopy upper bound}

\begin{proof}
Let $U_A(\tau)$ denote unitary evolution under the noiseless signal and control Hamiltonian. When depolarizing noise is present, signal interrogation of duration $\tau$ implements the channel
\begin{equation}
    \rho(\tau)=e^{-\gamma\tau}U_A(\tau)\rho(0)U_A^\dagger(\tau)+(1-e^{-\gamma\tau})\frac{\mathds{1}}{2}.
    \label{eq:depolarizing_exact_shot}
\end{equation}
When a traceless observable like $\langle X\rangle$ or $\langle Y \rangle$ is measured, its expectation value is therefore multiplied by $e^{-\gamma\tau}$ relative to its noiseless value, independently of the pulse sequence. Using this simple observation, we analyze how minor modifications to the RQS protocol achieve the stated time complexity.

First consider $\gamma \lesssim \Delta_\omega$, where we run either Gaussian randomized spectroscopy or high-precision spectroscopy. For the former, when $\gamma\sqrt N\lesssim1$, a Gaussian response can still be compiled in time $\widetilde O(\sqrt N)$ with its outcome $\pm 1$ probabilities only modified by a constant factor, so the noiseless argument is unchanged and gives
\begin{equation}
    T=\widetilde O\left(\sqrt N\frac{\kappa^2}{\epsilon^2}\right).
\end{equation}
When $\gamma\sqrt N\gtrsim1$, we modify the protocol by reducing every sampled Gaussian response $g$ by the factor $1/(\gamma\sqrt N)$. Running the Gaussian compiler (Algorithm \ref{algo:gaussian_response_compiler}) with this modification and applying its performance guarantee gives that each compiler query uses time
\begin{equation}
    \widetilde O\left(\frac{1}{\gamma}+\frac{1}{\Delta_\omega}\right)=\widetilde O\left(\frac{1}{\gamma}\right).
\end{equation}
Writing the compiled response as $g/(\gamma\sqrt N)$, Gaussian integration gives
\begin{equation}
    \E_g\left[g\sin\left(\frac{g^TA}{\gamma\sqrt N}\right)\right]=\frac{1}{\gamma\sqrt N}\exp\left(-\frac{\|A\|_2^2}{2N\gamma^2}\right)A.
\end{equation}
The exponential factor is estimated by the first $r_0$ shots of Algorithm~\ref{algo:randomized_spectroscopy_coarse}, and the number of shots needed for this calibration stage are unchanged. However, multiplying by a factor of $\gamma\sqrt{N}$ to recover an unbiased estimate of $A$ multiplies the estimator variance by $N\gamma^2$ relative to the noiseless case. Consequently, the required number of shots increases from $\widetilde O(\kappa^2/\epsilon^2)$ to $\widetilde O(N\gamma^2\kappa^2/\epsilon^2)$. Since each shot now lasts for time $\widetilde O(1/\gamma)$, the total cost is
\begin{equation}
    T=\widetilde O\left(N\gamma\frac{\kappa^2}{\epsilon^2}\right).
\end{equation}
Combining the noiseless and noisy regimes gives
\begin{equation}
    T=\widetilde O\left((\sqrt N+N\gamma)\frac{\kappa^2}{\epsilon^2}\right).
    \label{eq:noisy_gaussian_rqs}
\end{equation}
Next consider the noisy cost of high-precision randomized spectroscopy. Let $H$ denote the final gain of Algorithm~\ref{algo:randomized_spectroscopy_fine}; suppressing logarithmic factors, $H=\Theta(\kappa/\epsilon)$. If $\gamma H\lesssim1$, all phase-unwrapping measurements have are unchanged up to constant factor reductions in contrast between $+1$ and $-1$ measurement outcome probabilities, so the cost remains $\widetilde O(NH)$.

If $\gamma H\gtrsim1$, we modify the protocol by stopping the phase amplification process at gain $O(1/\gamma)$ instead of $O(\kappa/\epsilon)$. Because $\gamma\lesssim\Delta_\omega$, compiling such a measurement still takes $O(1/\gamma)$ time using the deterministic pulse compiler (Algorithm \ref{algo:deterministic_response_compiler}). Running phase unwrapping (Algorithm~\ref{algo:phase_unwrapping}) to this gain estimates each coordinate to accuracy $O(\gamma)$. The final measurement at the same gain also has error $O(\gamma)$, so averaging $O((\gamma H)^2)$ independent measurements reduces its error to $O(1/H)$. The total cost per coordinate of $A$ is therefore
\begin{equation}
    O\left((\gamma H)^2\frac{1}{\gamma}\right)=O(\gamma H^2).
\end{equation} 
Summing over the $2N$ coordinates and including the weak-noise case gives
\begin{equation}
    T_{\rm high}=\widetilde O\left(N\left[H+\gamma H^2\right]\right)=\widetilde O\left(\frac{N\kappa}{\epsilon}\left[1+\gamma\frac{\kappa}{\epsilon}\right]\right).
    \label{eq:noisy_high_gain_rqs}
\end{equation}

For $\gamma\gtrsim\Delta_\omega$, we simply run reusable Qdyne spectroscopy, and proposition~\ref{prop:large_noise_rqs} gives
\begin{equation}
    T_{\rm Qdyne}=\widetilde O\left((N\Delta_\omega+\gamma)\frac{\kappa^2}{\epsilon^2}\right).
\end{equation}
Taking the better of \eqref{eq:noisy_gaussian_rqs} and \eqref{eq:noisy_high_gain_rqs} for $\gamma\lesssim\Delta_\omega$, and switching to reusable Qdyne spectroscopy for $\gamma\gtrsim\Delta_\omega$, proves the proposition.
\end{proof}

\subsection{Deferred proofs for quantum computational spectroscopy}\label{sec:deferred_qcs_proofs}
\subsubsection{Proof of correctness for phase oracle transduction}\label{sec:deferred_oracle_transduction}
\begin{proof}[Proof of Algorithm \ref{alg:qcs_phase_oracle} guarantee]
We prove the guarantee in three steps. First, we show that the sensing intervals define a Gram matrix whose conditioning factor depends only on $\Delta_\omega$ and the bandwidth, so all filter vectors $w_k$ are well-behaved. We then average the calls to the elementary transduction procedure over the sampled sensing times and harmonics to approximate the arcsin and recover $\lambda A^TX$, carefully tracking the accumulated approximation errors. Finally, we show that sensor depolarization acts as a mean-zero random momentum displacement with the stated covariance.

We begin with the response vector $v_\tau(t)$. Directly evaluating equation \eqref{eq:modulated_v_r} shows that each pair of entries $[v_\tau(t)]_j$ and $[v_\tau(t)]_{N+j}$ contain the components $d_j\cos(\omega_j t)$ and $d_j\sin(\omega_j t)$ respectively, with
\begin{equation}
d_j = \frac{1}{2}\left[
\operatorname{sinc}\left(\frac{(\omega_j-\omega_c)\tau}{2}\right)
+
\operatorname{sinc}\left(\frac{(\omega_j+\omega_c)\tau}{2}\right)
\right].    
\end{equation}

The choice of $J$ ensures $\tau(\omega_N-\omega_1)=O(1)$. The first sinc factor is therefore bounded below by a positive constant, while the second is bounded below uniformly for every value of $\omega_c$. Consequently $\|v_\tau(t)\|_2\leq\sqrt N$, and the Gram matrices can be rewritten as $G_k = \frac{1}{W_0}D\left(\int_{u_k}^{u_k + W_0} f(t)f(t)^T dt\right)D$, where $D = \textnormal{diag}(d_1,...d_N, d_1, ..., d_N)$.

Consider the complex Gram matrix based on the term in brackets, neglecting $D$ for now. For two distinct signed frequencies $\nu,\nu'\in \{\omega_1,..\omega_N,-\omega_1,...,-\omega_N\}$, the integral term is bounded as
\begin{equation}
\left|
\frac{1}{W_0}\int_{u_k}^{u_k+W_0}e^{i(\nu-\nu')t}dt\right|\leq
\frac{2}{W_0|\nu-\nu'|}.    
\end{equation}
Note that for any $\nu\neq \nu'$, $|\nu-\nu'| \geq \Theta(\Delta_\omega)$. Then, fixing a particular $\nu$ and summing over all $\nu'\neq \nu$
\begin{equation}
    \sum_{\nu\neq \nu'} \frac{2}{W_0|\nu-\nu'|} \leq \frac{C}{W_0\Delta_\omega} \sum_{r=1}^{2N}\frac{1}{r} = O\left(\frac{\log 2N}{W_0\Delta}\right) \ .
\end{equation}
Using $W_0 = \Theta(\log(2N)/\Delta_\omega)$, we can choose an appropriate universal constant in $W_0$ so that this sum is bounded by a constant of our choosing, say $1/4$. This is a bound on every off-diagonal row sum of the Gram matrix. Meanwhile, each diagonal entry is $1$. Gershgorin's circle theorem then tells us that every eigenvalue of the complex Gram matrix lies within magnitude $1/4$ of $1$. Returning to the real basis and using the fact that each entry of $D$ is bounded below by a constant, it follows that for any constant $C'<1$ we can choose our parameters appropriately such that $C'\mathds{1} \preceq G_k \preceq \mathds{1}$. 

For concreteness, let us take $C' = 1/64$. It follows immediately from $w_k=G_k^{-1}v_\tau$ that
\begin{equation}
    \mathbb{E}_{t_k}\left[w_kv_\tau(t_k)^T\right]=I,
\qquad
\mathbb{E}_{t_k}\left[w_kw_k^T\right]=G_k^{-1},
\qquad
\|w_k\|_2\leq64\sqrt N.
\end{equation}
Next, consider one call to Algorithm~\ref{alg:qcs_transduction}, conditioned on fixed values of the randomly sampled $t_k$ and $\ell_k$. Since $\theta_{c,I_k}=\tau A^Tv_\tau(t_k)$, the elementary transduction guarantee says that this call differs by at most $O((N+\gamma^2)\tau^2)$ in diamond norm from evolution for time $\tau$ under the generator
\begin{equation}
    \mathcal{L}_{k,t_k,\ell_k}=-\frac{i\alpha A^Tv_\tau(t_k)}{2}[\sin O_k,\cdot]+\frac{\gamma}{2}\left(\sin O_k(\cdot)\sin O_k+\cos O_k(\cdot)\cos O_k-\mathcal I\right).
\end{equation}
Because $\|A\|_2\leq1$, $\|v_\tau(t_k)\|_2\leq\sqrt N$, and $\|\sin O_k\|,\|\cos O_k\|\leq1$, its diamond norm is bounded by
\begin{equation}
    \|\mathcal{L}_{k,t_k,\ell_k}\|_\diamond\leq\alpha|A^Tv_\tau(t_k)|+\gamma\leq\sqrt N+\gamma.
\end{equation}
Therefore, whenever $(\sqrt N+\gamma)\tau\leq1$, expanding the corresponding channel to first order gives
\begin{equation}
    e^{\tau\mathcal{L}_{k,t_k,\ell_k}}=\mathcal I+\tau\mathcal{L}_{k,t_k,\ell_k}+\mathcal R_{k,t_k,\ell_k},\qquad \|\mathcal R_{k,t_k,\ell_k}\|_\diamond\leq c(N+\gamma^2)\tau^2.
\end{equation}
Averaging this equation over $t_k$ and $\ell_k$ gives
\begin{equation}
    \mathbb E_{t_k,\ell_k}\left[e^{\tau\mathcal{L}_{k,t_k,\ell_k}}\right]=\mathcal I+\tau\mathbb E_{t_k,\ell_k}\left[\mathcal{L}_{k,t_k,\ell_k}\right]+O\left((N+\gamma^2)\tau^2\right).
\end{equation}
On the other hand, expanding the channel generated by the averaged generator we actually want to implement gives
\begin{equation}
    \exp\left(\tau\mathbb E_{t_k,\ell_k}\left[\mathcal{L}_{k,t_k,\ell_k}\right]\right)=\mathcal I+\tau\mathbb E_{t_k,\ell_k}\left[\mathcal{L}_{k,t_k,\ell_k}\right]+O\left((N+\gamma^2)\tau^2\right).
\end{equation}
Thus, averaging over the random choices $t_k$ and $\ell_k$ before each call to the elementary transducer changes the $k$th step, relative to evolution under the desired generator, by at most $O((N+\gamma^2)\tau^2)$ in diamond norm. This is of the same order as the error from Algorithm~\ref{alg:qcs_transduction}, so up to the same precision scale our algorithm implements the desired evolution in expectation.

Next we control the total error over all $J$ steps. Because the choices $(t_k,\ell_k)$ are independent between different calls, the average of the complete $J$-step channel is the composition of the $J$ independently averaged channels. For example, for two consecutive calls,
\begin{align}
    \mathbb E_{t_1,\ell_1,t_2,\ell_2}\left[\mathcal T_{2,t_2,\ell_2}\circ\mathcal T_{1,t_1,\ell_1}\right]
    &=\mathbb E_{t_2,\ell_2}\left[\mathcal T_{2,t_2,\ell_2}\circ\mathbb E_{t_1,\ell_1}[\mathcal T_{1,t_1,\ell_1}]\right]\\
    &=\mathbb E_{t_2,\ell_2}[\mathcal T_{2,t_2,\ell_2}]\circ\mathbb E_{t_1,\ell_1}[\mathcal T_{1,t_1,\ell_1}],
\end{align}
and iterating this identity gives the same result for all $J$ calls. The diamond-norm error between two compositions of channels is at most the sum of the errors in the individual factors, so the total accumulated error is
\begin{equation}
O\left(J(N+\gamma^2)\tau^2\right)=O\left(\frac{(N+\gamma^2)T^2}{J}\right)=O(\eta),
\end{equation}
where we used $\tau=T/J$ and the chosen value of $J$. Finally, note that this choice of $J$ guarantees the condition $\tau \leq c/(\sqrt{N}+\gamma)$ required by Algorithm \ref{alg:qcs_transduction}. For $\eta\leq1$, $J\gtrsim1+(N+\gamma^2)T^2$, and since $(\sqrt N+\gamma)^2\leq2(N+\gamma^2)$, it follows that $(\sqrt N+\gamma)T/J=O(1)$. Increasing the fixed constant in the definition of $J$ if necessary therefore ensures $(\sqrt N+\gamma)\tau\leq1$.

We next identify the phase factor realized by the averaged channel. For fixed $t_k$, the random choice of harmonic gives
\begin{equation}
\mathbb{E}_{\ell_k}[\sin O_k] =
\frac{1}{s_0}
q_b\left(\frac{w_k^TX}{C}\right).    
\end{equation}
All of these operators are functions of the mutually commuting position operators $X_\mu$, so their phases accumulate directly. The total coherent phase is therefore
\begin{equation}  
    \Phi_A(X)
    =
    \frac{\alpha\tau}{2s_0}
    \sum_{k=1}^J
    \mathbb{E}_{t_k}\left[
    (A^Tv_\tau(t_k))
    q_b\left(\frac{w_k^TX}{C}\right)
    \right]. \label{eq:phi_A(X)}
\end{equation}
If $q_b(u)$ is replaced by $u$, 
\begin{align}
    \Phi_A(X)
    &=
    \frac{\alpha\tau}{2s_0C}
    \sum_{k=1}^J
    A^T\mathbb{E}_{t_k}\left[
    v_\tau(t_k)w_k^T
    \right]X \\
    &=
    \frac{\alpha T}{2s_0C}A^TX
    =
    \lambda A^TX.
\end{align}
It remains to control the approximation $q_b(u)\simeq u$ on the actual memory state. We first show that the initial memory state is sufficiently concentrated in position space that the argument $w_k^TX/C$ lies in the interval $[-1/2,1/2]$ with high probability, and control the approximation error on the cases where it lies outside the interval. 

For one $D$-dimensional memory register, odd moments of $X_D$ vanish in the vacuum state because $X_D$ only couples neighboring Fock levels so an odd number of raising and lowering operators never returns to $\ket{0}$. Meanwhile, a well-known fact in the continuous-variable setting is $\bra{0}X^{2r}\ket{0} = (2r-1)!!/2^r$, and upon truncating do our $D$-dimensional subspace this probability remains an upper bound on $\bra{0}X_D^{2r}\ket{0}$.  Expanding the exponential below and using $(2r-1)!!=(2r)!/(2^rr!)$ then gives
\begin{equation}
\langle0|e^{sX_D}|0\rangle\leq\sum_{r=0}^{\infty}\frac{s^{2r}}{(2r)!}\frac{(2r-1)!!}{2^r}=\sum_{r=0}^{\infty}\frac{(s^2/4)^r}{r!}=e^{s^2/4}.
\end{equation}
This expression factorizes across all memory registers since $X_D$ operators commute with each other, giving
\begin{equation}
    \left\langle\Psi_0\right|e^{s w_k^TX}\left|\Psi_0\right\rangle=\prod_{\mu=1}^{2N}\langle0|e^{s(w_k)_\mu X_D}|0\rangle\leq e^{s^2\|w_k\|_2^2/4}.
\end{equation}
Applying exponential Markov's inequality and optimizing over $s>0$ therefore yields
\begin{equation}
    \Pr_{\Psi_0}\left[|w_k^TX|\geq r\right]\leq2\exp\left(-\frac{r^2}{\|w_k\|_2^2}\right).
\end{equation}
In particular, since $\|w_k\|_2\leq64\sqrt N$,
\begin{equation}
    \Pr_{\Psi_0}\left[\left|\frac{w_k^TX}{C}\right|>\frac12\right]\leq2\exp\left(-c\frac{C^2}{N}\right)
\end{equation}
for a numerical constant $c>0$. The same tail bound also gives
\begin{equation}
    \Big\langle\Psi_0\Big|\left(1+\frac{|w_k^TX|}{C}\right)^2\mathbf{1}_{\{|w_k^TX|>C/2\}}\Big|\Psi_0\Big\rangle\leq c'\exp\left(-c\frac{C^2}{N}\right).
\end{equation}
Since $|q_b(u)|\leq\sum_{m=0}^b\binom{2m}{m}/(4^m(2m+1))\leq\pi/2$, this last inequality controls the error when $|w_k^TX/C|>1/2$. Taking $C=\Theta(\sqrt{N\log((1+T)/\eta)})$ with a sufficiently large numerical constant makes the resulting state-norm contribution at most a fixed constant times $\eta/T$.

In all other cases $|w_k^TX/C|\leq1/2$, so we can directly bound the truncation of the arcsine series. Since $\arcsin(\sin u)=u$ on this interval,
\begin{equation}
    u-q_b(u)=\sum_{m=b+1}^{\infty}\frac{\binom{2m}{m}}{4^m(2m+1)}\sin^{2m+1}(u).
\end{equation}
Using $\binom{2m}{m}\leq4^m$ and $|\sin u|\leq1/2$ gives
\begin{equation}
    |u-q_b(u)|\leq\sum_{m=b+1}^{\infty}\left(\frac12\right)^{2m+1}=\frac16\,4^{-b}\leq4^{-b}.
\end{equation}
Thus choosing $b=\Theta(\log((1+T)/\eta))$ makes the error associated with power series truncation also at most a fixed constant times $\eta/T$.

We now have a bound on $\|(q_b(w^TX/C) - w^TX/C)\ket{\Psi_0}\|_2$. Applying Cauchy-Schwarz to equation \eqref{eq:phi_A(X)},
\begin{equation}
    \left\|\left(\Phi_A(X)-\lambda A^TX\right)\ket{\Psi_0}\right\|_2\leq\frac{\alpha\tau}{2s_0}\sum_{k=1}^J\left(\mathbb E_{t_k}[(A^Tv_\tau(t_k))^2]\right)^{1/2}\left(\mathbb E_{t_k}\left[\left\|\left(q_b\left(\frac{w_k^TX}{C}\right)-\frac{w_k^TX}{C}\right)\ket{\Psi_0}\right\|_2^2\right]\right)^{1/2}=O(\eta),
\end{equation}
because each second factor is $O(\eta/T)$ and $J\tau=T$. This establishes the algorithmic guarantee in the noiseless setting.

Finally consider the effect of continuous sensor depolarization. Using $\sin O\,\rho\,\sin O+\cos O\,\rho\,\cos O=\frac{1}{2}\left(e^{iO}\rho e^{-iO}+e^{-iO}\rho e^{iO}\right)$, the memory noise generator in each call to Algorithm \ref{alg:qcs_transduction} can be written as
\begin{equation}
    \frac{\gamma}{4}\left(e^{iO_k}\rho e^{-iO_k}+e^{-iO_k}\rho e^{iO_k}-2\rho\right).
\end{equation}
This has a simple probabilistic interpretation. During an infinitesimal interval $dt$, with probability $\gamma dt/4$ the memory receives the phase kick $e^{iO_k}$, with probability $\gamma dt/4$ it receives the opposite phase kick $e^{-iO_k}$, and otherwise no kick occurs. It follows that during one call of duration $\tau$, the number of such kicks is Poisson distributed with mean $\gamma\tau/2$, and each kick independently has sign $\pm 1$ with equal probability. Since the total active sensing time is $J\tau=T$, the total number of memory kicks over the complete protocol has mean $\gamma T/2$. 

For the $k$th call, recall $O_k=\operatorname{sgn}(a_{\ell_k})\frac{\ell_k}{C}w_k^TX$. A single memory kick therefore adds the phase $\pm\operatorname{sgn}(a_{\ell_k})\ell_k w_k^TX/C$. The overall sign is symmetric, so the factor $\operatorname{sgn}(a_{\ell_k})$ can be neglected. If $\zeta$ denotes the total vector multiplying $X$ as a result of all such kicks, the effective phase oracle is consequently displaced from phase $\lambda A$ to phase $\lambda A+\zeta$. Defining $\xi=\zeta/\lambda$ and using $\lambda=\alpha T/(2s_0C)$, one kick in the $k$th call contributes
\begin{equation}
    \pm\frac{2s_0\ell_k}{\alpha T}w_k
\end{equation}
to $\xi$. Since the signs are equally likely, every jump has zero mean and hence $\mathbb E[\xi]=0$. We next compute its covariance. Conditioned on $t_k$ and $\ell_k$, the expected number of kicks during the $k$th call is $\gamma\tau/2$. Since the kicks are mean-zero and independent, their contribution to the covariance of $\xi$ is therefore
\begin{equation}
    \frac{\gamma\tau}{2}\left(\frac{2s_0\ell_k}{\alpha T}\right)^2w_kw_k^T.
\end{equation}
Averaging over the choice of $\ell$ gives
\begin{equation}
    \mathbb E_{\ell_k}[\ell_k^2]=\sum_\ell\frac{|a_\ell|}{s_0}\ell^2=\frac{c_2}{s_0},\qquad c_2=\sum_\ell|a_\ell|\ell^2.
\end{equation}
Then averaging over $t_k$ and using $\mathbb E_{t_k}[w_kw_k^T]=G_k^{-1}$, the covariance contribution of the $k$th call is
\begin{equation}
    \frac{2\gamma s_0c_2\tau}{\alpha^2T^2}G_k^{-1}.
\end{equation}
The memory kicks in the different calls are independent, so their covariance matrices add. Using $\tau=T/J$ gives
\begin{equation}
    \operatorname{Cov}(\xi)=\frac{2\gamma s_0c_2}{\alpha^2T}\frac{1}{J}\sum_{k=1}^JG_k^{-1}.
\end{equation}
We have already shown that $G_k^{-1}\preceq64I$. Moreover, the harmonic expansion of $q_b$ gives $c_2=\operatorname{poly}(b)$, while $s_0\leq\pi/2$. Since $b=\Theta(\log((1+T)/\eta))$, 
\begin{equation}
    \|\operatorname{Cov}(\xi)\|_{\rm op}=\widetilde O\left(\frac{\gamma}{\alpha^2T}\right).
\end{equation}

All coherent signal phases and all memory kicks are functions of the commuting position operators $X_\mu$. They therefore commute with one another, so for a fixed realization of $\xi$ the effective memory state is obtained simply by adding the two phase displacements. Up to the approximation errors already bounded above, the averaged memory state is consequently
\begin{equation}
    \mathbb E_{\xi}\left[e^{-i\lambda(A+\xi)^TX}\ket{\Psi_0}\!\bra{\Psi_0}e^{i\lambda(A+\xi)^TX}\right].
\end{equation}
Let $\Phi_A(X)$ denote the coherent phase actually produced by the truncated arcsine construction. Because $\Phi_A(X)$, $A^TX$, and $\xi^TX$ all commute, unitary invariance gives, for every fixed $\xi$,
\begin{align}
    &\left\|e^{-i(\lambda\xi^TX+\Phi_A(X))}\ket{\Psi_0}-e^{-i\lambda(A+\xi)^TX}\ket{\Psi_0}\right\|_2 =\left\|\left(e^{-i(\Phi_A(X)-\lambda A^TX)}-\mathds{1}\right)\ket{\Psi_0}\right\|_2 \\
    &\qquad \leq\left\|\left(\Phi_A(X)-\lambda A^TX\right)\ket{\Psi_0}\right\|_2=O(\eta),
\end{align}
where the inequality follows from $|e^{-ix}-1|\leq|x|$ applied in the common eigenbasis of the position operators. The trace distance between two pure states is at most their Euclidean distance, and trace distance is convex under averaging over $\xi$. Hence the same $O(\eta)$ bound holds for the corresponding averaged density matrices.

Every error source is thus $O(\eta)$; choosing constants appropriately proves the stated guarantee.
\end{proof}

For completeness, we note that every unitary $U_O$ has a simple circuit implementation of polynomial size. Since $O$ is always proportional to $X$, we can write $O\ket{x} = o(x)\ket{x}$. Each value of $o(x)$ is known exactly, and there are $O(ND)$ of them. There is a polynomial-complexity unitary $V:\ket{x}\ket{0}\rightarrow \ket{x}\ket{o(x)}$, since the map is basic arithmetic that can be implemented with polynomially many Toffoli gates.
Then, note
\begin{equation}
    U_O = \sum_x \ketbra{x}{x} \otimes R_y(\pi/2-o(x))
\end{equation}
If the memory state is $\ket{\Psi} = \sum_x c_x\ket{x}$, simply introduce a temporary register $\ket{0}_{\rm temp}$ and apply $V$, so $\sum_x c_x\ket{x}\ket{0}_{\rm temp} \longmapsto \sum_x c_x\ket{x}\ket{o(x)}_{\rm temp}$. Then introduce a third register in $\ket{0}$, act with $R_y(\pi/2)$, and apply controlled-$R_{-y}$ between the temporary and new registers. This creates $\sum_x c_x \ket{x}\ket{o(x)}_{\rm temp}R_y(\pi/2-o(x))\ket{0}$. Finally, uncompute $V$ and discard the temporary register to obtain $\sum_x c_x \ket{x}R_y(\pi/2-o(x))\ket{0}$. This requires $\widetilde{O}(ND)$ operations; recall that $D$ is polynomial in our parameters of interest.

\subsubsection{Proof of guarantee for reusable quantum computational spectroscopy} \label{sec:qcs_reusable_proof_guarantee}
Here we prove the algorithmic guarantee of Algorithm \ref{alg:qcs_reusable}. We need one technical lemma before stating the proof.
\begin{lemma}\label{lemma:truncation_bound}
    Let $\Pi_D = \sum_{n=0}^{D-1} \ketbra{n}{n}$ denote the projector onto the first $D$ Fock levels of an oscillator, and let $X = (a+a^\dagger)/\sqrt{2}$ denote the continuous-variable position operator. Let $X_D$ denote the truncated position operator. Then for any $|u| \leq \sqrt{D/2}$, there is a positive constant $c$ such that
    \begin{align}
       &\|e^{-iuX_D}\ket{0} - \Pi_D e^{-iuX}\ket{0}\|_2    \leq e^{-cD} \\ 
       &\|(\mathds{1} - \Pi_D )e^{-iuX}\ket{0}\|_2    \leq e^{-cD} \ .
    \end{align}
\end{lemma}
\begin{proof}
    Let $\ket{\psi_D(u)} = e^{-iuX_D}\ket{0}$ and $\ket{\phi_D(u)} = \Pi_De^{-iuX}\ket{0}$. Then
    \begin{equation}
        \frac{d}{du}\ket{\phi_D(u)} = -i\Pi_DX e^{-iuX} \ket{0} = -iX_D\ket{\phi_D(u)} - i\Pi_D X(\mathds{1}-\Pi_D)e^{-iuX}\ket 0 \ ,
    \end{equation}
    where we inserted $\Pi_D + (\mathds{1} - \Pi_D)$. Meanwhile, $\frac{d}{du}\ket{\psi_D(u)} = -iX_De^{-iuX_D}\ket{0}$. Then applying Duhamel's formula, 
    \begin{equation}
        \ket{\phi_D(u)} - \ket{\psi_D(u)} = -i \int_0^u e^{-i(u-v)X_D}\Pi_D X (\mathds 1 - \Pi_D) e^{-ivX}\ket{0} \ dv\ .
    \end{equation}
    Taking the norm and applying unitarity,
    \begin{equation}
        \| \ket{\phi_D(u)} - \ket{\psi_D(u)}\|_2 \leq \int_0^{|u|}\|\Pi_D X(\mathds 1 - \Pi_D)e^{-ivX}\ket{0}\|_2 \ dv \ .
    \end{equation}
    Since $\Pi_DX(\mathds1 - \Pi_D) = \sqrt{D/2}\ketbra{D-1}{D}$, the integrand becomes
    \begin{equation}
        \|\Pi_DX(\mathds1 - \Pi_D) e^{-ivX}\ket{0}\|_2 = \sqrt{D/2}|\bra{D}e^{-ivX}\ket{0}| \ .
    \end{equation}
    The entire expression is not an exact oscillator, so standard identities for coherent states and bounding the integrand by its supremum give us 
    \begin{equation}
        \| \ket{\phi_D(u)} - \ket{\psi_D(u)}\|_2  \leq |u|\sqrt{D/2}\sup_{|v|\leq |u|} e^{-v^2/4} \frac{(|v|/\sqrt{2})^D}{\sqrt{D!}} \ .
    \end{equation}
    Now assuming $u^2\leq D/2$ allows us to conclude that the supremum is achieved at $|v| = |u|$ because $e^{-v^2/4}v^D$ is increasing for $v^2<2D$. Moreover, $|u|/\sqrt{2}\leq \sqrt D/2$, so $\frac{(|u|/\sqrt{2})^D}{\sqrt{D!}}\leq \frac{D^{D/2}2^{-D}}{\sqrt{D!}}$. By Stirling, this is bounded by $(\sqrt{e}/2)^D$ which, because $\sqrt{e}/2<1$, is $\leq e^{-\Omega(D)}$. The first claimed inequality follows. The second follows by using the fact that $e^{-iuX}\ket{0}$ is a coherent state, for which occupation is a Poisson distribution, so a Chernoff bound for Poisson random variables yields the second inequality from the first.
\end{proof}

\begin{proof}[Proof of correctness for Algorithm \ref{alg:qcs_reusable}]
We now prove that total time $\widetilde{O}(T_{\rm batch})$ enables the algorithmic guarantee. We first analyze one batch of $L$ queries to Algorithm \ref{alg:qcs_linear_readout}, under the assumption that the phase oracle readout distribution follows the ideal oscillator model from equation \eqref{eq:ideal_oscillator_dist}; we will later bound the error incurred by this approximation. Under the ideal model, a phase-oracle interrogation of active time $t$ followed by linear readout produces a vector $U$ with mean $A$ and covariance
\begin{equation}
    \operatorname{Cov}(U)\preceq\widetilde O\left(\frac{N}{\alpha^2t^2}+\frac{\gamma}{\alpha^2t}\right)\mathds{1}_{2N}.
\end{equation}
The first term is the projection noise of the memory readout, since $h^{-2}=\widetilde O(N/(\alpha^2t^2))$, while the second is the random shift generated by sensor depolarization whose covariance was quantified in Algorithm \ref{alg:qcs_phase_oracle}. The $L$ vectors in one batch are independent, so applying the error-cancellation bound for multilinear observables (Lemma~\ref{lem:independent_nonlinear_estimation}) gives, for every $\|K\|_\infty\leq\kappa$,
\begin{equation}
    \E\left[\left|K[U_1,\ldots,U_L]-K[A^{\otimes L}]\right|^2\right]\leq C_L\kappa^2\sum_{q=1}^L(2N)^{q-1}\left(\frac{N}{\alpha^2t^2}+\frac{\gamma}{\alpha^2t}\right)^q,
\end{equation}
where logarithmic factors have been absorbed into $C_L$. We want the right hand side to be bounded by $\epsilon^2$. Writing $\sigma^2 \coloneqq \frac{N}{\alpha^2t^2}+\frac{\gamma}{\alpha^2t}$, we thus want $\sum_{q=1}^L (2N)^{q-1}\sigma^{2q} \leq \epsilon^2/\kappa^2$.

Consider first the regime $\epsilon/\kappa\leq(2N)^{-1/2}$. Here, if we make $\sigma^2 \leq c_L(\epsilon/\kappa)^2$ for an appropriate constant $c_L$, our bound is obtained. This is seen as follows. For each $q$, imposing $\sigma^2 \leq c_L(\epsilon/\kappa)^2$ makes each term in the sum at most $\kappa^2{L\choose q}(2N)^{q-1}c_L^q\left(\frac{\epsilon}{\kappa}\right)^{2q} = \left[2N\left(\frac{\epsilon}{\kappa}\right)^2\right]^{q-1}\left(\frac{\epsilon}{\kappa}\right)^2 c_L^q\kappa^2{L\choose q}$. Using $\epsilon/\kappa \leq 1/\sqrt{2N}$, we have that the sum over $q$ is bounded as $\sum_{q=1}^L {L\choose q} c_L^q \epsilon^2$. Choosing $c_L$ to be a sufficiently small constant depending only on $L$, this expression becomes $\leq \epsilon^2$.

Expanding $\sigma^2 \leq c_L(\epsilon/\kappa)^2$, we find
\begin{equation}
    t\gtrsim\max\left\{\frac{\sqrt{2N}\kappa}{\alpha\epsilon},\frac{\gamma\kappa^2}{\alpha^2\epsilon^2}\right\}
\end{equation}
so the algorithmic guarantee holds in the fine-precision regime. Now consider the standard-precision regime $\epsilon/\kappa\geq(2N)^{-1/2}$. It is instead sufficient to require
\begin{equation}
    \frac{N}{\alpha^2t^2}+\frac{\gamma}{\alpha^2t}\lesssim_L(2N)^{-(L-1)/L}\left(\frac{\epsilon}{\kappa}\right)^{2/L}.
\end{equation}
This condition forces the $q=L$ term to be of size $O_L(\epsilon^2/\kappa^2)$, while the condition $\epsilon/\kappa\geq(2N)^{-1/2}$ also forces the $q=1$ term to be $O_L(\epsilon^2/\kappa^2)$. Because all terms between $q=1$ and $q=L$ form a geometric sequence, they all obey the same bound, making the whole sum $O_L(\epsilon^2)$. Solving the inequality for $t$, separately for the two terms, gives
\begin{equation}
    t\gtrsim_L\max\left\{\frac{(2N)^{1-\frac{1}{2L}}}{\alpha}\left(\frac{\kappa}{\epsilon}\right)^{1/L},\frac{\gamma(2N)^{1-\frac1L}}{\alpha^2}\left(\frac{\kappa}{\epsilon}\right)^{2/L}\right\}.
\end{equation}
This proves that the chosen time $T_{\rm batch}$ is sufficient up to an $L$-dependent constant.

Now we account for the finite dimension of each memory register and approximation error in the phase oracle. Recall that in Algorithm \ref{alg:qcs_phase_oracle}, the applied phase is $e^{-i\lambda(A+\xi)^T X}$; applying the quarter-turn operation $R_D$ then prepares the memory state $\ket{\lambda(A+\xi)/\sqrt{2}}$. We require that the error incurred from reading out this displacement amplitude is well-approximated by equation \eqref{eq:ideal_oscillator_dist}; to this end, we must show that a typical amplitude fits within the memory dimension $D$ specified in equation \eqref{eq:qcs_combined_cutoff}. 

Algorithm \ref{alg:qcs_phase_oracle} takes $\lambda = O(\alpha T/2\sqrt{N})$ for interrogation time $t$. We set $t = T_{\rm batch}/L$, and for any index $\mu = 1,...,2N$ we have $|\lambda A_\mu|^2 \leq \Tilde{O}(\alpha^2 t^2/N)$. Next consider the noise term $\lambda\xi$, which we know has covariance $\Tilde O(\lambda^2 \gamma/(\alpha^2 T))$. Plugging in the bound on $\lambda^2$ gives
\begin{equation}
    \text{Cov}(\lambda\xi) = \Tilde{O}\left(\frac{\gamma t}{N}\right)\mathds{1}_{2N} \ .
\end{equation}
Each coordinate of $\zeta=\lambda \xi$ is a sum of random kicks whose total number is a Poisson random variable with mean $\gamma t/2$, and the magnitude of each kick is $\ell w/C$, where $\ell$ is the randomly sampled harmonic, $C$ is the chosen normalization, and $w$ is the corresponding response vector, all defined in Algorithm \ref{alg:qcs_phase_oracle}. Using the bounds in the subsequent proof, it follows that $\left|\frac{\ell w}{C}\right| \leq O(\sqrt{b})$, where $b = O(\log((1+t)/\eta))$ was the power at which we truncated the arcsin power series.  This is a bound on the magnitude of each Poisson jump. Combining this with our bound on the variance of each $\zeta_\mu$, and treating $\zeta_\mu$ as the sum of each Poisson random variable, Bernstein's inequality gives us for any $x>0$,
\begin{equation}
    \textnormal{Pr}(|\zeta_\mu| > x)\leq 2\exp\left(-c\min\left\{\frac{Nx^2}{\gamma t}, \frac{x}{\sqrt{b}}\right\}\right) \ .
\end{equation}
If we require that $\zeta_\mu$, for all $2N$ coordinates and $L$ repetitions, should fall within a given magnitude with probability at least $1-p$, this implies taking
\begin{equation}
    x = C\log\left(\frac{4NL}{p}\right)\left[\sqrt{\frac{\gamma t}{N}} + \sqrt{b}\right] \ .
\end{equation}
We take $p=0.01$ for concreteness. Adding back $\lambda A_\mu$, it follows that the coordinates of the memory displacement satisfy, for each one of the $L$ batches and with probability at least $0.99$,
\begin{equation}
    \max_{i\in \{1...L\},\  \mu\in \{1,...,2N\}} |\lambda A_\mu + \zeta_\mu^{(i)}| \leq \Tilde{O}_L\left(\lambda + \sqrt{\frac{\gamma t}{N}} + \sqrt{b} \right) \ .
\end{equation}
Plugging in our bounds on each term and equation \eqref{eq:qcs_combined_cutoff}, it follows that with probability at least $0.99$, every one of the $2NL$ coordinates satisfies
\begin{equation}
    |\lambda A_\mu + \zeta_\mu^{(i)}|^2 \leq D/2 \ .
\end{equation}
Here we obtain the factor of $1/2$ by choosing the constant in $D$ sufficiently large. For the remainder of the proof, we condition on this event. Now we show that the state of the memory register is well-approximated by an ideal oscillator state. Let $\ket{\psi_{D,\mu}^{(i)}}$ denote the actual state of the memory register $\mu$ in round $i\in \{1,...,L\}$. Then $\|\ket{\psi_{D, \mu}^{(i)}} - \Pi_D\ket{\psi_\mu^{(i)}}\| \leq e^{-cD}$; applying this to all $2NL$ memory registers and telescoping the difference gives us $\|\ket{\Psi_D} - \Pi_D^{\otimes 2NL} \ket{\Psi}\|_2 \leq 2NLe^{-cD}$, where we use $\ket{\Psi_D}$ to denote the joint state of the $2NL$ registers. 

Next we compute how this trace distance approximation error carries over to our observable estimates. For the $i^{\rm th}$ round, let $X^{(i)}$ denote the $2N$-component vector of oscillator position operators and $X_D^{(i)}$ its finite-dimensional counterpart. Recall that we normalized each $X_D$ measurement in Algorithm \ref{alg:qcs_linear_readout} by $\lambda$; as such, for notational convenience, define $\Tilde{X}_D^{(i)} = X_D^{(i)}/\lambda$ and likewise $\Tilde{X}^{(i)} = X^{(i)}/\lambda$. Then, note that measuring all $\Tilde{X}_D^{(i)}$ and contracting them against the tensor $K$ is equivalent to measuring the operator-valued contraction $K[\Tilde{X}_D^{(1)}, ..., \Tilde{X}_D^{(L)}]$. Because $X_D=\Pi_DX\Pi_D$, we can express this observable in terms of ideal oscillator operators,
\begin{equation}
   K[\Tilde{X}_D^{(1)}, ..., \Tilde{X}_D^{(L)}]=
    \Pi_D^{\otimes 2NL}
    K[\Tilde{X}^{(1)}, ..., \Tilde{X}^{(L)}]
    \Pi_D^{\otimes 2NL}.
\end{equation}
It follows that
\begin{align}
    \left[K[\Tilde{X}_D^{(1)}, ..., \Tilde{X}_D^{(L)}]- K[A^{\otimes L}]
    \right]^2\leq\Pi_D^{\otimes 2NL}
    \left[ K[\Tilde{X}^{(1)}, ..., \Tilde{X}^{(L)}] - K[A^{\otimes L}]\right]^2
    \Pi_D^{\otimes 2NL},
    \label{eq:qcs_compression_bound}
\end{align}
where the inequality follows by omitting a subtracted cross-term that is positive semidefinite. We now replace the finite-dimensional state with an ideal oscillator state while controlling the error; our previous ideal oscillator bound will then enter. Since $\|X_D\|_{\infty}\leq\sqrt{2D}$ and $\|K\|_\infty\leq\kappa$ (where we overload notation and take the infinity norm to mean the maximum contraction against Euclidean-normalized vector for both operators and tensors),
\begin{equation}
    \left\|K[\Tilde{X}_D^{(1)}, ..., \Tilde{X}_D^{(L)}]- K[A^{\otimes L}]\right\|_\infty
    \leq
    \kappa + \kappa\left( \frac{\sqrt{2ND}}{h}\right)^L.
\end{equation}
Combining this operator norm bound with $\|\ket{\Psi_D} - \Pi_D^{\otimes 2NL} \ket{\Psi}\|_2 \leq 2NLe^{-cD}$, 
\begin{align}
    \left|\bra{\Psi_D}
    \left[ K[\Tilde{X}_D^{(1)}, ..., \Tilde{X}_D^{(L)}]- K[A^{\otimes L}]
    \right]^2\ket{\Psi_D} - 
    \bra{\Psi}\Pi_D^{\otimes 2NL}\left[K[\Tilde{X}_D^{(1)}, ..., \Tilde{X}_D^{(L)}]- K[A^{\otimes L}]
    \right]^2
    \Pi_D^{\otimes 2NL}
    \ket{\Psi}
    \right| \\
    \leq
    C_L\kappa^2 NL\left( 1+\frac{ND}{h^2}\right)^Le^{-cD}.
    \label{eq:qcs_state_replacement_error}
\end{align}
We have successfully replaced the ideal operator with the truncated operator being measured. We next need to replace the projector ideal oscillator state with the true finite-dimensional state $\ket{\Psi_D}$. Using Equation~\eqref{eq:qcs_compression_bound} and writing $\Pi_D^{\otimes 2NL}\ket{\Psi} = \ket{\Psi} - (\mathds{1}- \Pi_D^{\otimes 2NL})\ket{\psi}$,
\begin{align}
     \bra{\Psi}\Pi_D^{\otimes 2NL}&\left[K[\Tilde{X}_D^{(1)}, ..., \Tilde{X}_D^{(L)}]- K[A^{\otimes L}]
    \right]^2\Pi_D^{\otimes 2NL}
    \ket{\Psi} \leq\left\|\left[K[\Tilde{X}^{(1)}, ..., \Tilde{X}^{(L)}]-K[A^{\otimes L}]
    \right]\Pi_D^{\otimes 2NL}\ket{\Psi}\right\|_2^2 \\
    &\leq
    2\left(\left\|\left[K[\Tilde{X}^{(1)}, ..., \Tilde{X}^{(L)}]
    -K[A^{\otimes L}]\right]\ket{\Psi}\right\|_2^2 +
    \left\|\left[K[\Tilde{X}^{(1)}, ..., \Tilde{X}^{(L)}]
    -
    K[A^{\otimes L}]
    \right]
    \left(\mathds{1}-\Pi_D^{\otimes 2NL}\right)
    \ket{\Psi}
    \right\|_2^2\right).
    \label{eq:qcs_projected_ideal_error}
\end{align}
The first term was shown above to be bounded by $O(\epsilon^2)$. We now show that the second is negligible. Let $\hat n_\mu^{(i)}$ be the occupation operator of mode $\mu$ in copy $i$. Since $X^2+P^2=2\hat n+\mathds{1}$, we have $X^2\leq2\hat n+\mathds{1}$. Together with $\|K\|_\infty \leq \kappa$, we have the operator inequality
\begin{equation}
   K[\Tilde{X}^{(1)}, ..., \Tilde{X}^{(L)}]^2
    \preceq \frac{\kappa^2}{h^{2L}} \prod_{i=1}^L\left(
    \sum_{\mu=1}^{2N}\hat n_\mu^{(i)}+N
    \right).
    \label{eq:qcs_number_operator_bound}
\end{equation}
Here the factors belonging to each of the $L$ different copies commute, so the inequalities may be multiplied. Now recall that we conditioned on $    |\lambda A_\mu + \zeta_\mu^{(i)}|^2 \leq D/2$, so every oscillator mode is a coherent state whose mean occupation is at most $D/4$. Thus if $n$ denotes the occupation number of any one mode, then $n$ is Poisson with mean at most $D/4$. For every integer $r$,
\begin{align}
    \E\left[(n+1)^r\mathds{1}_{n\geq D}\right]
    &\leq
    2^{-D}\E\left[(n+1)^r2^n\right]
    \nonumber\\
    &=
    2^{-D}e^\nu
    \E_{m\sim{\rm Pois}(2\nu)}\left[(m+1)^r\right]
    \nonumber\\
    &\leq
    C_r(D+1)^r e^{-cD},
    \label{eq:qcs_weighted_poisson_tail}
\end{align}
where $\nu\leq D/4$. In the second line we used the elementary Poisson identity obtained by multiplying its probability mass function by $2^n$. The event selected by $\mathds{1}-\Pi_D^{\otimes 2NL}$ is that at least one of the $2NL$ occupations is at least $D$. Applying a union bound and Equation~\eqref{eq:qcs_weighted_poisson_tail} to Equation~\eqref{eq:qcs_number_operator_bound} therefore gives
\begin{align}
    \left\|K[\Tilde{X}^{(1)}, ..., \Tilde{X}^{(L)}]
    \left(\mathds{1}-\Pi_D^{\otimes 2NL}\right)
    \ket{\Psi}
    \right\|_2^2\leq C_L\kappa^2\frac{N^{L+1}(D+1)^L}{h^{2L}}e^{-cD}.
    \label{eq:qcs_polynomial_tail}
\end{align}
The contribution from the scalar $K[A^{\otimes L}]$ is smaller, since $|K[A^{\otimes L}]|\leq\kappa$ and $\left\| \left(\mathds{1}-\Pi_D^{\otimes 2NL}\right)\ket{\Psi}\right\|_2^2\leq 2NL e^{-cD}$. Hence
\begin{align}
   \left\|\left[K[\Tilde{X}^{(1)}, ..., \Tilde{X}^{(L)}]
    -K[A^{\otimes L}]
    \right]
    \left(\mathds{1}-\Pi_D^{\otimes 2NL}\right)
    \ket{\Psi}
    \right\|_2^2 \leq C_L\kappa^2\left[1+\frac{N^{L+1}(D+1)^L}{h^{2L}}\right]e^{-cD}.
    \label{eq:qcs_full_tail_error}
\end{align}

Both Equations~\eqref{eq:qcs_state_replacement_error} and \eqref{eq:qcs_full_tail_error} are a polynomial prefactor times an exponential in $D$, recalling that $h^{-2}=\widetilde O\left(\frac{N}{\alpha^2t^2}\right)$. Choosing the logarithmic factor in $D$ to be appropriately large, we can make this error at most $\epsilon^2/100$ (where the factor of $100$ is for concreteness below). This only contributes factors doubly logarithmic in $N, 1/\epsilon$ to the number of register qubits. 

Finally, we can count the total estimation error. Choose the numerical constant in $T_{\rm batch}$ so that the ideal-oscillator calculation above gives
\begin{equation}
    \E\left[
    \left|
    K[U_1,\ldots,U_L]
    -
    K[A^{\otimes L}]
    \right|^2
    \right]
    \leq
    \frac{2\epsilon^2}{25}.
\end{equation}
Combining Equations~\eqref{eq:qcs_state_replacement_error} and~\eqref{eq:qcs_projected_ideal_error}, and conditioning on the displacement event of probability at least $0.99$, gives a finite-register mean-squared error bounded by $[\epsilon^2/100+2(2\epsilon^2/25+\epsilon^2/100)]/0.99=19\epsilon^2/99<\epsilon^2/5$. Markov's inequality therefore gives
\begin{equation}
    \Pr\left[
    \left|
    Z_j^{(r)}
    -
    K_j[A^{\otimes L}]
    \right|>\epsilon
    \right]
    \leq
    \frac1{5}.
\end{equation}
The cutoff event itself fails with probability at most $0.01$. Moreover, each batch calls Algorithm \ref{alg:qcs_phase_oracle} $L$ times with precision parameter $\eta$, so the resulting trace-distance error from an ideal phase oracle call is at most $L\eta$ per batch. Converting to total variation means that the probability of the event $\left|Z_j^{(r)}- K_j[A^{\otimes L}]\right|>\epsilon$ changes by at most $L\eta$. Choosing $\eta = 9/(100L)$ means that with probability at least $7/10$, each of the $R$ batches successfully estimates each observable. 

Finally, the $R$ composite batches are independent. For any fixed $j$, the median can fail only if at least half of the $R$ batches fail for $K_j$. A Chernoff bound therefore gives
\begin{equation}
    \Pr\left[
    \left|
    \widehat F_j-K_j[A^{\otimes L}]
    \right|>\epsilon
    \right]
    \leq
    e^{-cR}.
\end{equation}
Taking $R=C_L\log(2M/\delta)$ with a sufficiently large constant and union bounding over $j=1,\ldots,M$ proves the required simultaneous success probability, since each round succeeds with probability at least $7/10 > 1/2$. The outer loop only runs a logarithmic number of times, so the interrogation time guarantee follows.
\end{proof}
\subsubsection{Proof of guarantees for matched linear QCS}
\label{sec:deferred_matched_qcs}

Here we prove the guarantee of Algorithm~\ref{alg:qcs_matched_oracle}, which follows its counterpart for standard QCS, and we only detail the steps that change.

\begin{proof}[Proof of the guarantee of Algorithm~\ref{alg:qcs_matched_oracle}]
The properties~\eqref{eq:qcs_matched_filter_properties} of the normalized filter replace the corresponding properties of $w_k$ in Section~\ref{sec:deferred_oracle_transduction}. We first verify them, and then show that the linearization of the arcsine series and the computation of the noise covariance follow.

Since $\mathds{1}/64\preceq G_k\preceq\mathds{1}$, we have $\E_k[v_\tau q_k^T]=U$ and $\E_k[q_kq_k^T]=U^TG_k^{-1}U\preceq64\,\mathds{1}_r$. The weight satisfies $\beta_k\geq\sqrt r$ pointwise, and Jensen's inequality gives $\E_k\beta_k\leq(r+\E_k\|q_k\|_2^2)^{1/2}\leq\sqrt{65r}$, so $\sqrt r\leq\E_k\beta_k\leq\sqrt{65r}$. The first property follows from $\|q_k\|_2\leq\beta_k$, and the second from the cancellation of $\beta_k$ between the sampling density and $\widetilde q_k$. For the third property, one factor of $\beta_k\geq\sqrt r$ remains in the denominator, so
\begin{equation}
    \widetilde{\mathbb E}_k[\widetilde q_k\widetilde q_k^T]=\frac{\E_k\beta_k}{65r}\E_k\left[\frac{q_kq_k^T}{\beta_k}\right]\preceq\frac{64\,\E_k\beta_k}{65r\sqrt r}\mathds{1}_r\preceq\frac{8}{r}\mathds{1}_r.
\end{equation}

As in Equation~\eqref{eq:phi_A(X)}, the coherent phase is 
\begin{equation}
\Phi(X)=\frac{\alpha\tau}{2s_0}\sum_k\widetilde{\mathbb E}_k[(A^Tv_\tau(t_k))\,q_b(\widetilde q_k(t_k)^TX/C_0)],    
\end{equation}
and replacing $q_b(u)$ by $u$ gives exactly $\lambda(U^TA)^TX$ with $\lambda=\alpha T/(2s_0\sqrt{65r}\,C_0)$, by the second property. Since $\|\widetilde q_k\|_2\leq1$, the vacuum tail bound of Section~\ref{sec:deferred_oracle_transduction} for $\widetilde q_k(t)^TX$ is independent of $N$ and $r$. The truncation and tail estimates there therefore give 
\begin{equation}
\|(q_b(\widetilde q_k(t)^TX/C_0)-\widetilde q_k(t)^TX/C_0)\ket{\Psi_0}\|_2\leq c'(4^{-b}+e^{-cC_0^2})    
\end{equation}
for every $t$. Using $|A^Tv_\tau(t)|\leq\sqrt N$ and summing over the $J$ calls, the error of the coherent phase on the vacuum is at most $\frac{\alpha T}{2s_0}\sqrt N\,c'(4^{-b}+e^{-cC_0^2})$. This is $O(\eta)$ for $b$ and $C_0^2$ of order $\log(N(1+T)/\eta)$, which is the only place where $N$ enters $C_0$.

Finally, each noise kick $e^{\pm iO_k}$ displaces the memory momenta by $\pm\ell_k\widetilde q_k(t_k)/C_0$, contributing a symmetric jump 
\begin{equation}
\pm\ell_k\widetilde q_k(t_k)/(C_0\lambda)=\pm2s_0\sqrt{65r}\,\ell_k\widetilde q_k(t_k)/(\alpha T)    
\end{equation}
to $\xi$ at rate $\gamma/2$. The covariance of a compound Poisson sum is its rate times the second moment of one jump, so the third property gives
\begin{equation}
    \operatorname{Cov}(\xi)=\sum_{k=1}^J\frac{\gamma\tau}{2}\,\frac{4\cdot65\,r\,s_0^2}{\alpha^2T^2}\,\E[\ell_k^2]\,\widetilde{\mathbb E}_k[\widetilde q_k\widetilde q_k^T]\preceq\frac{16\cdot65\,\gamma s_0c_2}{\alpha^2T}\mathds{1}_r,
\end{equation}
where $c_2=\sum_\ell|a_\ell|\ell^2=\operatorname{poly}(b)$. Every jump has norm at most $2s_0\sqrt{65r}(2b+1)/(\alpha T)=\widetilde O(\sqrt r/(\alpha T))$. The remaining steps are identical to those for Algorithm~\ref{alg:qcs_phase_oracle}.
\end{proof}

\begin{proof}[Proof of Proposition~\ref{prop:qcs_target_linear}]
In the ideal oscillator model, linear readout returns $V=U^TA+\xi+G$, with $G$ Gaussian of covariance $\mathds{1}_r/(2\lambda^2)$, so $(U^TK_j)^TV$ is unbiased with variance at most $\kappa^2\widetilde O(r/(\alpha^2T^2)+\gamma/(\alpha^2T))$. The stated choice of $T$ makes this at most $2\epsilon^2/25$. The finite registers, the phase-oracle error, and the median over repetitions are handled as in Section~\ref{sec:qcs_reusable_proof_guarantee} with $2N$ replaced by $r$. For the cutoff, each coordinate of $\lambda\xi$ is a sum of jumps of size $\widetilde O(1)$ with variance $\widetilde O(\gamma T/r)$. Together with $\lambda^2=\widetilde\Theta(\alpha^2T^2/r)$ and $\alpha^2T\gtrsim\gamma$, this gives $D=\widetilde O(1+\alpha^2T^2/r)$.
\end{proof}

\noindent Next we establish the correctness of nonlinear matched QCS.

\subsubsection{Proof of guarantee for nonlinear matched quantum computational spectroscopy} \label{sec:deferred_qcs_target_specific}
\begin{proof}[Proof of correctness for Algorithm~\ref{alg:qcs_coherent}]
We give the performance guarantee for $U=\mathds{1}_{2N}$, i.e. transduction without matching to any $r$-dimensional subspace. Obtaining the guarantee for $U \in \mathbb R^{2N\times r}$ then follows directly. 

Without a matched subspace projector, the proof differs from that of Algorithm~\ref{alg:qcs_reusable} only in the ideal-oscillator mean-squared error and in the final implementation of the nonlinear measurement. We first prove the coherent variance bound and show that Equation~\eqref{eq:qcs_coherent_batch_time} makes it $O(\epsilon^2)$. The finite-dimensional cutoff for the memory, approximation error in the phase-oracle, and joint success probability fro taking the median can then be inherited from the previous proof.

Consider one phase-oracle call of duration $t=T_{\rm batch}/\lceil L/2\rceil$. In the ideal oscillator model, after applying $R_D$ the memory is a mixture of coherent states $\ket{h(A+\xi)}$, where
$$h^{-2}=\widetilde O\left(\frac{N}{\alpha^2t^2}\right),\qquad \|\Sigma^{(j)}\|_{\rm op}=\widetilde O\left(\frac{\gamma}{\alpha^2t}\right).$$
Here $\Sigma^{(j)}=\E[\xi^{(j)}(\xi^{(j)})^T]$. The displacement $\xi^{(j)}$ is generated by a Poisson number of independent memory kicks induced by sensor depolarization, with mean $\gamma t/2$. Let $v$ denote the displacement of the memory induced by each jump. The sign of $v$ is uniformly random, so $\E[\xi^{(j)}]=0$ and all odd moments of $\xi^{(j)}$ vanish.

For a sum of jumps of $\pm v$, whose total number is distributed according to a Poisson distribution of mean $\gamma t/2$, the covariance and the non-Gaussian part of the fourth moment are:
$$\Sigma^{(j)}=\frac{\gamma t}{2}\E_v[vv^T],\qquad \text{fourth cumulant}=\frac{\gamma t}{2}\E_v[v^{\otimes4}].$$
The phase-oracle construction gives $\|v\|_2^2=\widetilde O(N/(\alpha^2t^2))$. Thus, for every matrix $B$,
\begin{align}
    \frac{\gamma t}{2}\E_v[|v^TBv|^2] &\leq \left(\max_v\|v\|_2^2\right)\frac{\gamma t}{2}\E_v[\|Bv\|_2^2] = \left(\max_v\|v\|_2^2\right)\operatorname{tr}(B^\dagger B\Sigma^{(j)}) \nonumber\\
    &\leq \widetilde O\left(\frac{\gamma N}{\alpha^4t^3}\right)\|B\|_F^2.
    \label{eq:qcs_coherent_fourth_cumulant}
\end{align}
The fourth cumulant is no larger than the covariance multiplied by the squared size of one memory kick.

We now bound the fluctuations of one normally ordered pair. Temporarily replace $a_D$ by the full oscillator ladder operator $a$. Using $a_\mu a_\rho^\dagger=a_\rho^\dagger a_\mu+\delta_{\mu\rho}$ and $a_\mu\ket{h(A+\xi)}=h(A_\mu+\xi_\mu)\ket{h(A+\xi)}$ gives
\begin{equation}
    \E\left[(Q^{(j)}_{\mu\nu}-A_\mu A_\nu)^\dagger(Q^{(j)}_{\rho\sigma}-A_\rho A_\sigma)\right]=\operatorname{Cov}\left((A_\mu+\xi_\mu^{(j)})(A_\nu+\xi_\nu^{(j)}),(A_\rho+\xi_\rho^{(j)})(A_\sigma+\xi_\sigma^{(j)})\right)+\frac{\delta_{\mu\rho}}{h^2}\left(A_\nu A_\sigma+\Sigma_{\nu\sigma}^{(j)}\right).
    \label{eq:qcs_coherent_pair_covariance}
\end{equation}
Viewed as a positive covariance matrix on the ordered pair of indices $(\mu,\nu)$, we then have the operator inequality
\begin{equation}
    \E\left[(Q^{(j)}-AA^T)^\dagger\otimes(Q^{(j)}-AA^T)\right]\preceq\widetilde O\left[\left(\frac{\gamma}{\alpha^2t}+h^{-2}\right)\left(\mathds{1}\otimes AA^T+AA^T\otimes\mathds{1}\right)+\left(\frac{\gamma^2}{\alpha^4t^2}+\frac{\gamma h^{-2}}{\alpha^2t}+\frac{\gamma N}{\alpha^4t^3}\right)\mathds{1}\right].
    \label{eq:qcs_coherent_pair_bound}
\end{equation}
The first term contracts one of the two tensor indices with $A$ and leaves the other free. It then consists of four terms of the form $A_\mu A_\rho \Sigma_{\mu\rho}$; using the bound on the covariance matrix yields the $\gamma/(\alpha^2 t)$ term. It also consists of the term $\delta_{\mu\rho}A_\mu A_\rho/h^2$, which contributes the $h^{-2}$ term. Then the second term leaves both tensor indices free, so it consists of terms bilinear in $\Sigma$, proportional to $\Sigma/h^2$, and the fourth cumulant of $\xi^{(j)}$. These produce the $\left(\frac{\gamma^2}{\alpha^4t^2}+\frac{\gamma h^{-2}}{\alpha^2t}+\frac{\gamma N}{\alpha^4t^3}\right)$ factor. 

Substituting $h^{-2}=\widetilde O(N/(\alpha^2t^2))$ into Equation~\eqref{eq:qcs_coherent_pair_bound}, the contribution of each memory copy to the final variance is consequently bounded by
\begin{equation}
    1+\widetilde O_L\left(\frac{\gamma(2N)}{\alpha^2t}+\frac{(2N)^2}{\alpha^2t^2}+\frac{\gamma^2(2N)^2}{\alpha^4t^2}+\frac{\gamma(2N)^3}{\alpha^4t^3}\right) \ .
\end{equation}
This follows because the variance of one copy is given by equation \eqref{eq:qcs_coherent_pair_bound} multiplied by a factor of $2N$, for the dimension of the contracted vector. This is at most
\begin{equation}
    \left(1+C_L\frac{\gamma(2N)}{\alpha^2t}\right)\left(1+C_L\frac{\gamma(2N)}{\alpha^2t}+C_L\frac{(2N)^2}{\alpha^2t^2}\right).
    \label{eq:qcs_coherent_pair_factor}
\end{equation}
If $L$ is odd, the leftover linear observable satisfies
$$\E[(W_\mu-A_\mu)(W_\nu-A_\nu)]=\Sigma_{\mu\nu}^{(j)}+\frac{\delta_{\mu\nu}}{4h^2},$$
so its contribution to the variance is bounded by the second factor in Equation~\eqref{eq:qcs_coherent_pair_factor}.

There are $\lfloor L/2\rfloor$ such contributions which multiply, because each memory state is an independent phase oracle acquisition. Factoring out $\kappa^2/(2N)$ gives
\begin{equation}
    \operatorname{Var}(O_K)\leq\frac{\widetilde O_L(\kappa^2)}{2N}\left[\left(1+\frac{\gamma(2N)}{\alpha^2t}\right)^{\lfloor L/2\rfloor}\left(1+\frac{\gamma(2N)}{\alpha^2t}+\frac{(2N)^2}{\alpha^2t^2}\right)^{\lceil L/2\rceil}-1\right].
    \label{eq:qcs_coherent_variance}
\end{equation}
Importantly, by using normally-ordered observables we only incurred a single term $A_\nu A_\rho$, which allows our exponent to be $\lceil L/2\rceil$ rather than $L$ and enables the improvement over the reusable protocol. Now we can show that the choice of interrogation time allows the variance to be bounded as $O(\epsilon^2)$.

When $\epsilon/\kappa\leq(2N)^{-1/2}$, it is sufficient to make both $\gamma(2N)/(\alpha^2t)$ and $(2N)^2/(\alpha^2t^2)$ at most a sufficiently small fixed-$L$ multiple of $(2N)(\epsilon/\kappa)^2$. Equation~\eqref{eq:qcs_coherent_variance} then gives $\operatorname{Var}(O_K)=O_L(\epsilon^2)$, which requires
$$t\gtrsim_L\max\left\{\frac{\sqrt{N}\kappa}{\alpha\epsilon},\frac{\gamma\kappa^2}{\alpha^2\epsilon^2}\right\}.$$

Now suppose $\epsilon/\kappa\geq(2N)^{-1/2}$ and $\gamma\lesssim\alpha$. From Equation~\eqref{eq:qcs_coherent_variance}, it is sufficient to require
\begin{equation}
    \left(1+\frac{\gamma(2N)}{\alpha^2t}\right)^{\lfloor L/2\rfloor}\left(1+\frac{\gamma(2N)}{\alpha^2t}+\frac{(2N)^2}{\alpha^2t^2}\right)^{\lceil L/2\rceil}-1\lesssim_L(2N)\left(\frac{\epsilon}{\kappa}\right)^2.
    \label{eq:qcs_coherent_coarse_requirement}
\end{equation}
For nonnegative $x,y$,
$$(1+x)^{\lfloor L/2\rfloor}(1+x+y)^{\lceil L/2\rceil}-1\leq C_L\left(x+y+x^L+y^{\lceil L/2\rceil}+x^{\lfloor L/2\rfloor}y^{\lceil L/2\rceil}\right).$$
This follows because if $x,y\leq1$, every nonconstant monomial is bounded by $x+y$; if $x\geq\max\{1,y\}$ the polynomial is $O_L(x^L)$; and if $y\geq\max\{1,x\}$ it is $O_L(y^{\lceil L/2\rceil}+x^{\lfloor L/2\rfloor}y^{\lceil L/2\rceil})$.

Since $(2N)(\epsilon/\kappa)^2\geq1$, controlling the three highest-degree terms also controls the two linear terms. Equation~\eqref{eq:qcs_coherent_coarse_requirement} therefore follows if
\begin{equation}
    \left(\frac{(2N)^2}{\alpha^2t^2}\right)^{\lceil L/2\rceil},\qquad \left(\frac{\gamma(2N)}{\alpha^2t}\right)^L,\qquad \left(\frac{\gamma(2N)}{\alpha^2t}\right)^{\lfloor L/2\rfloor}\left(\frac{(2N)^2}{\alpha^2t^2}\right)^{\lceil L/2\rceil}
    \label{eq:qcs_coherent_coarse_terms}
\end{equation}
are each at most a fixed-$L$ multiple of $(2N)(\epsilon/\kappa)^2$. Solving these three requirements gives, respectively,
\begin{align}
    t&\gtrsim_L\frac{2N}{\alpha}\left(\sqrt{2N}\frac{\epsilon}{\kappa}\right)^{-1/\lceil L/2\rceil},\qquad t\gtrsim_L\frac{2N}{\alpha}\frac{\gamma}{\alpha}\left(\sqrt{2N}\frac{\epsilon}{\kappa}\right)^{-2/L}, \nonumber\\
    t&\gtrsim_L\frac{2N}{\alpha}\left(\frac{\gamma}{\alpha}\right)^{\frac{\lfloor L/2\rfloor}{\lfloor L/2\rfloor+2\lceil L/2\rceil}}\left(\sqrt{2N}\frac{\epsilon}{\kappa}\right)^{-\frac{2}{\lfloor L/2\rfloor+2\lceil L/2\rceil}}.
    \label{eq:qcs_coherent_exact_coarse_time}
\end{align}
We now bound these three requirements by the simpler expression used in the algorithm. Because $\sqrt{2N}\epsilon/\kappa\geq1$,
$$\left(\sqrt{2N}\frac{\epsilon}{\kappa}\right)^{-1/\lceil L/2\rceil}\leq\left(\sqrt{2N}\frac{\epsilon}{\kappa}\right)^{-2/(L+1)}.$$
Because also $\gamma/\alpha\lesssim1$,
$$\frac{\gamma}{\alpha}\left(\sqrt{2N}\frac{\epsilon}{\kappa}\right)^{-2/L}\leq\left(\frac{\gamma}{\alpha}\right)^{1/3}\left(\sqrt{2N}\frac{\epsilon}{\kappa}\right)^{-4/[3(L+1)]}.$$
If $L$ is even, the mixed requirement in Equation~\eqref{eq:qcs_coherent_exact_coarse_time} is $(\gamma/\alpha)^{1/3}(\sqrt{2N}\epsilon/\kappa)^{-4/(3L)}$, and is bounded by the same expression. If $L$ is odd, it is the weighted geometric mean
\begin{equation}
    \left[\left(\sqrt{2N}\frac{\epsilon}{\kappa}\right)^{-2/(L+1)}\right]^{\frac{4}{3L+1}}\left[\left(\frac{\gamma}{\alpha}\right)^{1/3}\left(\sqrt{2N}\frac{\epsilon}{\kappa}\right)^{-4/[3(L+1)]}\right]^{\frac{3(L-1)}{3L+1}},
\end{equation}
and is therefore bounded by the larger of the same two quantities. Consequently it is sufficient to take
\begin{equation}
    t=\widetilde O_L\left[\max\left\{\frac{N^{L/(L+1)}}{\alpha}\left(\frac{\kappa}{\epsilon}\right)^{2/(L+1)},\frac{\gamma^{1/3}N^{1-\frac{2}{3(L+1)}}}{\alpha^{4/3}}\left(\frac{\kappa}{\epsilon}\right)^{\frac{4}{3(L+1)}}\right\}\right],
\end{equation}
where factors of two are absorbed into the fixed constants.

From this point through the conversion to a constant-success finite-register batch, the proof proceeds identically to the proof of Algorithm~\ref{alg:qcs_reusable} upon choosing the constants in $T_{\rm batch}$ sufficiently large. Apply that error budget with accuracy $\epsilon/2$, take $\eta_{\rm out}=\epsilon/10$, and make the total readout failure probability, including simulation error, at most $1/100$. The reported outcome is then $\epsilon$-accurate with probability at least $7/10-1/100>1/2$, so the same median argument applies. Since we rerun the inner loop for each of the $M$ observables, the total interrogation time is $\widetilde O_L(MT_{\rm batch}^{\rm coh})$.

For matched nonlinear observables with subspace dimension $r < 2N$, let $U\in\mathbb R^{2N\times r}$ span a common support of the specified tensors, and replace $A$ by $U^TA$ and $K$ by $\widetilde K=K[U,\ldots,U]$. Algorithm~\ref{alg:qcs_matched_oracle} produces the same effective memory state and displacement-noise law, with $h^{-2}=\widetilde O(r/(\alpha^2t^2))$ and $\|v\|_2^2=\widetilde O(r/(\alpha^2t^2))$. Since $\|U^TA\|_2\leq1$ and $\|\widetilde K\|_\infty=\|K\|_\infty$, the entire variance analysis and finite-register argument above carry over with every dimension-dependent factor $2N$ replaced by $r$, proving the matched nonlinear guarantee.

\end{proof}

\subsubsection{Proof of guarantees for QCS with qubit phases}
\label{sec:deferred_qubit_qcs}

We first record the noise properties of the qubit phase oracle and fix the parameters of Algorithm~\ref{alg:qcs_qubit_reusable_readout}, and then prove Proposition~\ref{prop:qcs_qubit_memory_noisy}. We write the proofs for the full spectrum; for the matched version, replace $2N$ by $r$ and $A$ by $U^TA$.

As discussed in Section~\ref{sec:qcs_qubit_memory}, the proof of Algorithm~\ref{alg:qcs_phase_oracle} applies with $X$ replaced by $Z_M$ and $\ket{\Psi_0}$ by $\ket{+}^{\otimes2N}$. The moment generating function bound differs from the vacuum bound only by a factor of two in the exponent, which is absorbed into $C$. Consider an acquisition of duration $t$, which has gain $2\lambda=\alpha t/(s_0C)$. Each noise kick $e^{\pm iO_k}$ is a product of single-qubit rotations $e^{\pm i\operatorname{sgn}(a_{\ell_k})\ell_k(w_k)_\mu Z_\mu/C}$, which shift the relative phase of qubit $\mu$ by $\mp2\operatorname{sgn}(a_{\ell_k})\ell_k(w_k)_\mu/C$. The phase noise $2\lambda\xi$ is therefore a sum of independent symmetric jumps, which occur at rate $\gamma/2$ during call $k$ and are distributed as $\pm2\ell_kw_k(t_k)/C$. Write $\Delta_k$ for an independent copy of one such jump, so that $4\lambda^2\operatorname{Cov}(\xi)=\sum_k\frac{\gamma\tau}{2}\E[\Delta_k\Delta_k^T]$, and let $\Delta_{\max}=\widetilde O(1)$ bound the norm of every jump. Then $4\lambda^2\|\operatorname{Cov}(\xi)\|_{\rm op}=\widetilde O(\gamma t/N)$. The noise reduces the visibility of qubit $\mu$ to the known value
\begin{equation}
    \chi_\mu=\E\left[e^{2i\lambda\xi_\mu}\right]=\exp\left(\sum_{k=1}^J\frac{\gamma\tau}{2}\E\left[\cos\Delta_{k,\mu}-1\right]\right)\geq\exp\left(-2\lambda^2\operatorname{Var}(\xi_\mu)\right),
\end{equation}
which is real because the jumps are symmetric. Throughout, we restrict every acquisition to durations satisfying $(4+\Delta_{\max}^2)\,4\lambda^2\|\operatorname{Cov}(\xi)\|_{\rm op}\leq1$. This permits $t=\widetilde\Theta(N/\gamma)$ and guarantees $\chi_\mu\geq e^{-1/8}>3/4$. Let $\sigma$ be the target standard deviation of each output vector, chosen as in Section~\ref{sec:qcs_reusable_proof_guarantee}. In Algorithm~\ref{alg:qcs_qubit_reusable_readout}, we take $H$ to be the largest gain of the form $2^J/4$ that is at most $1/\sigma$ and whose acquisitions satisfy this restriction, $R=\lceil C/(\sigma^2H^2)\rceil$, and $\epsilon_{\rm inv}$ such that $\sqrt{2N}\epsilon_{\rm inv}/H\leq\epsilon/(8L\kappa)$.

\begin{lemma}
\label{lem:noisy_qubit_score_covariance}
Fix a reference $a_0$ with $\|H(A-a_0)\|_\infty\leq1/8$, independent of the acquisitions at gain $H$ in Algorithm~\ref{alg:qcs_qubit_reusable_readout}, and write $\theta=H(A-a_0)$. The scores $W$ of one register satisfy $\E[W]=\sin\theta$ coordinatewise and $\operatorname{Cov}(W)\preceq C\mathds{1}_{2N}$. Each output $S$ of Algorithm~\ref{algo:sine_inversion} satisfies $|\E[S_\mu]-\theta_\mu|\leq\epsilon_{\rm inv}$ and $|S_\mu|\leq\pi/6$ for every $\mu$, and $\operatorname{Cov}(S)\preceq C\mathds{1}_{2N}$.
\end{lemma}

\begin{proof}
The coordinatewise statements follow as in the proof of Proposition~\ref{prop:high_precision_randomized_spectroscopy}, so we only need to bound the covariances. These are nontrivial because all qubits of a register share the same noise phases $H\xi$. We use the Poincaré inequality for compound Poisson sums, which follows from the Efron--Stein inequality applied to increasingly fine partitions of the sensing time. It states that every bounded function $F$ satisfies
\begin{equation}
    \Var(F(H\xi))\leq\sum_{k=1}^J\frac{\gamma\tau}{2}\E\left[\left(F(H\xi+\Delta_k)-F(H\xi)\right)^2\right].
\end{equation}
Fix a unit vector $u$ and take $F(x)=\sum_\mu u_\mu\sin(\theta_\mu+x_\mu)$. By Taylor's theorem, $F(x+\Delta)-F(x)$ is the sum of the linear term $\sum_\mu u_\mu\cos(\theta_\mu+x_\mu)\Delta_\mu$ and a remainder of magnitude at most $\sum_\mu|u_\mu|\Delta_\mu^2/2$, which Cauchy--Schwarz bounds by $\frac{\Delta_{\max}}{2}(\sum_\mu u_\mu^2\Delta_\mu^2)^{1/2}$. After weighting by the jump rates, the square of the linear term contributes at most $H^2\|\operatorname{Cov}(\xi)\|_{\rm op}$ and the square of the remainder at most $\frac{\Delta_{\max}^2}{4}H^2\|\operatorname{Cov}(\xi)\|_{\rm op}$. Under the duration restriction this gives
\begin{equation}
    \left\|\operatorname{Cov}\big(\sin(\theta+H\xi)\big)\right\|_{\rm op}\leq\left(2+\frac{\Delta_{\max}^2}{2}\right)H^2\|\operatorname{Cov}(\xi)\|_{\rm op}\leq\frac12.
\end{equation}
Conditional on the noise, the outcomes of different qubits are independent with variance at most one. The law of total covariance and $\chi_\mu>3/4$ therefore give $\operatorname{Cov}(W)\preceq C\mathds{1}_{2N}$.

For $S$, recall that Algorithm~\ref{algo:sine_inversion} multiplies clipped means $Q_g$ of independent groups of scores. For logarithmically large groups, Hoeffding's inequality makes clipping rare enough that $\operatorname{Cov}(Q_g)\preceq\mathds{1}_{2N}/16$ and $|\E[Q_{g,\mu}]|\leq1/4$. For independent random vectors $Y$ and $Z$, $\operatorname{Cov}(Y\odot Z)=\operatorname{Cov}(Y)\odot\E[ZZ^T]+\E[Y]\E[Y]^T\odot\operatorname{Cov}(Z)$, where $\odot$ denotes the entrywise product. For positive semidefinite matrices, $\|M_1\odot M_2\|_{\rm op}\leq\max_\mu(M_2)_{\mu\mu}\|M_1\|_{\rm op}$. Since every diagonal entry of $\E[Q_gQ_g^T]$ is at most $1/8$, the operator norm $v_d$ of the covariance of $Q_1\odot\cdots\odot Q_d$ satisfies $v_d\leq v_{d-1}/8+16^{-d}$, and hence $v_d\leq d\,8^{-(d-1)}/16$. The output $S=\sum_jc_jQ_1\odot\cdots\odot Q_{2j+1}$ is a sum of correlated terms. For every unit vector $u$, the triangle inequality for standard deviations gives $\Var(u^TS)^{1/2}\leq\sum_jc_jv_{2j+1}^{1/2}$, which is bounded by a numerical constant, so $\operatorname{Cov}(S)\preceq C\mathds{1}_{2N}$.
\end{proof}

\begin{proof}[Proof of Proposition~\ref{prop:qcs_qubit_memory_noisy}]
We first show that the reference is accurate despite the reduced visibility, then apply Lemma~\ref{lem:noisy_qubit_score_covariance} and the error cancellation argument of Section~\ref{sec:qcs_reusable_proof_guarantee}, and finally count the interrogation time.

For a register at gain $h_j$, the $X$ and $Y$ outcomes of qubit $\mu$ have means $\chi_\mu\cos(h_jA_\mu)$ and $\chi_\mu\sin(h_jA_\mu)$, with $\chi_\mu>3/4$. When all empirical means are within $1/16$ of these values, which fails with probability at most $\delta/4$ by the argument in the proof of Proposition~\ref{prop:high_precision_randomized_spectroscopy}, the vector $\overline X_{\mu,j}+i\overline Y_{\mu,j}$ is within $\sqrt2/16$ of $\chi_\mu e^{ih_jA_\mu}$. Its argument is then within $\arcsin(\sqrt2/12)<1/8$ of $h_jA_\mu$ modulo $2\pi$, which is the guarantee at each scale used in that proof. The visibility shortens the vector without moving its argument, so the same induction gives $\|H(A-a_0)\|_\infty\leq1/8$.

Conditional on the reference, Lemma~\ref{lem:noisy_qubit_score_covariance} shows that each $U_i$ has bias $\|\E[U_i\mid a_0]-A\|_2\leq\sqrt{2N}\epsilon_{\rm inv}/H\leq\epsilon/(8L\kappa)$ and covariance $\operatorname{Cov}(U_i\mid a_0)\preceq C\mathds{1}_{2N}/(RH^2)\preceq\sigma^2\mathds{1}_{2N}$. Moreover, $U_1,\ldots,U_L$ are conditionally independent because they use disjoint registers. Lemma~\ref{lem:independent_nonlinear_estimation}, applied conditionally on $a_0$, then gives each contraction $K_j[U_1,\ldots,U_L]$ constant success probability, and the median over batches proceeds as in Section~\ref{sec:qcs_reusable_proof_guarantee}.

An acquisition at gain $h$ costs interrogation time $s_0Ch/\alpha$. The reference therefore costs at most $4r_{\rm p}s_0CH/\alpha$, since $\sum_jh_j\leq2H$, and the residual stage costs $LRms_0CH/\alpha$. If the duration restriction does not limit $H$, then $H>1/(2\sigma)$ and $R=O(1)$, so a batch costs $\widetilde O_L(\sqrt N/(\alpha\sigma))$. Otherwise each acquisition lasts $\widetilde\Theta(N/\gamma)$, so $H^2=\widetilde\Theta(\alpha^2N/\gamma^2)$, and the $R$ acquisitions cost $\widetilde O_L(RN/\gamma)=\widetilde O_L(\gamma/(\alpha^2\sigma^2))$. Since $r_{\rm p}$ and $m$ are logarithmic, substituting $\sigma$ reproduces $T_{\rm batch}$. Each register is measured immediately after it is prepared, so only one register of $2N$ memory qubits is held at a time.
\end{proof}

\subsubsection{Proof of guarantees for matched QCS with one memory qubit}
\label{sec:deferred_one_qubit_matched}

\begin{proof}[Proof of Proposition~\ref{prop:jc_single_qubit_matched_oracle}]
We show that each run at gain $h_j$ writes the phase $h_jx$, with $x=K^TA/\kappa$, into the memory up to a deterministic error of at most $1/64$. We then show that this error fits within the tolerance of phase unwrapping, and finally sum the interrogation and wall-clock times.

On a window $I$ of length $W_0=\Theta(\log(2N)/\Delta_\omega)$, the separation of the signed frequencies makes the absolute row sums of $G_I-\mathds{1}/2$ at most $1/4$, as in Section~\ref{sec:deferred_oracle_transduction}. The Neumann series then gives $\|G_I^{-1}\|_{1\rightarrow1}\leq4$ and $G_I\succeq\mathds{1}/4$. Since $K$ has at most $2m$ nonzero entries and every entry of $f(t)$ has magnitude at most one, 
\begin{equation}
    |q_I(t)|\leq4\|K\|_1/\kappa\leq4\sqrt{2m}.
\end{equation}
The definition of $G_I$ gives $\frac{1}{W_0}\int_Iq_If\,dt=K/\kappa$, and Cauchy--Schwarz with $\frac{1}{W_0}\int_Iq_I^2\,dt=K^TG_I^{-1}K/\kappa^2\leq4$ gives $\frac{1}{W_0}\int_I|q_I|\,dt\leq2$. Consequently, a run at gain $h_j$ interrogates the signal for $\sum_{\mathcal C}d_{\mathcal C}\leq2h_j$. Moreover, since $T_j\geq4\sqrt{2m}\,h_j$, every duration satisfies $d_{\mathcal C}\leq|\mathcal C|$, so each controlled sensing interval fits within its cell.

During the controlled sensing interval in cell $\mathcal C$, the sensor is in a basis state on which $Z_s$ acts as $\sigma_{\mathcal C}Z_M$, so the interval applies $e^{-i\sigma_{\mathcal C}\theta_{\mathcal C}Z_M/2}$ to the memory exactly. Let $y(t)=\sigma_{\mathcal C}$ while the sensor senses in cell $\mathcal C$ and $y(t)=0$ otherwise. A run then leaves the memory in $(\ket0+e^{i\phi}\ket1)/\sqrt2$ with $\phi=A^T\int y(t)f(t)\,dt$, whereas the ideal modulation $h_jq_I(t)/T_j$ would give \begin{equation}
\frac{h_j}{T_j}A^T\int q_I(t)f(t)\,dt=h_jx    
\end{equation}
On every cell, $\int_{\mathcal C}(y-h_jq_I/T_j)\,dt=0$ and $|y-h_jq_I/T_j|\leq2$. As in the proof of Algorithm~\ref{algo:discretize_response}, we may subtract $f$ at the center of each cell and use $\|f'(t)\|_2\leq\omega_N\sqrt N$ and $\|A\|_2\leq1$. Summing over cells gives 
\begin{equation}
    |\phi-h_jx|\leq\omega_N\sqrt N\,T_j\Delta t\leq1/64 .
\end{equation}

The $X$ and $Y$ outcomes of a run at gain $h_j$ therefore have means within $1/64$ of $\cos(h_jx)$ and $\sin(h_jx)$. When all empirical means are within $1/16$ of their expectations, which fails with probability at most $\delta$ for the chosen $r_{\rm p}$ by Hoeffding's inequality and a union bound over the $2(J+1)$ means, $\overline X_j+i\overline Y_j$ is within $\sqrt2(1/16+1/64)<0.12$ of $e^{ih_jx}$. Its argument is then within $\arcsin(0.12)<1/8$ of $h_jx$ modulo $2\pi$. This is the guarantee at each scale used in the proof of Proposition~\ref{prop:high_precision_randomized_spectroscopy}, whose induction gives $|\widehat\phi_J-h_Jx|<1/8$. Since $h_J\geq\kappa/(8\epsilon)$, it follows that 
\begin{equation}
    |\widehat F_K-K^TA|=\kappa|\widehat\phi_J/h_J-x|\leq\epsilon.
\end{equation}

Finally, $\sum_jh_j\leq2h_J\leq\kappa/(2\epsilon)$. The total interrogation time is therefore at most $\sum_j2r_{\rm p}\cdot2h_j\leq2r_{\rm p}\kappa/\epsilon$. The wall-clock time is at most 
\begin{align}
    \sum_j2r_{\rm p}T_j\leq\sum_j2r_{\rm p}(4\sqrt{2m}\,h_j+W_0)\\\leq4\sqrt{2m}\,r_{\rm p}\kappa/\epsilon+2r_{\rm p}(J+1)W_0
\end{align} 
Since $r_{\rm p}$ and $J$ are logarithmic, this gives Equation~\eqref{eq:qcs_one_qubit_cost}.
\end{proof}
We now bound the interrogation time under stronger noise. With $r=1$, we have $U=K/\kappa$, and $O_k=\arcsin(\widetilde q_k(t_k))Z_M$ satisfies $\sin O_k=\widetilde q_k(t_k)Z_M$. Hence, 
\begin{equation}
\mathcal Q_{O_k}(\rho)=(1-\widetilde q_k(t_k)^2)\rho+\widetilde q_k(t_k)^2Z_M\rho Z_M    
\end{equation}
exactly. The comparison of each call with its averaged generator is unchanged from the proof of Algorithm~\ref{alg:qcs_matched_oracle}. By the second property in Equation~\eqref{eq:qcs_matched_filter_properties}, the averaged generator is exactly $-\frac{i\alpha\tau}{2\sqrt{65}}x[Z_M,\cdot]$, with no truncation or tail error, and the averaged noise generator is $\frac{\gamma\tau}{2}\widetilde{\mathbb E}_k[\widetilde q_k^2](Z_M(\cdot)Z_M-\mathcal I)$. These generators commute. A memory qubit prepared in $\ket+$ therefore ends with relative phase $2\lambda x$, where $\lambda=\alpha T/(2\sqrt{65})$, and with contrast
\begin{equation}
    \exp(-\gamma\sum_k\tau\widetilde{\mathbb E}_k[\widetilde q_k^2])\geq e^{-\gamma T},
\end{equation}
since $|\widetilde q_k|\leq1$. Capping each acquisition at $T\leq1/(8\gamma)$ keeps the contrast above $3/4$. The proof of Proposition~\ref{prop:qcs_qubit_memory_noisy} for a single coordinate then gives interrogation time $\widetilde O(\max\{\kappa/(\alpha\epsilon),\gamma\kappa^2/(\alpha^2\epsilon^2)\}\log(1/\delta))$, which depends on neither $N$ nor $m$.%
}{%
  \typeout{AQS: appendices_body.tex not found -- appendix material omitted.}%
}
\end{document}